\documentclass[graybox,envcountchap,openany]{svmono-modified}

\usepackage{type1cm}         
\usepackage{makeidx}         
\usepackage{graphicx}
\usepackage{amsmath}
\usepackage{amssymb}
\usepackage{placeins}        
\usepackage{multicol}        
\usepackage[bottom]{footmisc}
\usepackage{braket}
\usepackage{comment}
\usepackage{hyperref}
\usepackage{pgfplots}
\usepackage[lofdepth]{subfig}
\usepackage{bbold}
\usepackage{empheq}
\usepackage{booktabs}
\usepackage{ntheorem}
\usepackage{mdframed}      
\usepackage{bm,bbm}  
\usepackage{todonotes}
\usepackage{mathrsfs}
\usepackage{longtable}
\usepackage{caption}
\usepackage{amscd}
\usepackage{multirow}
\usepackage{tikz,circledsteps,circuitikz}
\usetikzlibrary{quantikz2}

\newcommand{\bq}{\begin{equation}}
\newcommand{\eq}{\end{equation}}
\newcommand{\bqn}{\begin{equation*}}
\newcommand{\eqn}{\end{equation*}}
\newcommand{\bqali}{\begin{equation}\begin{aligned}}
\newcommand{\eqali}{\end{aligned}\end{equation}}
\newcommand{\bqnali}{\begin{equation*}\begin{aligned}}
\newcommand{\eqnali}{\end{aligned}\end{equation*}}
\newcommand{\com}[2]{\left[ {#1},{#2} \right]}
\newcommand{\acom}[2]{\left\{ {#1},{#2} \right\}}

\newcommand{\Tr}[1]{\operatorname{Tr}\left[#1\right]}

\newcommand{\TR}[2]{\operatorname{Tr}^{\text{\tiny(#1)}}\left[#2\right]}
\newcommand{\x}{{\bf x}}

\newcommand{\kB}{k_\text{\tiny B}}

\newcommand{\id}{\operatorname{id}}
\newcommand{\hatHc}{\hat H_\text{\tiny C}}
\newcommand{\hatHm}{\hat H_\text{\tiny M}}
\newcommand{\tikzcircle}[2][red,fill=red]{\tikz[baseline=-0.5ex]\draw[#1,radius=#2] (0,0) circle ;}%

\theoremstyle{break}
\theoremheaderfont{\bfseries}
\newmdtheoremenv[%
leftmargin=40,%
rightmargin=40,
backgroundcolor=gray!10,%
innertopmargin=10pt,%
ntheorem]{myexample}{Example}[chapter]

\theoremstyle{break}
\theoremheaderfont{\bfseries}
\newmdtheoremenv[%
leftmargin=40,%
rightmargin=40,
backgroundcolor=gray!10,%
innertopmargin=10pt,%
ntheorem]{myexercise}{Exercise}[chapter]

\theoremstyle{break}
\theoremheaderfont{\bfseries}
\newmdtheoremenv[%
leftmargin=40,%
rightmargin=40,
backgroundcolor=gray!10,%
innertopmargin=10pt,%
ntheorem]{myrecall}{Recall}[chapter]

\usepackage{datetime}

\newdateformat{monthyeardate}{%
  \monthname[\THEMONTH] \THEYEAR}

\newcommand{\version}{1.0 -- \monthyeardate\today}

\makeindex             % used for the subject index

\newcommand{\tindex}[1]{\index{#1}}

\graphicspath{{/Figs/}}

\begin{document}

\author{Matteo Carlesso}
\title{Lecture Notes on Quantum Algorithms in Open Quantum Systems}
\subtitle{Department of Physics, University of Trieste}
\maketitle

\frontmatter%%%%%%%%%%%%%%%%%%%%%%%%%%%%%%%%%%%%%%%%%%%%%%%%%%%%%%

\chapter*{Preface}
\setcounter{page}{1}

These lecture notes aim to provide a clear and comprehensive introduction to using open quantum system theory for quantum algorithms. They are based on various sources, including research papers and textbooks, which can be found in literature. While I have focused on clarity and consistency, there is always space for improvement. I will  work on expanding these notes and welcome your constructive feedback and comments. Feel free to send them to \href{mailto:matteo.carlesso@units.it}{matteo.carlesso@units.it}. \\

In the \textit{Suggested Bibliography} reported below, the reader can find the list of references I considered to prepare these notes. Some of these are unpublished (please contact me directly for more info).
I apologise in advance to all the numerous authors whose contributions I did not cite. The field is vast and the intent of these notes is not to serve as a comprehensive review article. I have certainly not done justice to the literature.\\

These notes will be used as a basis for some courses that I held at the Department of Physics of the University of Trieste. Specifically, these are
\begin{itemize}
    \item \textit{357SM - Quantum Algorithms in Opens Quantum Systems (Spring 2024)}
    \item \textit{987DF - Quantum Computing Algorithms (Autumn 2024)}
\end{itemize}

\section*{Acknowledgments}
For the preparation of these lecture notes, I am in great debt with several people. Among these Angelo Bassi, Francesco Cesa, Giulio Crognaletti, Giovanni di Bartolomeo, Sandro Donadi and Michele Vischi for inputs, discussion and support.\\

I also acknowledge the financial support from the National Quantum Science and Technology Institute  through the PNRR MUR project PE0000023-NQSTI, the University of Trieste and the Italian National Institute for Nuclear Physics (INFN).

\vfill
\noindent \textcopyright \textit{Matteo Carlesso$^{1,2}$, \monthyeardate\today, Trieste}\\
$^1$Department of Physics, University of Trieste, Strada Costiera 11, 34151 Trieste, Italy\\
$^2$Istituto Nazionale di Fisica Nucleare, Trieste Section, Via Valerio 2, 34127 Trieste, Italy\\

\chapter*{Suggested Bibliography}

\section*{Open Quantum Systems}

Although not explicitly discussed in these notes, a part of the above mentioned courses regards the theory of open quantum systems. For this part of the courses, I suggest the following literature. 

\begin{itemize}
\item Angelo Bassi. Lecture Notes on Advanced Quantum Mechanics. Unpublished.
\item Heinz-Peter Breuer, Francesco Petruccione. The theory of open quantum systems. Oxford, 2022.
\item Mahn-Soo Choi. A Quantum Computation Workbook. Springer, 2022.
\item Roberto Floreanini. Lecture Notes on Quantum Information. Unpublished.
    \item Daniel A. Lidar. Lecture Notes on Theory of Open Quantum Systems. arXiv preprint arXiv:1902.00967.
    \item Ahsan Nazir. Lecture notes on open quantum systems. \url{https://www.yumpu.com/en/document/view/8219582/lecture-notes-on-open-quantum-systems-workspace}.
    \item Stefano Olivares. Lecture Notes on Quantum Computing. University of Milano, 2020.
\end{itemize}

\section*{Quantum Computation}
\begin{itemize}
\item Giuliano Benenti, Giulio Casati, Davide Rossini, and Giuliano Strini. Principles of quantum computation and information: a comprehensive textbook. World Scientific, 2019.
\item Mahn-Soo Choi. A Quantum Computation Workbook. Springer, 2022.
     \item Christoph Dittel. Lecture Notes on Quantum information theory. arXiv preprint arXiv:2311.12442.
    \item  Ronald de Wolf. Lecture Notes on Quantum Computing. arXiv preprint arXiv:1907.09415.
    \item Aram W Harrow, Avinatan Hassidim, and Seth Lloyd. Quantum algorithm for linear systems of equations. Physical Review Letters \textbf{103}, 150502, (2009).
    \item Anton Frisk Kockum, Ariadna Soro, Laura García-Álvarez, Pontus Vikstål, Tom Douce, Göran Johansson, and Giulia Ferrini. Lecture Notes on Quantum Computing. arXiv preprint arXiv:2311.08445.
    \item Lin Lin. Lecture notes on quantum algorithms for scientific computation. arXiv preprint arXiv:2201.08309.
    \item M. A. Nielsen and I. L. Chuang
Quantum Computation and Quantum Information. Cambridge University Press 2010.
    \item Stefano Olivares. Lecture Notes on Quantum Computing. University of Milano, 2020.
    \item Oswaldo Zapata. A Short Introduction to Quantum Computing for Physicists. arXiv preprint arXiv:2306.09388.
    
\end{itemize}

\section*{Quantum Error Correction}
\begin{itemize}
\item Giuliano Benenti, Giulio Casati, Davide Rossini, and Giuliano Strini. Principles of quantum computation and information: a comprehensive textbook. World Scientific, 2019.
\item Mahn-Soo Choi. A Quantum Computation Workbook. Springer, 2022.
    \item Andrew N. Cleland. An introduction to the surface code. SciPost Phys. Lect. Notes 49 (2022).
    \item  Ronald de Wolf. Lecture Notes on Quantum Computing. arXiv preprint arXiv:1907.09415.
    \item Sandro Donadi. Lecture Notes on Quantum Algorithms in Open Quantum Systems. Unpublished.
    \item Ekert, A and Hosgood, T and Kay, A and Macchiavello, C. Introduction to Quantum Information Science. \url{https://qubit.guide}.
        \item Austin G. Fowler, Matteo Mariantoni, John M. Martinis, and Andrew N. Cleland. Surface codes: Towards practical
large-scale quantum computation. Phys. Rev. A, 86:032324, 2012.
    \item Steven M. Girvin. Introduction to quantum error correction and fault tolerance. SciPost Phys. Lect. Notes 70 (2023).
    \item Stefano Olivares. Lecture Notes on Quantum Computing. University of Milano, 2020.
    \item Andrew M. Steane. A Tutorial on Quantum Error Correction. IOS Press 2006.
      \item Oswaldo Zapata. A Short Introduction to Quantum Computing for Physicists. arXiv preprint arXiv:2306.09388.
\end{itemize}

\section*{Dynamical Decoupling and Quantum Error Mitigation}
\begin{itemize}
\item Zhenyu Cai, Ryan Babbush, Simon C. Benjamin, Suguru Endo, William J. Huggins, Ying Li, Jarrod R. McClean, and
Thomas E. O’Brien. Quantum error mitigation. Review Modern Physics \textbf{95},045005  (2023).
\item Sandro Donadi. Lecture Notes on Quantum Algorithms in Open Quantum Systems. Unpublished.
\item Tudor Giurgica-Tiron, Yousef Hindy, Ryan LaRose, Andrea Mari, and William J. Zeng. Digital zero noise extrapolation
for quantum error mitigation. In 2020 IEEE International Conference on Quantum Computing and Engineering (QCE).
IEEE,  2020.

\end{itemize}

\tableofcontents

\mainmatter%%%%%%%%%%%%%%%%%%%%%%%%%%%%%%%%%%%%%%%%%%%%%%%%%%%%%%%

\chapter{Circuit model for quantum computation}

In quantum computation, the basic ingredients are qubits and gates. The composition of different gates acting on a series of qubits is what we called an algorithm. Here we introduce quantum gates and some algorithms.

\section{Qubit gates}

The single qubit algebra can be described in terms of the identity $\hat{\mathbb{1}}$ and Pauli $\hat \sigma_{x}$, $\hat \sigma_{y}$ and $\hat \sigma_{z}$ operators. All single qubit gates are a linear composition of these. In particular, they can be visualised as rotations of the state $\ket\psi$ on the Bloch sphere. The three elementary rotations by an angle $\theta$ around the Cartesian axes are defined as $\hat R^j(\theta)=e^{-i\theta \hat\sigma_j/2}$ for $j=x,y,z$. In particular, in the computational basis, which is the one mainly used in quantum computation, one has
\begin{equation}
    \begin{aligned}
         R^x(\theta)&=
         \begin{pmatrix}
             \cos(\theta/2) & -i \sin(\theta/2)\\
             -i \sin(\theta/2) & \cos(\theta/2)
         \end{pmatrix},\\
        R^y(\theta)&=
         \begin{pmatrix}
             \cos(\theta/2) & -\sin(\theta/2)\\
              \sin(\theta/2) & \cos(\theta/2)
         \end{pmatrix},\\
        R^z(\theta)&=
         \begin{pmatrix}
             \exp({-i \theta/2})&0\\
             0&\exp({i\theta/2})
         \end{pmatrix}.
    \end{aligned}
\end{equation}
Then, the rotation of an angle $\theta$ around the unit axis ${\bf n}$ is given by
\begin{equation}
    \hat R^{\bf n}(\theta)=e^{-i\theta {\bf n}\cdot\hat {\boldsymbol \sigma}/2}=\cos(\theta/2)\hat{\mathbb 1}-i\sin{(\theta/2)}{\bf n}\cdot\hat {\boldsymbol \sigma}.
\end{equation}
Beside the rotations, there are six important single-qubit gates that are standard. These are $X$, $Y$, $Z$ and
\begin{equation}
    H=\frac{1}{\sqrt{2}}
    \begin{pmatrix}
        1&1\\
        1&-1
    \end{pmatrix},
    \quad
    S=
    \begin{pmatrix}
        1&0\\
        0&i
    \end{pmatrix},
    \quad
    T=
    \begin{pmatrix}
        1&0\\
        0&e^{i\pi/4}
    \end{pmatrix}.
\end{equation}
In particular, $X$, $Y$ and $Z$ are respectively the Pauli operators $\hat \sigma_x$, $\hat \sigma_y$ and $\hat \sigma_z$ represented in the computational basis, and $H$ is known as the Hadamard gate. \\

Eventually, the state of the qubit is measured.  In particular, this is always the measurement of $\hat \sigma_z$ and one always obtains one of the two discrete outcomes: ``0'' or ``1''. Given the generic state $\ket\psi=\alpha\ket0+\beta\ket1$, with $\alpha$ and $\beta$ being complex and $|\alpha|^2+|\beta|^2=1$, then one has a probability $p_0=|\alpha|^2$ to have the outcome ``0'' and $p_1=|\beta|^2$ to have the outcome ``1''. 

\begin{myexample}
The gate $X$ flips states. Indeed,
\begin{equation}
    X\begin{pmatrix}
        \alpha\\
        \beta
    \end{pmatrix}=
\begin{pmatrix}
    0&1\\
    1&0
\end{pmatrix}       
        \begin{pmatrix}
        \alpha\\
        \beta
    \end{pmatrix}=     
    \begin{pmatrix}
        \beta\\
        \alpha
    \end{pmatrix}.
\end{equation}
\end{myexample}

\begin{myexample}
    The Hadamard gate $H$ generates uniform superpositions. In particular, one has
    \begin{equation}
    \begin{aligned}
        &H
        \begin{pmatrix}
            1\\0
        \end{pmatrix}=
        \frac{1}{\sqrt{2}}
    \begin{pmatrix}
        1&1\\
        1&-1
    \end{pmatrix}
    \begin{pmatrix}
            1\\0
        \end{pmatrix}=\frac{1}{\sqrt{2}}
        \begin{pmatrix}
            1\\1
        \end{pmatrix},\\
                &H
        \begin{pmatrix}
            0\\1
        \end{pmatrix}=
        \frac{1}{\sqrt{2}}
    \begin{pmatrix}
        1&1\\
        1&-1
    \end{pmatrix}
    \begin{pmatrix}
            0\\1
        \end{pmatrix}=\frac{1}{\sqrt{2}}
        \begin{pmatrix}
            1\\-1
        \end{pmatrix}.
        \end{aligned}
    \end{equation}
    Namely, one has
    \begin{equation}
        \hat H\ket{0}=\ket{+},\quad\text{and}  \quad    \hat H\ket{0}=\ket{-},
    \end{equation}
where $\ket\pm=(\ket0\pm\ket1)/\sqrt{2}$. Notably, the Hadamard gate maps the basis of $\hat \sigma_z$ in that of $\hat \sigma_x$, and back.

\end{myexample}

\begin{myexercise}\label{ex.Hadamardasrot}
    Express the Hadamard gate as a rotation.
\end{myexercise}

\begin{myexercise}\label{ex.3rotations}
    Prove that, given two fixed non-parallel  normalised vectors ${\bf n}$ and ${\bf m}$, any unitary single qubit gate $\hat U$ can be expressed as
    \begin{equation}
        \hat U=e^{i\alpha}\hat R^{\bf n}(\beta)\hat R^{\bf m}(\gamma)\hat R^{\bf n}(\delta),
    \end{equation}
    with $\alpha, \beta, \gamma, \delta \in \mathbb R$.
\end{myexercise}

It is common to represent quantum circuits with diagrams with the time running from left to right, where lines correspond to qubits and boxes to gates. For example, the following diagram 
\begin{equation}
\begin{quantikz}
\ket0&\gate{H}&\gate{R_Z(\theta)}& \meter{}
\end{quantikz}    
\end{equation}
corresponds to the following logical consecutive operations
\begin{itemize}
    \item[0)] Prepare the qubit in the ground state $\ket0$.
    \item[1)] Apply the Hadamard gate $H$.
    \item[2)] Apply a rotation of an angle $\theta$ around the $z$ axis.
    \item[3)] Measure the state of the qubit.
\end{itemize}

When one is working with more than one qubit, there is the need to construct the representation of the states the common computational basis. In the case of two qubits, the basis is given by $\set{\ket{00},\ket{01},\ket{10},\ket{11}}$, whose representation in the common computational basis is 
\begin{equation}\label{eq.representation2qubit}
    \ket{00}\sim
    \begin{pmatrix}
        1\\0\\0\\0
    \end{pmatrix},
    \quad
      \ket{01}\sim
    \begin{pmatrix}
        0\\1\\0\\0
    \end{pmatrix},
    \quad
      \ket{10}\sim
    \begin{pmatrix}
        0\\0\\1\\0
    \end{pmatrix},
    \quad
      \ket{11}\sim
    \begin{pmatrix}
        0\\0\\0\\1
    \end{pmatrix},
\end{equation}
where the symbol $\sim$ indicates that the state $\ket \psi$ was represented on the computational basis. This is constructed through the tensor product, i.e.
\begin{equation}
    \ket{\psi\phi}\sim
    \begin{pmatrix}
    \psi_1\\
    \psi_2
    \end{pmatrix}
    \otimes
    \begin{pmatrix}
    \phi_1\\
    \phi_2
    \end{pmatrix}
    =\begin{pmatrix}
        \psi_1\phi_1\\
        \psi_1\phi_2\\
        \psi_2\phi_1\\
        \psi_2\phi_2
    \end{pmatrix}.
\end{equation}

Owning the computational representation, we can introduce some 2-qubit gates. One of the most useful among these gates is the CNOT or control-NOT gate:
\begin{equation}
    CNOT=
    \begin{pmatrix}
        1&0&0&0\\
        0&1&0&0\\
        0&0&0&1\\
        0&0&1&0
    \end{pmatrix},
\end{equation}
and is represented as
\begin{equation}
\begin{quantikz}
    & \ctrl{1} &  \\
    & \targ{}  & 
\end{quantikz}
\quad\text{or equivalentely as}\quad
\begin{quantikz}
    & \ctrl{1} &  \\
    & \gate{X}  & 
\end{quantikz}
\end{equation}
It acts on a target qubit (qubit 1) in a way that depends on the state of a control qubit (qubit 0). Namely, it applies an $X$ gate to the qubit 1 if the state of qubit 0 is 1, otherwise it does not change the state:
\begin{equation}
    CNOT\ket{00}=\ket{00},\quad CNOT\ket{01}=\ket{01},\quad CNOT\ket{10}=\ket{11},\quad CNOT\ket{11}=\ket{10}.
\end{equation}
\begin{myexercise}\label{ex.CNOT.entanglement}
    Prove that CNOT can generate entanglement.
\end{myexercise}

A second important 2-qubit gate is the SWAP, which swaps the state between two qubits. Namely
\begin{equation}
    SWAP \ket a\otimes\ket b=\ket b\otimes \ket a.
\end{equation}
A SWAP operation can be constructed using a concatenation of CNOT gates. In particular:
\begin{equation}
    SWAP\sim
\begin{quantikz}
&\swap{1}&\midstick[2,brackets=none]{=}&\gate[2,swap]{} &\midstick[2,brackets=none]{=}& \ctrl{1} & \targ{} & \ctrl{1} &  \\
&\targX{}&& &  & \targ{}  &    \ctrl{-1}    & \targ{}  & 
\end{quantikz}
\end{equation}

Similarly as the CNOT, one can construct a controlled unitary gate, where the state of the control qubit determines if a unitary gate $\hat U$ is applied to the target qubit:
\begin{equation}
    C(U)\sim
    \begin{pmatrix}
        1&0&0&0\\
        0&1&0&0\\
        0&0&U_{00}&U_{01}\\
        0&0&U_{10}&U_{11}
    \end{pmatrix}
    \sim
    \begin{quantikz}
        &\ctrl{1}&\\
        &\gate{U}&
    \end{quantikz}
\end{equation}
where $U_{ij}$ are the matrix elements of $\hat U$.

\subsection{Hadamard test}\label{hadamard.test}\tindex{Hadamard test}

The Hadamard test is a useful tool for computing expectation values of a unitary, black-box operator $\hat U$ with respect to a state $\ket \psi$, which can be in principle a multi-qubit state. Since in general $\hat U$ is not Hermitian, one measures independently the real and immaginary part of $\braket{\psi|\hat U|\psi}$. 

The circuit for the real Hadamard test is
\begin{equation}\label{circuit.hadamardim0}
\begin{quantikz}
    \ket0&\gate{H}&\ctrl{1}&         \gate{H}   &\meter{}\\
    \ket\psi&\qwbundle{}&\gate{U}&              &
\end{quantikz}    
\end{equation}
and it performs as follows. The first step is to generate a superposition in the first qubit (qubit 0):
\begin{equation}
    \ket0\ket\psi\xrightarrow[]{\hat H\otimes\hat {\mathbb 1}}
    \tfrac{1}{\sqrt{2}}(\ket0+\ket1)\ket\psi.
\end{equation}
Then, we entangle the qubits with the $C(U)$ gate:
\begin{equation}
    \tfrac{1}{\sqrt{2}}(\ket0+\ket1)\ket\psi\xrightarrow[]{C(U)}
    \tfrac{1}{\sqrt{2}}(\ket0\ket\psi+\ket1\hat U\ket\psi),
\end{equation}
and apply the Hadamard gate to qubit 0:
\begin{equation}
    \xrightarrow[]{\hat H\otimes\hat {\mathbb 1}}
    \tfrac12\left[(\ket0+\ket1)\ket\psi+(\ket0-\ket1)\hat U\ket\psi\right]=\tfrac12\left[\ket0(\hat{\mathbb 1}+\hat U)\ket\psi+\ket1(\hat{\mathbb 1}-\hat U)\ket\psi\right].
\end{equation}
Finally, one measures qubit 0, and the probability of finding the qubit in $\ket 0$ is
\begin{equation}\label{circuit.hadamardim.result0}
    P(\ket0)=\tfrac14\braket{\psi|\left(   \hat{\mathbb 1}+\hat U^\dag \right)\left(   \hat{\mathbb 1}+\hat U \right)|\psi}=\tfrac12\left(1+\Re\braket{\psi|\hat U|\psi}\right).
\end{equation}
Thus, by measuring only one qubit (qubit 0) one has an indication of the real part of $\braket{\psi|\hat U|\psi}$. To estimate the imaginary part, the circuit is modified as follows:
\begin{equation}\label{circuit.hadamardim}
    \begin{quantikz}
    \ket0&\gate{H}&\gate{S^\dag}&\ctrl{1}&         \gate{H}   &\meter{}\\
    \ket\psi&\qwbundle{}&&\gate{U}&              &
\end{quantikz}    
\end{equation}
Then, the state before the measurement is
\begin{equation}
    \tfrac12\left[\ket0(\hat{\mathbb 1}-i\hat U)\ket\psi+\ket1(\hat{\mathbb 1}+i\hat U)\ket\psi\right],
\end{equation}
and correspondingly one has
\begin{equation}\label{circuit.hadamardim.result}
    \tilde P(\ket0)=\tfrac12\left(1+\Im\braket{\psi|\hat U|\psi}\right).
\end{equation}

Notably, to well characterise these probabilities, there is the need to run the protocol several times to construct a statistics.

\begin{myexercise}\label{ex.S.im.U}
    Prove that the circuit in Eq.~\eqref{circuit.hadamardim} provides the result in Eq.~\eqref{circuit.hadamardim.result}. 
\end{myexercise}

\section{No-cloning theorem}

For different computational reasons, one would like to create an independent and identical copy of an arbitrary state with a unitary operation. Nevertheless, the following theorem prevents it.

\begin{theorem}[No-cloning]\tindex{No-cloning theorem}\label{th:no-cloning}

Consider two quantum systems $\mathcal A$ and $\mathcal B$ with corresponding Hilbert spaces of the same dimensions $\mathbb H_{\mathcal A}$ and $\mathbb H_{\mathcal B}$. Then, it is not possible to construct a unitary operation $\hat U$ acting on $\mathbb H_{\mathcal A}\otimes\mathbb H_{\mathcal B}$ that copies an arbitrary state of $\mathcal A$ over an initial, reference state of $\mathcal B$. Namely, $\nexists \hat U$ such that
\begin{equation}\label{eq.cloning}
    \hat U\ket{\psi}\ket{e}=\ket{\psi}\ket\psi,
\end{equation}
where $\ket\psi$ is an arbitrary state and $\ket{e}$ is a reference state.\\

Proof. A simple proof goes as follows. Suppose there exists $\hat U$ such that described in Eq.~\eqref{eq.cloning}. Then, one considers the scalar product between the state $\ket{\psi,e}=\ket\psi\ket e$ and $\ket{\phi,e}=\ket\phi\ket e$, where $\ket \phi$ is a second arbitrary state. This gives
\begin{equation}
    \braket{\phi,e|\psi, e}=\braket{\phi|\psi}\braket{e|e}.
\end{equation}
Exploiting the unitarity of $\hat U$ we have
\begin{equation}
    \braket{\phi,e|\psi, e}=\braket{\phi,e|\hat U^\dag\hat U|\psi, e}
\end{equation}
Now, we apply Eq.~\eqref{eq.cloning} to both these states:
\begin{equation}
\braket{\phi,e|\hat U^\dag\hat U|\psi, e}=\braket{\phi,\phi|\psi, \psi}=\braket{\phi|\psi}^2.
\end{equation}
By putting together the last three expressions we find
\begin{equation}
    \braket{\phi|\psi}\braket{e|e}=\braket{\phi|\psi}^2,
\end{equation}
which holds true only if $\braket{\phi|\psi}=0$ or $\ket \phi=e^{i\alpha(\phi,\psi)}\ket\psi$ with $\alpha(\phi,\psi)$ being a phase possibly depending on the two input states. In both cases, one does not allow for full arbitrariness, thus proving the no-cloning theorem. 

\end{theorem}

Importantly for the quantum computation field, the no-cloning theorem prevents the employment of classical error correction techniques on quantum states. One needs to employ quantum error corrections, which will be subject of Chapter \ref{ch.errorcorrection}, that effectively circumvent the no-cloning theorem.

\section{Dense coding}\tindex{Dense coding}

An interesting quantum algorithm is that of dense coding. Suppose Alice has two classical bits $x$ and $y$ that wants to communicate (securely) to Bob, and can do it only via a single qubit. The following protocol allows for it. It assumes to have two qubits on which six operations are performed:
\begin{equation}
\begin{quantikz}
\ket{0}&\gate{H}&\ctrl{1}\slice{share}&\gate{U}\slice{send}&\ctrl{1}&\gate{H}&    \meter{}  \\
 \ket{0}&&\targ{}&&\targ{}&&\meter{}
\end{quantikz}
\end{equation}
The first operation is to prepare an initial entangled state
\begin{equation}
    \ket{0}\ket{0}\xrightarrow[]{\hat H\otimes \hat{\mathbb 1}}\tfrac{1}{\sqrt{2}}(\ket0+\ket1)\ket0\xrightarrow[]{CNOT}\tfrac{1}{\sqrt{2}}(\ket0\ket0+\ket1\ket1)=\ket{\psi_+}.
\end{equation}
The second operation is to share the state among Alice (qubit 0) and Bob (qubit 1).
Then, the third operation is the encoding: Alice encodes the state of $(x,y)$ in the operation performed with the gate $\hat U$:
\begin{equation*}
    \begin{array}{ c | c }
    x,y & \hat U \\
    \hline
 0,0&\hat{\mathbb 1}\otimes\hat{\mathbb 1}\\
 0,1&\hat \sigma_x\otimes\hat{\mathbb 1}\\
 1,0&\hat \sigma_z\otimes\hat{\mathbb 1}\\
 1,1&i\hat \sigma_y\otimes\hat{\mathbb 1}
  \end{array}
\end{equation*}
This leads to
\begin{equation}
    \ket{\psi_+}\xrightarrow[]{\hat U\otimes\hat{\mathbb 1}}
    \begin{cases}
        \tfrac{1}{\sqrt{2}}(\ket0\ket0+\ket1\ket1)=\ket{\psi_+},&\text{if }(x,y)=(0,0),\\
        \tfrac{1}{\sqrt{2}}(\ket1\ket0+\ket0\ket1)=\ket{\phi_+},&\text{if }(x,y)=(0,1),\\
        \tfrac{1}{\sqrt{2}}(\ket0\ket0-\ket1\ket1)=\ket{\psi_-},&\text{if }(x,y)=(1,0),\\
        \tfrac{1}{\sqrt{2}}(\ket0\ket1-\ket1\ket0)=\ket{\phi_-},&\text{if }(x,y)=(1,1),
    \end{cases}
\end{equation}
where $\ket{\psi_\pm}$ and $\ket{\phi_\pm}$ are the four Bell states (fully entangled state, being a basis of the common Hilbert space). 
The fourth operation consists in Alice sending the qubit 0 to Bob. Any operation performed on the two qubits by Bob is now fully local.
The fifth operation is the decoding. Bob applies the last two operations (CNOT and $\hat H\otimes \hat{\mathbb 1}$) which together form the inverse operation of the encoding:
\begin{equation}\label{eq.decoding}
    \begin{aligned}
        \ket{\psi_+}&\to\ket0\ket0,\\
        \ket{\phi_+}&\to\ket0\ket1,\\
        \ket{\psi_-}&\to\ket1\ket0,\\
        \ket{\phi_-}&\to\ket1\ket1.
    \end{aligned}
\end{equation}
The last operation is to Bob to measure the state of both  qubits, which will identify which bits Alice encoded in her qubit.

\section{Quantum teleportation}\tindex{Quantum teleportation}

An application of the dense coding protocol is the quantum teleportation, that allows for sending a generic state $\ket\psi=\alpha\ket0+\beta\ket1$ from Alice to Bob without knowing a priory the state. The protocol is based on the use of three qubits and six operations:
\begin{equation}
    \begin{quantikz}
        \ket\psi&&\slice{share}&\ctrl{1}&\gate{H}&\meter{}\slice{classical communication}&\setwiretype{c}\wire[d][2]{c}\\
        \ket0&\gate{H}&\ctrl{1}&\targ{}&&\meter{}&\setwiretype{c}\wire[d][1]{c}\\
        \ket1&&\targ{}&&&&\gate{U}&\ket\psi
    \end{quantikz}
\end{equation}
The first operation is to prepare an entangled state between qubit 1 and 2 (similarly as in the dense coding protocol):

\begin{equation}
    \ket{\psi01}\xrightarrow[]{\hat{\mathbb 1}\otimes\hat H\otimes\hat{\mathbb 1}}
    \ket\psi\tfrac1{\sqrt{2}}(\ket0+\ket1)\ket1\xrightarrow[]{\hat{\mathbb 1}\otimes CNOT}
    \ket\psi\tfrac1{\sqrt{2}}(\ket{01}+\ket{1 0})=\ket\psi\ket{\phi_+}.
\end{equation}
Then, the second operation is to  shared the qubits among Alice (qubit 0 and 1) and Bob (qubit 2).
The third operation consists in applying a decoding operation (see dense coding) to the first two qubits. Namely, the decoding operation acts as in Eq.~\eqref{eq.decoding}. Thus, owning that $\ket\psi=\alpha\ket0+\beta\ket1$, one obtains
\begin{equation}
    \begin{aligned}
      \ket\psi\ket{\phi_+}&\xrightarrow[]{CNOT\otimes\hat{\mathbb 1}}
      \tfrac{1}{\sqrt{2}}\left[\alpha\ket0(\ket{01}+\ket{10})+\beta\ket1(\ket{11}+\ket{00})\right]\\
      &\xrightarrow[]{\hat H\otimes\hat{\mathbb 1}\otimes\hat{\mathbb 1}}\tfrac12\left[ \ket{00}(\alpha\ket1+\beta\ket0)+\ket{01}(\alpha\ket0+\beta\ket1)+\ket{10}(\alpha\ket1-\beta\ket0)+\ket{11}(\alpha\ket0-\beta\ket1)
      \right].
    \end{aligned}
\end{equation}
The fourth operation consists in Alice measuring her qubits. There are 4 possible couples, and thus four possible collapses (according to the measurement postulate of quantum mechanics). These are $\ket{00}, \ket{01},\ket{10}$ and $\ket{11}$ with probability $1/4$ each. The fundamental point of the protocol is that the collapse of the state of the first 2 qubit implies that of the last qubit, being in Bob's hands. In particular, if Alice measures the couple $(0,0)$, then qubit 2 collapses in $\alpha\ket1+\beta\ket0$; and similarly for the other three measurement outcomes. 
The fifth operation is the classical communication of the outcomes of the measurement to Bob. Consequently, the sixth operation is a unitary operation $\hat U$ on qubit 2 that depends on the outcomes ($q_0$ and $q_1$) of the measurement:
\begin{equation}
        \begin{array}{c| c |c| c }
    q_0&q_1 &\ket{q_2}& \hat U \\
    \hline
 0&0&\alpha\ket1+\beta\ket0&\hat \sigma_x\\
 0&1&\alpha\ket0+\beta\ket1&\hat{\mathbb 1}\\
 1&0&\alpha\ket1-\beta\ket0&i\hat \sigma_y\\
 1&1&\alpha\ket0-\beta\ket1&\hat \sigma_z
  \end{array}
\end{equation}
where $\ket{q_2}$ is the state on which qubit 2 has collapsed after the measurement.
By applying the unitary we obtain
\begin{equation}
    \begin{aligned}
        \alpha\ket1+\beta\ket0\xrightarrow[\phantom{i\hat\sigma_x}]{\hat\sigma_x}\ket\psi,\\
    \alpha\ket0+\beta\ket1\xrightarrow[\phantom{i\hat\sigma_x}]{\hat{\mathbb 1}}\ket\psi,\\
        \alpha\ket1-\beta\ket0\xrightarrow[\phantom{i\hat\sigma_x}]{i\hat \sigma_y}\ket\psi,\\
        \alpha\ket0-\beta\ket1\xrightarrow[\phantom{i\hat\sigma_x}]{\hat \sigma_z}\ket\psi.
    \end{aligned}
\end{equation}
In such a way, Bob retrieves the state $\ket\psi$ without that neither Bob or Alice had measure it. 

We notice that there is a strong difference with the case studied in the no cloning theorem. Here, one needs to measure two qubits to perform the protocol: this a fundamentally different procedure with respect to a unitary operation.

\section{Quantum Phase estimation}\index{Quantum phase estimation}
\label{sec.QPE}

The framework of quantum phase estimation (QPE) is the following. Consider a unitary operation $\hat U$ where the state $\ket \psi$ is one of its eigenstates. In particular, one has
\begin{equation}\label{eq.phase.estim.0}
    \hat U\ket\psi=e^{2\pi i \varphi}\ket \psi.
\end{equation}
Then, the task is to determine the phase $\varphi$ with a certain given precision. 

\subsection{Single-qubit quantum phase estimation}

The Hadamard test described in Sec.~\ref{hadamard.test} can be used to implement a single qubit phase estimation. Indeed,
 from Eq.~\eqref{eq.phase.estim.0} one gets that
\begin{equation}
    \braket{\psi|\hat U|\psi}=e^{2\pi i \varphi}.
\end{equation}
Then,
by merging with Eq.~\eqref{circuit.hadamardim.result0} one has
\begin{equation}
    P(\ket0)=\tfrac12(1+\cos(2\pi\varphi)),
\end{equation}
which implies
\begin{equation}\label{eq.qpe.1qubit}
    \varphi=\pm\frac{\arccos\left(1-2P(\ket0)\right)}{2\pi}+2\pi k,
\end{equation}
where $k\in\mathbb N$.
Notice that such a circuit cannot distinguish the sign of $\varphi$. Conversely, using both Eq.~\eqref{circuit.hadamardim.result0} and Eq.~\eqref{circuit.hadamardim.result}, one has
\begin{equation}
    \varphi=\arctan\left(\frac{1-2P(\ket0)}{1-2\tilde P(\ket0)}\right).
\end{equation}

Now, for the sake of simplicity, let us restrict to the case of $\varphi\in[0,1[$. Suppose we would like to estimate the value of $\varphi$ with a single run of the circuit in Eq.~\eqref{circuit.hadamardim0}. Then, if the outcome is $+1$ (i.e., the state collapses on $\ket0$), we have $P(\ket0)=1$. Conversely, with the outcome being $-1$ we have $P(\ket0)=0$. Then, by employing Eq.~\eqref{eq.qpe.1qubit} we obtain
\begin{equation}
        \begin{array}{c| c |c|c}
    \text{outcome}&P(\ket0) &\bar\varphi&\varphi_v \\
    \hline
 +1&1&0&[0,1/2[\\
 -1&0&1/2&[1/2,1[
  \end{array}
\end{equation}
where $\bar\varphi$ gives the best estimation for the real value of the phase $\varphi_v$.
Since there are no other possible outcomes with a single run, the phase is estimated with an error $\epsilon=1/2$, namely $\varphi_v\in[\bar\varphi,\bar\varphi+\epsilon[$. This is a really low accuracy for a deterministic algorithm. To improve this accuracy, one should run the algorithm several times (namely, a number of times that scales as $\mathcal O(1/\epsilon^2)$, where $\epsilon$ is the target error bound), or consider alternative methods, as the N-qubit quantum phase estimation described below.

\subsection{Kitaev's method for single-qubit quantum phase estimation}

In the fixed point representation, a natural number $k$ can be represented with a real number $\varphi\in[0,1[$ by employing $d$ bits, i.e.
\begin{equation}
    \varphi=(.\varphi_{d-1}\dots\varphi_0),
\end{equation}
where $\varphi_k\in\{0,1\}$, as far as $k\leq 2^d-1$.

\begin{myexample}
    To make an explicit example of the fixed point representation, the value of $k=41$ corresponds to the $d=6$ bit's string $[101001]$ and can be represented with $\varphi=0.640625$ being equivalent to $(.101001)$. Indeed,  by employing the following expression with the string $\varphi=(.\varphi_{d-1}\dots\varphi_0)=(.101001)$ one has
    \begin{equation}
        \sum_{i=0}^{d-1}\varphi_i 2^{i-d}=\varphi_52^{-1}+\varphi_42^{-2}+\varphi_32^{-3}+\varphi_22^{-4}+\varphi_12^{-5}+\varphi_02^{-6}=2^{-1}+2^{-3}+2^{-6}=0.640625.
        \end{equation}
        Such a value, when multiplied by $2^6$ gives exactly 41.
\end{myexample}

In the simplest scenario of $d=1$, one has $\varphi=(.\varphi_0)$ with $\varphi_0\in\{0,1\}$. Thus, when performing once the real Hadamard test, one has $P(\ket0)=1$ if $\varphi_0=0$ (i.e., $\bar\varphi=0$), and $P(\ket 0)=0$ if $\varphi_0=1$ (i.e., $\bar\varphi=1/2$).

Next, we consider the case of $d$ bits, where $\varphi=(.0\dots0\varphi_0)$. Here, the first $d$ bits are 0 and the last one is $\varphi_0$. To determined the value of $\varphi_0$ one needs to reach a precision of $\epsilon<2^{-d}$. This would require $\mathcal O(1/\epsilon^2)=\mathcal O(2^{2d})$ repeated applications of the single-qubit quantum phase estimation, or number of queries to $\hat U$.
The observation from Kitaev’s method is that if we can have access to $\hat U^j$ for a suitable power $j$, then the number of queries to $\hat U$ can be reduced. 
If one substitutes $\hat U^j$ to $\hat U$, with the corresponding circuit being
\begin{equation}\label{circuit.hadamardim1}
\begin{quantikz}
    \ket0&\gate{H}&\ctrl{1}&         \gate{H}   &\meter{}\\
    \ket\psi&\qwbundle{}&\gate{U^j}&              &
\end{quantikz}    
\end{equation}
then the probability changes in 
\begin{equation}
    P(\ket0)=\tfrac{1}{2}(1+\cos(2\pi j \varphi)).
\end{equation}
Importantly, every time one multiplies a number by a factor 2, the bits in the fixed point representation are shifted to the left. To make an example, 
\begin{equation}
    2\times(.00\varphi_0)=(.0\varphi_0).
\end{equation}
Then, one has that $2^{d-1}\varphi=2^{d-1}(.0\dots0\varphi_0)=(.\varphi_0)$.
Thus, applying the circuit in Eq.~\eqref{circuit.hadamardim1} with $j=d-1$ to estimate $(.0\dots0\varphi_0)$ is equivalent to apply the circuit in Eq.~\eqref{circuit.hadamardim0} to estimate $(.\varphi_0)$. 

This idea can be extended to general phases with $d$ bits, i.e.~$\varphi=(.\varphi_{d-1}\dots\varphi_0)$. Indeed, one has
\begin{equation}
\hat Ue^{2\pi i \varphi}\ket\psi=\hat Ue^{2\pi i (.\varphi_{d-1}\dots\varphi_0)}\ket\psi=e^{2\pi i (\varphi_{d-1}.\varphi_{d-2}\dots\varphi_0)}\ket\psi=e^{2\pi i \varphi_{d-1}}e^{2\pi i (.\varphi_{d-2}\dots\varphi_0)}\ket\psi,
\end{equation}
but $e^{2\pi i \varphi_{d-1}}=1$ independently from the value of $\varphi_{d-1}$. Thus
\begin{equation}
\hat Ue^{2\pi i \varphi}\ket\psi=e^{2\pi i (.\varphi_{d-2}\dots\varphi_0)}\ket\psi,
\end{equation}
i.e.~the application of $\hat U$ shifts the bits and allows the evaluation of the first bit after the decimal point.

\subsection{$n$-qubit quantum phase estimation}

Notably,  both the previous algorithms necessitate an important classical post-processing. Employing $n$ ancillary qubits allow the reduction of such post-processing. This is based on the application of the Inverse Quantum Fourier Transform $\hat F^\dag$.

\begin{myrecall}[Quantum Fourier transform]
    The discrete Fourier transform of a $N$-component vector with complex components $\set{f(0),\dots,f(N-1)}$ is a new complex vector $\set{\tilde f(0),\dots,\tilde f(N-1)}$, defined as
    \begin{equation}
        F(f(j),k)=\tilde f(k)=\tfrac{1}{\sqrt{N}}\sum_{j=0}^{N-1}e^{2\pi i j k /N}f(j).
    \end{equation}
The Quantum Fourier transform (QFT) acts similarly: it acts as the unitary operator $\hat F$  on a quantum register of $n$ qubits, where $N=2^n$, in the computational basis as
\begin{equation}\label{eq.fourier.transf}
    \hat F\ket j=\tfrac{1}{\sqrt{2^n}}\sum_{k=0}^{2^n-1} e^{2\pi i j k /2^n}\ket k,
\end{equation}
where $\ket{j}=\ket{j_{n-1}\dots j_0}$ and $\ket{k}=\ket{k_{n-1}\dots k_0}$. Namely, the application of the quantum Fourier transform $\hat F$ to the state $\ket{j}=\ket{j_{n-1}\dots j_0}$ gives
\begin{equation}\label{eq.fourier.transf1}
    \hat F\ket j=\tfrac{1}{\sqrt{2^n}}\left(\ket0+e^{2\pi i (0.j_0)}\ket1\right)\left(\ket0+e^{2\pi i (0.j_1j_0)}\ket1\right)\dots\left(\ket0+e^{2\pi i (0.j_{n-1}\dots j_0)}\ket1\right).
\end{equation}

    In the case of a superposition $\ket \psi=\sum_j f(j)\ket j$, one has
    \begin{equation}
        \ket{\tilde \psi}=\hat F\ket \psi=\sum_{k=0}^{2^n-1}\tilde f(k)\ket k,
    \end{equation}
    where the coefficients $\tilde f(k)$ are the discrete Fourier transform of the coeficients $f(j)$.

    The inverse quantum Fourier transform $\hat F^\dag$ acts as
    \begin{equation}
        \hat F^\dag\ket j=\tfrac{1}{\sqrt{2^n}}\sum_{j=0}^{2^n-1} e^{-2\pi i j k /2^n}\ket k,
    \end{equation}
    in a completely similar way as Eq.~\eqref{eq.fourier.transf} but with negative phases.
\end{myrecall}

\begin{myexample}
The application of the quantum Fourier transform $\hat F$ to the state $\ket j=\ket{10}=\ket{j_1=1,j_0=0}$ gives
\begin{equation}
\begin{aligned}
    \hat F\ket j&=\tfrac{1}{2}\left(\ket0+e^{2\pi i (0.j_0)}\ket1\right)\left(\ket0+e^{2\pi i (0.j_1j_0)}\ket1\right),\\
    &=\tfrac{1}{\sqrt{2}}(\ket0+\ket1)\tfrac{1}{\sqrt{2}}(\ket0-\ket1).
\end{aligned}
\end{equation}

\end{myexample}

The algorithm implementing the (standard) quantum phase estimation uses a first register of $n$ ancillary qubits and a second register of which we want to compute the phase. The first register is initially prepared in the $\ket0$ state for all the qubits. The circuit implementing the algorithm is the following
\begin{equation}
    \begin{quantikz}
    \lstick[5]{First register,\\$n$ qubits}    \ket0&\gate{H}&&&&\ \dots\ &\ctrl{5}\slice{end first step}&\gate[5]{F^\dag}&\meter{}
\\
\vdots
\\
        \ket0&\gate{H}&&&\ctrl{3}&\ \dots\ &&&\meter{}
\\
        \ket0&\gate{H}&&\ctrl{2}&&\ \dots\ &&&\meter{}
\\
        \ket0&\gate{H}&\ctrl{1}&&&\ \dots\ &&&\meter{}
\\
      \lstick[1]{Second register,\\$t$ qubits}  \ket\psi&\qwbundle{}&\gate{U^{2^0}}&\gate{U^{2^1}}&\gate{U^{2^2}}&\ \dots\ &\gate{U^{2^{n-1}}}&&\ket\psi
    \end{quantikz}
\end{equation}
In particular, the state of the first register after the end of the first part of the algorithm (see red dashed line) reads
\begin{equation}
    \tfrac{1}{\sqrt{2^n}}\left(\ket0+e^{2\pi i (2^{n-1}\varphi)}\ket1\right)\dots\left(\ket0+e^{2\pi i (2^{0}\varphi)}\ket1\right).
\end{equation}
Now, by considering the binary representation of $\varphi=(\varphi_{n-1}\dots\varphi_0)$, the latter expression becomes
\begin{equation}
    \tfrac{1}{\sqrt{2^n}}\left(\ket0+e^{2\pi i (0.\varphi_0)}\ket1\right)\left(\ket0+e^{2\pi i (0.\varphi_1\varphi_0)}\ket1\right)\dots\left(\ket0+e^{2\pi i (0.\varphi_{n-1}\dots \varphi_0)}\ket1\right),
\end{equation}
which is exactly equal to $\hat F\ket j$ in Eq.~\eqref{eq.fourier.transf1} for $\ket j=\ket\varphi$. Thus, applying the inverse Fourier transform $\hat F^\dag$ one gets $\ket \varphi$, which is then measured.

\section{Harrow-Hassidim-Lloyd algorithm}
\index{Harrow-Hassidim-Lloyd}

The Harrow-Hassidim-Lloyd (HHL) algorithm allows for the resolution of linear system problems on a quantum computer. To be precise, the problem to be solved is described as finding the $N_b$ complex entries of $\x$ that solve the following problem \label{lin2022lecture}
\begin{equation}
    A\x={\bf b},
\end{equation}
where $A$ is an hermitian and non-singular $N_b\times N_b$ matrix and ${\bf b}$ is a $N_b$ vector, both defined on $\mathbb C$. Classically, the solution is given by
\begin{equation}
    \x=A^{-1}{\bf b}.
\end{equation}
The question is then how one can implement this on a quantum computer.

First, let us assume that the entries of ${\bf b}$ are such that $||{\bf b}||=1$. Then, ${\bf b}$ can be stored in a $n_b$-qubit state $\ket b$, through the following mapping:
\begin{equation}
    {\bf b}=
    \begin{pmatrix}
        b_0\\
        \vdots\\
        b_{N_b-1}
    \end{pmatrix}
    \leftrightarrow
    b_0\ket0+\dots + b_{N_b-1}\ket{N_b-1}=\ket b,
\end{equation}
where $N_b=2^{n_b}$. For example, this can be done via a unitary operation $\hat U_{\bf b}$.
Now, we define $\ket x=\hat A^{-1}\ket b$, where $\hat A$ in the computational representation gives the classical matrix $A$. Notably, the state $\ket x$ needs to be normalised to be stored in a quantum register. Thus, one has
\begin{equation}
    \ket x=\frac{\hat A^{-1}\ket b}{||\hat A^{-1}\ket b||},
\end{equation}
where the normalisation problem can be tackled in a second moment.

Consider the spectral decomposition of $\hat A$:
\begin{equation}
    \hat A\ket{v_j}=\lambda_j \ket{v_j},
\end{equation}
where $\lambda_j$ and $\ket{v_j}$ are respectively the eigeinvalues and eigeinstates of $\hat A$. We also assume that the ordering of the eigeinvalues is such that
\begin{equation}
    0<\lambda_0\leq\dots\leq\lambda_{N_b-1}<1.
\end{equation}
In general this will not be the case, but one can remap the problem in order to fall within this case. We also assume that all the $N_b$ eigeinvalues have an exact $d$-bit representation. 

By applying what in Sec.~\ref{sec.QPE}, we can query $\hat A$ via an unitary operation $\hat U=e^{2\pi i \hat A}$ using QPE.  For example, suppose $\ket b=\ket{v_j}$, then we have
\begin{equation}
\hat U_\text{\tiny QPE}\ket{0}^{\otimes d}\ket{v_j}=\ket{\lambda_j}\ket{v_j}.    
\end{equation}
In particular, the (not-normalised) solution of the linear system problem would be
\begin{equation}
    \hat A^{-1}\ket b=\hat A^{-1}\ket{v_j}=\tfrac{1}{\lambda_j}\ket{v_j}.
\end{equation}
More generally, one can decompose the state $\ket b$ on the basis of $\hat A$, i.e.~
\begin{equation}\label{eq.HHL.b}
\ket b=\sum_{j=0}^{2^{n_b}-1}\beta_j\ket{v_j},
\end{equation}
where $\beta_j$ are a linear combination of $b_j$. Then the QPE procedure gives
\begin{equation}
    \hat U_\text{\tiny QPE}\ket0^{\otimes d}\ket b=\sum_j\beta_j\ket{\lambda_j}\ket{v_j},
\end{equation}
and the solution of the problem is given by
\begin{equation}\label{eq.aimHHL}
   \hat A^{-1}\ket b=\sum_{j=0}^{2^{n_b}-1}\frac{\beta_j}{\lambda_j}\ket{v_j}. 
\end{equation}
The aim of the HHL algorithm is to generate the normalised version of the state in Eq.~\eqref{eq.aimHHL} from the general state $\ket b$ as shown in Eq.~\eqref{eq.HHL.b}.

The algorithm works with three registers. The first one is an ancillary register made of a single qubit, the second is also an ancillary register but made of $d$ qubits, the third register is made of $n_b$ qubits and will encode the solution of the problem.  The HHL circuit is the following
\begin{equation}
\begin{quantikz}
    \ket0\phantom{_{\otimes n}} &               &                        &                               &\gate{R} \slice{$\ket{\Psi_5}$} &\meter{} \slice{$\ket{\Psi_6}$}  \\
    \ket0^{\otimes d}           &\qwbundle{d}   &                        &\gate[2]{U_\text{\tiny QPE}} \slice{$\ket{\Psi_4}$}  &\ctrl{-1} &           &\gate[2]{U^{-1}_\text{\tiny QPE}}\slice{$\ket{\Psi_9}$}&\ket0^{\otimes d}\\
    \ket0^{\otimes n_b}         &\qwbundle{n_b} &\gate{\hat U_{\bf b}}  \slice{$\ket{\Psi_1}$} &                                &          &           &                                 &\ket x
\end{quantikz}
\end{equation}
The algorithm works as the following. Initially, all the qubits are prepared in $\ket0$:
\begin{equation}
    \ket{\Psi_0}=\ket0\ket0^{\otimes d}\ket{0}^{\otimes n_b},
\end{equation}
then the information about ${\bf b}$ is encoded in the last register:
\begin{equation}
    \ket{\Psi_1}=\hat{\mathbb 1}\otimes\hat{\mathbb 1}^{\otimes d}\otimes\hat U_{\bf b}  \ket{\Psi_0}=\ket0\ket0^{\otimes d}\ket{b}.
\end{equation}
We apply the QPE procedure, which is here broke down in the corresponding three steps. The first is the application of the Hadamard gate:
\begin{equation}
    \ket{\Psi_2}=\hat{\mathbb 1}\otimes \hat H^{\otimes d}\otimes \hat{\mathbb 1}\ket{\Psi_1}=\ket0\tfrac{1}{2^{d/2}}(\ket0+\ket1)^{\otimes d}\ket{b}.
\end{equation}
This is followed by the controlled unitary $\hat U^j$:
\begin{equation}
    \ket{\Psi_3}=\hat{\mathbb 1}\otimes C(U^j)\ket{\Psi_2}=\ket0\tfrac{1}{2^{d/2}}\sum_{k=0}^{2^d-1}e^{2\pi i k \varphi}\ket k\ket b,
\end{equation}
where $\hat U\ket b=e^{2\pi i \varphi}\ket b$ with $\varphi\in[0,1[$. Finally, we apply the inverse Fourier transform to the second register
\begin{equation}
\begin{aligned}
\ket{\Psi_4}&=    \hat{\mathbb 1}\otimes \hat F^\dag\otimes \hat{\mathbb 1}^{\otimes n_b}\ket{\Psi_3},\\
&=\ket0\tfrac{1}{2^{d/2}}\sum_{k=0}^{2^d-1}e^{2\pi i k \varphi}\hat F^\dag\ket k\ket b,     \\
&=\ket0\tfrac{1}{2^{d}}\sum_{k=0}^{2^d-1}e^{2\pi i k \varphi}\sum_{y=0}^{2^d-1}e^{-2\pi i y k/2^d}
\ket y\ket b.
\end{aligned}
\end{equation}
However, one has that
\begin{equation}
    \sum_{k=0}^{2^d-1}e^{2\pi i k (\varphi-y/2^d)}=
    \begin{cases}
        \sum_{k=0}^{2^d-1}e^0=2^d, &\text{if }\varphi=y/2^d,\\
        0, &\text{if }\varphi\neq y/2^d,\\
    \end{cases}
\end{equation}
meaning that the $k$ sum selects the value of $y=\varphi 2^d$. Thus, 
\begin{equation}
    \ket{\Psi_4}=\ket0\ket{\varphi 2^d}\ket b.
\end{equation}
In general, $\ket b$ is in a superposition of $\ket{v_j}$, then 
\begin{equation}
    \hat U\ket{v_j}=e^{2\pi i \hat A}\ket{v_j}=e^{2\pi i \lambda_j}\ket{v_j}.
\end{equation}
Then, the entire QPE gate maps
\begin{equation}
  \ket{\Psi_1}=\ket0\ket0^{\otimes d}\sum_{j=0}^{2^{n_b}-1}\beta_j\ket{v_j}\xrightarrow[]{U_\text{\tiny QPE}}\ket{\Psi_4}=\ket0\sum_{j=0}^{2^{n_b}-1}\beta_j\ket{\lambda_j 2^d}\ket{v_j}.
\end{equation}
We apply a controlled rotation on the first register, such that
\begin{equation}
    \ket{\Psi_5}=C(R)\otimes\hat{\mathbb 1}^{\otimes n_b}\ket{\Psi_4}=\sum_{j=0}^{2^{n_b}-1}\beta_j\left(\sqrt{1-\tfrac{C^2}{\lambda_j^2}}\ket0+\tfrac{C}{\lambda_j}\ket1\right)\ket{\lambda_j2^d}\ket{v_j},
\end{equation}
where $C\in \mathbb R$ is an arbitrary constant. At this point we perform the measurement of the first register. If the outcome is $+1$ and the state collapses in $\ket0$ then we discard the run; if the outcome is $-1$ with the state collapsed in $\ket1$ then we retain the run. To increase the probabilities of having the outcome $-1$, we make $C$ as large as possible. After the collapse of the first register in $\ket 1$, the state of the second and third register is
\begin{equation}
    \ket{\Psi_6}=\frac{1}{\left(\sum_{j=0}^{2^{n_b}-1}|\beta_j/\lambda_j|^2\right)^{1/2}}\sum_{j=0}^{2^{n_b}-1}\tfrac{\beta_j}{\lambda_j}\ket{\lambda_j 2^d}\ket{v_j},
\end{equation}
where we exploited that $C\in\mathbb R$. Now, we apply the inverse QPE, which has also three steps. The first is the application of the QFT:
\begin{equation}
\begin{aligned}
    \ket{\Psi_7}&=\hat F\otimes\hat{\mathbb 1}^{\otimes n_b}\ket{\Psi_6},\\
    &=\frac{1}{\left(\sum_{j=0}^{2^{n_b}-1}|\beta_j/\lambda_j|^2\right)^{1/2}}\sum_{j=0}^{2^{n_b}-1}\tfrac{\beta_j}{\lambda_j}\hat F\ket{\lambda_j 2^d}\ket{v_j},\\
    &=\frac{1}{\left(\sum_{j=0}^{2^{n_b}-1}|\beta_j/\lambda_j|^2\right)^{1/2}}\sum_{j=0}^{2^{n_b}-1}\tfrac{\beta_j}{\lambda_j}
\tfrac{1}{2^{d/2}}    \sum_{y=0}^{2^d-1}e^{2\pi i y (\lambda_j 2^d)/2^d}\ket y
    \ket{v_j}.
\end{aligned}
\end{equation}
Then, we apply the controlled unitary $C(U^{-j})$, which gives
\begin{equation}
\begin{aligned}
    \ket{\Psi_8}&=C(U^{-j})\ket{\Psi_7},\\
    &=\frac{1}{\left(\sum_{j=0}^{2^{n_b}-1}|\beta_j/\lambda_j|^2\right)^{1/2}}\sum_{j=0}^{2^{n_b}-1}\tfrac{\beta_j}{\lambda_j}
\tfrac{1}{2^{d/2}}    \sum_{y=0}^{2^d-1}e^{2\pi i y \lambda_j}\ket y
    e^{-2\pi i \lambda_j y}\ket{v_j},
    \end{aligned}
\end{equation}
where the two phases cancel and thus
\begin{equation}
        \ket{\Psi_8}=\tfrac{1}{2^{d/2}}    \sum_{y=0}^{2^d-1}\ket y\sum_{j=0}^{2^{n_b}-1}\tfrac{\beta_j}{\lambda_j}
\frac{1}{\left(\sum_{j=0}^{2^{n_b}-1}|\beta_j/\lambda_j|^2\right)^{1/2}}\ket{v_j}.
\end{equation}
Finally, the application of Hadamard's gates on the second register gives
\begin{equation}
        \ket{\Psi_9}=\hat H^{\otimes d}\otimes\hat{\mathbb 1}^{\otimes n_b}\ket{\Psi_8}=\ket0^{\otimes d}\sum_{j=0}^{2^{n_b}-1}\tfrac{\beta_j}{\lambda_j}
\frac{1}{\left(\sum_{j=0}^{2^{n_b}-1}|\beta_j/\lambda_j|^2\right)^{1/2}}\ket{v_j},
\end{equation}
where the third register is exactly in the form in Eq.~\eqref{eq.aimHHL} after the proper normalisation. Thus, 
\begin{equation}
    \ket{\Psi_9}=\ket0^{\otimes d}\ket x,
\end{equation}
embeds the solution of the linear system $A\x={\bf b}$.

\chapter{Variational Quantum Algorithms}
\index{Variational quantum algorithms}

This class of algorithms employs a quantum and a classical computer to solve some optimisation problems. The quantum computer performs the quantum evolution of a state with respect to an Hamiltonian that is transformed, say from $\hat H_0$ to $\hat H_1$. The classical computer determines how such a transformation should take place employing the variational principle. Typically, the problem is to map the state from the ground state of $\hat H_0$ to that of $\hat H_1$, whose ground state is unknown. Thus, one wants to have a well-known $\hat H_0$. This is often taken as that of the Ising model.

\section{The Ising model}\tindex{Ising model}
In a combinatorial optimisation problem, one has a string of $n$ bits and wants to optimise a particular problem. The problem is mapped in a minimisation (or maximisation) of a cost function $C:\set{0,1}^n\to\mathbb R$. Notably, the maximisation problem can be obtained from the minimisation one buy a minus sign: $C\to-C$. 

To solve a combinatorial optimisation problem via a quantum algorithm, one needs to encode the problem onto a quantum system. 
In the following, we show how the Ising Hamiltonian can be used to embed such an optimisation problem.

The Ising model was developed to study the phase transition in magnetic materials. It consists in $n$ spins that can be coupled via long-range interactions. The corresponding Hamiltonian is
\begin{equation}\label{eq.Hising}
     \hatHc=-\sum_{i=1}^n h_i\hat \sigma_z^{(i)}-\sum_{1\leq i<j\leq n}J_{ij}\hat \sigma_z^{(i)}\hat \sigma_z^{(j)},
\end{equation}
where $h_i$ are the single spin magnetic fields describing the single spin evolution and $J_{ij}$ the spin-spin couplings. The choice of the latter encodes if the spins are encouraged to be aligned (ferromegnetic phase) or anti-aligned (antiferromagnetic phase). Since only $\hat \sigma_z$ are appearing in $\hatHc$, then its spectral decomposition can be expressed in the computational basis:
\begin{equation}
    \hatHc=\sum_{z=0}^{2^n-1}C(z)\ket z\bra z,
\end{equation}
where $C(z)$ is the energy of the specific spin configuration $\ket z$. Then, by properly mapping a combinatorial problem in the choice of $\set{h_i}$ and $\set{J_{ij}}$, one can find the optimal solution by minimizing the energy, i.e.~by finding the configuration $\ket z$ that corresponds to minimal energy (or cost) $C(z)$.

\section{Mapping combinatorial optimisation problems into the Ising model}

Many problems can be mapped in the form in Eq.~\eqref{eq.Hising}, and hence solve with a quantum computer, by choosing the appropriate values of $\set{h_i}$ and $\set{J_{ij}}$. Here we consider some explicit examples.\\

\noindent {\bf Subset sum problem.} Given an integer number $m$ (total value) and a set of $N$ positive and negative integers $n=\set{n_1,\dots,n_N}$, which is the subset of the latter integers whose sum gives $m$?

\begin{myexample}
    Consider the case of $m=7$ and $n=\set{-5,-3,1,4,9}$. The subset $\set{-3,1,9}$ solves the problem: $-3+1+9=7=m$.
\end{myexample}

\begin{myexercise}
    Consider the case of $m=13$ and $n=\set{-3,2,8,4,20}$. Show that the corresponding subset sum problem has no solution.
\end{myexercise}

The subset sum problem can be framed as an energy minimisation problem as follows. Consider the sum
 $   \sum_{i=1}^N n_i z_i-m$,
where $n_i$ are the elements of $n$ and $z_i\in\set{0,1}$ are weights that select or not the corresponding element $n_i$ in the sum (effectively, this is the way to select a specific subsection).
We define $\mathcal E(z)$ as the square of such a sum:
\begin{equation}
    \mathcal E(z)=\mathcal E(z_1,\dots,z_N)=\left(\sum_{i=1}^N n_i z_i-m\right)^2.
\end{equation}
Then, if there is a subset  solving the problem, one has that exists a value of $z=\set{z_1,\dots,z_N}$ such that $\mathcal E(z)=0$. Conversely, if all the possible values of $z$ give $\mathcal E(z)\neq 0$, then there is no subset that can solve the subset sum problem. One can already see that the $z$ corresponding to the solution of the problem is the one minimising $\mathcal E(z)$. We now show the connection with the Ising model. We introduce the classical spins $s_i=\pm 1$, which will be employed in place of the weights $z_i$. Namely, one uses
\begin{equation}
    z_i=\tfrac12(1-s_i),
\end{equation}
so that $s_i=+1$ (spin up) corresponds to $z_i=0$ and $s_i=-1$ (spin down) to $z_i=1$. We define the corresponding classical Hamiltonian
\begin{equation}
\begin{aligned}
    \mathcal H(s_1,\dots,s_N)&=\left(\sum_{i=1}^N n_i \tfrac12(1-s_i)-m\right)^2,\\
    &=\tfrac14\sum_{i,j=1}^N n_in_j s_is_j-\sum_{i=1}^N\left(\tfrac12\sum_{j=1}^Nn_j-m\right)n_is_i+\left(\tfrac12\sum_{i=1}^Nn_i-m\right)^2,
\end{aligned}
\end{equation}
where the last term is independent from $s_i$ and thus is a negligible constant of the problem. After having defined 
\begin{equation}
    J_{ij}=-\frac{n_in_j}{4},\quad\text{and}\quad h_i=\left(\tfrac12\sum_{j=1}^Nn_j-m\right)n_i,
\end{equation}
the Hamiltonian becomes
\begin{equation}\label{eq.H.subsetsum}
      \mathcal H(s_1,\dots,s_N)=-\sum_{1\leq i<j\leq n}J_{ij}s_is_j-\sum_{i=1}^n h_i s_i+\text{const},
\end{equation}
where 
\begin{equation}
    \text{const}=\left(\tfrac12\sum_{i=1}^Nn_i-m\right)^2-\sum_{i=1}^NJ_{ii}s_i^2,
\end{equation}
is $s_i$ independent since $s_i^2=1$ for any value of $i$. To solve the problem on a quantum computer, one quantises the Hamiltonian in Eq.~\eqref{eq.H.subsetsum} by substituting $s_i\to\hat\sigma_z^{(i)}$ and gets Eq.~\eqref{eq.Hising}.\\

\noindent {\bf Number partitioning problem.} Another combinatory problem that can be mapped in an Ising Hamiltonian is the number partitioning problem. It asks if a set of $N$ integers $\set{n_1,\dots,n_N}$ can be partitioned in two subsets such that the sum of the elements in the individual subsets is equal.

\begin{myexample}
    Consider the set $n=\set{1,2,3,4,6,10}$. In such a case, one can consider the case of $\set{1,2,4,6}$ and $\set{3,10}$, whose individual sums are both equal to 13.
\end{myexample}

The classical Hamiltonian for this problem can be straightforwardly constructed as
\begin{equation}
    \mathcal H(s_1,\dots,s_N)=\left(\sum_{i=1}^Nn_is_i\right)^2,
\end{equation}
with $s_i=\pm 1$. Clearly, the solution $s=\set{s_1,\dots,s_N}$ is such that $\mathcal{H}(s)=0$. Expanding the square, we find 
\begin{equation}\label{eq.partitioning}
    \mathcal H(s)=-\sum_{1\leq i<j\leq N}J_{ij}s_is_j-\Tr{J_{ij}},
\end{equation}
where 
\begin{equation}
    J_{ij}=-\frac{n_in_j}{2},\quad\text{and}\quad \Tr{J_{ij}}=\sum_{i=1}^NJ_{ii}s_i^2.
\end{equation}
The classical Hamiltonian in Eq.~\eqref{eq.partitioning} can be quantised and one obtains that in Eq.~\eqref{eq.Hising} with no need to introduce the magnetic fields, i.e.~$h_i=0$.

\section{Adiabatic Theorem}
\index{Adiabatic theorem}
Adiabatic quantum computation is based on the adiabatic theorem. The latter considers the case of a time dependent Hamiltonian, that changes from $\hat H_0$ at time $t=0$ to $\hat H_1$ at time $t=\tau$. We also assume that the two Hamiltonians do not commute, i.e.$[\hat H_0,\hat H_1]\neq0$. The theorem
states that a system prepared in the $n$-th eigeinstate of $\hat H_0$ goes in the $n$-th eigeinstate of $\hat H_1$ if the  transformation is made slowly enough, i.e.~adiabatically.
The application to quantum computation then is to take an initial Hamiltonian with a ground state that can be easily prepared and then adiabatically change the Hamiltonian to that of the problem one wants to optimise. If the system is initially in the ground state of $\hat H_0$, then will remain in the ground state of the target Hamiltonian $\hat H_1$ and it will encode the solution of the optimisation problem.

The proof of the adiabatic theorem is the following. Consider the instantaneous spectralisation of a time-dependent Hamiltonian $\hat H(t)$, which is
\begin{equation}\label{eq.Hspectral.adiab}
    \hat H(t)\ket{n(t)}=E_n(t)\ket{n(t)},
\end{equation}
where $E_n(t)$ and $\ket{n(t)}$ are respectively the corresponding instantaneous eigeinvalues and eigeinstates.  
Given a state $\ket{\psi(t)}$ at time $t$, one can always express it as a superposition of the instantaneous eigeinstates as
\begin{equation}\label{eq.state.adiab}
    \ket{\psi(t)}=\sum_nc_n(t)\ket{n(t)},
\end{equation}
where
\begin{equation}
    c_n(t)=\braket{n(t)|\psi(t)},
\end{equation}
determine the probabilities $P_n(t)=|c_n(t)|^2$ of being in $\ket{n(t)}$ at time $t$. The evolution of $c_n(t)$ can be determined via
\begin{equation}
\begin{aligned}
    \dot c_n(t)&=\braket{\dot n(t)|\psi(t)}+\braket{n(t)|\dot\psi(t)},\\
    &=\braket{\dot n(t)|\psi(t)}-\tfrac i\hbar\braket{n(t)|\hat H(t)|\psi(t)},\\
    &=\braket{\dot n(t)|\psi(t)}-\tfrac i\hbar E_n(t)\braket{n(t)|\psi(t)},\\
\end{aligned}
\end{equation}
where we defined $\ket{\dot n(t)}=\tfrac{\D}{\D t}\ket{n(t)}$, and  we applied the Schr\"odinger equation and applied the Hamiltonian to its eigeinstate. Then, the imposing Eq.~\eqref{eq.state.adiab}, we get
\begin{equation}\label{eq.cn-cm}
        \dot c_n(t)=   \sum_mc_m(t)\braket{\dot n(t)|m(t)}-\tfrac i\hbar E_n(t)c_n(t),
\end{equation}
which determines a system of coupled differential equations. In complete generality, the evolution of $c_n(t)$ depends on $c_m(t)$ for all values of $m$. To determine the first term of Eq.~\eqref{eq.cn-cm}, we consider the time derivative of Eq.~\eqref{eq.Hspectral.adiab} with $\ket{n(t)}$ subsituted with $\ket{m(t)}$ and projecting it on $\bra{n(t)}$. This gives
\begin{equation}
    \braket{n(t)|\tfrac{\D}{\D t}\hat H(t)|m(t)}+\braket{n(t)|\hat H|\dot m(t)}=\dot E_m(t)\delta_{nm}+E_m(t)\braket{n(t)|\dot m(t)},
    \end{equation}
which can be recasted as
\begin{equation}
    \left(E_n(t)-E_m(t)\right)\braket{n(t)|\dot m(t)}=\dot E_m(t)\delta_{nm}-\braket{n(t)|\tfrac{\D}{\D t}\hat H(t)|m(t)}.
\end{equation}
For $m\neq n$, one then has
\begin{equation}
    \braket{\dot n(t)| m(t)}=\frac{\braket{n(t)|\tfrac{\D}{\D t}\hat H(t)|m(t)}}{\left(E_n(t)-E_m(t)\right)},
\end{equation}
where we exploited that $\braket{\dot n(t)| m(t)}=-\braket{ n(t)| \dot m(t)}$. Thus, by separating the case of $m=n$ and $m\neq n$ in Eq.~\eqref{eq.cn-cm}, we have
\begin{equation}\label{eq.cn-cm1}
     \dot c_n(t)=\left(\braket{\dot n(t)|n(t)}-\tfrac i\hbar E_n(t)\right)c_n(t)+\sum_{m\neq n}c_m(t)\frac{\braket{n(t)|\tfrac{\D}{\D t}\hat H(t)|m(t)}}{\left(E_n(t)-E_m(t)\right)}.
\end{equation}
In the limit where the Hamiltonian $\hat H(t)$ changes slowly enough, i.e.~for $\braket{n(t)|\tfrac{\D}{\D t}\hat H(t)|m(t)}\ll\left(E_n(t)-E_m(t)\right)$ for all $n$ and $m$, then one can neglect the last term in Eq.~\eqref{eq.cn-cm1}. This is the so-called adiabatic approximation, which gives the following solutions
\begin{equation}
    c_n(t)=e^{i\theta_n(t)}e^{i\gamma_n(t)}c_n(0),
\end{equation}
where we defined
\begin{equation}
    \theta_n(t)=-\tfrac 1\hbar\int_0^t\D s\,E_n(s),\quad \text{and}\quad \gamma_n(t)=-i\int_0^t\D s\,\braket{\dot n(s)|n(s)}.
\end{equation}
In particular, $\gamma_n(t)\in \mathbb R$ is known as the Berry phase.

Importantly, under the adiabatic approximation, one has that the probabilities evolve as
\begin{equation}
    P_n(t)=|c_n(t)|^2=|c_n(0)|^2=P_n(0),
\end{equation}
which is the final proof of the theorem.

\begin{remark}
It is important to understand the limits in which the adiabatic approximation is valid. To prove it in complete generality, one should require that the time-scale $\tau$ of the transformation is such that 
\begin{equation}
    \tau\gg \max_{n\neq m}\max_{0\leq t\leq \tau}\left|\frac{\braket{n(t)|\tfrac{\D}{\D t}\hat H(t)|m(t)}}{\left(E_n(t)-E_m(t)\right)}\right|.
\end{equation}
For the perspective of quantum computation, one can restrict to the case of $n=0$ and $m=1$. This is the case where the system is initially prepared in the ground state $n=0$ and one does not want a jump in the first excited state $m=1$. In such a case, the approximation is valid if
\begin{equation}
\tau\gg \max_{0\leq t\leq \tau}\left|\frac{\braket{\psi_0(t)|\tfrac{\D}{\D t}\hat H(t)|\psi_1(t)}}{\left(E_0(t)-E_1(t)\right)}\right|.
\end{equation}
Notably, the more the energy gap $E_1-E_0$ closes, the larger value of $\tau$ one has to consider. In the case of a linear transition between the initial $\hat H_0$ and final Hamiltonian $\hat H_1$ (i.e.~$\hat H(t)=(1-t/\tau)\hat H_0+t/\tau \hat H_1$), a necessary condition for keeping the energy gap open is that $[\hat H_0,\hat H_1]\neq0$. Figure \ref{fig:energy_qva} represents graphically how the gap should remain open during the Hamiltonian change so that the initial state being the ground state of $\hat H_0$ is mapped to the ground state of $\hat H_1$, which encodes the solution of the problem.

\begin{figure}[h]
    \centering
    \includegraphics[width=0.7\linewidth]{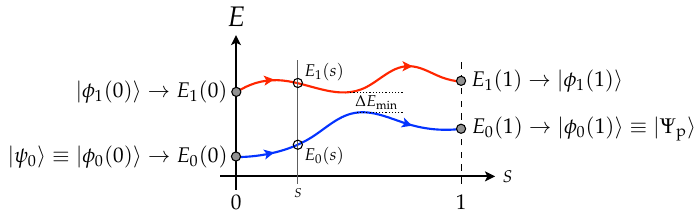}
    \caption{Graphical representation of how the energy levels of $\hat H(t)$ change in time. As long as the minimum energy gap $\Delta E_\text{min}$ is finite, one can employ the adiabatic theorem to go from the ground state of $\hat H_0$ (here denoted as $\ket{\psi_0}$) to that of $\hat H_1$ ($\ket{\Psi_\text{p}}$). Here, we used the parameter $s$ to parametrise the time flow: $t=s\tau$. 
    }
    \label{fig:energy_qva}
\end{figure}
\end{remark}

\section{Quantum Annealing}\tindex{Quantum annealing}

The quantum annealing is an heuristic quantum algorithm based on the adiabatic theorem. It aims at solving hard combinatorial optimisation problems using the Ising Hamiltonian as a target Hamiltonian.

The algorithm exploits an Hamiltonian transformation of the form
\begin{equation}
    \hat H(t)=\left(1-s(t)\right)\hat H_0+s(t) \hatHc,
\end{equation}
where $s(t)$ is a suitable smooth function of time with $s(0)=0$ and $s(\tau)=1$, and $\hatHc$ is the Ising Hamiltonian
in Eq.~\eqref{eq.Hising} whose ground state encodes the solution of the problem. Moreover,  one requires that $H_0$ has a non-degenerate ground state that is easy to prepare and that $[\hat H_0,\hat H_1]\neq0$. A simple choice is
\begin{equation}\label{eq.Hm.annealing}
    \hat H_0=-\sum_{i=1}^n \hat \sigma_x^{(i)},
\end{equation}
which has $\ket +^{\otimes n}$ as ground state. This can be easily prepared via Hadamard gates: $\ket +^{\otimes n}=\hat H^{\otimes n}\ket{0}^{\otimes n}$.

If the quantum annealing time $\tau$ is not sufficiently long (i.e. the transformation is not sufficiently adiabatic), which is essentially always the case, then one reach a  state $\ket{\psi(\tau)}$, which has a probability $p$ of being the solution  of the problem. Such a probability (of success) is given by $p=|\braket{z_\text{sol}|\psi(\tau)}|^2$, where $\ket{z_\text{sol}}$ is the state encoding the exact solution. To obtain the solution with a 99\% certainty, one has to repeat the annealing procedure $m$ times. Indeed, 
\begin{equation}
    P_\text{succ}^m=1-(1-p)^m=0.99.
\end{equation}
The corresponding total time required is given by
\begin{equation}
    T_{99\%}=m \tau=\frac{\ln(1-0.99)}{\ln(1-p)}\tau
\end{equation}

A strong challenge for the quantum annealing is the full connectivity of the qubits. Indeed, in a quantum computer, the qubits interactions, which are parameterised by $J_{ij}$, are typically null beyond nearest-neighbour sites. This strongly limits the scaling of universally annealing where one can suitably tune all the values of $J_{ij}$.

\section{Quantum Approximate Optimisation Algorithm (QAOA)} \tindex{Quantum Approximate Optimisation Algorithm}

The Quantum Approximate Optimisation Algorithm (QAOA) is a hybrid quantum-classical algorithm that allows for optimising a cost function and finding an approximated solution. It is an application of the adiabatic theorem, similarly as the quantum annealing, which is run on a quantum computer, while a classical computer optimises the cost function.

We start from a quantum annealing Hamiltonian of the form 
\begin{equation}\label{eq.Ht.qaoa}
    \hat H(t)=(1-s(t))\hatHm+s(t)\hatHc,
\end{equation}
where $s(t)$ is an arbitrary function such that $s(0)=0$ and $s(\tau)=1$ with $\tau$ being the total time of the algorithm. The initial Hamiltonian $\hatHm$ is such that its ground state can be prepared easily. $\hatHc$ is instead the cost Hamiltonian whose ground state encodes the solution to the problem. The QAOA is based on the observation that the best way to implement the annealing Hamiltonian in Eq.~\eqref{eq.Ht.qaoa} is a Trotter procedure. Namely, this is to consider the unitary evolution with respect to $\hat H(t)$ and decompose it in small time steps. Then, we have
\begin{equation}
    \hat U(\tau)=\operatorname{T} \exp\left[-\tfrac{i}{\hbar}\int_0^\tau\D t\,\hat H(t)\right]\simeq \prod_{k=1}^p \exp\left[-\tfrac i\hbar \hat H(k\Delta t)\Delta t\right],
\end{equation}
where $\operatorname{T}$ indicates the time ordering, one assumes a large number of steps $p\gg1$ of length $\Delta=\tau/p$. Owing that for $[\hat A,\hat B]\neq 0$ one has
\begin{equation}
    e^{i(\hat A+\hat B)\Delta t}=  e^{i\hat A\Delta t}  e^{i\hat B\Delta t}+\mathcal O((\Delta t)^2),
\end{equation}
and since
we require that $[\hatHc,\hatHm]\neq 0$, one has that at each time step the following approximation is valid to the order $(\Delta t)^2$:
\begin{equation}
    \hat U(\tau)\simeq\prod_{k=1}^p\exp\left[-\tfrac i\hbar \left(1-s(k \Delta t)\right) \hatHm\Delta t\right]\exp\left[-\tfrac i\hbar s(k \Delta t) \hatHc\Delta t\right].
\end{equation}
Now, the key idea of QAOA is to redefine the time dependence in the following way:
\begin{equation}
    \tfrac 1\hbar\left(1-s(k \Delta t)\right)\Delta t\to \beta_k,\quad\text{and}\quad \tfrac 1\hbar s(k \Delta t)\Delta t\to \gamma_k.
\end{equation}
Thus, we have 
\begin{equation}
     \hat U(\tau)\simeq\prod_{k=1}^p\exp\left[- i\beta_k \hatHm \right]\exp\left[- i\gamma_k \hatHc \right],
\end{equation}
where the parameters ${\bf \beta}=(\beta_1,\dots,\beta_p)$ and ${\bf \gamma}=(\gamma_1,\dots,\gamma_p)$ become the variational parameters to be optimised. Crucial difference with respect to the quantum annealing case is that one optimises over a set of $2p$ parameters instead of a fixed time segments. Finally, one constructs the variational state
\begin{equation}
    \ket{{\bf \gamma},{\bf \beta}}=\prod_{k=1}^pe^{- i\beta_k \hatHm}e^{- i\gamma_k \hatHc}\ket{\text{init}},
\end{equation}
where the initial state $\ket{\text{init}}$ is the ground state of $\hatHm$. In the case of $\hatHm$ being equal to Eq.~\eqref{eq.Hm.annealing}, one has
\begin{equation}
    \ket{\text{init}}=\hat H^{\otimes n}\ket 0^{\otimes n}.
\end{equation}
In the computational basis, the variational state reads
\begin{equation}
    \ket{{\bf \gamma},{\bf \beta}}= \sum_{z=0}^{2^n-1}d_z({\bf \gamma},{\bf \beta})\ket z,
\end{equation}
where $d_z({\bf \gamma},{\bf \beta})$ defines the superposition in the $Z$ basis. Notably, since $\hatHc$ encodes in its ground state the solution of the problem, one needs to minimise the expectation value of $\hatHc$ computed on the variational state. Namely
\begin{equation}
    E_p({\bf \gamma},{\bf \beta})=\braket{{\bf \gamma},{\bf \beta}|\hatHc|{\bf \gamma},{\bf \beta}}=\sum_{z=0}^{2^n-1}P_z({\bf \gamma},{\bf \beta})C(z),
\end{equation}
where $P_z({\bf \gamma},{\bf \beta})=|d_z({\bf \gamma},{\bf \beta})|^2$ is the probability of having the $\ket z$ state and $C(z)=\braket{z|\hatHc|z}$ is the corresponding cost.
The best $({\bf \gamma},{\bf \beta})$ are such that 
\begin{equation}
    ({\bf \gamma}^*,{\bf \beta}^*)=\arg \min_{{\bf \gamma},{\bf \beta}}E_p({\bf \gamma},{\bf \beta}).
\end{equation}
Such an optimisation is performed on classical computer (classical optimiser). The circuit representation of the QAOA is 
\begin{equation}\label{circuit.QAOA}
\begin{quantikz}
\ket0&\qwbundle{n}&\gate{\rotatebox[origin=c]{90}{$H^{\otimes n}$}}
\gategroup[1,steps=6,style={dashed,rounded
    corners,fill=blue!20, inner
    xsep=2pt},background,label style={label
    position=above,anchor=north,yshift=0.2cm}]{{\sc
    quantum computer}}&\gate{\rotatebox[origin=c]{90}{$e^{-i\gamma_1 H_\text{\tiny C}}$}}&\gate{\rotatebox[origin=c]{90}{$e^{-i\beta_1 H_\text{\tiny M}}$}}&\, \dots\,  &\gate{\rotatebox[origin=c]{90}{$e^{-i\gamma_p H_\text{\tiny C}}$}}&\gate{\rotatebox[origin=c]{90}{$e^{-i\beta_p H_\text{\tiny M}}$}}&\meter{}&\setwiretype{c}&\gate[style={rounded
    corners}]{\text{Compute: }E_p({\bf \gamma},{\bf \beta})}\gategroup[3,steps=3,style={dashed,rounded corners,fill=red!20, inner xsep=2pt},background,label style={label position=above,anchor=north,yshift=0.2cm}]{{\sc classical optimiser}}\wire[d]{c}\setwiretype{n}\\
    \setwiretype{n}&&&&&\,\dots\,&&&&&\gate[2,style={rounded
    corners}]{\text{Gives the minimum?}}\wire[d]{c}&\gate[style={rounded
    corners}]{\text{Yes}}\setwiretype{c}&\gate[style={rounded
    corners}]{\text{Stop}}\\
    \setwiretype{n}&&&&&\,\dots\,&&&&&&\gate[style={rounded
    corners}]{\text{No}}\setwiretype{c}&\gate[style={rounded
    corners}]{\text{New $(\gamma,\beta)$}}\wire[d]{c}\\
    \setwiretype{n}&&&\wire[u][3]{c}&\wire[u][3]{c}\setwiretype{c}&&\wire[u][3]{c}&\wire[u][3]{c}&&&&&
\end{quantikz}
\end{equation}

\begin{myexercise}
    Derive the explicit expression of the cost fucntion $C(z)$ in terms of the coefficients $J_{ij}$ and $h_i$.
\end{myexercise}

The single $k$ step of the QAOA, when considering $\hatHm$ as in Eq.~\eqref{eq.Hm.annealing} and $\hatHc$ being the Ising Hamiltonian, is implemented as the following. First we consider $\hatHm$,
\begin{equation}
    \hatHm=-\sum_{i=1}^n \hat \sigma_x^{(i)}.
\end{equation}
Then, the corresponding unitary acts independently on each qubit
\begin{equation}
    e^{-i\beta_k\hatHm}=e^{i\beta_k\sum_{i=1}^n\hat \sigma_x^{(i)}}=\prod_{i=1}^ne^{i\beta_k\hat \sigma_x^{(i)}}.
\end{equation}
Then, the corresponding action can be implemented with a rotation on the single $i$-th qubit. The circuit implementing it is
\begin{equation}
    \begin{quantikz}
        &\gate{R_X(-2\beta_k)}&
    \end{quantikz}
\end{equation}
Indeed a rotation around ${\bf n}$ by an angle $\theta$ is defined a $\hat R^{\bf n}(\theta)=e^{-i\theta {\bf n}\cdot\hat {\boldsymbol \sigma}/2}$. The implementation of the unitary related to $\hatHc$ can be divided in two steps, indeed the two terms of $\hatHc$ in Eq.~\eqref{eq.Hising} commute. Then, one writes
\begin{equation}
    e^{-i\gamma_k\hatHc}=e^{i\gamma_k\left(\sum_{i=1}^n h_i\hat \sigma_z^{(i)}+\sum_{1\leq i<j\leq n}J_{ij}\hat \sigma_z^{(i)}\hat \sigma_z^{(j)}\right)}=\prod_{1\leq i<j\leq n} e^{i\gamma_kJ_{ij}\hat \sigma_z^{(i)}\hat \sigma_z^{(j)}}\prod_{i=1}^ne^{i\gamma_k h_i\hat \sigma_z^{(i)}}.
\end{equation}
Here, the single qubits factors act as rotations, with a circuit being
\begin{equation}
        \begin{quantikz}
        &\gate{R_Z(-2\gamma_k h_i)}&
    \end{quantikz}
\end{equation}
On the other hand, the two qubits interactions are 2-local gates, which can be implemented via a rotation between two CNOT gates. Namely, the corresponding circuit will read
\begin{equation}
        \begin{quantikz}
        &\ctrl{1}&&\ctrl{1}&\\
        &\targ{}&\gate{R_Z(-2\gamma_k J_{ij})}&\targ{}&
    \end{quantikz}
\end{equation}
thus becoming very easy to be implemented.

\section{Variational Quantum Eigensolver (VQE)}

Similarly as the QAOA, the Variational Quantum Eigeinsolver (VQE) is an heuristic approach to solve combinatorial optimisation problem that exploits a combination of quantum computation and classical optimisation. In particular, the QAOA can be seen as a specific implementation of the VQE algorithm. 

The algorithm is designed to solve problems that can be stated as finding the ground state energy $E_0$ of $n$ qubit Hamiltonian. Namely, to find the configuration corresponding to the state $\ket{\Psi_0}$  that
\begin{equation}
    \hat H\ket{\Psi_0}=E_0\ket{\Psi_0}.
\end{equation}
The generality with respect to  QAOA comes in the form of the cost Hamiltonian $\hatHc$. Indeed, one assumes for it the most general form, which is
\begin{equation}
    \hatHc=\sum_\alpha h_\alpha\hat P_\alpha=\sum_\alpha h_\alpha\bigotimes_{j=1}^n\hat \sigma_{\alpha_j}^{(j)},
\end{equation}
where $h_\alpha$ are coefficients and the $\hat P_\alpha$ are called Pauli strings. The latter are product of $n$ single-qubit Pauli matrices (including the identity). Thus, compared to QAOA (which exploits the Ising model), this Hamiltonian is not limited to two qubit interactions only, but can consider $n$ qubit interactions. This is particularly relevant when considering more complex systems where the Ising model fails to describe the entire complexity of the problem.

Then, the steps of VQE are the following:
\begin{itemize}
    \item[1.] Map the problem in a cost Hamiltonian $\hatHc$ so that the solution is embedded in its ground state.
    \item[2.] Prepare the initial state as the ground state of $\hat H$.
    \item[3.] Generate the trial state $\ket{\Psi(\theta)}$, which is determined by a set of parameters $\theta$.
    \item[4.] Measure the expectation values of the Pauli strings in the Hamiltonian, i.e. $\braket{\Psi(\theta)|\hat P_\alpha|\Psi(\theta)}$. This is the end of the computation on the quantum computer
    \item[5.] Compute the corresponding energy, i.e. $E(\theta)=\sum_\alpha h_\alpha \braket{\Psi(\theta)|\hat P_\alpha|\Psi(\theta)}$
    \item[6.] Update or accept the values of $\theta$ based on the result.
    \item[7.] If updated, one goes back to point 2.
\end{itemize}

Notably, when searching for the ground state energy of the cost Hamiltonian, there are several pitfalls that the update step must deal with. For example, the parameter landscape may have local minima where one does not want to remain stacked. 
\chapter{Noisy Intermediate-Scale Quantum (NISQ) computation}

For quantum algorithms to function properly, we need to ensure that the basic units of quantum information, the qubits, are as reliable as the bits in classical computers. These qubits must be shielded from environmental noise that can disrupt their states, while still have to be controlled by external agents. This control involves making the qubits entangle and eventually measuring their states to extract the outputs of the quantum computation. Technically, it is feasible to minimise the impact of noise without compromising the quantum information process through the development of Quantum Error Correction (QEC) and Quantum Error Mitigation (QEM) protocols, see Chap.~\ref{ch.errorcorrection}.

Many quantum algorithms that come with guaranteed performance need millions of physical qubits to effectively use QEC methods. It could take decades to build fault-tolerant quantum computers capable of reaching this scale. Presently, quantum devices typically have around 100-1000 physical qubits, often referred to as noisy intermediate-scale quantum (NISQ)\index{Noisy intermediate-scale quantum} devices. These devices lack error correction and are imperfect, yet in the NISQ era, the aim is to maximize the quantum computational capabilities of current devices, while working on techniques for fault-tolerant quantum computation.\\

\begin{figure}
    \centering
    \includegraphics[width=0.48\linewidth]{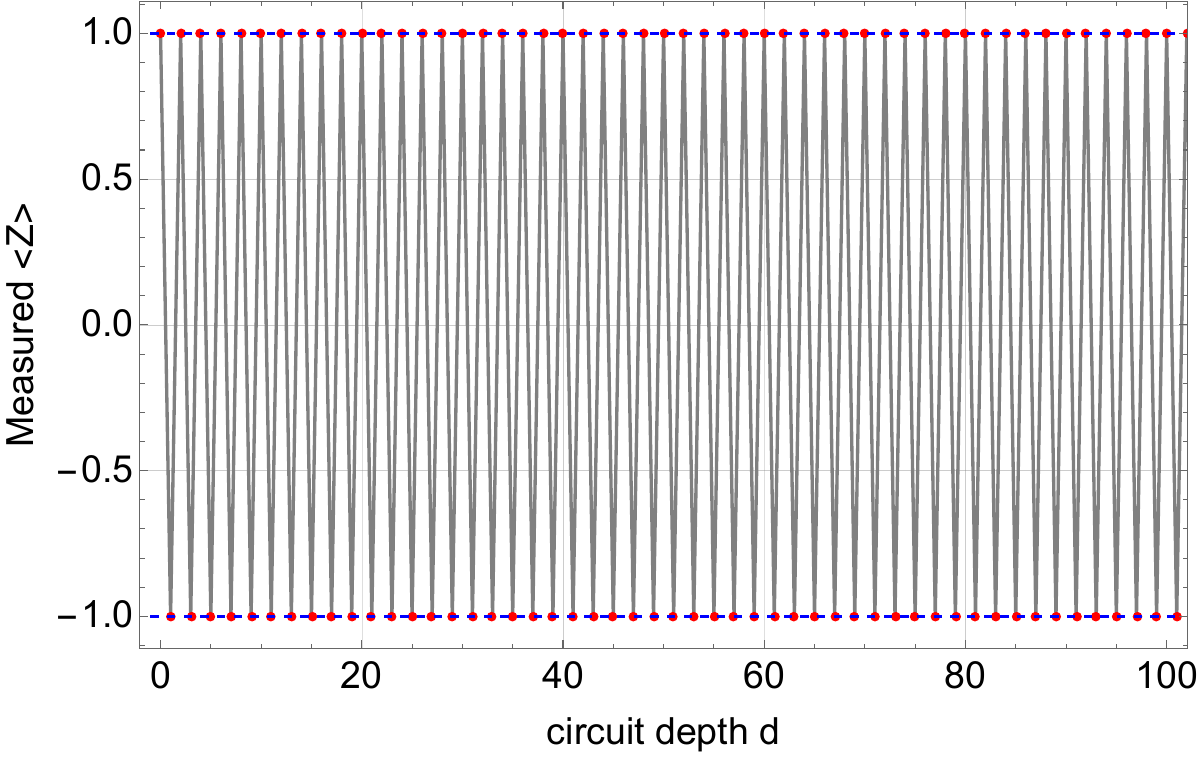}\hfill \includegraphics[width=0.48\linewidth]{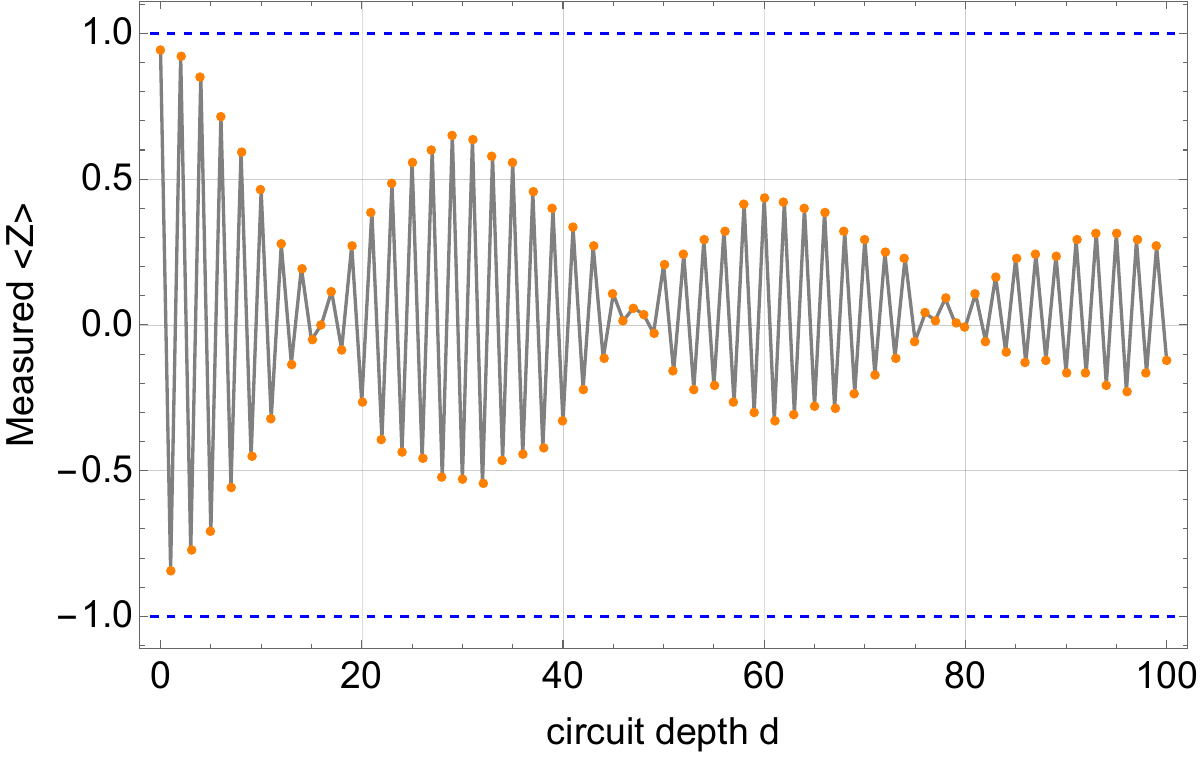}
    \caption{Expectation value of the polarisation $\braket{Z}$ for the circuit in Eq.~\eqref{eq.Z.noiseless} with respect to the depth $d$ of the circuit: (left panel) in the case of no noises; (right panel) when noises are accounted.}
    \label{fig.Z.noiseless}
\end{figure}

Here, we study how noises impact quantum circuits. The following circuit will be considered as a basis of the study:
\begin{equation}\label{eq.Z.noiseless}
    \begin{quantikz}
        \ket0&\gate{X^d}\slice{$\ket\psi$}&\meter{}
    \end{quantikz}
\end{equation}
where the gate $X$ is repeated $d$ times. The value of $d$ is also called the depth of the circuit. The state before the measurement, when no noise is considered, is given by
\begin{equation}
    \ket\psi=(i \hat R_x(\theta=\pi))^d\ket{0}=i^d\left[\cos(d \pi/2)\ket0-\sin(d \pi/2)\ket 1\right],
\end{equation}
indeed one has that the $X$ gate can be realised as a rotation of an angle $\pi$ around the $x$ axis: $\hat \sigma_x=i \hat R_x(\pi)$. Notably, the factor $i^d$ is just a negligible global phase. The 
expectation value of the polarisation is
\begin{equation}
    \braket{Z}=\braket{\psi|\hat\sigma_z|\psi}=\cos^2(d \pi/2)-\sin^2(d \pi/2)=\cos(d\pi),
\end{equation}
which is shown in the left panel of Fig.~\ref{fig.Z.noiseless}. Clearly, the value of $\braket{Z}$ jumps from $+1$ to $-1$ depending on the value of $d$.
However, when we perform such a simple experiment the result is quite different due to the noises and errors acting on the system. This is represented in  the right panel Fig.~\ref{fig.Z.noiseless}.
Such a result account for the presence of different noises and errors. These are listed and studied below.

\section{Miscalibrated gates}

\begin{figure}
    \centering
    \includegraphics[width=0.48\linewidth]{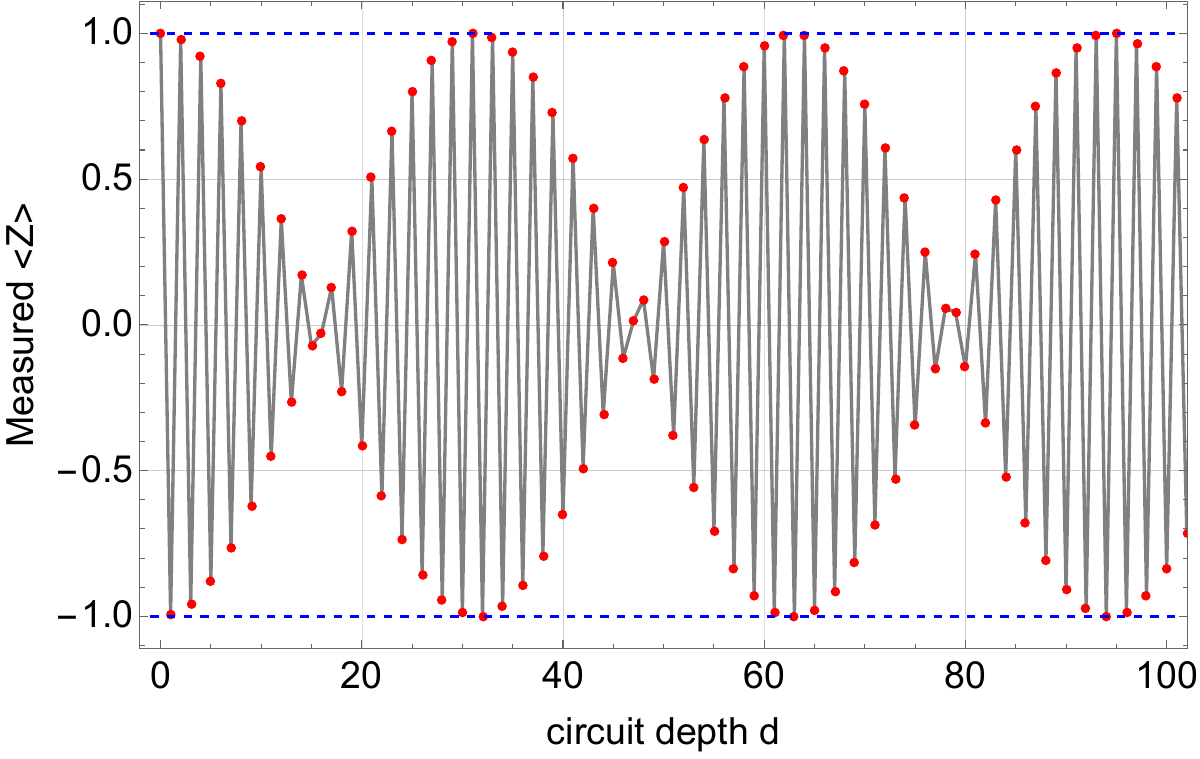}\hfill\includegraphics[width=0.475\linewidth]{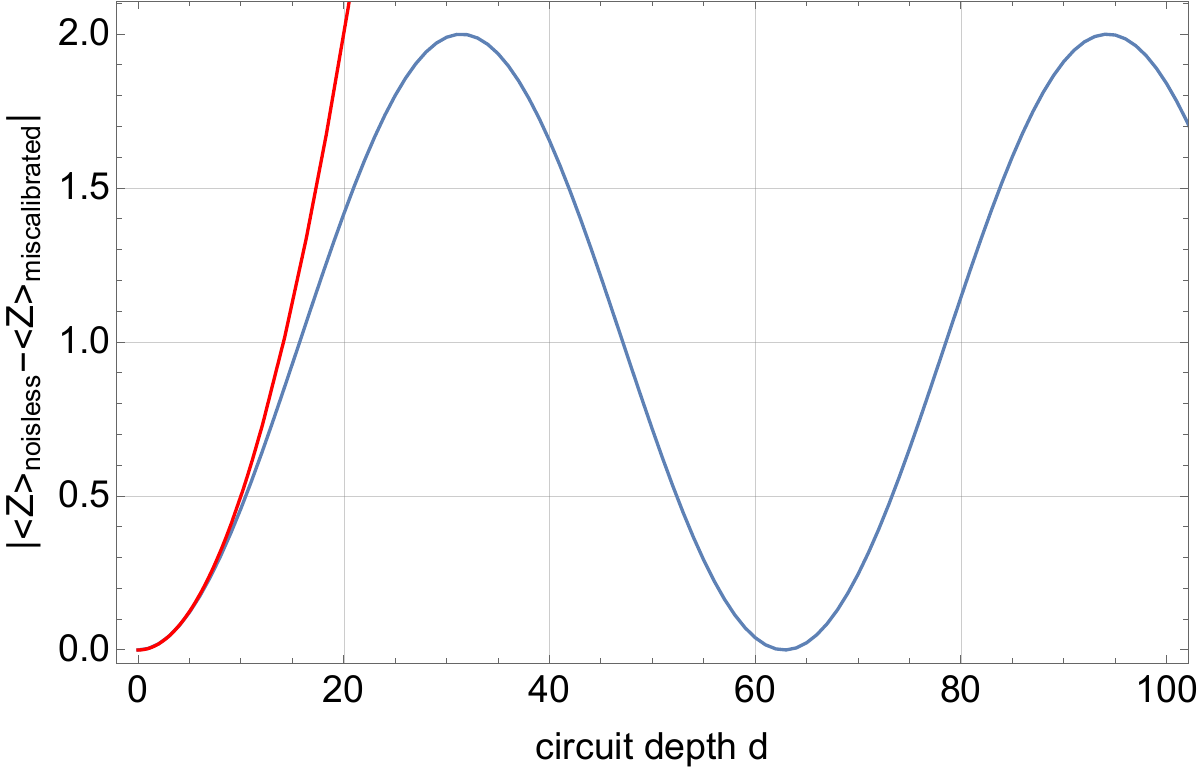}
    \caption{(Left panel) Expectation value (red dots) of the polarisation $\braket{Z}$ for the circuit in Eq.~\eqref{eq.Z.noiseless} with respect to the depth $d$ of the circuit when miscalibrated gates are considered. (Right panel) Difference with respect to the noiseless case (blue line) and the small $\epsilon$ and $d$ expansion (red line).  Here we considered $\epsilon=0.1$.}
    \label{fig.Z.miscalibrated}
\end{figure}

As we saw previously, the gate $X$ can be performed as a rotation of an angle $\pi$ around the $x$ axis. Now, let us suppose that the gate is systematic miscalibrated. Specifically, one performs a rotation of an angle of $\pi+\epsilon$ in place of only $\pi$. Then, we have that the actual gate $\tilde X$ is given by
\begin{equation}
    \begin{quantikz}
        &\gate{\tilde X}&\midstick[1,brackets=none]{=}&\gate{X}&\gate{iR_x(\epsilon)}&
   \end{quantikz}
\end{equation}
indeed one has that 
\begin{equation}
  R_x( \pi+\epsilon)= R_x( \pi) R_x( \epsilon).
\end{equation}
Then, when running the circuit with $d$ repetitions, one simply has
\begin{equation}\label{circ.miscalibrated}
    \begin{quantikz}
        \ket0&\gate{\tilde X^d}&\meter{}&\wireoverride{n} \midstick[1,brackets=none]{=} &\wireoverride{n}\ket0&\gate{X^d}&\gate{iR_x(d\epsilon)}\slice{$\ket\psi$}&\meter{}
   \end{quantikz}
\end{equation}
where
\begin{equation}
    \ket\psi=i^d\left[\cos\left(d \tfrac{(\pi+\epsilon)}{2}\right)\ket0-i\sin\left(d \tfrac{(\pi+\epsilon)}{2}\right)\ket 1\right].
\end{equation}
Correspondingly, one has that the expectation value for the polarisation is
\begin{equation}
    \braket{Z}=\cos^2\left(d \tfrac{(\pi+\epsilon)}{2}\right)-\sin^2\left(d \tfrac{(\pi+\epsilon)}{2}\right)=\cos(d(\pi+\epsilon)),
\end{equation}
which is shown in the left panel of Fig.~\ref{fig.Z.miscalibrated}. 
For small values of $\epsilon$ and $d$, one performs an error of
\begin{equation}
|    \braket{Z}_\text{noiseless}-\braket{Z}_\text{miscalibrated}|\sim \tfrac12 d^2\epsilon^2,
\end{equation}
which scales quadratically with the miscalibration $\epsilon$. As it is shown in the right panel of Fig.~\ref{fig.Z.miscalibrated}, the difference with respect to the noiseless result can be substantial. Indeed, for values of $d$ such that $d\epsilon\sim (2n+1) \pi$, with $n\in \mathbb N$, we have that
\begin{equation}
    \braket{Z}_\text{miscalibrated}=-\braket{Z}_\text{noiseless}.
\end{equation} 
This means that the error completely inverts the output signal.

\section{Projection noise and sampling error}

Consider the following trivial circuit
\begin{equation}\label{circ.sampling}
    \begin{quantikz}
        \ket \psi&\meter{}
    \end{quantikz}
\end{equation}
whose possible values of the polarisation are $z=+1$, meaning that the state after the collapse is $\ket0$, and $z=-1$, with corresponding state $\ket1$. 

\begin{figure}
    \centering
    \includegraphics[width=0.7\linewidth]{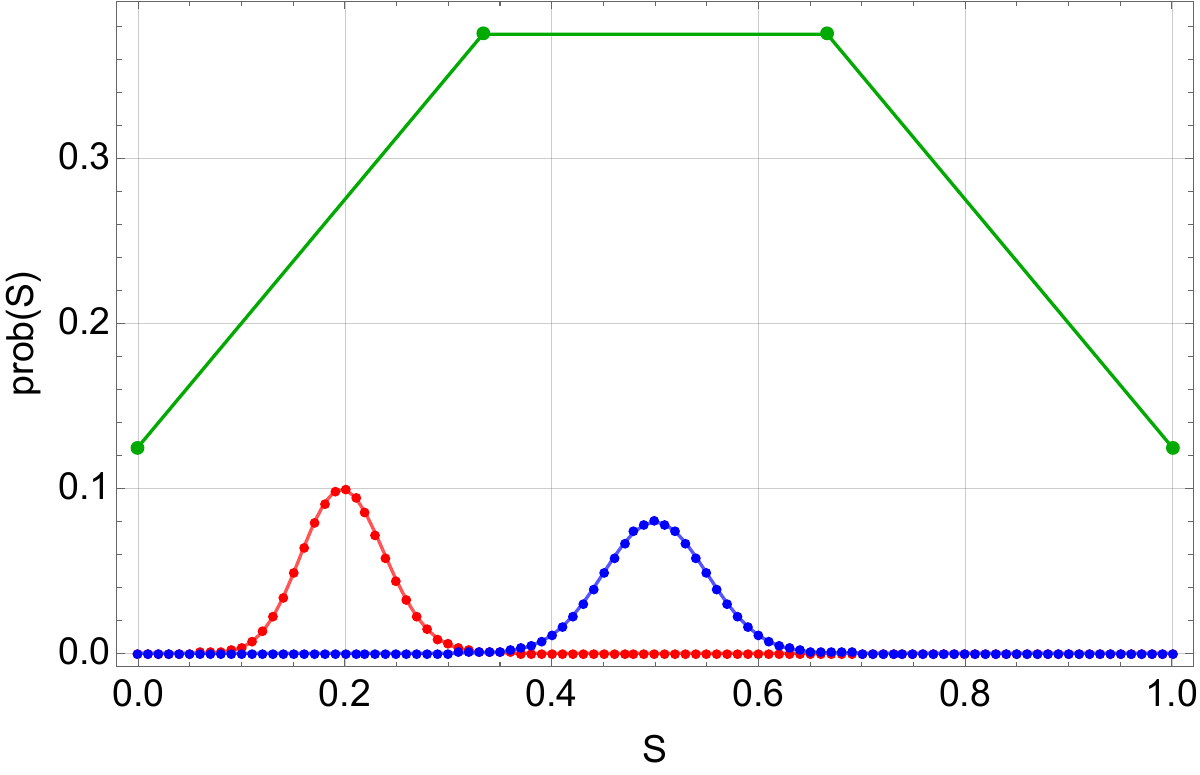}
    \caption{Binomial distribution followed by the sampling mean $S$ for $N=100$ samplings with $\mathbb p=0.2$ (red plot) and $\mathbb p=0.5$ (blue plot), and for $N=3$ with $\mathbb p=0.5$ (green plot). }
    \label{fig.binomialsampling}
\end{figure}

\begin{figure}
    \centering
    \includegraphics[width=0.48\linewidth]{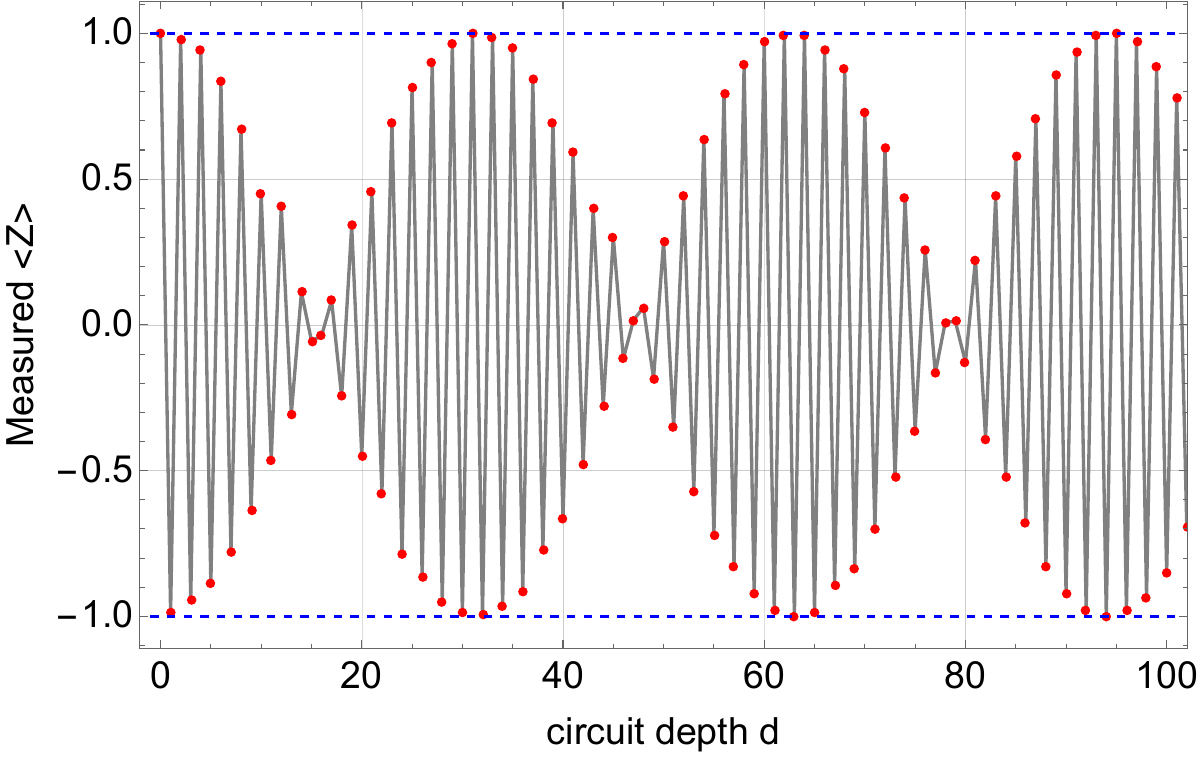}    \includegraphics[width=0.48\linewidth]{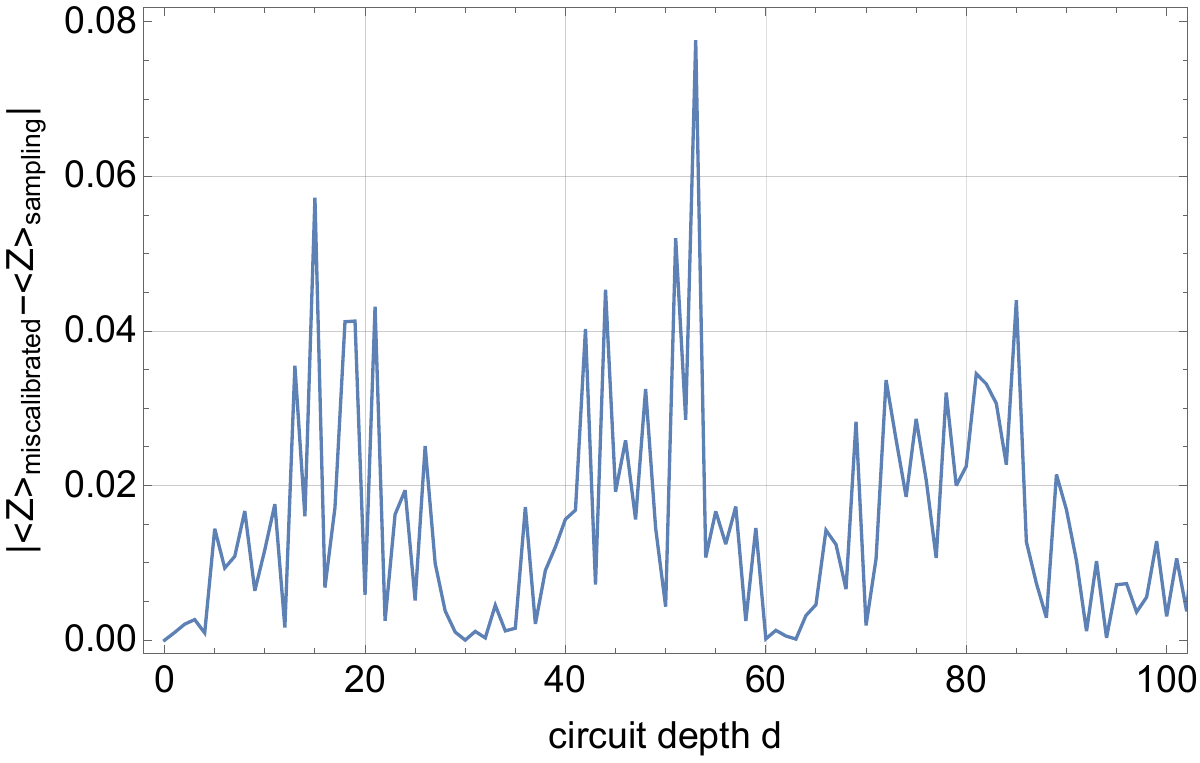}
    \caption{(Left panel) Expectation value (red dots) of the polarisation $\braket{Z}$ for the circuit in Eq.~\eqref{eq.Z.noiseless} with respect to the depth $d$ of the circuit when miscalibrated gates and error sampling are considered. (Right panel) Difference with respect to the miscalibrated case.  Here we considered $\epsilon=0.1$ and $N=10$.}
    \label{fig.Z.sampling}
\end{figure}

Let us define a measurement operator $\hat M$ that indicates if the state of the qubit is in the $\ket1$ state. Such an operator can be constructed as 
\begin{equation}
\begin{aligned}
        \hat M&=\ket1\bra1,\\
        &=1 \ket1\bra1+0\ket0\bra0,\\
        &=\sum_{m=0,1} m\hat\Pi_m,
\end{aligned}
\end{equation}
where $m$ are the eigeinvalues of $\hat M$ and $\hat \Pi_m=\ket m\bra m$, with $\sum_m\hat \Pi_m=\hat {\mathbb 1}$. Namely, the expectation value of $\hat M$ indicates the probability $\mathbb p(m=1)$ that the state $\ket \psi$ is equal to $\ket 1$:
\begin{equation}
    \braket{\psi|\hat M|\psi}=\braket{\psi|1}\braket{1|\psi}=|\braket{1|\psi}|^2=\mathbb p(m=1).
\end{equation}
Correspondingly, one has $\mathbb p(m=0)=1-\mathbb p(m=1)$. To ease the notation, in the following we use $\mathbb p=\mathbb p(m=1)$. The outcomes of the operator $\hat M$ are distributed via a binomial distribution $\mathcal B(\mathbb p)$. To be more explicit, let us consider the repetition of the above circuit $N$ times, meaning that we have $N$ samplings of the protocol. Then, 
we can construct a series of outputs: $\set{m_n}_{n=1}^N=\set{0,0,1,0,1,1,0,\dots}$, where each entry is a random variable. This corresponds in a series of states into which the system has collapsed after the measurement: $\set{\ket0,\ket0,\ket1,\ket0,\ket1,\ket1,\ket0,\dots}$. Suppose we have $N=3$, then we have a total of $2^N=2^3$ outputs $(m_1,m_2,m_3)$. Namely
\begin{equation}
        \begin{array}{c| c |c }
(m_1,m_2,m_3)&(\ket{m_1},\ket{m_2},\ket{m_3})&S\\
    \midrule
(0,0,0)&(\ket0,\ket0,\ket0)&0\\
(0,0,1)&(\ket0,\ket0,\ket1)&1/3\\
(0,1,0)&(\ket0,\ket1,\ket0)&1/3\\
(0,1,1)&(\ket0,\ket1,\ket1)&2/3\\
(1,0,0)&(\ket1,\ket0,\ket0)&1/3\\
(1,0,1)&(\ket1,\ket0,\ket1)&2/3\\
(1,1,0)&(\ket1,\ket1,\ket0)&2/3\\
(1,1,1)&(\ket1,\ket1,\ket1)&1
  \end{array}
\end{equation}
where we also computed the corresponding sampling mean $S$, which is defined as
\begin{equation}
    S=\frac{1}{N}\sum_{n=1}^Nm_n.
\end{equation}
The statement then is that the sampling mean $S$, which is also a random variable, is distributed as a binomial distribution $\mathcal B(\mathbb p)$, with mean $\mathbb E[S]=\mathbb p$ and variance $\mathbb V[S]=\mathbb p(1- \mathbb p)/N$. Fig.~\ref{fig.binomialsampling} shows some examples.
Now, by assuming that the state $\ket\psi$ in Eq.~\eqref{circ.sampling} is that one has before the measurement in Eq.~\eqref{circ.miscalibrated}, i.e.
\begin{equation}
        \ket\psi=i^d\left[\cos\left(d \tfrac{(\pi+\epsilon)}{2}\right)\ket0-i\sin\left(d \tfrac{(\pi+\epsilon)}{2}\right)\ket 1\right],
\end{equation}
then the corresponding distribution $\mathcal B(\mathbb p_d)$ depends on the probability of being in the state $\ket1$ after a depth $d$ of the circuit. This is given by
\begin{equation}\label{eq.sampling.pd}
\mathbb p_d=\sin^2\left(d\tfrac{(\pi+\epsilon)}{2}\right).
\end{equation}
The corresponding expectation value of the polarisation is shown in Fig.~\ref{fig.Z.sampling}, where we also report its difference with respect to the case shown in Fig.~\ref{fig.Z.miscalibrated}.

\section{Measurement error}

\begin{figure}
    \centering
    \includegraphics[width=0.4\linewidth]{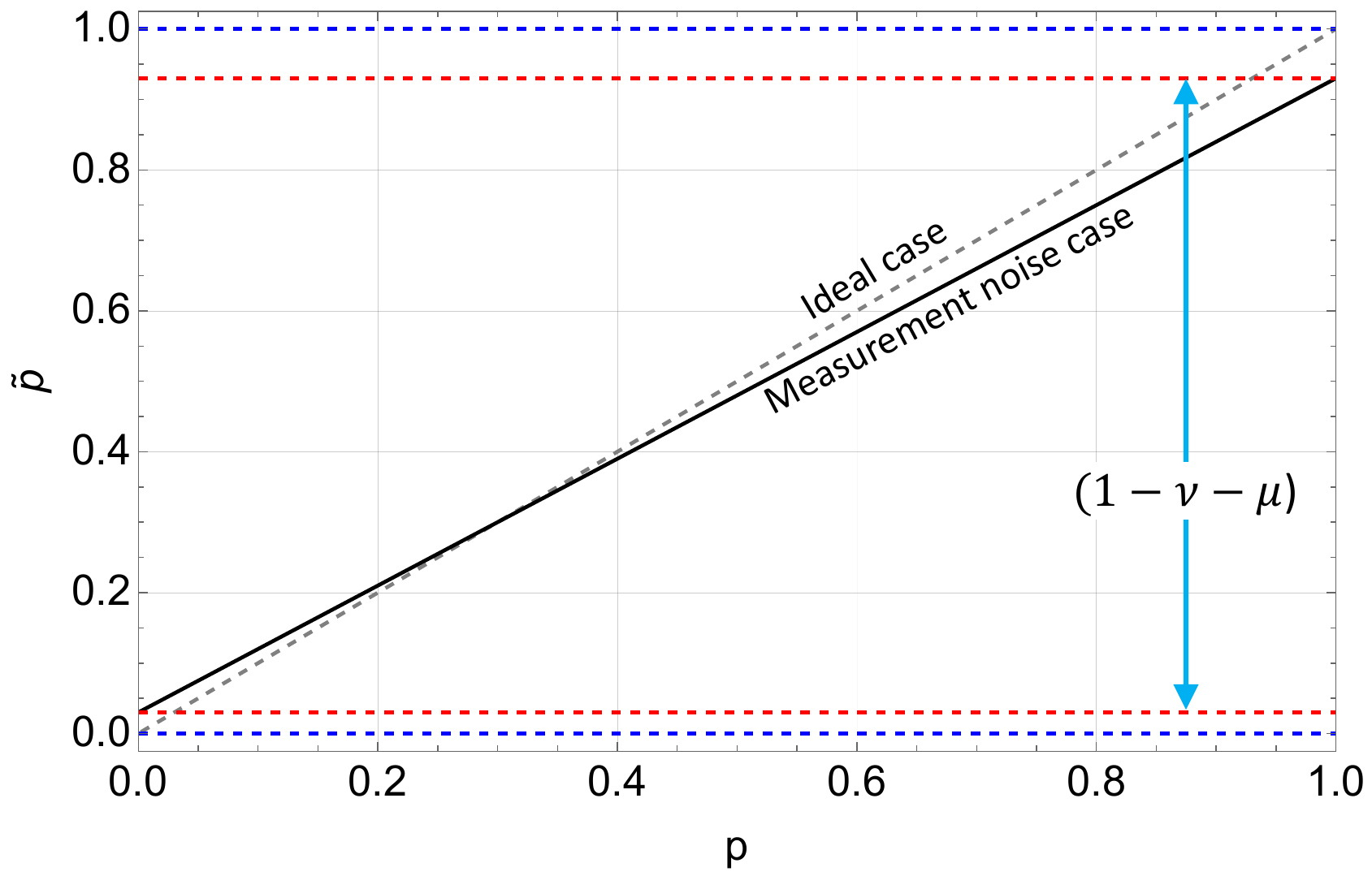} \\
    \vspace{0.1cm}
    \includegraphics[width=0.48\linewidth]{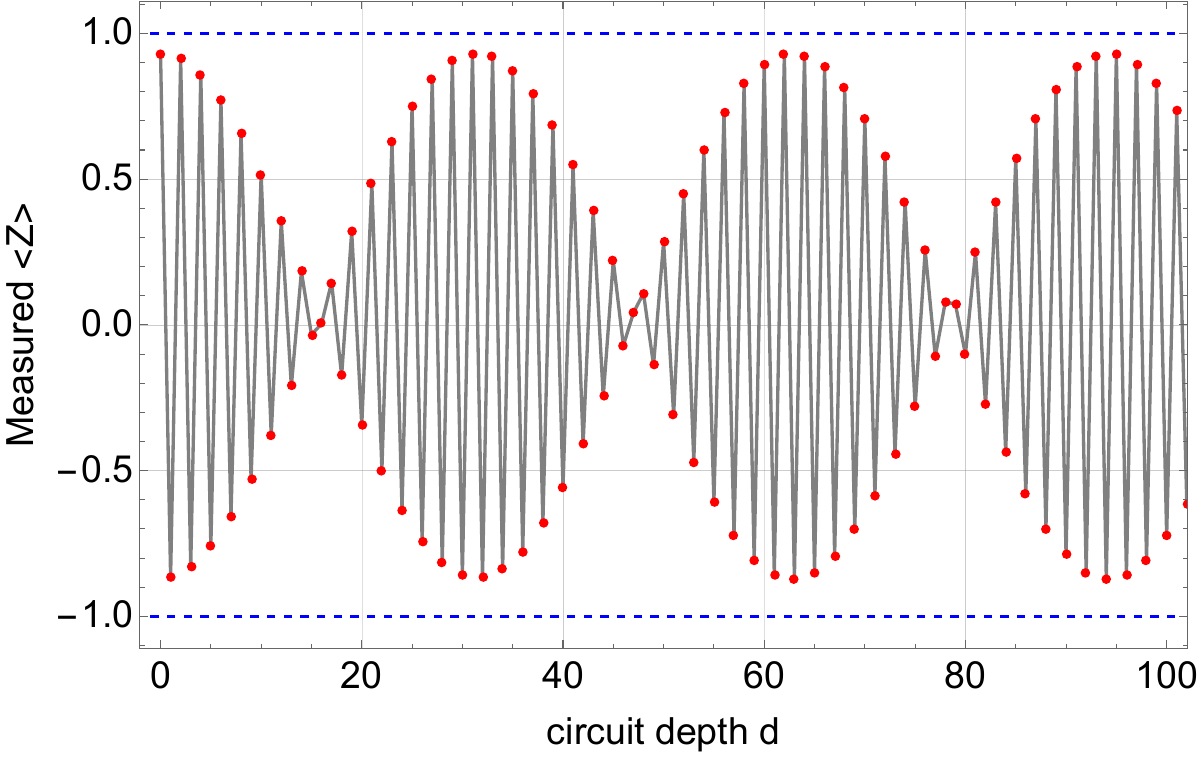}    \includegraphics[width=0.48\linewidth]{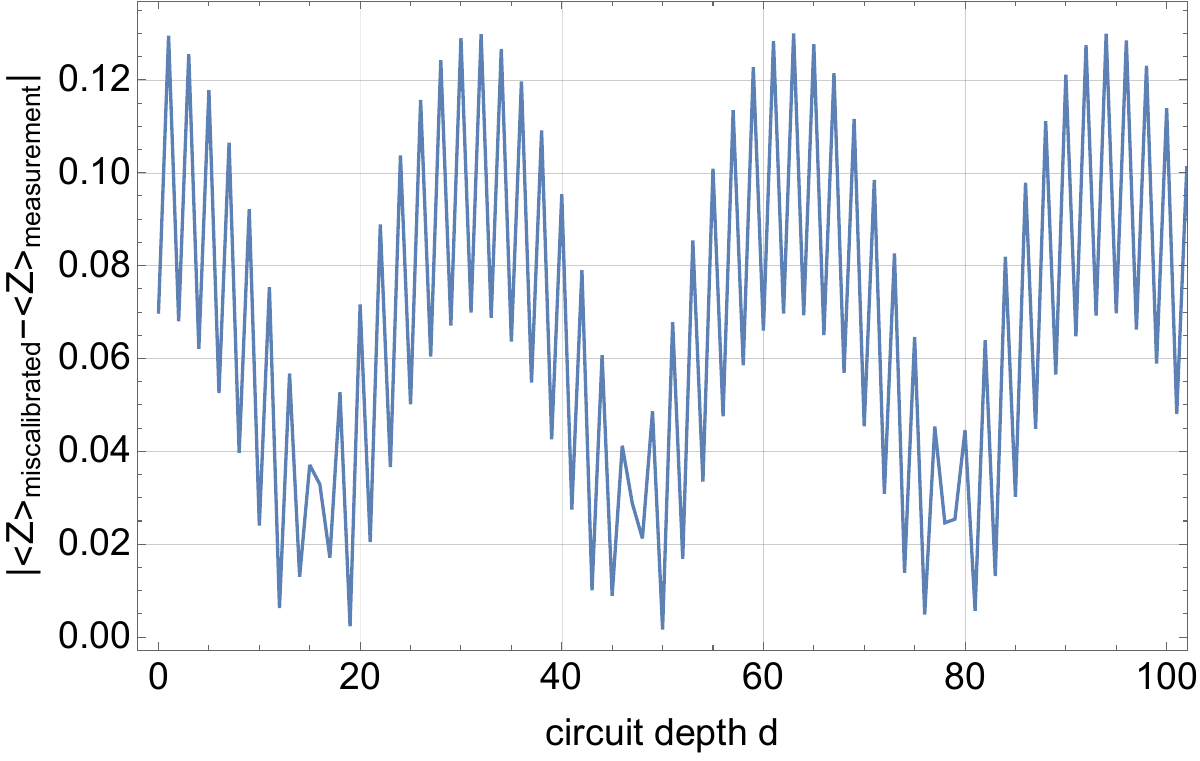}
    \caption{(Top panel) Representation of Eq.~\eqref{eq.PP}. (Bottom left panel) Expectation value (red dots) of the polarisation $\braket{Z}$ for the circuit in Eq.~\eqref{eq.Z.noiseless} with respect to the depth $d$ of the circuit when miscalibrated gates and measurement errors are considered. (Bottom right panel) Difference with respect to the miscalibrated case.  Here we considered $\epsilon=0.1$, $\nu=0.07$ and $\mu=0.03$.}
    \label{fig.PPtilde}
\end{figure}

Another source of errors is related to the performance of the measurement apparatus. To be explicit, consider the following trivial circuit
\begin{equation}
    \begin{quantikz}
        \ket0&\gate{U}\slice{$\ket\psi$}&\meter{}
    \end{quantikz}
\end{equation}
Suppose that is the  state $\ket\psi=\ket 0$ to be fed to the measurement apparatus. Then, if the apparatus is perfectly set, it will give as the polarisation value $z=+1$ corresponding to the output $m=0$. However, there might be errors and there could be a non-null probability $\mu$ that the measurement apparatus gives as an output $m=1$. Similarly, if $\ket\psi=\ket1$, one could have a probability $\nu$ to have $m=0$. The following scheme represents the supposed outcome and the actual outcome with the corresponding probabilities:
\begin{equation}\label{diag.supposed.actual}
\text{supposed $m$}=\ 
\begin{tikzcd}
{ 0 \ } &&&& \ 0 \\
	\\
	{ 1 \ } &&&&\  1
	\arrow["\mu"{description, pos=0.3}, shorten <=1pt, shorten >=1pt, from=1-1, to=3-5]
	\arrow["{(1-\nu)}"{description}, shorten <=11pt, shorten >=11pt, from=3-1, to=3-5]
	\arrow["\nu"{description, pos=0.3}, shorten <=1pt, shorten >=1pt, from=3-1, to=1-5]
	\arrow["{(1-\mu)}"{description}, shorten <=11pt, shorten >=11pt, from=1-1, to=1-5]
\end{tikzcd}
\ =\text{actual $\tilde m$.}
\end{equation}
We can define the probability vectors for the supposed and actual outcomes, which respectively are
\begin{equation}\label{eq.vectorPP}
    \mathbb P_M=
    \begin{pmatrix}
        1-\mathbb p\\
        \mathbb p
    \end{pmatrix}
    \quad\text{and}\quad
        \mathbb P_{\tilde M}=
    \begin{pmatrix}
        1-\tilde{\mathbb p}\\
        \tilde{\mathbb p}
    \end{pmatrix},
\end{equation}
where $\mathbb p=\mathbb p(m=1)$ and $\tilde{\mathbb p}=\tilde{\mathbb p}(\tilde m=1)$ are  the probability of having the outcome $m=1$ in the supposed and actual case respectively. Such vectors are related by the matrix $A$, which quantifies the performances of measurement apparatus: $\mathbb P_{\tilde M}=A\mathbb P_{M}$, where
\begin{equation}
    A=
    \begin{pmatrix}
        P(\tilde m=0|m=0)&P(\tilde m=0|m=1)\\
        P(\tilde m=1|m=0)&P(\tilde m=1|m=1)
    \end{pmatrix},
\end{equation}
where $P(\tilde m|m)$ is the conditional probability of having the actual outcome $\tilde m$ given the supposed outcome $m$. The diagram in Eq.~\eqref{diag.supposed.actual} sets the entries of $A$ to
\begin{equation}
    A=
    \begin{pmatrix}
        1-\mu&\nu\\
        \mu&1-\nu
    \end{pmatrix}.
\end{equation}
By merging the latter and Eq.~\eqref{eq.vectorPP}, we find
\begin{equation}\label{eq.PP}
    \tilde{\mathbb p}=\mathbb p+\mu-(\nu+\mu)\mathbb p,
\end{equation}
which is represented in Fig.\ref{fig.PPtilde}.
To recalibrate the measurement apparatus, i.e.~to quantify experimentally the values of $\mu$ and $\nu$ one would need to run the following two trivial circuits:
\begin{equation}
    \begin{quantikz}
        \ket0&\meter{}\\
       \ket1&\meter{} 
    \end{quantikz}
\end{equation}
They should  give 100\% of the time the outcomes $m=0$ and $m=1$ respectively. Variations of such percentage will characterise the values of $\mu$ and $\nu$.
The bottom panels in Fig.~\ref{fig.PPtilde} show the corresponding polarisation and difference with the miscalibrated case.

\subsection{Environmental noise}

\begin{figure}
    \centering
  \includegraphics[width=0.48\linewidth]{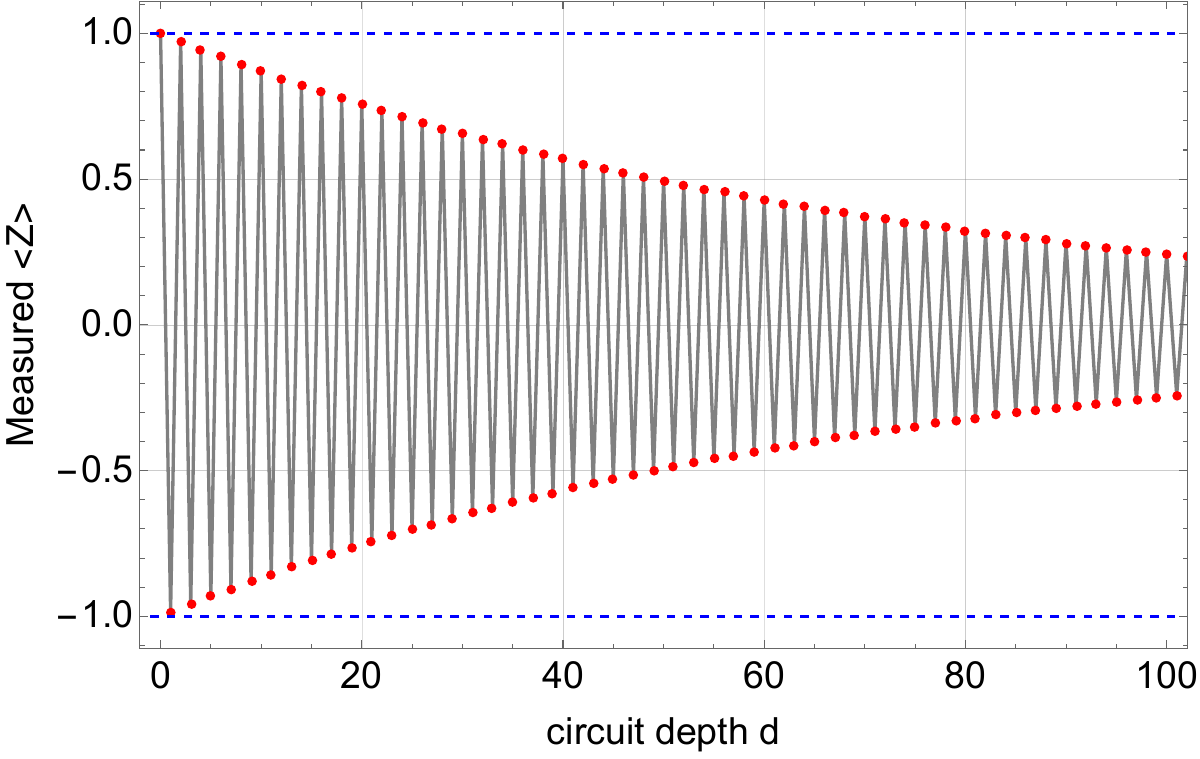}  \includegraphics[width=0.48\linewidth]{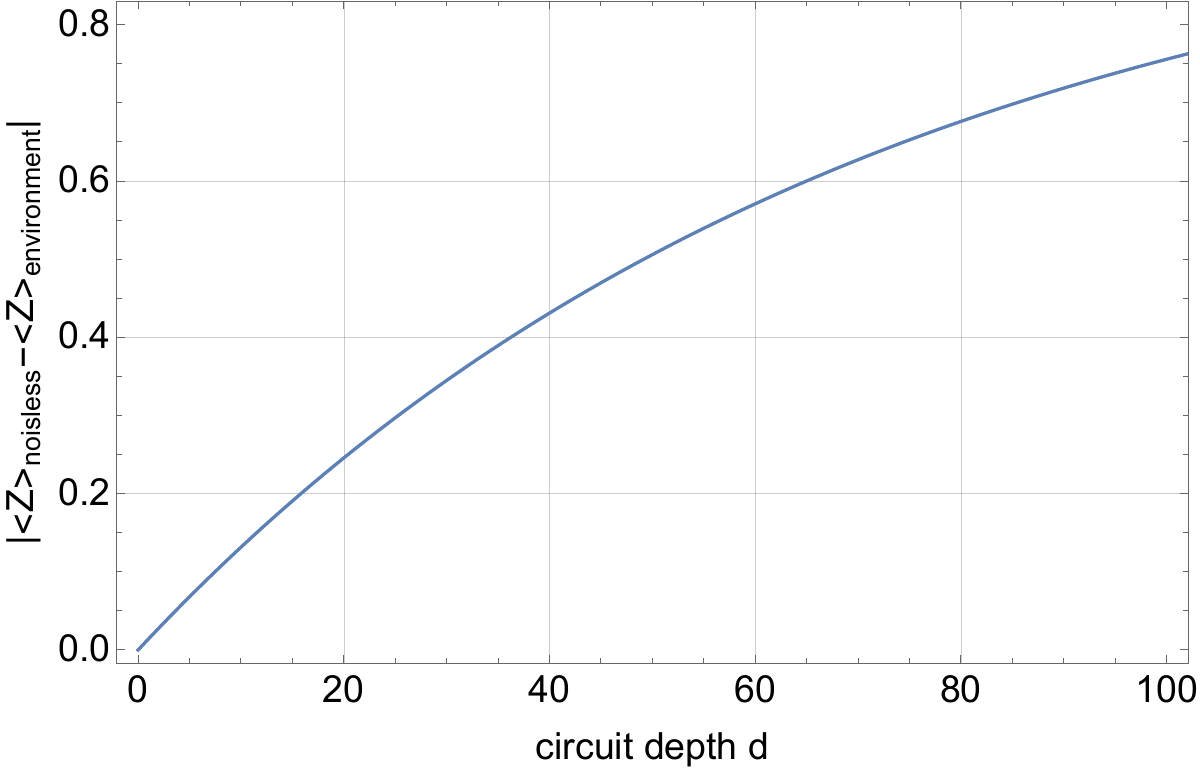}
    \caption{(Left panel) Expectation value (red dots) of the polarisation $\braket{Z}$ for the circuit in Eq.~\eqref{eq.Z.noiseless} with respect to the depth $d$ of the circuit when the environment noise considered. (Right panel) Difference with respect to the noiseless case.  Here we considered $\mathbb p_\text{E}=0.007$.}
    \label{fig.Z.environment}
\end{figure}

We now dwell in the most interesting source of noise, which is the one due to the external environment. In a circuit-based representation of the evolution, one now has
\begin{equation}
    \begin{quantikz}
        \ket0&\gate{\tilde G}&\meter{}
    \end{quantikz}
\end{equation}
where $\tilde G$ represents the noisy version of the gate $G$. There are different possible ways how one can account for the noise action. Here, we will consider the following way. Every gate $G$ of the noiseless case is substituted by $\tilde G$, where an extra gate, say $\mathcal E$, is added with a probability $\mathbb p_\text{E}$. Namely,
\begin{equation}
        \begin{array}{c|c|c }
\text{Case}&\text{Probability}&\text{Effective circuit}\\
    \hline
\text{A}&1-\mathbb p_\text{E}&\begin{quantikz}
    \ket0&\gate{\mathbb 1}\gategroup[1,steps=2,style={dashed,rounded
    corners,fill=blue!20, inner
    xsep=2pt},background,label style={label
    position=above,anchor=north,yshift=0.4cm}]{$\tilde G_\text{A}$}&\gate{G}&\meter{}
\end{quantikz}
\\
\midrule
\text{B}&\mathbb p_\text{E}&\begin{quantikz}
    \ket0&\gate{\mathcal E}\gategroup[1,steps=2,style={dashed,rounded
    corners,fill=blue!20, inner
    xsep=2pt},background,label style={label
    position=above,anchor=north,yshift=0.4cm}]{$\tilde G_\text{B}$}&\gate{G}&\meter{}
\end{quantikz}
  \end{array}
\end{equation}
For the sake of simplicity, we will consider the case of $\mathcal E=X$ and $G=X$. With this choice, in the case A, one has $\tilde G=\mathbb 1 X=X$, and the gate is properly implemented. While the case B, one has $\tilde G=XX=\mathbb 1$, which nullify the action of the original gate. Now, in the ideal case ($\mathcal E=\mathbb 1$ always), the state before the measurement is $\ket\psi=\ket1$, so the outcome is
\begin{equation}
    \braket{m}=\Tr{\hat M\hat \rho}=1,
\end{equation}
indeed, the probability of having $m=1$ is $\mathbb p(m=1)=1$. In the case with the environmental noise, one has
\begin{equation}
\begin{aligned}
    \braket{m}&=\sum_{m=0,1}m\, \mathbb p(m),\\
    &=(1-\mathbb p_\text{E})\time 1+\mathbb p_\text{E}\times 0,\\
    &=(1-\mathbb p_\text{E}).
\end{aligned}
\end{equation}
In particular, this result can be constructed as follows. Starting from the initial state $\ket0$, we construct the corresponding initial statistical operator $\hat\rho_0$. Then, with a probability $\mathbb p_\text{A}=(1-\mathbb p_\text{E})$ evolves as in the case A, and with a probability $\mathbb p_\text{B}=\mathbb p_\text{E}$ as in the case B. Namely, one has
\begin{equation}
    \hat\rho=\mathbb p_\text{A}\hat \rho_\text{A}+\mathbb p_\text{B}\hat \rho_\text{B},
\end{equation}
where
\begin{equation}
        \hat \rho_\text{B}=\hat \sigma_x\hat \rho_0\hat\sigma_x,\quad\text{and}\quad
        \hat \rho_\text{B}=\hat {\mathbb 1}\hat \rho_0\hat{\mathbb 1}.
\end{equation}
This gives
\begin{equation}
    \hat\rho=(1-\mathbb p_\text{E})\ket1\bra1+\mathbb p_\text{E}\ket0\bra0,
\end{equation}
which, in the computational basis, is represented as
\begin{equation}
 \rho=\begin{pmatrix}
     \mathbb p_\text{E}&0\\
     0&(1-\mathbb p_\text{E})
 \end{pmatrix}.
\end{equation}
The corresponding expectation value of the measurement operator $\hat M$ is
\begin{equation}
\begin{aligned}
    \braket{m}&=\Tr{\hat M\hat \rho},\\
    &=\Tr{
\begin{pmatrix}
    0&0\\
    0&1
\end{pmatrix} 
    \begin{pmatrix}
     \mathbb p_\text{E}&0\\
     0&(1-\mathbb p_\text{E})
 \end{pmatrix}},\\
 &=\Tr{
    \begin{pmatrix}
     0&0\\
     0&(1-\mathbb p_\text{E})
 \end{pmatrix}},\\
 &=(1-\mathbb p_\text{E}),
 \end{aligned}
\end{equation}
as expected. Conversely, the polarisation is
\begin{equation}
    \braket{Z}=(2\mathbb p_\text{E}-1).
\end{equation}
   Applying the noisy gate $d$ times, we find
\begin{equation}\label{eq.Z.environ}
    \braket{Z}=(2\mathbb p_\text{E}-1)^d,
\end{equation}
which is represented in Fig.~\ref{fig.Z.environment}.
\begin{myexercise}
    Verify the relation in Eq.~\eqref{eq.Z.environ}.
\end{myexercise}

\subsection{Global noise action}

The last step of this section is to put together the different noises and errors discussed here. Namely, when one has a circuit, to account for the error, such a circuit needs to be substituted as represented
\begin{equation}
    \begin{quantikz}
     \ket0&\gate{X}&\meter{}&\wireoverride{n} \midstick[1,brackets=none]{$\rightarrow$} &\wireoverride{n}\ket0 &\push{\Big(}\gategroup[1,steps=7,style={dashed,rounded
    corners,fill=blue!20, inner
    xsep=2pt},background,label style={label
    position=above,anchor=north,yshift=0.4cm}]{}&\gate{\mathcal E}&\gate{X}&\gate{iR_x(\epsilon)}&\push{\Big)^d}&\gate[style={rounded
    corners}]{\text{Sampling}}&\gate[style={rounded
    corners}]{\text{Measurement}}&\meter{}
    \end{quantikz}
\end{equation}
The action of the environment noise and miscalibration lead the qubit in the state 
\begin{equation}
    \hat \rho=(1-\mathbb p_\text{E})\ket{\psi_\text{A}}\bra{\psi_\text{A}}+\mathbb p_\text{E}\ket{\psi_\text{B}}\bra{\psi_\text{B}},
\end{equation}
where
\begin{equation}
\begin{aligned}
    \ket{\psi_\text{A}}&=\cos\left(d \tfrac{(\pi+\epsilon)}{2}\right)\ket0-i\sin\left(d \tfrac{(\pi+\epsilon)}{2}\right)\ket 1,\\
    \ket{\psi_\text{B}}&=\cos\left(d \tfrac{\epsilon}{2}\right)\ket0-i\sin\left(d \tfrac{\epsilon}{2}\right)\ket 1.
    \end{aligned}
\end{equation}
Specifically, 
such a state can be rewritten as
\begin{equation}
    \hat \rho=\left(\rho_{00}\ket0\bra0+\rho_{11}\ket1\bra1+\rho_{10}\ket1\bra0+\rho_{01}\ket0\bra1\right),
\end{equation}
where the $\rho_{00}$ and $\rho_{11}$ populations are fundamental in computing the expectation value of the polarisation:
\begin{equation}
\begin{aligned}
    \braket{Z}&=\rho_{00}-\rho_{11},\\
    &=(1-\mathbb p_\text{E})\cos\left(d (\pi+\epsilon)\right)+\mathbb p_\text{E} \cos\left(d \epsilon\right).
\end{aligned}
\end{equation}

On the other hand, to compute the effect of the sampling error, one needs only $ \rho_{11}$, which is equal to the probability of having $m=1$, i.e.~of being in the state $\ket1$. Thus, Eq.~\eqref{eq.sampling.pd} becomes 
\begin{equation}
    \mathbb p_d=(1-\mathbb p_\text{E})\frac{(1-\cos(d(\pi+\epsilon)))}{2}+\mathbb p_\text{E}\frac{(1-\cos(d\epsilon))}{2},
\end{equation}
with respect to which one constructs the statistics of the outcomes that are distributed following a binomial distribution $\mathcal B(\mathbb p_d)$. 

Finally, the measurement error can be accounted by mapping the sampling means $S_d$ in those accounting for diagram in Eq.~\eqref{diag.supposed.actual}. This consists in 
\begin{equation}
    S_d\rightarrow S_d+\mu-(\nu+\mu)S_d.
\end{equation}
The result of considering all the noises and errors is given by the right panel of Fig.~\ref{fig.Z.noiseless}.

\chapter{Quantum Error Correction}
\label{ch.errorcorrection}

\section{Quantum Error Correction}
\index{Quantum error correction}

Several techniques have been developed to correct errors due to the presence of noise. This is not only a considered issue in the quantum information context, but also in the classical one.
In both cases, the key ingredient is the redundancy.

\begin{figure}[h!]
    \centering
    \includegraphics[width=0.8\linewidth]{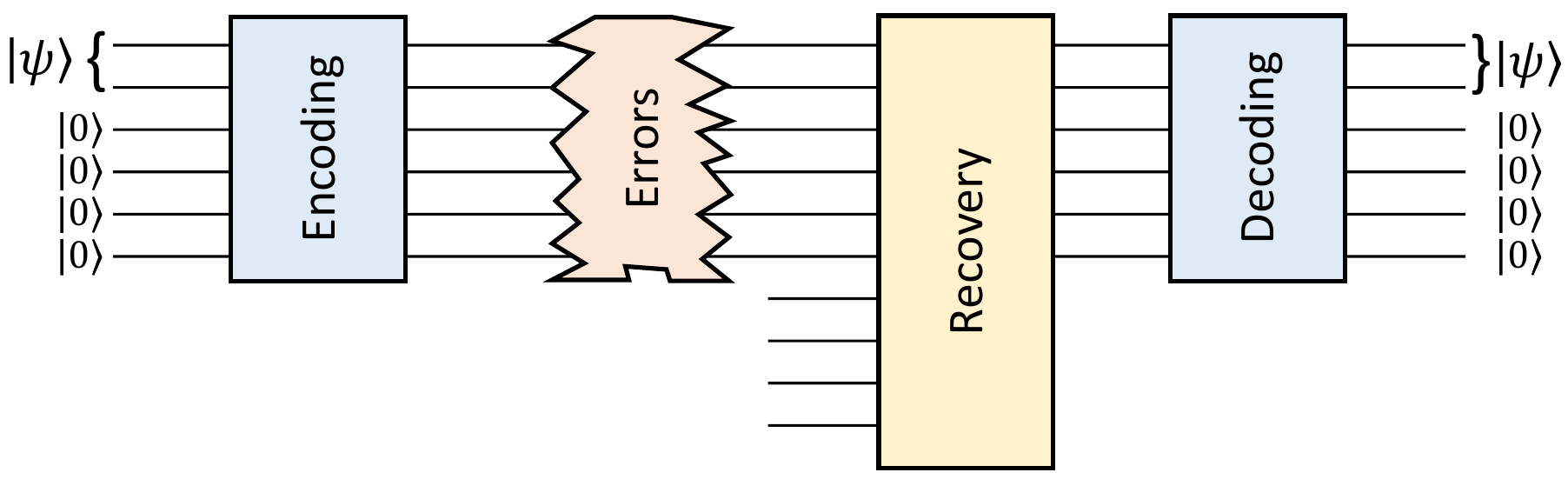}
    \caption{Schematic representation of the Quantum Error Correcting approach.}
    \label{fig:schema QEC}
\end{figure}

\subsection{Classical error correction}
\index{Classical error correction}

\begin{figure}
    \centering
    \includegraphics[width=0.5\linewidth]{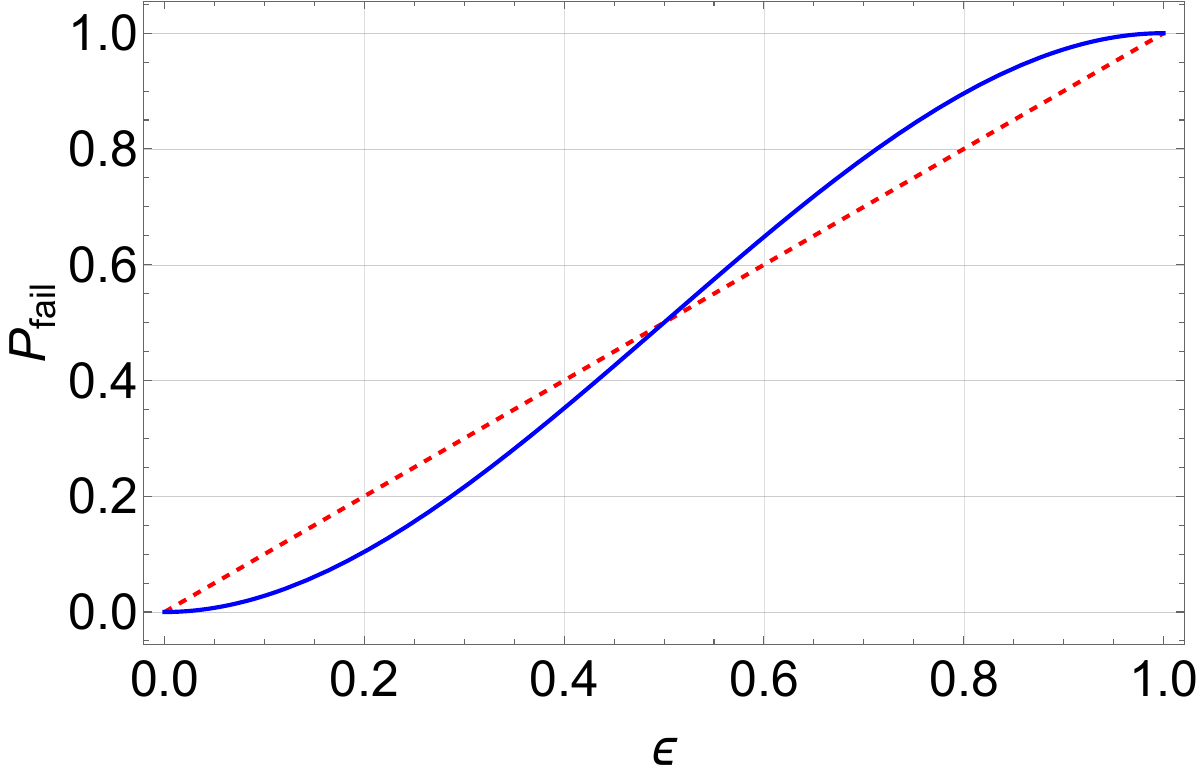}
    \caption{Probability of failing $P_\text{fail}$ for the single bit channel (dashed red line) and the three-bit classical correcting code (blue line).}
    \label{fig:pfailclassical}
\end{figure}

Consider the following example, that will make clear the usefulness of redundancies. We have Alice that wants to send a single bit information to Bob. The communication channel is not perfect: a bit-flip error noise can act on the bit with the following probabilities:
\begin{equation}\label{graph.bit.flip}
\text{Alice sends the bit}=\ 
\begin{tikzcd}
{ 0 \ } &&&& \ 0 \\
	\\
	{ 1 \ } &&&&\  1
	\arrow["\epsilon"{description, pos=0.3}, shorten <=1pt, shorten >=1pt, from=1-1, to=3-5]
	\arrow["{(1-\epsilon)}"{description}, shorten <=11pt, shorten >=11pt, from=3-1, to=3-5]
	\arrow["\epsilon"{description, pos=0.3}, shorten <=1pt, shorten >=1pt, from=3-1, to=1-5]
	\arrow["{(1-\epsilon)}"{description}, shorten <=11pt, shorten >=11pt, from=1-1, to=1-5]
\end{tikzcd}
\ =\text{is the bit the Bob receives.}
\end{equation}
With this scheme, the protocol has a probability of failing that is
\begin{equation}
    P_\text{fail}=\epsilon,
\end{equation}
which is equal to the probability $\epsilon$ of the bit being flipped. The idea of error correction is to suitably modify the protocol so one can recover the wanted information, with a failing probability being 
\begin{equation}\label{eq.pfail.epsilon}
       P_\text{fail}<\epsilon.
\end{equation}
Specifically, what Alice does is to send the bit string 000 in place of only the single bit 0. This operation is called encoding, and in this specific case, one encodes the classical information in the following way
\begin{equation}
\begin{aligned}
        0&\to000,\\
        1&\to111.
\end{aligned}
\end{equation}
Then, Bob receives three bits and needs to decode the information. This is performed via a majority voting. For example, let us assume the second bit is flipped while the other remain untouched: Bob receives 010, and  the majority voting gives
\begin{equation}
    010\to0.
\end{equation}
This is a decoding of the classical information. Considering all the possible three bits strings ($A,B,C$) that Bob can receive if Alice sends 000, with the corresponding probabilities $p(A,B,C)$, we have 
\begin{equation}
\text{Alice encodes 0 in 000}\to\text{Bob receives}\quad
        \begin{array}{c| c| c c  }
(A,B,C)&p(A,B,C)&\text{decoded bit}\\
 \cline{1-3}
(0,0,0)&(1-\epsilon)^3&0&\multirow{2}{*}{At most 1 error}\\
(0,0,1)&\epsilon(1-\epsilon)^2&0\\
(0,1,0)&\epsilon(1-\epsilon)^2&0&\multirow{2}{*}{Majority voting works}\\
(1,0,0)&\epsilon(1-\epsilon)^2&0\\
 \cline{1-3}
(0,1,1)&\epsilon^2(1-\epsilon)&1&\multirow{2}{*}{2 or more errors}\\
(1,0,1)&\epsilon^2(1-\epsilon)&1\\
(1,1,0)&\epsilon^2(1-\epsilon)&1&\multirow{2}{*}{Majority voting fails}\\
(1,1,1)&\epsilon^3&1
  \end{array}
\end{equation}
The probability that the protocol fails is given by the sum of the probabilities of the failing cases:
\begin{equation}
    P_\text{fail}=3\times\epsilon^2(1-\epsilon)+e^3=3\epsilon^2-2\epsilon^3,
\end{equation}
which is reported in Fig.~\ref{fig:pfailclassical}.

Specifically, for $0\leq\epsilon<1/2$, we have that Eq.~\eqref{eq.pfail.epsilon} is satisfied. This means that by using this protocol one has less probability to fail ($p_\text{fail}$) rather than using a single bit ($\epsilon$). Therefore, the redundancy is a good approach to reduce errors, as long as Eq.~\eqref{eq.pfail.epsilon} is satisfied.

\subsection{Quantum information context}

The direct application of redundancy in the quantum information context encounters some important, although not insurmountable, difficulties. 

\begin{itemize}
    \item[-] The no-cloning theorem (see Sec.~\ref{th:no-cloning}) does not allow to create copies of an unknown quantum state. This means that Alice cannot generate $\ket\psi\ket\psi\ket\psi$ to protect an unknown state $\ket\psi$.
    \item[-] The second difficulty comes in how the classical error correcting code operates: one measures the state of the bit and applies a correcting operation accordingly. In the quantum case, the measurement operation would destroy the coherence of the state and thus the information encoded in the state. To be more quantitative, take the generic state $\ket\psi=\alpha\ket0+\beta\ket1$. After the measurement, the state collapses in $\ket0$ or in $\ket 1$ with the respective probabilities. However, one cannot reconstruct the coherence of the state after its collapse.
    \item[-] In classical information, the only possible error is a bit-flip: $0\to1$ and $1\to0$. Conversely, in quantum information the class of possible noises is far wider. For example, the phase-flip maps $\alpha\ket0+\beta\ket1\to \alpha\ket0-\beta\ket1$, and this does not have a classical counterpart. Moreover, in the quantum context, one can also have infinitesimal errors that can accumulate as the depth of the algorithm increases. An example can be $\alpha\ket0+\beta\ket1\to \alpha\ket0+\hat R^x(\epsilon)\beta\ket1$, where $\hat R^x(\epsilon)$ is a rotation of an infinitesimal angle $\epsilon $ around the $x$ axis.

\end{itemize}

\subsection{The 3-qubit bit-flip code}\label{QEC.bifflip}\index{3-qubit bit-flip code}

The first Quantum Error Correction (QEC) code we see is that correcting bit-flip errors. This is the counterpart of that seen in the classical information context.

Suppose Alice wants to send the generic state $\ket\psi=\alpha\ket0+\beta\ket1$ to Bob via a bit-flip noisy channel. We assume that the noise acts independently on each of the qubits that Alice sends. This is an important assumption for the QEC codes we will see here. We assume that the noise leaves the qubit untouched with a probability $(1-\epsilon)$, while it applies $\hat \sigma_x$ with a probability $\epsilon$. Indeed, $\hat \sigma_x\ket0=\ket1$ and $\hat \sigma_x\ket1=\ket0$. This is essentially the quantum version of what shown in Eq.~\eqref{graph.bit.flip}.

To protect the information from bit-flip errors, Alice employs the following encoding:
\begin{equation}
\begin{aligned}
    \ket0&\to\ket{0_\text{L}}=\ket0\ket0\ket0,    \\
    \ket1&\to\ket{1_\text{L}}=\ket1\ket1\ket1 ,   
\end{aligned}
\end{equation}
where $\ket0$ and $\ket 1$ are physical qubits, while $\ket{0_\text{L}}$ and $\ket{1_\text{L}}$ are logical qubits. Then, the generic state $\ket\psi$ is encoded in
\begin{equation}
    \ket\psi=\alpha\ket0+\beta\ket1\to\alpha\ket{0_\text{L}}+\beta\ket{1_\text{L}}=\alpha\ket{000}+\beta\ket{111},
\end{equation}
with the notation $\ket{q_1q_2q_3}=\ket{q_1}\ket{q_2}\ket{q_3}$. This encoding can be implemented via the following circuit
\begin{equation}\label{circ.bit.flip.encoding}
    \begin{tikzcd}
        \ket\psi&\ctrl{1}&\ctrl{2}&\rstick[3]{$\ket{\Psi_1}$}\\
        \ket 0&\targ{}&&\\
        \ket 0&&\targ{}&
    \end{tikzcd}
\end{equation}
Indeed, we have
\begin{equation}
    \ket{\psi00}=\alpha\ket{000}+\beta\ket{100}\xrightarrow[]{CNOT\otimes\hat{\mathbb 1}}\alpha\ket{000}+\beta\ket{110}\xrightarrow[]{\hat{\mathbb 1}\otimes CNOT}\alpha\ket{000}+\beta\ket{111}=\ket{\Psi_1}.
\end{equation}
We underline that the entangled state $\ket{\Psi_1}$ is not equal to $\ket\psi\ket\psi\ket\psi$, so the no-cloning theorem is not violated.

Now, the state $\ket{\Psi_1}$ is sent to Bob via the noisy channel. Bob receives one of the following states $\ket{\Psi_2}$ with the respective probabilities $p(\ket{\Psi_2})$:
\begin{equation}
\text{Alice encodes $\ket\psi$ in $\ket{\Psi_1}$}\to\text{Bob receives}\quad
        \begin{array}{c |c c  }
\ket{\Psi_2}&p(\ket{\Psi_2})\\
 \cline{1-2}
\alpha\ket{000}+\beta\ket{111}&(1-\epsilon)^3&\multirow{4}{*}{At most 1 error}\\
\alpha\ket{001}+\beta\ket{110}&\epsilon(1-\epsilon)^2\\
\alpha\ket{010}+\beta\ket{101}&\epsilon(1-\epsilon)^2\\
\alpha\ket{100}+\beta\ket{011}&\epsilon(1-\epsilon)^2\\
 \cline{1-2}
\alpha\ket{011}+\beta\ket{100}&\epsilon^2(1-\epsilon)&\multirow{4}{*}{2 or more errors}\\
\alpha\ket{101}+\beta\ket{010}&\epsilon^2(1-\epsilon)\\
\alpha\ket{110}+\beta\ket{001}&\epsilon^2(1-\epsilon)\\
\alpha\ket{111}+\beta\ket{000}&\epsilon^3
  \end{array}
\end{equation}
Let us suppose that Bob receives the state $\ket{\Psi_2}=\alpha\ket{100}+\beta\ket{011}$. To correct the bit-flip, Bob would be tempted to perform a simultaneous measurement of the spin of the three qubits, i.e.~$\hat \sigma_z^{(1)}\hat \sigma_z^{(2)}\hat \sigma_z^{(3)}$. Such an operation would give as an outcome $100$ with probability $|\alpha|^2$ and $011$ with probability $|\beta|^2$. Then, by applying a majority voting, Bob would understand that the first qubit is flipped. However, the coherence of the state is lost. Indeed, the measurement of the spin of the qubits collapses the state. To solve the problem, one needs to perform the so-called error syndrome. In particular, Bob employs two ancillary qubits that are prepared in the $\ket0$ state and coupled to the qubits carrying the encoded state. The circuit implementing the correction then is
\begin{equation}
    \begin{tikzcd}
        \lstick[3]{$\ket{\Psi_2}$}  &\ctrl{3}   &           &\ctrl{4}   &           &                       &\gate[3]{\hat U}       \\     
                                    &           &\ctrl{2}   &           &           &                       &                       \\
                                    &           &           &           &\ctrl{2}   &                       &                       \\
        \ket0                       &\targ{}    &\targ{}    &           &           &\meter{}\wire[r][1]["x_0"{above,pos=0.5}]{c}&\wire[u][1]{c}      \\
        \ket0                       &           &           &\targ{}    &\targ{}    &\meter{}\wire[r][1]["x_1"{above,pos=0.5}]{c}&\wire[u][1]{c}      
    \end{tikzcd}
\end{equation}
To be quantitative, starting from $\ket{\Psi_2}=\alpha\ket{100}+\beta\ket{011}$, we have
\begin{equation}\begin{aligned}
    \ket{\Psi_2}\ket{00}&=\alpha\ket{10000}+\beta\ket{01100}\xrightarrow[]{CNOT_{1,4}}\alpha\ket{10010}+\beta\ket{01100},\\
    &\xrightarrow[]{CNOT_{2,4}}\alpha\ket{10010}+\beta\ket{01110},\\
    &\xrightarrow[]{CNOT_{1,5}}\alpha\ket{10011}+\beta\ket{01110},\\
    &\xrightarrow[]{CNOT_{3,4}}\alpha\ket{10010}+\beta\ket{01111},\\
    &=(\alpha\ket{100}+\beta\ket{011})\ket{11}.
\end{aligned}
\end{equation}
Fundamentally, the last two qubits are not entangled with the first three. Thus, the measurement on the last two qubits does not impose the collapse of the first three. After such a measurement, Bob has the outcomes $x_0=1$ and $x_1=1$. In particular, $x_0=1$ indicates that one among the first and the second qubit has flipped. Similarly, $x_1=1$ indicates that one among the first and the third qubit has flipped. Then, Bob knows, under the assumption of single qubit errors, that the first qubit has flipped and can apply $\hat U=\hat \sigma_x^{(1)}$ to flip it back. 

In general, Bob will apply the following unitary operations to correct the errors:
\begin{equation}
    \begin{array}{cc|c}
         x_0&x_1&\hat U  \\
         \midrule
         0&0&\hat{\mathbb 1}\\
         0&1&\hat \sigma_x^{(3)}\\
         1&0&\hat \sigma_x^{(2)}\\
         1&1&\hat \sigma_x^{(1)}\\         
    \end{array}
\end{equation}
After having applied the correction, Bob gets the state $\ket{\Psi_3}$ with the following probabilities
\begin{equation}
        \begin{array}{c | c| c  }
\ket{\Psi_2}&\ket{\Psi_3}&p(\ket{\Psi_3})\\
     \midrule
\alpha\ket{000}+\beta\ket{111}&\alpha\ket{000}+\beta\ket{111}&(1-\epsilon)^3\\
\alpha\ket{001}+\beta\ket{110}&\alpha\ket{000}+\beta\ket{111}&\epsilon(1-\epsilon)^2\\
\alpha\ket{010}+\beta\ket{101}&\alpha\ket{000}+\beta\ket{111}&\epsilon(1-\epsilon)^2\\
\alpha\ket{100}+\beta\ket{011}&\alpha\ket{000}+\beta\ket{111}&\epsilon(1-\epsilon)^2\\
     \midrule
\alpha\ket{011}+\beta\ket{100}&\alpha\ket{111}+\beta\ket{000}&\epsilon^2(1-\epsilon)\\
\alpha\ket{101}+\beta\ket{010}&\alpha\ket{111}+\beta\ket{000}&\epsilon^2(1-\epsilon)\\
\alpha\ket{110}+\beta\ket{001}&\alpha\ket{111}+\beta\ket{000}&\epsilon^2(1-\epsilon)\\
\alpha\ket{111}+\beta\ket{000}&\alpha\ket{111}+\beta\ket{000}&\epsilon^3
  \end{array}
\end{equation}
Finally, Bob applies the following decoding circuit
\begin{equation}
    \begin{tikzcd}
          \lstick[3]{$\ket{\Psi_2}$}  &\ctrl{2}   &\ctrl{1}   &\ket{\phi}\\ 
                               &           &\targ{}    &\ket0\\
                             &\targ{}      &           &\ket0
    \end{tikzcd}
\end{equation}
which is the inverse of the circuit in Eq.~\eqref{circ.bit.flip.encoding}. The final state $\ket\phi$ is
\begin{equation}\begin{aligned}
    \ket\phi&=\alpha\ket0+\beta\ket1=\ket\psi,\quad\quad\,\text{with probability}\quad p=(1-\epsilon)^3+3\epsilon(1-\epsilon)^2,\\
    \ket\phi&=\alpha\ket1+\beta\ket0=\hat\sigma_x\ket\psi,\quad\text{with probability}\quad p=3\epsilon^2(1-\epsilon)+\epsilon^3.
\end{aligned}\end{equation}
Thus, the failing probability of this QEC code is
\begin{equation}
    P_\text{fail}=3\epsilon^2-2\epsilon^3,
\end{equation}
which is the same as in the classical correcting algorithm seen previously and it is plotted in Fig.~\ref{fig:pfailclassical}.
Notably, Bob does not learn anything about the weights $\alpha$ and $\beta$ thought this QEC code. 
The coherence of the state remains intact.

\subsection{The 3-qubit phase-flip code}\index{3-qubit phase-flip code}

Let us consider the case of the phase-flip noise, where the following error generates with a probability $\epsilon$:
\begin{equation}
    \begin{aligned}
        \ket0\to&\ \hat\sigma_z\ket0=\ket0,\\
        \ket1\to&\ \hat\sigma_z\ket1=-\ket1.
    \end{aligned}
\end{equation}
Consequently, one has
\begin{equation}
    \ket\psi=\alpha\ket0+\beta\ket1\to\alpha\ket0-\beta\ket1.
\end{equation}
Notably, this noise does not have a classical counterpart. Since the error is imprinted in the relative phase between $\ket0$ and $\ket1$, the bit-flip QEC code developed in Sec.~\ref{QEC.bifflip} does not correct this type of errors. However, a phase-flip error in the computational basis $\set{\ket0,\ket1}$ corresponds to a bit-flip error in the $\set{\ket+,\ket-}$ basis. Indeed,
\begin{equation}
    \begin{aligned}
      &\hat\sigma_z\ket+=\ket-,\\
      &\hat\sigma_z\ket-=\ket+.
    \end{aligned}
\end{equation}
Then, by simply adding a Hadamard gate, one changes basis and thus is able to employ the bit-flip QEC code to correct phase-flip errors. This is done both in the encoding and the decoding parts of the code. The encoding circuit is then 
\begin{equation}
    \begin{tikzcd}
        \ket\psi&\ctrl{1}&\ctrl{2}&\gate{H}&\rstick[3]{$\ket{\Psi_1}$}\\
        \ket 0&\targ{}&&\gate{H}&\\
        \ket 0&&\targ{}&\gate{H}&
    \end{tikzcd}
\end{equation}
while the deconding becomes
\begin{equation}
    \begin{tikzcd}
          \lstick[3]{$\ket{\Psi_2}$}   &\gate{H} &\ctrl{2}   &\ctrl{1}   &\ket{\phi}\\ 
                            &\gate{H}   &           &\targ{}    &\ket0\\
                        &\gate{H}     &\targ{}      &           &\ket0
    \end{tikzcd}
\end{equation}
The remaining parts of the code remain identical.

\subsection{The 9-qubit Shor code}\index{9-qubit Shor code}

We saw how the 3-qubit bit-flip and phase flip QEC codes can correct respectively bit-flip and phase flip errors. Here, we show that concatenating these two codes, one can protects for generic single qubit errors. Indeed, consider the situation of a single qubit initially prepared in the state $\ket\psi$. Suppose it is coupled to the surrounding enviroment, whose state is initially $\ket{e}$, and that the latter entangles with the system. Such a transformation is described as
\begin{equation}
\ket\psi\ket e\to c_0 \hat{\mathbb 1}\ket\psi\ket{e_0}+c_1\hat \sigma_x\ket\psi\ket{e_1}+c_2\hat \sigma_y\ket\psi\ket{e_2}+c_3\hat \sigma_z\ket\psi\ket{e_3},
\end{equation}
where $c_i$ are suitable constants, and $\ket{e_i}$ are states of the environment. Then, the state of the system is transformed via the application of the four Pauli operators. Here, $\hat\sigma_0=\hat {\mathbb 1}$ does not imply any change in the state, so no error needs to be corrected. The errors due to $\hat \sigma_x$ and $\hat \sigma_z$ are respectively corrected via bit-flip and phase-flip QEC codes. It remains that due to $\hat \sigma_y$. However, one can notice that, since the Pauli matrices form a Lie algebra, one can express $\hat \sigma_y$ in terms of $\hat \sigma_x$ and $\hat \sigma_z$. Namely, $\hat \sigma_y=i\hat \sigma_x\hat \sigma_z$. Then, one needs only to correct two consecutive errors (phase-flip and then bit-flip) to correct a bit-phase flip. The following QEC code is sufficient to perform such a correction.

The encoding of the 9-qubit Shore code is given by
\begin{equation}
\begin{aligned}
\ket{0}&\to\ket{0_\text{L}}=\frac{1}{\sqrt{8}}\left(\ket{000}+\ket{111}\right)\left(\ket{000}+\ket{111}\right)\left(\ket{000}+\ket{111}\right),\\
\ket{1}&\to\ket{1_\text{L}}=\frac{1}{\sqrt{8}}\left(\ket{000}-\ket{111}\right)\left(\ket{000}-\ket{111}\right)\left(\ket{000}-\ket{111}\right).
\end{aligned}
\end{equation}
This implies the following encoding for a generic state $\ket\psi$
\begin{equation}\label{eq.9qubit.encoding}
\ket\psi\to\frac{\alpha}{\sqrt{8}}\left(\ket{000}+\ket{111}\right)\left(\ket{000}+\ket{111}\right)\left(\ket{000}+\ket{111}\right)+\frac{\beta}{\sqrt{8}}\left(\ket{000}-\ket{111}\right)\left(\ket{000}-\ket{111}\right)\left(\ket{000}-\ket{111}\right).
\end{equation}
The encoding is implemented via the following circuit
\begin{equation}\label{circ.9qubit.encoding}
\begin{tikzcd}
\ket\psi	&\ctrl{3}	&\ctrl{6}	&\gate{H}	&\ctrl{1}	&\ctrl{2}	&\\
\ket0	&		&		&			&\targ{}	&		&\\
\ket0	&		&		&			&		&\targ{}	&\\
\ket0	&\targ{}	&		&\gate{H}	&\ctrl{1}	&\ctrl{2}	&\\
\ket0	&		&		&			&\targ{}	&		&\\
\ket0	&		&		&			&		&\targ{}	&\\
\ket0	&		&\targ{}	&\gate{H}	&\ctrl{1}	&\ctrl{2}	&\\
\ket0	&		&		&			&\targ{}	&		&\\
\ket0	&		&		&			&		&\targ{}	&
\end{tikzcd}
\end{equation}

The action of the first two CNOT gates and three Hadamard in Eq.~\eqref{circ.9qubit.encoding} is to map the qubits 1, 4 and 7 as follows:
\begin{equation}
\begin{aligned}
\ket{\psi 00}&=\alpha\ket{000}+\beta\ket{100},\\
&\to\alpha\ket{000}+\beta\ket{110},\\
&\to\alpha\ket{000}+\beta\ket{111},\\
&\to\alpha\ket{+++}+\beta\ket{- - -}.
\end{aligned}
\end{equation} 
Namely, they perform the encoding for the phase-flip QEC code:
\begin{equation}
\begin{aligned}
\ket0&\to\ket{+++},\\
\ket1&\to\ket{- - -}.
\end{aligned}
\end{equation}
Then, every $\ket{+}$ and $\ket{-}$ state in these qubits is further encoded with the last CNOT gates. Specifically, one has
\begin{equation}
\begin{aligned}
\ket{+00}&=\frac{1}{\sqrt{2}}\left(\ket{000}+\ket{100}	\right),\\
&\to\frac{1}{\sqrt{2}}\left(\ket{000}+\ket{110}	\right),\\
&\to\frac{1}{\sqrt{2}}\left(\ket{000}+\ket{111}	\right),\\
\end{aligned}
\end{equation}
and 
\begin{equation}
\begin{aligned}
\ket{-00}&=\frac{1}{\sqrt{2}}\left(\ket{000}-\ket{100}	\right),\\
&\to\frac{1}{\sqrt{2}}\left(\ket{000}-\ket{110}	\right),\\
&\to\frac{1}{\sqrt{2}}\left(\ket{000}-\ket{111}	\right).
\end{aligned}
\end{equation}
These, effectively perform the encoding for the bit-flip QEC code.
Such an encoding combines the phase-flip and the bit-flip encoding.

To extract the error syndrome, one employs a collective measurement, similarly as for the bit-flip. In particular, 8 ancillary qubits are employed to construct the following circuit
\begin{equation}
\begin{tikzcd}
&\ctrl{3}&&\ctrl{4}&&\gate{H}&\ctrl{1}&\ctrl{1}&\gate{H}&\\
&&\ctrl{2}&&&\gate{H}&\ctrl{1}&\ctrl{1}&\gate{H}&\\
&&&&\ctrl{2}&\gate{H}&\ctrl{3}&\ctrl{8}&\gate{H}&\\
\ket0&\targ{}&\targ{}&&&&&&&\meter{d_0}\\
\ket0&&&\targ{}&\targ{}&&&&&\meter{d_1}\\
&\ctrl{3}&&\ctrl{4}&&\gate{H}&\ctrl{1}&&\gate{H}&\\
&&\ctrl{2}&&&\gate{H}&\ctrl{1}&&\gate{H}&\\
&&&&\ctrl{2}&\gate{H}&\ctrl{8}&&\gate{H}&\\
\ket0&\targ{}&\targ{}&&&&&&&\meter{d_2}\\
\ket0&&&\targ{}&\targ{}&&&&&\meter{d_3}\\
&\ctrl{3}&&\ctrl{4}&&\gate{H}&&\ctrl{1}&\gate{H}&\\
&&\ctrl{2}&&&\gate{H}&&\ctrl{1}&\gate{H}&\\
&&&&\ctrl{2}&\gate{H}&&\ctrl{4}&\gate{H}&\\
\ket0&\targ{}&\targ{}&&&&&&&\meter{d_4}\\
\ket0&&&\targ{}&\targ{}&&&&&\meter{d_5}\\
\ket0&&&&&&\targ{}&&&\meter{d_6}\\
\ket0&&&&&&&\targ{}&&\meter{d_7}
\end{tikzcd}
\end{equation}
Here, the outcomes $(d_0,d_1)$, $(d_2,d_3)$ and $(d_4,d_5)$ respectively indicate bit-flip errors within the first, second and third block of three physical qubits. Specifically, for the first block, one employs exactly what described in Sec.~\ref{QEC.bifflip}. 

The outcomes $(d_6,d_7)$ are instead used to detect phase-flip errors of the logical state encoded with the three blocks. The collective measurements to do this are
\begin{equation}
    \begin{aligned}
        \hat \sigma_x^{(1)}\hat \sigma_x^{(2)}\hat \sigma_x^{(3)}\hat \sigma_x^{(4)}\hat \sigma_x^{(5)}\hat \sigma_x^{(6)},\\
        \hat \sigma_x^{(1)}\hat \sigma_x^{(2)}\hat \sigma_x^{(3)}\hat \sigma_x^{(7)}\hat \sigma_x^{(8)}\hat \sigma_x^{(9)},
    \end{aligned}
\end{equation}
which provide $d_6$ and $d_7$ respectively. If one gets, for example, $(d_6=-1,d_7=-1)$, then a phase flip occurred in the first block.

\subsection{On the redundancy and threshold}

As we saw, a fundamental step in the QEC codes is the redundancy of the state. Notably, there is no need in having exactly 3 copies. It can be extended to any $k$ copies, as long as $k>1$ is an odd number. What one wants is that the probability $P_\text{fail}$ that the QEC code fails  is smaller than the probability $\epsilon$ of an error occurring on a single physical qubit: $P_\text{fail}<\epsilon$.

\begin{figure}
    \centering
    \includegraphics[width=0.5\linewidth]{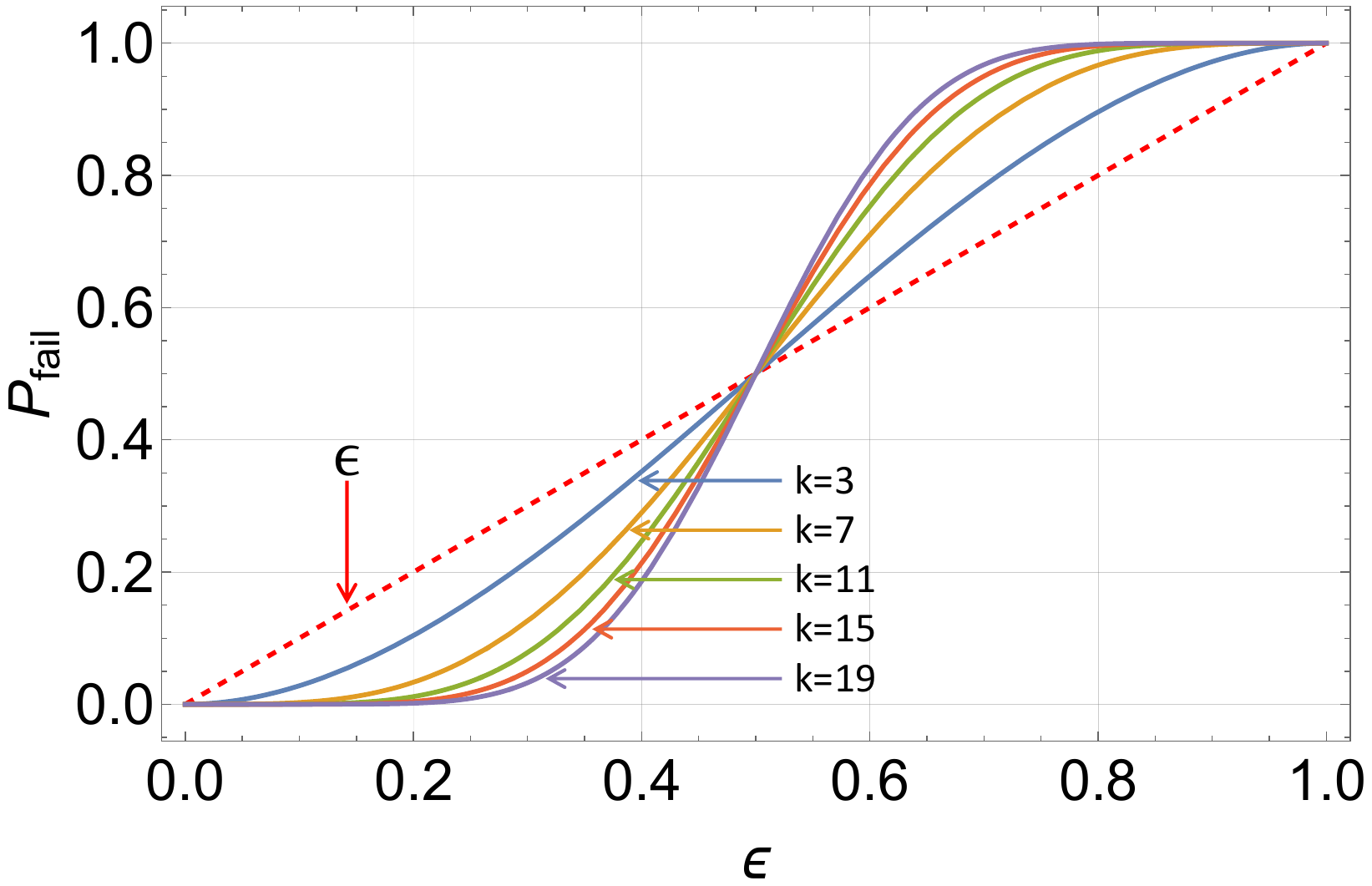}
    \caption{Probability of failing $P_\text{fail}$ for the single bit channel (dashed red line) with respect to a redundant encoding with $k$ physical qubits (continuous lines).}
    \label{fig.threshold}
\end{figure}

Consider the case of $k$ physical qubits encoding a single logical qubit. Given the probability $\epsilon$ of having an error on one of these qubits, that of having $j$ qubits with errors is given by
\begin{equation}
    p(j)=\epsilon^j(1-\epsilon)^{k-j},
\end{equation}
and there are 
\begin{equation}
\binom{k}{j}=\frac{k!}{(k-j)!j!},
\end{equation}
different possible combinations. Then, $P_\text{fail}$ is given by the sum over these when the faulty qubits are at least half of the total. This is
\begin{equation}\label{eq.Pfail.k}
    P_\text{fail}=\sum_{j=\tfrac{(k+1)}{2}}^k \binom{k}{j}\,\epsilon^j(1-\epsilon)^{k-j}.
\end{equation}
Namely, for $k=3$, one has
\begin{equation}
    P_\text{fail}=\sum_{j=\tfrac{(3+1)}{2}}^3 \binom{3}{j}\,\epsilon^j(1-\epsilon)^{3-j}=3\epsilon^2(1-\epsilon)+\epsilon^3.
\end{equation}
The behaviour of $P_\text{fail}$ for different values of $k$ is shown in  Fig.~\ref{fig.threshold}.

\begin{figure}[h]
    \centering
    \includegraphics[width=0.6\linewidth]{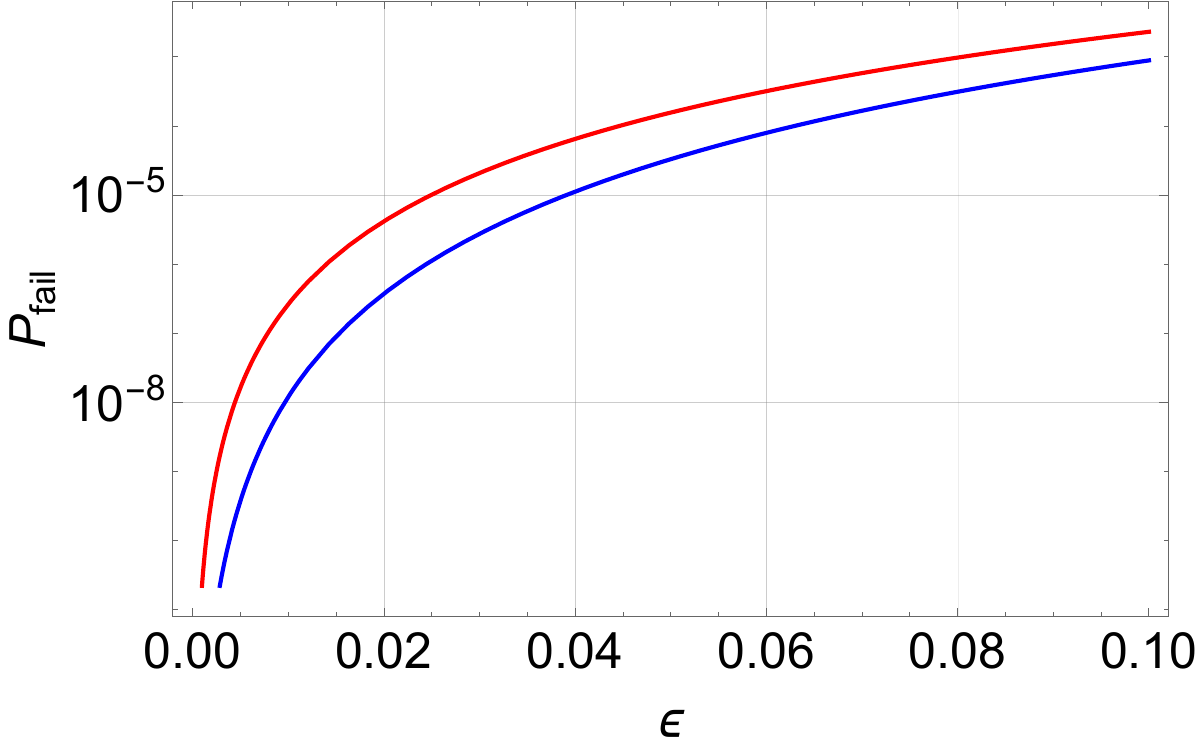}
    \caption{Comparison of the failing probabilities $P_\text{fail}$ for a single level encoding with 9 physical qubits (blue line) and that for a 2 level encoding each with 3 qubits (red line).}
    \label{fig:double encoding vs long encoding}
\end{figure}

However, one can consider an alternative approach. Instead of encoding a logical qubit just once in a large number of physical qubits, one can concatenate encodings. One encodes the logical qubit in different levels, where each level employs a small numeber of qubits. To be more explict, the following is the encoding of a single physical qubit in a 2 level encoding with three qubits each:
\begin{equation}
\ket0\xrightarrow[]{\text{first encoding}}\ket{000}\xrightarrow[]{\text{second encoding}}\ket{000}\ket{000}\ket{000},
\end{equation}
and similarly for $\ket1$. Now, with this encoding, the actual physical qubit is that in the highest level of encoding, and this is that directly suffering from the noise. Suppose there is a probability $\epsilon$ that an error occurs on this physical qubit. Then, at the level 1, the probability of failing, for example for the bit-flip QEC code, is
\begin{equation}
P_\text{fail,1}=3\epsilon^2-2\epsilon^3.
\end{equation}
This quantity is the probability that the noise corrupts a qubit at the level 1. Thus, when computing the probability of failing for the qubit at level 0, the actual logical qubit, $P_\text{fail,1}$ needs to be interpreted as the probability $\epsilon_1$ that an error occurs on the qubit at the level 1. Then, at level 0, one has that the failing probability is
\begin{equation}
\begin{aligned}
P_\text{fail,0}&=3P_\text{fail,1}^2-2P_\text{fail,1}^3,\\
&=3[3\epsilon^2-2\epsilon^3]^2-2[3\epsilon^2-2\epsilon^3]^3,\\
&=27 \epsilon^4-36\epsilon^5-42\epsilon^6+108\epsilon^7-72\epsilon^8+16\epsilon^9.
\end{aligned}
\end{equation}
The question is then which is the best encoding. Figure \ref{fig:double encoding vs long encoding}  compares the failing probabilities $P_\text{fail}$ for a single level encoding with 9 physical qubits (blue line), where Eq.~\eqref{eq.Pfail.k} gives
\begin{equation}
    P_\text{fail}=126 \epsilon^5-420\epsilon^6+540\epsilon^7-315\epsilon^8+70\epsilon^9,
\end{equation}
and that for a 2 level encoding each with 3 qubits (red line). This is a fair comparison, as both the approaches are employing the same number of physical qubits, i.e.~$n=9$.

To keep the discussion more general, suppose $p$ is the probability of failing for a qubit with no encoding (this is what we called $\epsilon$ until now). Then, the failing probability is
\begin{equation}
P_\text{fail}^{(0)}=p.
\end{equation}
Suppose that after one encoding the failing probability is
\begin{equation}
P_\text{fail}^{(1)}=c p^2,
\end{equation}
where $c$ is some suitable constant. In the case of the 3-qubit encoding, one had
\begin{equation}
P_\text{fail}=3p^2-2p^3\sim3p^2,
\end{equation}
for small values of $p$. After $2$ encodings, one has
\begin{equation}
P_\text{fail}^{(2)}=c (cp^2)^2=\frac{1}{c}(cp)^{2^2}.
\end{equation}
After $k$ encodings, one has
\begin{equation}
P_\text{fail}^{(k)}=p_\text{th}\left(\frac{p}{p_\text{th}}\right)^{2^k},
\end{equation}
where we defined the threshold probability as
\begin{equation}
p_\text{th}=\frac{1}{c}.
\end{equation}
Such a probability depends on varius parameters, among which the QEC code used, the physical components, the experimental implementation of the QEC protocol etc.

The threshold probability $p_\text{th}$ is fundamental due to the following theorem.

\begin{theorem}[Threshold theorem]\index{Threshold theorem}
If a threshold probability $p_\text{th}$ exists, then it is always possible to correct errors at a faster than they are created. It is sufficient to increase the level $k$ of encoding.

Proof. The proof is trivial. As long as the the occurrence of an error on the physical qubit $p=\epsilon$ is smaller than the threshold probability $p_\text{th}$, then the ratio
\begin{equation}
\frac{p}{p_\text{th}}<1,
\end{equation}
and thus the quantity
\begin{equation}
\left(\frac{p}{p_\text{th}}\right)^{2^k},
\end{equation}
can be made suitably small by simply increasing $k$.

\end{theorem}

The beauty of the threshold theorem is its simplicity. However, it also highlights a non-trivial problem, which is the necessity of employing a very large number of physical qubits. It naturally follows the question: How many physical qubits are necessary to quantum error correcting a faulty circuit?

Suppose we have $N$ components (this is the number of qubits times the number of gates). Suppose for each of these components one needs $R$ physical qubit accounting for QEC at the 1 level of encoding. Then, after $k$ levels of encoding there is a total of $N R^k$ qubits. Suppose we want that the entire circuit works with a failing probability $P_\text{fail, circuit}<\epsilon$, where $\epsilon$ is a given probability. Then, per component, we have
\begin{equation}
P_\text{fail}=p_\text{th}\left(\frac{p}{p_\text{th}}\right)^{2^k}<\frac{\epsilon}{N}.
\end{equation}
The question is then how many levels $k$ of encoding are necessary? Or equivalently, how many physical qubits $R^k$ per components are required? From the previous expression we obtain
\begin{equation}
2^k\sim\frac{\log_2\left(\frac{Np_\text{th}}{\epsilon}\right)}{\log_2\left(\frac{p_\text{th}}{p}\right)},
\end{equation}
which implies
\begin{equation}
R^k\sim\left(	\frac{\log_2\left(\frac{Np_\text{th}}{\epsilon}\right)}{\log_2\left(\frac{p_\text{th}}{p}\right)}\right)^{\log_2 R}.
\end{equation}
Thus, the size of the full circuit scales as 
\begin{equation}
N R^k\sim \operatorname{poly}\left(\log\frac{Np_\text{th}}{\epsilon}\right).
\end{equation}
This is the quantitative result of the threshold theorem.

\subsection{More layers of encoding or only more qubits}

Until now, we worked under the assumption of having only noises that act independently on the physical qubits. Let us suppose now a different kind of noise map, that correlates the noise on different qubits. Specifically, we consider a Kraus map acting on $N$ qubits, that reads
\begin{equation}\label{eq.strangeerror}
    \mathcal T[\hat \rho]=\frac{1}{N(N-1)}\sum_{i\neq j}^N\left[(1-p)\hat \rho+p\hat \sigma_x^{(i)}\hat \sigma_x^{(j)}\hat \rho\hat \sigma_x^{(j)}\hat \sigma_x^{(i)}\right],
\end{equation}
where there is a probability $p$ that the noise acts on the qubits and the prefactor is required to properly normalise the state.\\

First of all, let us introduce a graphical notation. Suppose we have three physical qubits, and the noise has acted on the first and second. Then, we denote this with the following scheme
\begin{equation}
\text{Before the noise}\quad
\begin{circuitikz}
\tikzstyle{every node}=[font=\normalsize]
\draw (10,17.75) to[short] (11,17.75);
\draw (10,18) to[short] (11,18);
\draw (10,18.25) to[short] (11,18.25);
\end{circuitikz}
\qquad\text{After the noise}
\quad
\begin{circuitikz}
\tikzstyle{every node}=[font=\normalsize]
\draw (10,17.75) to[short] (11,17.75);
\draw (10,18) to[short] (11,18);
\draw (10,18.25) to[short] (11,18.25);
\node [font=\normalsize] at (10.5,18.25) {X};
\node [font=\normalsize] at (10.5,18) {X};
\end{circuitikz}
\end{equation}

Now, suppose we are performing the encoding with $N=3$ qubits and $k=1$ layer of encoding. If there are no errors --- this happens with probability $(1-p)$ --- then physical qubits are
\begin{equation}
\begin{circuitikz}
\tikzstyle{every node}=[font=\normalsize]
\draw (10,17.75) to[short] (11,17.75);
\draw (10,18) to[short] (11,18);
\draw (10,18.25) to[short] (11,18.25);
\end{circuitikz}
\quad
\text{and correspond to a logical qubit}
\quad 
\begin{circuitikz}
\tikzstyle{every node}=[font=\normalsize]
\draw (10,18) to[short] (11,18);
\end{circuitikz}
\end{equation}
which is untouched by the noise. If there is one error, which corresponds to two qubits being affected with a probability $p$, then the physical qubits are in one of the following three states 
\begin{equation}\label{3qubit1error}
    \begin{circuitikz}
\tikzstyle{every node}=[font=\normalsize]
\draw (10,10.5) to[short] (11,10.5);
\draw (10,11) to[short] (11,11);
\draw (10,10.75) to[short] (11,10.75);
\node [font=\normalsize] at (10.5,11) {X};
\node [font=\normalsize] at (10.5,10.75) {X};
\draw (12.5,11) to[short] (13.5,11);
\draw (12.5,10.75) to[short] (13.5,10.75);
\draw (12.5,10.5) to[short] (13.5,10.5);
\draw (15,11) to[short] (16,11);
\draw (15,10.75) to[short] (16,10.75);
\draw (15,10.5) to[short] (16,10.5);
\node [font=\normalsize] at (13,10.75) {X};
\node [font=\normalsize] at (13,10.5) {X};
\node [font=\normalsize] at (15.5,11) {X};
\node [font=\normalsize] at (15.5,10.5) {X};
\draw [->, >=Stealth] (17.5,10.75) -- (18.75,10.75);
\draw (20,10.75) to[short] (21,10.75);
\node [font=\normalsize] at (20.5,10.75) {X};
\end{circuitikz}
\end{equation}
that correspond to a logical qubit that is affected to the noise. Here, we denote decodings with horizontal arrows. Thus, the protocol fails. 
These states contribute to the failing probability:
\begin{equation}
    P_\text{fail}\sim 3p,
\end{equation}
where the factor 3 is given by the number of equivalent states at the physical level, namely those represented graphically in the left side of Eq.~\eqref{3qubit1error}.

If there are two errors, which is a process involving a probability $p^2$ and four qubits, one has the following 9 states:
\begin{equation}
    \begin{circuitikz}
\tikzstyle{every node}=[font=\normalsize]
\draw (6.25,11) to[short] (7.25,11);
\draw (6.25,10.75) to[short] (7.25,10.75);
\draw (6.25,10.5) to[short] (7.25,10.5);
\draw (8.75,11) to[short] (9.75,11);
\draw (8.75,10.75) to[short] (9.75,10.75);
\draw (8.75,10.5) to[short] (9.75,10.5);
\draw (11.25,11) to[short] (12.25,11);
\draw (11.25,10.75) to[short] (12.25,10.75);
\draw (11.25,10.5) to[short] (12.25,10.5);
\draw (13.75,11) to[short] (14.75,11);
\draw (13.75,10.75) to[short] (14.75,10.75);
\draw (13.75,10.5) to[short] (14.75,10.5);
\draw (15,11) to[short] (16,11);
\draw (15,10.75) to[short] (16,10.75);
\draw (15,10.5) to[short] (16,10.5);
\draw (12.5,11) to[short] (13.5,11);
\draw (12.5,10.75) to[short] (13.5,10.75);
\draw (12.5,10.5) to[short] (13.5,10.5);
\draw (10,11) to[short] (11,11);
\draw (10,10.75) to[short] (11,10.75);
\draw (10,10.5) to[short] (11,10.5);
\draw (7.5,11) to[short] (8.5,11);
\draw (7.5,10.75) to[short] (8.5,10.75);
\draw (7.5,10.5) to[short] (8.5,10.5);
\draw (5,11) to[short] (6,11);
\draw (5,10.75) to[short] (6,10.75);
\draw (5,10.5) to[short] (6,10.5);
\node [font=\normalsize] at (5.25,11) {X};
\node [font=\normalsize] at (5.25,10.75) {X};
\node [font=\normalsize] at (5.75,11) {X};
\node [font=\normalsize] at (5.75,10.75) {X};
\node [font=\normalsize] at (6.5,11) {X};
\node [font=\normalsize] at (6.5,10.75) {X};
\node [font=\normalsize] at (7,11) {X};
\node [font=\normalsize] at (7,10.5) {X};
\node [font=\normalsize] at (7.75,11) {X};
\node [font=\normalsize] at (7.75,10.75) {X};
\node [font=\normalsize] at (8.25,10.75) {X};
\node [font=\normalsize] at (8.25,10.5) {X};
\node [font=\normalsize] at (9,11) {X};
\node [font=\normalsize] at (9,10.5) {X};
\node [font=\normalsize] at (9.5,11) {X};
\node [font=\normalsize] at (10.25,11) {X};
\node [font=\normalsize] at (10.25,10.5) {X};
\node [font=\normalsize] at (10.75,10.5) {X};
\node [font=\normalsize] at (11.5,10.5) {X};
\node [font=\normalsize] at (12,10.75) {X};
\node [font=\normalsize] at (12.75,10.75) {X};
\node [font=\normalsize] at (9.5,10.75) {X};
\node [font=\normalsize] at (10.75,11) {X};
\node [font=\normalsize] at (11.5,11) {X};
\node [font=\normalsize] at (12,10.5) {X};
\node [font=\normalsize] at (12.75,10.5) {X};
\node [font=\normalsize] at (14,10.75) {X};
\node [font=\normalsize] at (14,10.5) {X};
\node [font=\normalsize] at (15.25,10.75) {X};
\node [font=\normalsize] at (15.25,10.5) {X};
\node [font=\normalsize] at (13.25,11) {X};
\node [font=\normalsize] at (13.25,10.75) {X};
\node [font=\normalsize] at (14.5,11) {X};
\node [font=\normalsize] at (14.5,10.5) {X};
\node [font=\normalsize] at (15.75,10.75) {X};
\node [font=\normalsize] at (15.75,10.5) {X};
\end{circuitikz}
\end{equation}
which correspond to
\begin{equation}
    \begin{circuitikz}
\tikzstyle{every node}=[font=\normalsize]
\draw (6.25,11) to[short] (7.25,11);
\draw (6.25,10.75) to[short] (7.25,10.75);
\draw (6.25,10.5) to[short] (7.25,10.5);
\draw (8.75,11) to[short] (9.75,11);
\draw (8.75,10.75) to[short] (9.75,10.75);
\draw (8.75,10.5) to[short] (9.75,10.5);
\draw (11.25,11) to[short] (12.25,11);
\draw (11.25,10.75) to[short] (12.25,10.75);
\draw (11.25,10.5) to[short] (12.25,10.5);
\draw (13.75,11) to[short] (14.75,11);
\draw (13.75,10.75) to[short] (14.75,10.75);
\draw (13.75,10.5) to[short] (14.75,10.5);
\draw (15,11) to[short] (16,11);
\draw (15,10.75) to[short] (16,10.75);
\draw (15,10.5) to[short] (16,10.5);
\draw (12.5,11) to[short] (13.5,11);
\draw (12.5,10.75) to[short] (13.5,10.75);
\draw (12.5,10.5) to[short] (13.5,10.5);
\draw (10,11) to[short] (11,11);
\draw (10,10.75) to[short] (11,10.75);
\draw (10,10.5) to[short] (11,10.5);
\draw (7.5,11) to[short] (8.5,11);
\draw (7.5,10.75) to[short] (8.5,10.75);
\draw (7.5,10.5) to[short] (8.5,10.5);
\draw (5,11) to[short] (6,11);
\draw (5,10.75) to[short] (6,10.75);
\draw (5,10.5) to[short] (6,10.5);
\node [font=\normalsize] at (6.5,10.75) {X};
\node [font=\normalsize] at (7,10.5) {X};
\node [font=\normalsize] at (7.75,11) {X};
\node [font=\normalsize] at (8.25,10.5) {X};
\node [font=\normalsize] at (9,10.5) {X};
\node [font=\normalsize] at (12,10.75) {X};
\node [font=\normalsize] at (9.5,10.75) {X};
\node [font=\normalsize] at (11.5,11) {X};
\node [font=\normalsize] at (12.75,10.5) {X};
\node [font=\normalsize] at (14,10.75) {X};
\node [font=\normalsize] at (13.25,11) {X};
\node [font=\normalsize] at (14.5,11) {X};
\draw [->, >=Stealth] (5.5,10.25) -- (5.5,9.75);
\draw [->, >=Stealth] (6.75,10.25) -- (6.75,9.75);
\draw [->, >=Stealth] (8,10.25) -- (8,9.75);
\draw [->, >=Stealth] (9.25,10.25) -- (9.25,9.75);
\draw [->, >=Stealth] (10.5,10.25) -- (10.5,9.75);
\draw [->, >=Stealth] (11.75,10.25) -- (11.75,9.75);
\draw [->, >=Stealth] (13,10.25) -- (13,9.75);
\draw [->, >=Stealth] (14.25,10.25) -- (14.25,9.75);
\draw [->, >=Stealth] (15.5,10.25) -- (15.5,9.75);
\draw (5,9.5) to[short] (6,9.5);
\draw (6.25,9.5) to[short] (7.25,9.5);
\draw (7.5,9.5) to[short] (8.5,9.5);
\draw (8.75,9.5) to[short] (9.75,9.5);
\draw (10,9.5) to[short] (11,9.5);
\draw (11.25,9.5) to[short] (12.25,9.5);
\draw (12.5,9.5) to[short] (13.5,9.5);
\draw (13.75,9.5) to[short] (14.75,9.5);
\draw (15,9.5) to[short] (16,9.5);
\node [font=\normalsize] at (6.75,9.5) {X};
\node [font=\normalsize] at (8,9.5) {X};
\node [font=\normalsize] at (9.25,9.5) {X};
\node [font=\normalsize] at (11.75,9.5) {X};
\node [font=\normalsize] at (13,9.5) {X};
\node [font=\normalsize] at (14.25,9.5) {X};
\end{circuitikz}
\end{equation}
as the application of an bit-flip error twice on the same qubit correspond to not having an error, i.e.~$\hat\sigma_x^2=\hat{\mathbb{1}}$. In such a case, only some combination are faulty, while in others the errors have cancelled. Here, we denote equivalence with vertical arrows. These states will contribute to $P_\text{fail}$ with $+6p^2$. Thus, one gets
\begin{equation}
    P_\text{fail}=3p+6p^2+\dots\sim3p,
\end{equation}
where the $\dots$ indicate higher order errors. Nevertheless, is the lowest order term in $p$ that is the most significant, under the hypothesis of small error probabilities.\\

\begin{figure}[h]
    \centering
    \includegraphics[width=0.6\linewidth]{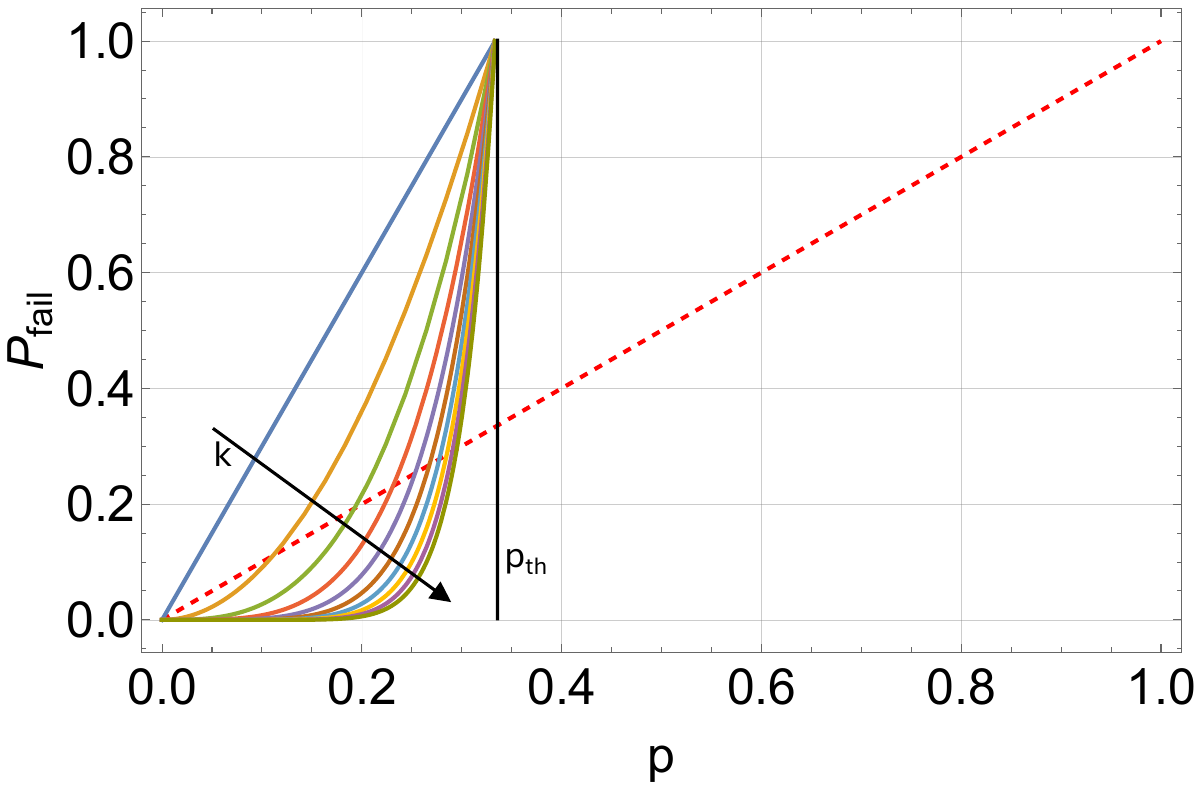}
    \caption{Comparison of the failing probabilities $P_\text{fail}$ for different $k$ layers of encoding with 3 qubits each. The arrow indicates the direction of increasing values of $k$, while the vertical black line indicates the threshold probability $p_\text{th}$.}
    \label{fig.3qubit-klayers}
\end{figure}

Consider now a double encoding $k=2$ with a total of $N=3^2=9$ qubits. In case of no errors, prob $=(1-p)$, we have
\begin{equation}
    \begin{circuitikz}
\tikzstyle{every node}=[font=\normalsize]
\draw (5,11) to[short] (6,11);
\draw (5,10.75) to[short] (6,10.75);
\draw (5,10.5) to[short] (6,10.5);
\draw (5,11.25) to[short] (6,11.25);
\draw (5,11.5) to[short] (6,11.5);
\draw (5,11.75) to[short] (6,11.75);
\draw (5,12.25) to[short] (6,12.25);
\draw (5,12.5) to[short] (6,12.5);
\draw (5,12) to[short] (6,12);
\draw [->, >=Stealth] (6.5,12.25) -- (7,12.25);
\draw [->, >=Stealth] (6.5,11.5) -- (7,11.5);
\draw [->, >=Stealth] (6.5,10.75) -- (7,10.75);
\draw (7.25,12.25) to[short] (8.25,12.25);
\draw (7.25,11.5) to[short] (8.25,11.5);
\draw (7.25,10.75) to[short] (8.25,10.75);
\draw [->, >=Stealth] (8.75,11.5) -- (9.25,11.5);
\node [font=\normalsize] at (5.5,13) {layer 2};
\node [font=\normalsize] at (7.75,13) {layer 1};
\node [font=\normalsize] at (10,13) {layer 0};
\draw (9.5,11.5) to[short] (10.5,11.5);
\end{circuitikz}
\end{equation}
which do not contribute to $P_\text{fail}$. If there is one error, prob $=p$, we have ${9\choose 2}=36$ states. Some states display two affected physical qubits in different layer-1 logical qubit,
\begin{equation}
    \begin{circuitikz}
\tikzstyle{every node}=[font=\normalsize]
\draw (5,11) to[short] (6,11);
\draw (5,10.75) to[short] (6,10.75);
\draw (5,10.5) to[short] (6,10.5);
\draw (5,11.25) to[short] (6,11.25);
\draw (5,11.5) to[short] (6,11.5);
\draw (5,11.75) to[short] (6,11.75);
\draw (5,12.25) to[short] (6,12.25);
\draw (5,12.5) to[short] (6,12.5);
\draw (5,12) to[short] (6,12);
\draw [->, >=Stealth] (6.5,12.25) -- (7,12.25);
\draw [->, >=Stealth] (6.5,11.5) -- (7,11.5);
\draw [->, >=Stealth] (6.5,10.75) -- (7,10.75);
\draw (7.25,12.25) to[short] (8.25,12.25);
\draw (7.25,11.5) to[short] (8.25,11.5);
\draw (7.25,10.75) to[short] (8.25,10.75);
\draw [->, >=Stealth] (8.75,11.5) -- (9.25,11.5);
\node [font=\normalsize] at (5.5,13) {layer 2};
\node [font=\normalsize] at (7.75,13) {layer 1};
\node [font=\normalsize] at (10,13) {layer 0};
\draw (9.5,11.5) to[short] (10.5,11.5);
\node [font=\normalsize] at (5.5,12.5) {X};
\node [font=\normalsize] at (5.5,11.25) {X};
\end{circuitikz}
\end{equation}
Some have two physical qubits in  the same layer-1 logical qubit. Thus, the corresponding layer-1 logical qubit fails, but the layer-0 logical qubit is still protected
\begin{equation}
    \begin{circuitikz}
\tikzstyle{every node}=[font=\normalsize]
\draw (5,11) to[short] (6,11);
\draw (5,10.75) to[short] (6,10.75);
\draw (5,10.5) to[short] (6,10.5);
\draw (5,11.25) to[short] (6,11.25);
\draw (5,11.5) to[short] (6,11.5);
\draw (5,11.75) to[short] (6,11.75);
\draw (5,12.25) to[short] (6,12.25);
\draw (5,12.5) to[short] (6,12.5);
\draw (5,12) to[short] (6,12);
\draw [->, >=Stealth] (6.5,12.25) -- (7,12.25);
\draw [->, >=Stealth] (6.5,11.5) -- (7,11.5);
\draw [->, >=Stealth] (6.5,10.75) -- (7,10.75);
\draw (7.25,12.25) to[short] (8.25,12.25);
\draw (7.25,11.5) to[short] (8.25,11.5);
\draw (7.25,10.75) to[short] (8.25,10.75);
\draw [->, >=Stealth] (8.75,11.5) -- (9.25,11.5);
\node [font=\normalsize] at (5.5,13) {layer 2};
\node [font=\normalsize] at (7.75,13) {layer 1};
\node [font=\normalsize] at (10,13) {layer 0};
\draw (9.5,11.5) to[short] (10.5,11.5);
\node [font=\normalsize] at (5.5,12.5) {X};
\node [font=\normalsize] at (5.5,12) {X};
\node [font=\normalsize] at (7.75,12.25) {X};
\end{circuitikz}
\end{equation}
This is the worst-case scenario with one error. If there are two errors, prob $=p^2$, the states that are relevant are those that have 2 layer-1 logical qubits affected. Namely, they reproduce the same graph as that in Eq.~\eqref{3qubit1error}. For example, of the form
\begin{equation}
    \begin{circuitikz}
\tikzstyle{every node}=[font=\normalsize]
\draw (5,11) to[short] (6,11);
\draw (5,10.75) to[short] (6,10.75);
\draw (5,10.5) to[short] (6,10.5);
\draw (5,11.25) to[short] (6,11.25);
\draw (5,11.5) to[short] (6,11.5);
\draw (5,11.75) to[short] (6,11.75);
\draw (5,12.25) to[short] (6,12.25);
\draw (5,12.5) to[short] (6,12.5);
\draw (5,12) to[short] (6,12);
\draw [->, >=Stealth] (6.5,12.25) -- (7,12.25);
\draw [->, >=Stealth] (6.5,11.5) -- (7,11.5);
\draw [->, >=Stealth] (6.5,10.75) -- (7,10.75);
\draw (7.25,12.25) to[short] (8.25,12.25);
\draw (7.25,11.5) to[short] (8.25,11.5);
\draw (7.25,10.75) to[short] (8.25,10.75);
\draw [->, >=Stealth] (8.75,11.5) -- (9.25,11.5);
\node [font=\normalsize] at (5.5,13) {layer 2};
\node [font=\normalsize] at (7.75,13) {layer 1};
\node [font=\normalsize] at (10,13) {layer 0};
\draw (9.5,11.5) to[short] (10.5,11.5);
\node [font=\normalsize] at (5.5,12.5) {X};
\node [font=\normalsize] at (5.5,12) {X};
\node [font=\normalsize] at (7.75,12.25) {X};
\node [font=\normalsize] at (5.5,10.75) {X};
\node [font=\normalsize] at (5.5,10.5) {X};
\node [font=\normalsize] at (7.75,10.75) {X};
\node [font=\normalsize] at (10,11.5) {X};
\end{circuitikz}
\end{equation}
At the layer-1, this correspond to a probability being $3p$. For each of the layer-1 affected logical qubits, one needs 2 layer-2 affected physical qubits, with associated a probability $3p$. Thus, one has
\begin{equation}
    P_\text{fail}=(3p)^2+\dots \sim 9p^2.
\end{equation}
With generic $k$ layers of encoding, one has
\begin{equation}
    P_\text{fail}=(3p)^k+\dots,
\end{equation}
which is graphically represented in Fig.~\ref{fig.3qubit-klayers}, where the lines are listed at increasing values of $k$ and condense towards the value of $p_\text{th}=1/3$.\\

Let us now consider the alternative. Instead of taking a large number of layers of encoding with just a few qubits per layer, we consider a large number of qubits on a single encoding layer.

For $N=3$, a single error, i.e.~having two physical qubits affected, is sufficient to make the encoding fail:
\begin{equation}
    \begin{circuitikz}
\tikzstyle{every node}=[font=\normalsize]
\draw (5,9.5) to[short] (6,9.5);
\draw (5,9.75) to[short] (6,9.75);
\draw (5,9.25) to[short] (6,9.25);
\draw [->, >=Stealth] (6.5,9.5) -- (7,9.5);
\draw (7.25,9.5) to[short] (8.25,9.5);
\node [font=\normalsize] at (6.75,10.25) {No error};
\node [font=\normalsize] at (11.75,10.25) {1 error};
\draw (10,9.75) to[short] (11,9.75);
\draw (10,9.5) to[short] (11,9.5);
\draw (10,9.25) to[short] (11,9.25);
\draw (12.25,9.5) to[short] (13.25,9.5);
\draw [->, >=Stealth] (11.5,9.5) -- (12,9.5);
\node [font=\normalsize] at (10.5,9.75) {X};
\node [font=\normalsize] at (10.5,9.25) {X};
\node [font=\normalsize] at (12.75,9.5) {X};
\end{circuitikz}
\end{equation}
where there are three different combinations (see Eq.~\eqref{3qubit1error}) that count. Thus, we have
\begin{equation}
    P_\text{fail}\sim3p
\end{equation}
For $N=5$, one requires two errors, with four qubits affected, to make the encoding fail. Indeed, 
\begin{equation}
    \begin{circuitikz}
\tikzstyle{every node}=[font=\normalsize]
\draw (5,9.5) to[short] (6,9.5);
\draw (5,9.75) to[short] (6,9.75);
\draw (5,9.25) to[short] (6,9.25);
\node [font=\normalsize] at (6.75,10.25) {No error};
\node [font=\normalsize] at (11.75,10.25) {1 error};
\draw (10,9.75) to[short] (11,9.75);
\draw (10,9.5) to[short] (11,9.5);
\draw (10,9.25) to[short] (11,9.25);
\node [font=\normalsize] at (10.5,9.75) {X};
\node [font=\normalsize] at (10.5,9.25) {X};
\draw [->, >=Stealth] (6.5,9.25) -- (7,9.25);
\draw [->, >=Stealth] (11.5,9.25) -- (12,9.25);
\draw (5,9) to[short] (6,9);
\draw (5,8.75) to[short] (6,8.75);
\draw (10,9) to[short] (11,9);
\draw (10,8.75) to[short] (11,8.75);
\draw (7.25,9.25) to[short] (8.25,9.25);
\draw (12.25,9.25) to[short] (13.25,9.25);
\node [font=\normalsize] at (16.75,10.25) {2 errors};
\draw (15,9) to[short] (16,9);
\draw (15,9.25) to[short] (16,9.25);
\draw (15,9.5) to[short] (16,9.5);
\draw (15,9.75) to[short] (16,9.75);
\draw (15,8.75) to[short] (16,8.75);
\draw (17.25,9.25) to[short] (18.25,9.25);
\draw [->, >=Stealth] (16.5,9.25) -- (17,9.25);
\node [font=\normalsize] at (15.5,9.75) {X};
\node [font=\normalsize] at (15.5,9.25) {X};
\node [font=\normalsize] at (15.5,9) {X};
\node [font=\normalsize] at (15.5,8.75) {X};
\node [font=\normalsize] at (17.75,9.25) {X};
\end{circuitikz}
\end{equation}
In such a case, the failing probability is given by
\begin{equation}
    P_\text{fail}=\frac12{5\choose2}{3\choose2}p^2+\dots\sim15p^2,
\end{equation}
where the first binomial chooses 2 qubits to affect among the available 5, the second binomial chooses 2  qubits among the remaining 3. The factor one-half acconts for the simmetry between the first error and the second one, i.e.~between the first couple of affected qubits and the second one.

\begin{figure}
    \centering
    \includegraphics[width=0.6\linewidth]{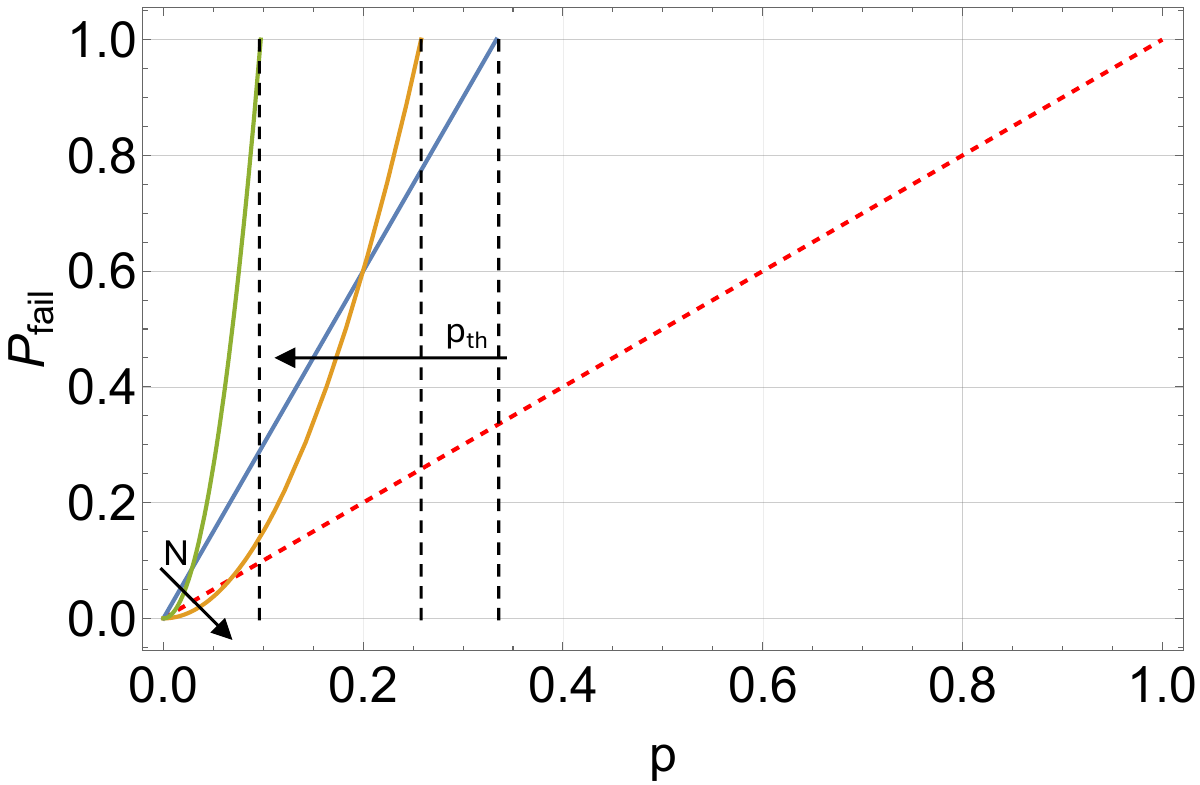}
    \caption{Comparison of the failing probabilities $P_\text{fail}$ for different values $N$ of the qubits with a single layer of encoding. The arrow indicates the direction of increasing values of $N$, while the vertical black dashed lines indicate how the curves cannot define a threshold probability.}
    \label{fig.Nqubit-klayers.pdf}
\end{figure}

Also for $N=7$, one requires 2 errors, i.e.~four affected qubits. Indeed
\begin{equation}
    \begin{circuitikz}
\tikzstyle{every node}=[font=\normalsize]
\draw (5,9.5) to[short] (6,9.5);
\draw (5,9.75) to[short] (6,9.75);
\draw (5,9.25) to[short] (6,9.25);
\node [font=\normalsize] at (6.75,10.25) {No error};
\node [font=\normalsize] at (11.75,10.25) {1 error};
\draw (10,9.75) to[short] (11,9.75);
\draw (10,9.5) to[short] (11,9.5);
\draw (10,9.25) to[short] (11,9.25);
\node [font=\normalsize] at (10.5,9.75) {X};
\node [font=\normalsize] at (10.5,9.25) {X};
\draw (5,9) to[short] (6,9);
\draw (5,8.75) to[short] (6,8.75);
\draw (10,9) to[short] (11,9);
\draw (10,8.75) to[short] (11,8.75);
\node [font=\normalsize] at (16.75,10.25) {2 errors};
\draw (15,9) to[short] (16,9);
\draw (15,9.25) to[short] (16,9.25);
\draw (15,9.5) to[short] (16,9.5);
\draw (15,9.75) to[short] (16,9.75);
\draw (15,8.75) to[short] (16,8.75);
\node [font=\normalsize] at (15.5,9.75) {X};
\node [font=\normalsize] at (15.5,9.25) {X};
\node [font=\normalsize] at (15.5,9) {X};
\node [font=\normalsize] at (15.5,8.75) {X};
\draw (7.25,9) to[short] (8.25,9);
\draw (12.25,9) to[short] (13.25,9);
\draw [->, >=Stealth] (6.5,9) -- (7,9);
\draw [->, >=Stealth] (11.5,9) -- (12,9);
\draw [->, >=Stealth] (16.5,9) -- (17,9);
\draw (17.25,9) to[short] (18.25,9);
\draw (5,8.5) to[short] (6,8.5);
\draw (5,8.25) to[short] (6,8.25);
\draw (10,8.5) to[short] (11,8.5);
\draw (10,8.25) to[short] (11,8.25);
\draw (15,8.5) to[short] (16,8.5);
\draw (15,8.25) to[short] (16,8.25);
\node [font=\normalsize] at (17.75,9) {X};
\end{circuitikz}
\end{equation}
In such a case, we get
\begin{equation}
    P_\text{fail}=\frac12{7\choose2}{5\choose2}p^2+\dots\sim105p^2.
\end{equation}
Figure \ref{fig.Nqubit-klayers.pdf} compares failing probabilities of the encodings with different values of $N$. As one can see the knee of the curves moves towards zero, meaning that the threshold does not exist.\\

Thus, for the error defined in Eq.~\eqref{eq.strangeerror}, employing several layers of encoding can allow for a fault-tolerant quantum computing, while employing only a single layer encoding with more qubits has strong limits.

\section{Stabiliser formalism}

The stabiliser formalism exploits the fact that there exist operations, namely stabilisers, that 
can be used to detect errors without changing the state of the logical qubit. While the single stabiliser can only tell if there was an error, without establishing which error, a specific set of stabilisers can identify the specific errors that occurred, and thus providing the information to correct it.

\subsection{Inverting quantum channels}

\begin{figure}[h]
    \centering
    \includegraphics[width=0.8\linewidth]{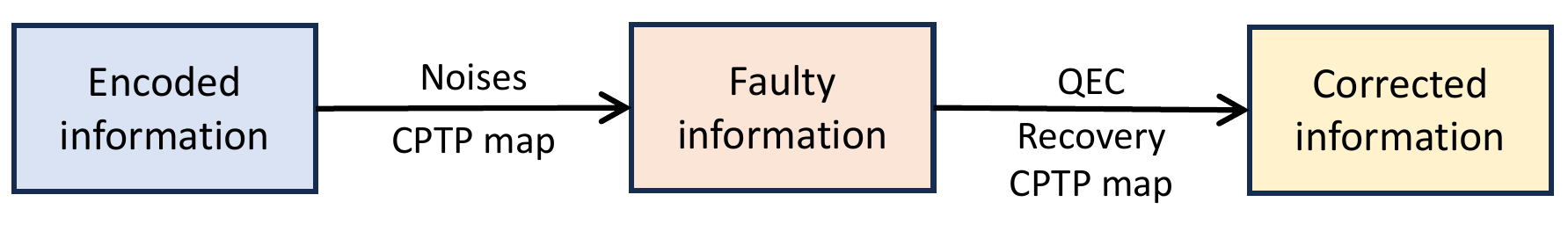}
    \caption{Schematic representation of the QEC scheme, where a recovery CPTP map is applied to recover the information as there were no errors.}
    \label{fig.QEC.scheme}
\end{figure}

The following scheme summarises the QEC philosophy. Given a qubit, this is encoded in a logical qubit made of a set of physical qubits. The interaction with the surrounding environment (or other faulty components of the physical circuit) leads to errors in the state. The action of these errors can be described in terms of a CPTP map. The QEC code applies a recovery CPTP map, which --- up to a certain probability --- gives back the same initial state, as there were no errors, cf.~Fig.~\ref{fig.QEC.scheme}.

The question is: when is this possible?\\

Consider two CPTP maps $\mathcal E$ and $\mathcal R$, which describe respectively the occurrence of environmental errors and the recovery map. Their action is
\begin{equation}
\mathcal B(\mathbb H)\xrightarrow[]{\mathcal E}\mathcal B(\mathbb H’)\xrightarrow[]{\mathcal R}\mathcal B(\mathbb H''),
\end{equation}
 where $\mathcal B(\mathbb H)$ is the space of all the linear operators acting on $\mathbb H$. We want to have no effects of the errors, i.e.~$\mathcal R\circ\mathcal E=\id$.

Now, we show that the following two statements are equivalent:
\begin{itemize}
\item[1)] Given a CPTP map $\mathcal E$, it exists another CPTP map $\mathcal R$ such that $\mathcal R\circ\mathcal E=\id$.

\item[2)] For any Kraus representation of $\mathcal E$, which is defined through the set of Kraus operators $\hat E_i$, then one has that
\begin{equation}
    \hat E_i^\dag\hat E_j=\mu_{ij}\hat {\mathbb 1},
\end{equation}
where $\mu_{ij}\in \mathbb C$ are the coefficients of the density matrix of the environment imposing the map $\mathcal E$. Namely,
\begin{equation}
    \hat \mu=\sum_{ij}\mu_{ij}\ket{e_i}\bra{e_j}.
\end{equation}
\end{itemize}

First, we prove that 1) implies 2). Consider $\ket\psi\in\mathbb H$, to which we apply the map $\mathcal E$. This gives
\begin{equation}
    \ket\psi\xrightarrow[]{\mathcal E}\sum_j \hat E_j\ket\psi\ket{e_j},
\end{equation}
where $\ket{e_j}$ is the state of the environment with which the system is entangling. Now, we apply the map $\mathcal R$:
\begin{equation}
    \xrightarrow[]{\mathcal R}\sum_{jk}\hat R_k\hat E_j\ket{\psi}\ket{e_j}\ket{a_k},
\end{equation}
where $\ket{a_k}$ is the state of an ancilla, whose interaction defines the map $\mathcal R$. Since we want that the map $\mathcal R$ works as a recovery map for the map $\mathcal E$, we have to impose that
\begin{equation}
    \sum_{jk}\hat R_k\hat E_j\ket{\psi}\ket{e_j}\ket{a_k}=\ket\psi\otimes(\dots),
\end{equation}
where $(\dots)$ is a suitable state of the environment and the ancilla. In this way, the entanglement between the system and the environment is transferred to the environment and the ancilla, with no correlation to the state of the system. This is possible only if
\begin{equation}
    \hat R_k\hat E_j=\alpha_{kj}\hat{\mathbb 1}.
\end{equation}
In such a case, one has that $(\dots)=\sum_{jk}\alpha_{kj}\ket{e_j}\ket{a_k}$ and 
\begin{equation}
\begin{aligned}
    \hat E_i^\dag\hat E_j&=\sum_k\hat E_i^\dag\hat R_k^\dag\hat R_k\hat E_j,\\
    &=\sum_k\alpha_{ki}^*\alpha_{kj}\hat {\mathbb 1},
\end{aligned}
\end{equation}
where we used that 
\begin{equation}
    \sum_k\hat R_k^\dag\hat R_k=\hat {\mathbb 1}.
\end{equation}
Then, we can define
\begin{equation}
    \mu_{ij}=\sum_k\alpha_{ki}^*\alpha_{kj}.
\end{equation}
Notably, we have that $\mu_{ji}^*=\mu_{ij}$, indeed
\begin{equation}
    \mu_{ji}^*\hat {\mathbb 1}=(\mu_{ij}\hat {\mathbb 1})^\dag=(\hat E_i^\dag\hat E_j)^\dag=\hat E_j^\dag\hat E_i=\mu_{ij}\hat {\mathbb 1}.
\end{equation}
Moreover, we have that 
\begin{equation}
    \sum_i\mu_{ii}\hat {\mathbb 1}=\sum_i\hat E_i^\dag\hat E_i=\hat {\mathbb 1},
\end{equation}
and that 
\begin{equation}
    \mu_{ii}\hat {\mathbb 1}=\hat E_i^\dag\hat E_i,
\end{equation}
is a positive operator, which implies that $\mu_{ii}>0$.
Thus, $\set{\mu_{ij}}_{ij}$ have all the properties to be the coefficients of a density matrix, and this proves that 1) implies 2).

We now prove that 2) implies 1). We start from 
\begin{equation}
    \hat E_i^\dag\hat E_j=\mu_{ij}\hat {\mathbb 1},
\end{equation}
and we diagonalise $\mu_{ij}$. This implies a change the Kraus operators according to $\hat E_i\to\hat F_i$ such that
\begin{equation}
 \hat F_i^\dag\hat F_j=\delta_{ij}p_i\hat {\mathbb 1},
\end{equation}
where $p_i>0$ by definition. We then introduce the isometries $\hat V_i$ that are related to $  \hat F_i$ via
\begin{equation}\label{eq.FonV}
    \hat F_i=\sqrt{p_i}\hat V_i.
\end{equation}
\begin{figure}[h]
    \centering
    \includegraphics[width=0.6\linewidth]{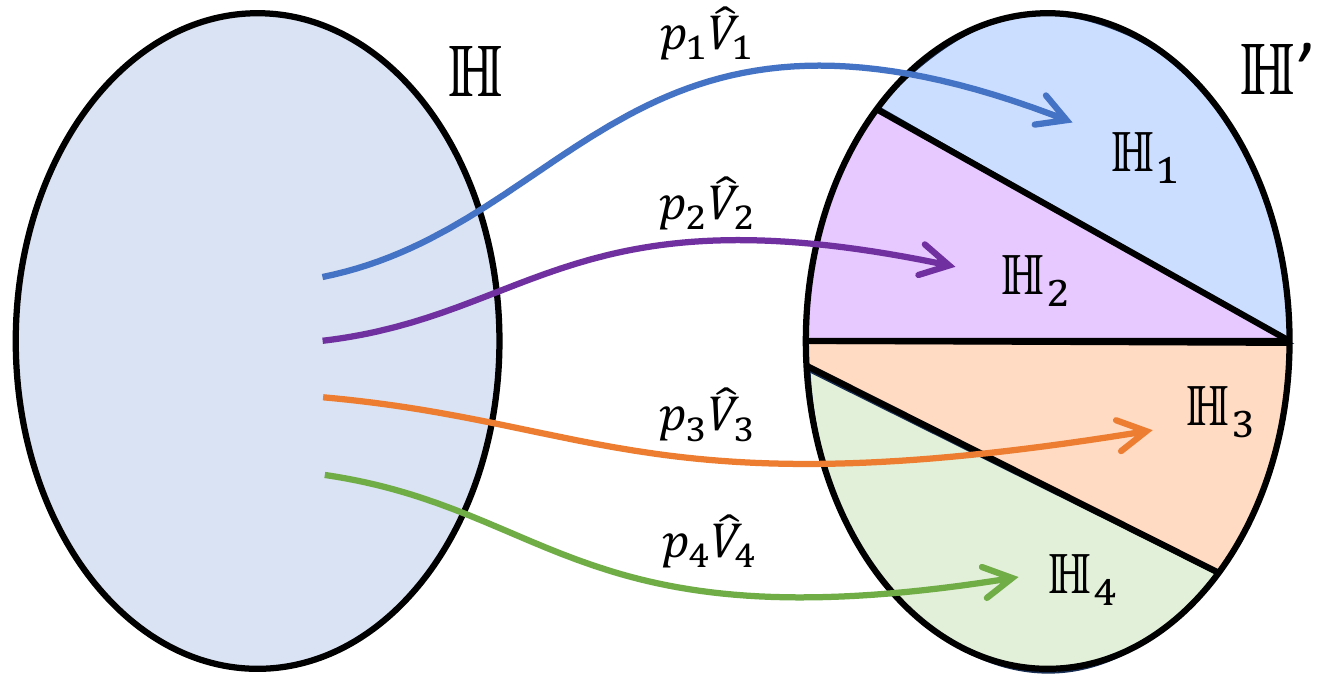}
    \caption{Graphical representation of the mapping between $\mathbb H$ and $\mathbb H'$.  }
    \label{fig.HsuH}
\end{figure}
Then, as represented in Fig.~\ref{fig.HsuH}, the map $\mathcal E$ is mapping $\mathbb H$ to different subspaces $\mathbb H_i'$ of $\mathbb H'$. Each of these mapping is performed with a probability $p_i$ by the operator $\hat V_i\propto\hat F_i$. Notably, since we diagonalised $\mu_{ij}$, the different operators $\hat V_i$ are orthogonal, and thus are also the subspaces $\mathbb H_i'$. Now, given this set of isometries, we can construct the recovery Kraus operators as
\begin{equation}
    \hat R_i=\hat V_i^\dag,
\end{equation}
which will act only on the corresponding subspace $\mathbb H_i'$, leaving the rest of Hilbert space $\mathbb H'$ untouched: indeed, for $i\neq j$ we have $\hat V_j^\dag F_j\propto\hat V_j^\dag V_j=0$. Finally, we compose the maps $\mathcal E$ and $\mathcal R$:
\begin{equation}
\begin{aligned}
    \hat \rho\xrightarrow[]{\mathcal E}&\mathcal E(\hat \rho)=\sum_ip_i\hat V_i\hat \rho\hat V_i^\dag,\\
    &\xrightarrow[]{\mathcal R}\mathcal R(\mathcal E(\hat \rho))=\sum_j\hat V_j^\dag\left(\sum_ip_i\hat V_i\hat \rho\hat V_i^\dag\right)\hat V_j,\\
    &=\sum_{ij}p_i\hat V_j^\dag\hat V_i\hat \rho\hat V_i^\dag\hat V_j,\\
    &=\sum_ip_i\hat \rho=\hat \rho,
\end{aligned}
\end{equation}
where we used that $\hat V_j^\dag\hat V_i=\delta_{ij}\hat{\mathbb 1}$.

\subsection{Correctable errors}

The generic scheme for QEC is the following. We are given $k$ qubits in an unknown state $\ket\psi\in\mathbb H$. We encode $\ket\psi$ in a larger number $n>k$ of qubits. These $n$ qubits are subject to errors, which are described in terms of a Kraus map $\mathcal E$ with the Kraus operators being $\hat E_i$ or equivalently $\hat V_i$, see Eq.~\eqref{eq.FonV}. The recovery protocol employs some extra $n’$ ancillary qubits to apply the QEC, which inverts the error Kraus map under certain conditions.

\begin{figure}[h]
    \centering
    \includegraphics[width=0.6\linewidth]{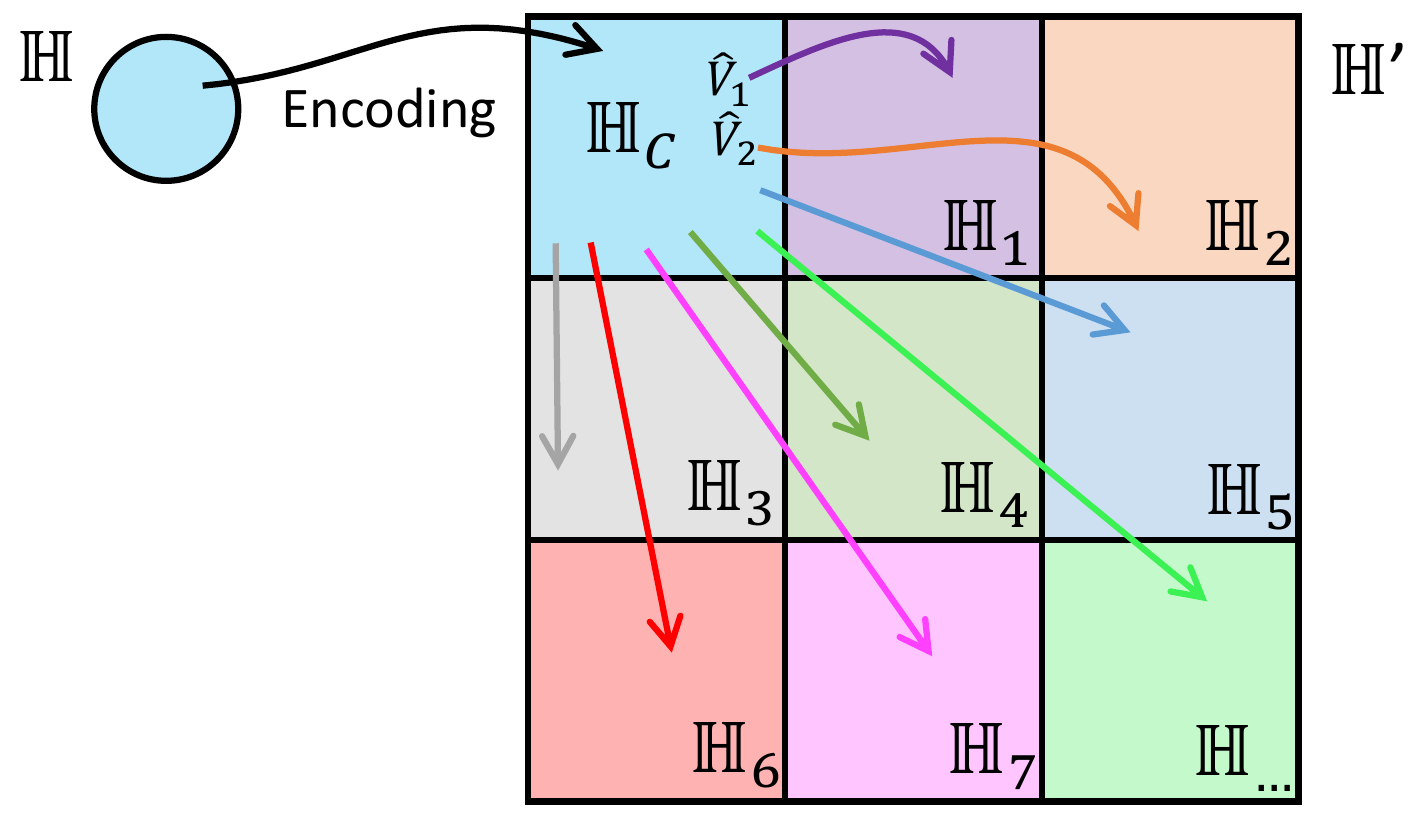}
    \caption{Graphical representation of the division of the Hilbert space $\mathbb H'$ in the code space $\mathbb H$ and error spaces $\mathbb H_i$, which are linked by the operators $\hat V_i$.}
    \label{fig.corr.errors}
\end{figure}

After the encoding, the relevant state will be $\ket{\psi’}\in\mathbb H_\text{C}$, which is a subspace of $\mathbb H’$ and it is called code space\index{Code space}. In particular, the entire Hilbert space $\mathbb H’$ is the union of  the code space $\mathbb H_\text{C}$ with the $\otimes_i \mathbb H’_i$. Here, the basis of each $\mathbb H’_i$ is obtained by applying the corresponding $\hat V_i$ to the basis of $\mathbb H_\text{C}$. Under this perspective, one can say that also the basis of $\mathbb H_\text{C}$ is generated in the same way, where the corresponding error operator is $\hat V_\text{C} =\hat {\mathbb 1}$. Now, since the subspace $\mathbb H’_i$ so constructed is orthogonal to  $\mathbb H_\text{C}$, then the error can be recovered. What is needed is a error syndrome measurement that identifies the subspace $\mathbb H’_i$ in which the state $\ket{\psi’}$ has been mapped. Given such a measurement, one can apply the corresponding Kraus recovery operator. This is graphically represented in Fig.~\ref{fig.corr.errors}

One can invert the error map if the corresponding operators $\hat E_i$ satisfy
\begin{equation}
\hat P\hat E_i^\dag\hat E_j\hat P=\mu_{ij}\hat P,
\end{equation}
where $\hat P$ is a projector on the code space $\mathbb H_\text{C}$. 

Notably, $\hat P$ acts as an identity operator if restricted on $\mathbb H_\text{C}$. This also implies that there exists a special set of Kraus operators, which allow to rewrite the error map, and thus also the recovery map, as a mixture of isometries. These are induced by a set of unitary operators $\hat U_i$ on the code space:
\begin{equation}
\hat V_i=\hat U_i\hat P,
\end{equation}
such that 
\begin{equation}
\hat P\hat U_i^\dag\hat U_j\hat P=\delta_{ij}\hat P.
\end{equation}
Then, one can map the state $\hat V_i\ket{\psi’}$ back to $\mathbb H_\text{C}$ by selecting the corresponding recovery Kraus operator $\hat R_i=\hat V_i^\dag$.

An important note is the following. Let be $\mathcal R$ the recovery map for $\mathcal E$. Then, one has
\begin{equation}
\begin{aligned}
\mathcal E(\hat \rho)&=\sum_i\hat E_i\hat \rho\hat E_i^\dag,\\
\mathcal R(\hat \rho)&=\sum_k\hat R_k\hat \rho\hat R_k^\dag,
\end{aligned}
\end{equation}
such that $\hat R_k\hat E_i=\alpha_{ki}\hat {\mathbb 1}$. We define the map $\mathcal D$ as
\begin{equation}
\mathcal D(\hat \rho)=\sum_i\hat D_i\hat \rho\hat D_i^\dag,
\end{equation}
with $\hat D_i$ appertaining to the span of $\set{\hat E_i}_i$, i.e.~$\hat D_i=\sum_j c_{ij}\hat E_j$. Then, the map $\mathcal D$  can be recovered with the same recovery map $\mathcal R$.

\begin{figure}[h]
    \centering
    \includegraphics[width=0.4\linewidth]{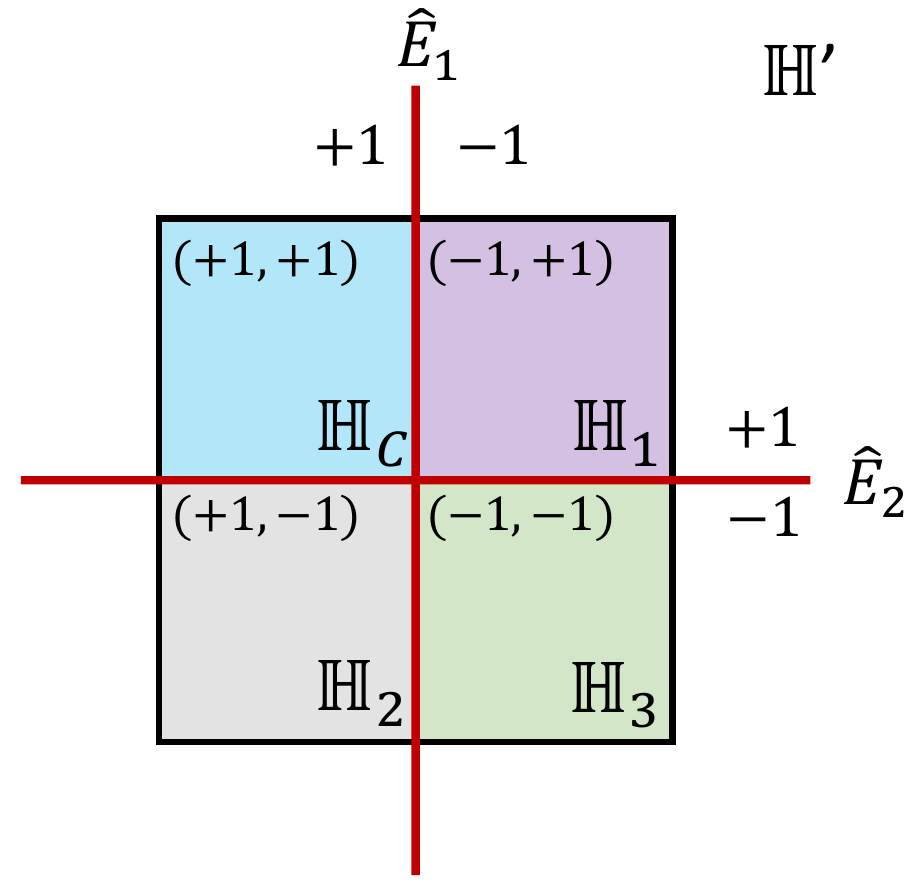}
    \caption{Division of the Hilbert space $\mathbb H'$ with respect to the subspaces defined by the eigeinvalues of $\hat E_1$ and $\hat E_2$.}
    \label{fig.2qubit}
\end{figure}

Consider the case of $n$ qubits . We want to construct the recovery map for errors due to the application of the Pauli operators. These are $\set{\hat {\mathbb 1},\hat \sigma_x,\hat \sigma_y,\hat \sigma_z}^{\otimes n}$, and they form the basis of $\mathcal B(\mathbb H^{\otimes n})$.
The operator $\hat {\mathbb 1}$ is the identity, so is not associated to any error. The operator $\hat \sigma_y=i\hat \sigma_x\hat \sigma_z$. So, one needs to construct the recovery map $\mathcal R$ that corrects only errors due to $\hat \sigma_x$ and  $\hat \sigma_z$. 
Now, for every Pauli operator different from the identity, we have two important properties: $\Tr{\hat \sigma_i}=0$ and $\hat \sigma_i^2=\hat {\mathbb 1}$. They imply that their eigenvalues are $\pm1$. Thus, one can divide the full Hilbert space $\mathbb H’=\mathbb H$, with  $\operatorname{dim}(\mathbb H)=2^{n}$, in two subspaces (of the same dimension), which are associated to the corresponding eigeinvalues, see Fig.~\ref{fig.2qubit}. Namely, given $\hat\sigma_i$ we have $\mathbb H_{\sigma_i=1}$ and $\mathbb H_{\sigma_i=-1}$, with $\operatorname{dim}(\mathbb H_i)=2^{n-1}$, whose union gives $\mathbb H’$. Suppose the code space $\mathbb H_\text{C}$ is defined in terms of the operators $\hat E_1$ and $\hat E_2$ as it follows: $\forall\ket\psi\in\mathbb H_\text{C}$, one has 
\begin{equation}
\hat E_1\ket{\psi}=+1\ket{\psi},\quad \text{and}\quad\hat E_2\ket{\psi}=+1\ket{\psi}.
\end{equation}
Any other combination identifies an error subspace $\mathbb H’_i$.

More in general, given the operators in $\mathcal B(\mathbb H)$, they are part of one of the following families:
\begin{itemize}
\item[-] They specify the subspaces we are going to divide $\mathbb H’$.
\item[-] They are responsible for errors. So they describe how a state moves from $\mathbb H_\text{C}$ to $\mathbb H’_i$.
\item[-] They are responsible for logical operations within the individual subspaces $\mathbb H_\text{C}$ and $\mathbb H_i$.
\end{itemize}

\subsection{Stabilisers}\index{Stabilisers}

For $n$ qubits, one has $4\times4^n$ Pauli operators. The first factor $4$ accounts for the four relevant phases $\set{+1,-1,+i,-i}$, while the rest count the operators being the basis of the operators acting on a $2^n$ dimensional space. We define as stabilisers  the elements of an abelian subgroup which is responsible of the division of $\mathbb H’$ in subspaces. Namely, there will be one code space $\mathbb H_\text{C}$ and the other are the error spaces.

\begin{figure}
    \centering
    \includegraphics[width=0.4\linewidth]{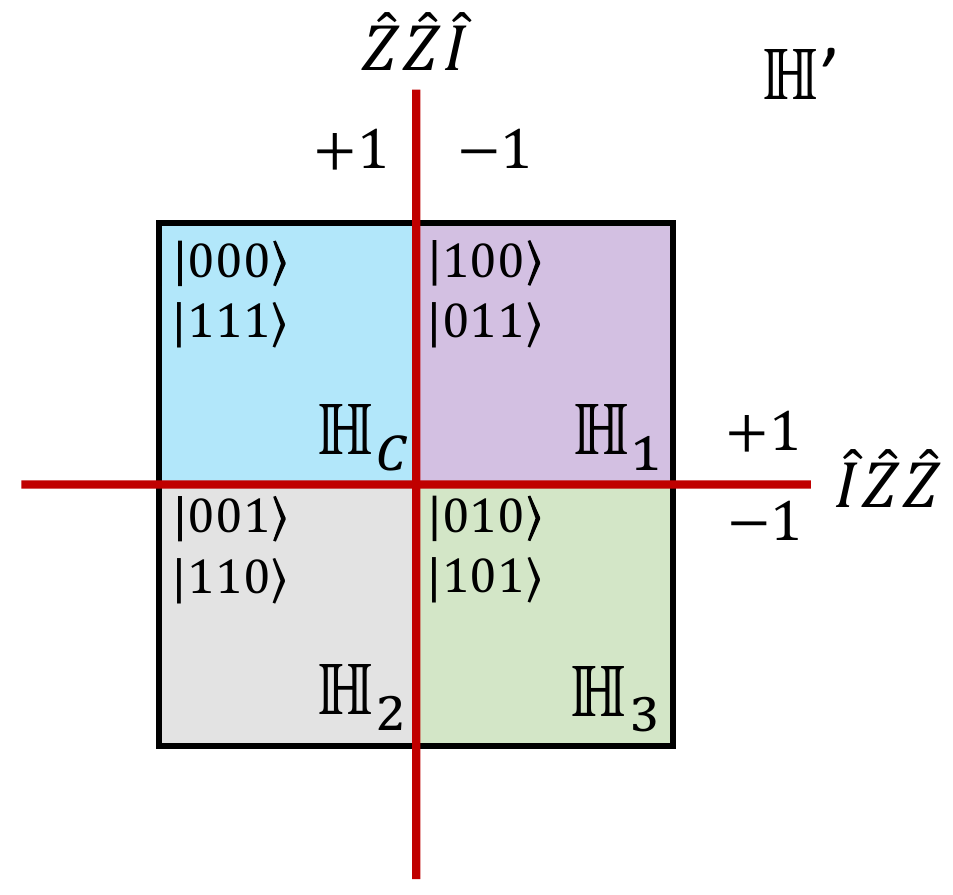}
    \caption{Division of the Hilbert space $\mathbb H'$ with respect to the subspaces defined by the eigeinvalues of $\hat \sigma_z\hat \sigma_z\hat {\mathbb 1}$ and $\hat {\mathbb 1}\hat \sigma_z\hat \sigma_z$.}
    \label{fig.3qubit}
\end{figure}

\begin{myexample}
Consider the case of $n=3$. The stabilisers are 
\begin{equation}
\text{(Stabilisers for $n=3$)}=\set{\hat {\mathbb 1}\hat {\mathbb 1}\hat {\mathbb 1},\hat \sigma_z\hat \sigma_z\hat {\mathbb 1},\hat {\mathbb 1}\hat \sigma_z\hat \sigma_z,\hat \sigma_z\hat {\mathbb 1}\hat \sigma_z}.
\end{equation}
Consider $\hat \sigma_z\hat \sigma_z\hat {\mathbb 1}$. It determines the parity of the first and the second qubit. Thus, it divides the Hilbert space $\mathbb H’$ in two parts, one associated to its $+1$ eigeinvalue and to its $-1$ eigeinvalue:
\begin{equation}
\begin{array}{ c| c  }
+1&-1\\
\midrule
\ket{000}&\ket{100}\\
\ket{111}&\ket{011}\\
\ket{001}&\ket{010}\\
\ket{110}&\ket{101}
\end{array}
\end{equation}
A similar division can be done considering the operator $\hat {\mathbb 1}\hat \sigma_z\hat \sigma_z$, for which we have
\begin{equation}
\begin{array}{ c| c  }
+1&-1\\
\midrule
\ket{000}&\ket{001}\\
\ket{111}&\ket{110}\\
\ket{100}&\ket{010}\\
\ket{011}&\ket{101}
\end{array}
\end{equation}
We notice that there are no other possible partitions of $\mathbb H’$. Indeed, the last non-trivial stabilisers is $\hat \sigma_z\hat {\mathbb 1}\hat \sigma_z$ can be expressed as the product of the other two:
\begin{equation}
\hat \sigma_z\hat {\mathbb 1}\hat \sigma_z=(\hat \sigma_z\hat {\mathbb 1}\hat \sigma_z)(\hat \sigma_z\hat \sigma_z\hat {\mathbb 1}).
\end{equation}
To be specific, the operators $\hat \sigma_z\hat {\mathbb 1}\hat \sigma_z$ and $\hat \sigma_z\hat \sigma_z\hat {\mathbb 1}$ are the generators of the abelian subgroup of the stabilisers. 

Now, we can define the code space $\mathbb H_\text{C}$ as that associated to the $+1$ eigeinvalues for all the stabilisers. Namely, this is the subspace of $\mathbb H’$ which is spanned by the +1 eigeinstates of all the generators of the abelian subgroup:
\begin{equation}
\mathbb H_\text{C}=\operatorname{span}(\ket{000},\ket{111}).
\end{equation}
The partitioning of $\mathbb H’$ is represented graphically in Fig.~\ref{fig.3qubit}.

\end{myexample}

Now, consider $\hat E_i$ being one of the $4\times4^n$ Pauli operators not being one of the stabilisers $S_k$. Now, since it is constructed as the product of single qubit Pauli operators, $\hat E_i$ can only commute or anticommute with the stabilisers $S_k$. 
\begin{itemize}
\item Assume that it commutes: $\com{\hat E_i}{\hat S_k}=0$. Then, we have that for any $\ket\psi\in\mathbb H_\text{C}$, it holds
\begin{equation}
\hat S_k\hat E_i\ket{\psi}=\hat E_i\hat S_k\ket{\psi}=\hat E_i\ket{\psi},
\end{equation}
where the last equality follows from the fact that $S_k$ is a stabiliser and thus acts as an identity on $\mathbb H_\text{C}$. Then, if $\hat E_i$ commutes stabiliser $\hat S_k$, it is associated to the eigeinvalue +1 of the latter. Indeed, the state $\ket{\phi_i}=\hat E_i\ket\psi$ is associated to the +1 eigeinvalue of $\hat S_k$.
\item Conversely, if it anticommutes: $\acom{\hat E_i}{\hat S_k}=0$, then
\begin{equation}
\hat S_k\hat E_i\ket{\psi}=-\hat E_i\hat S_k\ket{\psi}=-\hat E_i\ket{\psi}.
\end{equation}
In such a case, one says that $\hat E_i$ is associated to the -1 eigeinvalue of $S_k$.
\end{itemize}

\begin{myexample}
In the case of $n=3$ one has that the operators $\hat 	\sigma_x\hat {\mathbb 1}\hat {\mathbb 1}$ and $\hat 	\sigma_x\hat 	\sigma_x\hat 	\sigma_x$ are associated to the eigenvalues of $\hat \sigma_z\hat \sigma_z\hat {\mathbb 1}$ and $\hat {\mathbb 1}\hat \sigma_z\hat \sigma_z$ as
\begin{equation}
\begin{array}{c|c|c}
&\hat \sigma_z\hat \sigma_z\hat {\mathbb 1}&\hat {\mathbb 1}\hat \sigma_z\hat \sigma_z\\
\midrule
\hat 	\sigma_x\hat {\mathbb 1}\hat {\mathbb 1}&-1&+1\\
\hat 	\sigma_x\hat 	\sigma_x\hat 	\sigma_x&+1&+1
\end{array}
\end{equation}
Now, $\hat 	\sigma_x\hat 	\sigma_x\hat 	\sigma_x$ commutes with both the (generators of the subgroup of) stabilisers. Thus, it means that if
\begin{equation}
 \ket\psi\in\mathbb H_\text{C}, \quad\text{then}\quad  \hat 	\sigma_x\hat 	\sigma_x\hat 	\sigma_x\ket\psi\in\mathbb H_\text{C}.
\end{equation}
Namely, it is a normaliser (see below) and it acts as a logical $\hat \sigma_x$.

Conversely, $\hat 	\sigma_x\hat {\mathbb 1}\hat {\mathbb 1}$ anticommutes with $\hat \sigma_z\hat \sigma_z\hat {\mathbb 1}$. This means that given a state
\begin{equation}
 \ket\psi\in\mathbb H_\text{C}, \quad\text{then}\quad  \hat 	\sigma_x\hat {\mathbb 1}\hat {\mathbb 1}\ket\psi\in\mathbb H’_1\cup\mathbb H’_3,
\end{equation}
where $\mathbb H’_1$ and $\mathbb H’_3$ are respectively associated to the eigeinvalues $(-1,+1)$ and  $(-1,-1)$ of $(\hat \sigma_z\hat \sigma_z\hat {\mathbb 1},\hat {\mathbb 1}\hat \sigma_z\hat \sigma_z)$.
However, since $\hat 	\sigma_x\hat {\mathbb 1}\hat {\mathbb 1}$  commutes with $\hat {\mathbb 1}\hat \sigma_z\hat \sigma_z$, then $\hat 	\sigma_x\hat {\mathbb 1}\hat {\mathbb 1}\ket\psi\in\mathbb H’_1$.

Namely, the operator $\hat 	\sigma_x\hat {\mathbb 1}\hat {\mathbb 1}$ is one of the $\hat V_i$ errors that maps the states from the code space to the corresponding $\mathbb H’_i$.

\end{myexample}

If we have $k$ qubits that are encoded in $n$ qubits, with $n>k$, then we need $(n-k)$ generators from the stabilisers to define the partitions. By starting with $\operatorname{dim}(\mathbb H’)=2^n$, since each generator divides the Hilbert space in two parts, we have $2^{n-k}$ different subspaces of dimension $2^k$.

\subsection{Normalisers and Centralisers}\index{Normalisers}

There are Pauli operators that commute with all the elements of the stabilisers, but do not appartain to the stabilisers subgroup. These are the normalisers $\hat N_k$. They respect the partition of $\mathbb H$, meaning that they do not map states from the code space $\mathbb H_\text{C}$ to an error space $\mathbb H’_i$, and act non-trivially in the code space. They are defined via
\begin{equation}
\hat N_k\hat S_i\hat N_k^\dag=\hat S_j,
\end{equation}
where $\hat S_i$ are the stabilisers. If $i=j$ they are called centralisers, while for $i\neq j$ they are normalisers. In the particular case of the Pauli algebra, meaning that all the operators are generated by the product of Pauli operators, one has that the normalisers are centralisers. Indeed, 
\begin{equation}
\hat N_k\hat S_i\hat N_k^\dag=\pm\hat N_k\hat N_k^\dag\hat S_i=\pm\hat S_i,
\end{equation}
since Pauli operators can only commute or anticommute. However, given the stabilisers $\hat S_i$, the operator $-\hat S_i$ does not stabilise the code space. Thus, the $-$ sign cannot be accepted and one gets
\begin{equation}
\hat N_k\hat S_i\hat N_k^\dag=\hat S_i=\hat S_j.
\end{equation}
Thus, they are all centralisers.

\begin{myexample}
Consider the case of $n=3$. The operator $\hat\sigma_x\hat\sigma_x\hat\sigma_x$ acts as follows:
\begin{equation}
\begin{aligned}
\ket{000}&\xrightarrow[]{\hat\sigma_x}\ket{111},\\
\ket{111}&\xrightarrow[]{\hat\sigma_x}\ket{000}.
\end{aligned}
\end{equation}
Thus, it acts as a logical $\hat\sigma_x$. The same happens for $\hat\sigma_x\hat\sigma_x\hat\sigma_x\hat S_k$ for any stabiliser $\hat S_k$. Indeed, for $\ket\psi\in\mathbb H_\text{C}$, we have that 
\begin{equation}
\hat\sigma_x\hat\sigma_x\hat\sigma_x\hat S_k\ket\psi=\hat\sigma_x\hat\sigma_x\hat\sigma_x\ket\psi,
\end{equation}
since $\hat S_k$ acts as a logical identity on $\mathbb H_\text{C}$. Suppose we take the stabiliser $\hat S_k=\hat\sigma_z\hat\sigma_z\hat{\mathbb 1}$, then
\begin{equation}
(\hat\sigma_x\hat\sigma_x\hat\sigma_x)(\hat\sigma_z\hat\sigma_z\hat{\mathbb 1})=-\hat\sigma_y\hat\sigma_y\hat\sigma_x,
\end{equation}
which also acts as a logical $\hat \sigma_x$.

The normalisers contain $\hat\sigma_x\hat\sigma_x\hat\sigma_x$, $-\hat\sigma_y\hat\sigma_y\hat\sigma_y$, $\hat\sigma_z\hat\sigma_z\hat\sigma_z$, and all the products of these with all the stabilisers $\hat S_k$, which act as a logical identity. 
\begin{equation}
\begin{array}{c|c}
\text{Physical operation}&\text{Logical operation}\\
\midrule
\hat{\mathbb 1}\hat{\mathbb 1}\hat{\mathbb 1}&\hat{\mathbb 1}\\
\hat\sigma_z\hat\sigma_z\hat{\mathbb 1}&\hat{\mathbb 1}\\
\hat\sigma_z\hat{\mathbb 1}\hat\sigma_z&\hat{\mathbb 1}\\
\hat{\mathbb 1}\hat\sigma_z\hat\sigma_z&\hat{\mathbb 1}\\
\midrule
\hat\sigma_x\hat\sigma_x\hat\sigma_x&\hat\sigma_x\\
\hat\sigma_x\hat\sigma_x\hat\sigma_x\hat S_k&\hat\sigma_x\\
\midrule
-\hat\sigma_y\hat\sigma_y\hat\sigma_y&\hat\sigma_y\\
-\hat\sigma_y\hat\sigma_y\hat\sigma_y\hat S_k&\hat\sigma_y\\
\midrule
\hat\sigma_z\hat\sigma_z\hat\sigma_z&\hat\sigma_z\\
\hat\sigma_z\hat\sigma_z\hat\sigma_z\hat S_k&\hat\sigma_z
\end{array}
\end{equation}

\end{myexample}

\subsection{Stabiliser code}\label{sec.stabilisercode}

Consider the three qubits encoding a single logical qubit. The stabilisers are:
\begin{equation}
\hat{\mathbb 1}\hat{\mathbb 1}\hat{\mathbb 1},\quad\hat\sigma_z\hat\sigma_z\hat{\mathbb 1},\quad\hat\sigma_z\hat{\mathbb 1}\hat\sigma_z,\quad\hat{\mathbb 1}\hat\sigma_z\hat\sigma_z.
\end{equation}
Among the possible errors, there are some that are more and less likely to occur. Under the assumption of errors that act independently on the qubits, the error $\hat{\mathbb 1}\hat \sigma_x\hat{\mathbb 1}$ is more likely to occur than $\hat \sigma_x\hat \sigma_x\hat \sigma_x$. The first has weight 1 (only one operator different from the identity), while the second has weight 3.

Now, the question is which are the errors that can be corrected, and eventually how they can be corrected. As we already saw, the errors where only one of the qubits is modified can be corrected (see bit-flip, phase-flip and 9-qubit Shor QEC codes). These can be corrected via the application of the recovery operator $\hat R_k=\hat V_k^\dag$, so that
\begin{equation}
\hat R_k\hat V_k=\hat V_k^\dag\hat V_k=\hat{\mathbb 1}.
\end{equation}
However, the operator $\hat R_k$ can correct for a much wider class of operators. Indeed, given a state $\ket\psi\in\mathbb H_\text{C}$ and a stabiliser $\hat S_i$, one has
\begin{equation}
\hat R_k(\hat V_k\hat S_i)\ket\psi=\hat R_k\hat V_k\ket\psi=\ket\psi.
\end{equation}
Thus, $\hat R_k$ can correct also errors of the form of a correctable error multiplied by a stabiliser, i.e.~$\hat V_k\hat S_i$.

Conversely, an error in the class of normalisers which is not a stabiliser is a non-correctable error. Indeed, it acts non-trivially on the code space.

\section{Surface code}\index{Surface code}

\begin{figure}[h]
    \centering
    \includegraphics[width=0.7\linewidth]{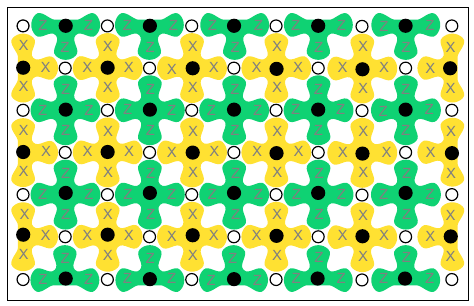}
    \caption{Graphical representation of the $L^2$ physical qubits (full and open circles) in the array generating the surface code through their interaction (green and yellow connections).}
    \label{fig:surfacecode}
\end{figure}

The surface code is a QEC code that is related to topology.
The idea is that a logical qubit is encoded in $L\times L $ physical qubits as in the layout presented in Fig.~\ref{fig:surfacecode}. The array they construct has to be considered with periodic boundary conditions.
The $L^2$ qubits are divided in two classes:
\begin{itemize}
    \item Half of the physical qubits are used as data qubits: they store quantum states $\ket{\psi_\text{L}}$ that will be used for computation. They are represented with open circles \tikzcircle[fill=white]{2pt}.
    \item Half of the physical qubits are called measurement qubits and they are employed as error detecting qubits. They are represented with full circles \tikzcircle[fill=black]{2pt}. There are two type of measurement qubits:
    \begin{itemize}
        \item Measure Z or Z-syndrome qubits, which are represented in green,
        \item Measure X or X-syndrome qubits, which are represented in yellow.
    \end{itemize}
\end{itemize}
Each data qubit is coupled to two X-syndrome and two Z-syndrome qubits. Each measurement qubit is coupled to four data qubits. These couplings are describe as the following. For the green block, i.e.~the Z-syndrome, we have 
\begin{equation}
\begin{gathered}
\includegraphics[width=0.2\linewidth]{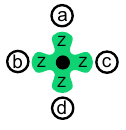}
\end{gathered}
=
    \begin{tikzcd}
        \lstick[4]{$\ket{\psi_{abcd}}$}                   &\ctrl{4}   &           &           &           &\\
                           &           &\ctrl{3}   &           &           &\\
                           &           &           &\ctrl{2}   &           &\\
                           &           &           &           &\ctrl{1}   &\\
          \ket{0_\text{\tikzcircle[fill=black]{2pt}}}      &\targ{}    &\targ{}    &\targ{}    &\targ{}    &\meter{}
    \end{tikzcd}
\end{equation}
where $ \ket{0_\text{\tikzcircle[fill=black]{2pt}}}$ indicates that the qubit \tikzcircle[fill=black]{3pt} has been initiallised in $\ket0$.
Similarly, for the yellow block one has 
\begin{equation}
\begin{gathered}
\includegraphics[width=0.2\linewidth]{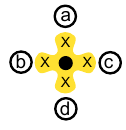}
\end{gathered}
=
    \begin{tikzcd}
        \lstick[4]{$\ket{\psi_{abcd}}$}   &                &\targ{}   &           &           &         &  &\\
                   &        &           &\targ{}   &           &       &    &\\
                    &       &           &           &\targ{}   &       &    &\\
                     &      &           &           &           &\targ{}&   &\\
          \ket{0_\text{\tikzcircle[fill=black]{2pt}}}  &\gate{H}    &\ctrl{-4}    &\ctrl{-3}    &\ctrl{-2}    &\ctrl{-1} &\gate{H}   &\meter{}
    \end{tikzcd}
\end{equation}
In such a way,  the measurement qubits are coupled to the data qubits.
These circuits are run in cycles, between one logical operation and the following one, so to keep track of the errors that occur in between.

To understand the logic of the surface code, let us focus on the case where one has only four physical qubits: 2 are data qubits and 2 are measurement qubits (one in X and one in Z).  The generators of the stabilisers are the operators $\hat X_a\hat X_b$ and $\hat Z_a\hat Z_b$. Here, we employ the notation $\hat Z_i$ to identify the $\hat \sigma_z^{(i)}$ Pauli operator acting on the $i$-th physical qubit. One can easily show that these operators commute, i.e. $\com{\hat X_a\hat X_b}{\hat Z_a\hat Z_b}=0$, although they do not at the level of single qubit, i.e.~$\com{\hat X_i}{\hat Z_i}\neq0$. Thus, they have common eigeinstates, which identify the division of the Hilbert space $\mathbb H'$ in the code and error subspaces (for the sake of simplicity, we will drop all the normalisation constants):
\begin{equation}
\begin{array}{ c| c|c  }
\ket\psi&\hat X_a\hat X_b&\hat Z_a\hat Z_b\\
\midrule
\ket{00}+\ket{11}&+1&+1\\
\ket{00}-\ket{11}&-1&+1\\
\ket{01}+\ket{10}&+1&-1\\
\ket{01}-\ket{10}&-1&-1
\end{array}
\end{equation}
The circuit that applies these two stabilisers is 
\begin{equation}\label{circ.surface2}
\begin{gathered}
\includegraphics[width=0.15\linewidth]{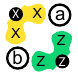}
\end{gathered}
=
    \begin{tikzcd}
          \lstick[1]{X}&\wireoverride{n}&\wireoverride{n}\ket{0_\text{\tikzcircle[fill=black]{2pt}}} &       \gate{H}         &\ctrl{1}   &    \ctrl{2}       &           &         & \gate{H} &\meter{}\\
          \lstick[1]{a}&\wireoverride{n}&\wireoverride{n}\lstick[2]{$\ket{\psi_{ab}}$}         &        &      \targ{}     &   &    \ctrl{2}       &       &    &&\rstick[2]{$\ket{\psi_{ab}'}$}\\
         \lstick[1]{b} &\wireoverride{n} &\wireoverride{n}                             &       &           &     \targ{}      &   &   \ctrl{1}    &    &&\\
          \lstick[1]{Z}&\wireoverride{n}&\wireoverride{n}\ket{0_\text{\tikzcircle[fill=black]{2pt}}}  &    &  &    &\targ{}     &\targ{}  &  &\meter{}
    \end{tikzcd}
\end{equation}
where the first qubit is the X-syndrome and the last is the Z-syndrome.
Considering the generic state $\ket{\psi_{ab}}$ for the qubits $a$ and $b$ being
\begin{equation}\label{eq.state0.surface2}
    \ket{\psi_{ab}}=A\ket{00}+B\ket{01}+C\ket{10}+D\ket{11},
\end{equation}
one can input this state in the circuit in Eq.~\eqref{circ.surface2} and, before the measurements, obtains that the total state reads
\begin{equation}\label{eq.state.abcd}
\begin{aligned}
    \ket{\Psi_{XabZ}}&=(A+D)\ket0(\ket{00}+\ket{11})\ket0\\
    &+(A-D)\ket1(\ket{00}-\ket{11})\ket0\\
    &+(B+C)\ket0(\ket{01}+\ket{10})\ket1\\
    &+(B-C)\ket1(\ket{01}-\ket{10})\ket1.
\end{aligned}
\end{equation}
It follows that, after the measurement of the X and Z-syndrome qubits, one obtains --- depending on the outcomes $\set{M_X,M_Z}$ --- the following states with the corresponding probabilities $P_{\ket{\psi_{ab}'}}$:
\begin{equation}
\begin{array}{ c| c|c  }
\set{M_X,M_Z}&\ket{\psi_{ab}'}&P_{\ket{\psi_{ab}'}}\\
\midrule
\set{+1,+1}&\ket{00}+\ket{11}&|A+D|^2\\
\set{-1,+1}&\ket{00}-\ket{11}&|A-D|^2\\
\set{+1,-1}&\ket{01}+\ket{10}&|B+C|^2\\
\set{-1,-1}&\ket{01}-\ket{10}&|B-C|^2
\end{array}
\end{equation}
After the collapse on one of these common eigenstates of $\hat X_a\hat X_b$ and $\hat Z_a\hat Z_b$, subsequent applications of the circuit in Eq.~\eqref{circ.surface2} will provide always --- in the assumption of no noise --- the same state. 

\begin{myexample}
Consider the state in Eq.~\eqref{eq.state.abcd} and suppose the first cycle (which acts effectively as an encoding) provides the measurements $\set{M_X=-1, M_Z=-1}$ and $\ket{\psi_{ab}'}=\ket{01}-\ket{10}$. Now, $\ket{\psi_{ab}'}$ is equal to $\ket{\psi_{ab}}$ in Eq.~\eqref{eq.state0.surface2} when setting $A=D=0$ and $B=-C=1$. The corresponding output state at the end of the second cycle before the measurement will be $\ket{\Psi_{XabZ}}=\ket1(\ket{01}-\ket{10})\ket1$. This has two important implications: 1) the state $\ket{\psi_{ab}'}$ remains untouched by the circuit, which is the implementation of the stabilisers: $\hat S_2\hat S_1\ket{\psi_{ab}'}=\ket{\psi_{ab}'}$; 2) also the output of the measurement    $\set{M_X=-1, M_Z=-1}$ remains the same.
\end{myexample}

This example shows that, without measuring directly $\ket{\psi_{ab}'}$, one can use the output of the measurements to infer the state of the data qubits. Turning the argument upside-down, an error occuring in the data qubits will be identified by the change in the outcomes of the measurements.

\begin{figure}[h]
    \centering
    \includegraphics[width=0.2\linewidth]{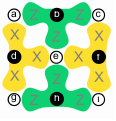}
    \caption{Graphical representation of the surface code with 5 data qubits and 4 measurement qubits.}
    \label{fig:reduced_surface}
\end{figure}

We notice that, in the case of 2 data qubits and 2 measurement qubits, one has
\begin{itemize}
    \item 4 degrees of freedom where to encode a logical state: there are 2 physical qubits having 2 dimensions. The total Hilbert space $\mathbb H'$ has $2\times2$ dimensions.
    \item 4 constrains from the stabilisers: we have 2 stabilisers and each divides $\mathbb H'$ in 2 subspaces.
\end{itemize}
Then, there are no free degrees of freedom where one can perform logical operations. In order to do that, one needs to impose the length of the array of physical qubits $L$ being an odd number $>1$. In such a way, one has 2 free degrees of freedom that can be employed. The minimal array is that having $L=3$ with 5 data qubits and 4 measurement qubits. Such an array in shown in Fig.~\ref{fig:reduced_surface}.
The circuit corresponding to the stabilisers application is given by
\begin{equation}\label{circ.surface5}
    \begin{tikzcd}
        \lstick[1]{a}&\wireoverride{n}&\wireoverride{n}&\wireoverride{n}\lstick[5]{$\ket{\psi_{acegi}}$}&&\ctrl{5}&&&\targ{}&&&&&&&&&&&\rstick[5]{$\ket{\psi_{acegi}'}$}\\
        \lstick[1]{c}&\wireoverride{n}&\wireoverride{n}&\wireoverride{n}&&&\ctrl{4}&&&&&\targ{}&&&&&&&&\\
        \lstick[1]{e}&\wireoverride{n}&\wireoverride{n}&\wireoverride{n}&&&&\ctrl{3}&&\targ{}&&&\targ{}&&\ctrl{6}&&&&&\\
        \lstick[1]{g}&\wireoverride{n}&\wireoverride{n}&\wireoverride{n}&&&&&&&\targ{}&&&&&\ctrl{5}&&&&\\
        \lstick[1]{i}&\wireoverride{n}&\wireoverride{n}&\wireoverride{n}&&&&&&&&&&\targ{}&&&\ctrl{4}&&&\\
        \lstick[1]{b,Z}&\wireoverride{n}&\wireoverride{n}&\wireoverride{n}\ket{0_\text{\tikzcircle[fill=black]{2pt}}}&&\targ{}&\targ{}&\targ{}&&&&&&&&&&&\meter{}\\
        \lstick[1]{d,X}&\wireoverride{n}&\wireoverride{n}&\wireoverride{n}\ket{0_\text{\tikzcircle[fill=black]{2pt}}}&\gate{H}&&&&\ctrl{-6}&\ctrl{-4}&\ctrl{-3}&&&&&&&\gate{H}&\meter{}\\
        \lstick[1]{f,X}&\wireoverride{n}&\wireoverride{n}&\wireoverride{n}\ket{0_\text{\tikzcircle[fill=black]{2pt}}}&\gate{H}&&&&&&&\ctrl{-6}&\ctrl{-5}&\ctrl{-3}&&&&\gate{H}&\meter{}\\
        \lstick[1]{h,Z}&\wireoverride{n}&\wireoverride{n}&\wireoverride{n}\ket{0_\text{\tikzcircle[fill=black]{2pt}}}&&&&&&&&&&&\targ{}&\targ{}&\targ{}&&\meter{}
    \end{tikzcd}
\end{equation}
Suppose we input the state $\ket{\psi_{acegi}}=\ket{00000}$. Then, the action of this circuit before the measurement is given by
\begin{equation}
    \ket{00000,0000}\xrightarrow[]{\text{circuit in Eq.~\eqref{circ.surface5}}}
    \begin{aligned}
&(\ket{00000}+\ket{10110}+\ket{01101}+\ket{11011})\ket{0000}\\
+&(\ket{00000}-\ket{10110}+\ket{01101}-\ket{11011})\ket{0100}\\
+&(\ket{00000}-\ket{10110}-\ket{01101}+\ket{11011})\ket{0010}\\
+&(\ket{00000}-\ket{10110}-\ket{01101}+\ket{11011})\ket{0110},
    \end{aligned}
\end{equation}
where we expressed the states in the form $\ket{\psi_{acegi},\phi_{bdfh}}$. Suppose the measurement  results in $\set{M_b,M_d,M_f,M_h}=\set{+1,+1,-1,+1}$, which corresponds to the measurement  state $\ket{\phi_{bdfh}}=\ket{0010}$. Then, the data state is given by 
\begin{equation}\label{eq.data.surface5}
    \ket{\text{data}}=\ket{\psi_{acegi}'}=\ket{00000}-\ket{10110}-\ket{01101}+\ket{11011},
\end{equation}
which remains untouched by subsequent applications of the circuit in Eq.~\eqref{circ.surface5}. This is an easy but lengthy computation if performed in terms of states. However, it becomes trivial and immediate if considering that the circuit corresponds to the application of the stabilisers $\hat S_4\hat S_3\hat S_2\hat S_1$ on the state $\ket{\text{data}}$, which has been already stabilised by the same circuit. Thus, $\hat S_4\hat S_3\hat S_2\hat S_1\ket{\text{data}}=\ket{\text{data}}$.

\subsection{Detecting errors}

There are several kinds of errors that can be detected with surface code.  For the sake of simplicity, we consider the case of the array with 5 data and 4 measurement qubits, and that the logical state is encoded in $\ket{\text{data}}$ shown in Eq.~\eqref{eq.data.surface5}.  The latter corresponds to the measurement state $\ket{0010}$, i.e.~to the measurement outcomes $\set{M_b,M_d,M_f,M_h}=\set{+1,+1,-1,+1}$. We construct the table of outcomes with respect to the number of cycles that are performed. In the case of no errors and no logical operations, such table reads
\begin{equation}
\begin{array}{c|c|c|c|c}
\text{\# cycles} &M_b&M_d&M_f&M_h  \\
\midrule
1 &+1&+1&-1&+1 \\
2 &+1&+1&-1&+1 \\
3 &+1&+1&-1&+1 \\
4 &+1&+1&-1&+1 \\
5 &+1&+1&-1&+1 \\
\vdots&\vdots&\vdots&\vdots&\vdots
    \end{array}
\end{equation}

We now introduce errors, that will appear exactly at the third cycle. There are also other relevant errors, but we only focuses on the following two kinds.
\begin{itemize}
    \item[1)] Errors on the measurement or on the syndrome qubits;

  These are the errors due to the erroneous output of a measurement $M_i$, or errors that are applied to the syndrome qubit. The latter will appear as the former.  Suppose we have an error on the measurement of the $f$ qubit. Then, the above table becomes
\begin{equation}
\begin{array}{c|c|c|c|c}
\text{\# cycles} &M_b&M_d&M_f&M_h  \\
\midrule
1 &+1&+1&-1&+1 \\
2 &+1&+1&-1&+1 \\
3 &+1&+1&\Circled[outer color=red]{+1}&+1 \\
4 &+1&+1&-1&+1 \\
5 &+1&+1&-1&+1 \\
\vdots&\vdots&\vdots&\vdots&\vdots
    \end{array}
\end{equation}
where we highlighted the difference between the two tables. The error is only momentaneous. Later applications of the cycle will erase the action of these kind of errors. Indeed, if the error is due to the random erroneous measurement outcome, then the following cycle will (most probably) provide the exact outcome. This will not be the case if there is a systematic error in the measurement process, which cannot be corrected. Conversely, if the error is caused by the application of an external action on the measurement qubit (say the surrounding environment acts with $\hat Z$ on the $f$ qubit), then the error vanishes in the next cycle since the qubit's state is initialised at the beginning of each cycle.\\

    \item[2)] Errors on the data qubits.
These are the errors on the data qubits that can be, for example, due to the surrounding environment, and that can corrupt the information encoded in the data state. It becomes then fundamental to being able to detect and account for such errors for the sake of computation.

Suppose we have a phase-flip error on the $e$ qubit. The state is then transformed as
\begin{equation}
    \ket{\psi'_{acegi}}\xrightarrow[]{\text{phase-flip error on qubit $e$}}\hat Z_e    \ket{\psi'_{acegi}}.
\end{equation}
Then, the Z-syndrome qubits will be unable to detect it. However, X-syndrome qubits $d$ and $f$ will detect the error: their coupling to the data qubit $e$ imposes the action of $\hat X_e$, which do not commute with $\hat Z_e$. Then, what happens is that the state of the X-syndrome qubits will flip. By supposing that the error occurs at the third cycle. Then, the table of outcomes becomes
\begin{equation}
\begin{array}{c|c|c|c|c}
\text{\# cycles} &M_b&M_d&M_f&M_h  \\
\midrule
1 &+1&+1&-1&+1 \\
2 &+1&+1&-1&+1 \\
3 &+1&\Circled[outer color=red]{-1}&\Circled[outer color=red]{+1}&+1 \\
4 &+1&\Circled[outer color=red]{-1}&\Circled[outer color=red]{+1}&+1 \\
5 &+1&\Circled[outer color=red]{-1}&\Circled[outer color=red]{+1}&+1 \\
\vdots&\vdots&\vdots&\vdots&\vdots
    \end{array}
\end{equation}
Importantly, the outcomes $M_d$ and $M_f$ change sign for any subsequent cycle (if no other errors or logical operations take place). This is how one can distinguish an error on the measurement and on the data qubits. The best way to account for this error is to employ a classical control software that will changes the sign of every subsequent measurement of that data qubit’s two adjacent X-syndrome qubits. \\

\end{itemize}

\begin{figure}[h]
    \centering
    \includegraphics[width=0.4\linewidth]{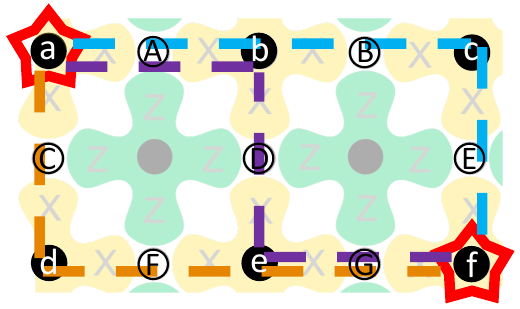}
    \caption{Graphical representation of $Z$ noises detected by the measurement qubits $a$ and $f$. Different paths (blue, purple and orange) can produce this error syndrome.}
    \label{fig:array erro surface}
\end{figure}

When one has an array of larger dimensions, then the situation is more complicated. For example, one might have that several data errors that form paths on the array. If this happens, the errors will be highlighted only by two syndrome qubits at the ends of the error path. An example is shown in Fig.~\ref{fig:array erro surface}, where the X-syndrome qubits $a$ and $f$ indicate that an error occurred. However, there is no indication about which path between $a$ and $f$ the Z-errors are covering.

This could be one among $ABE$ (blue path), $ADG$ (purpkle path) or $CFG$ (orange path). When accounting via classical control software for the errors, one needs to select one of these paths. The question is what happens if one selects the wrong path? The beauty of the surface code kicks in: as long as the error path and the selected one form a closed loop, the error is well accounted. This is shown with the following argument.  Suppose the error path is $ABE$, which is produced by $\hat Z_A\hat Z_B\hat Z_E$, and we select the path $CFG$ to be corrected, whose correction is given by $\hat Z_C\hat Z_F \hat Z_G$. This is however not a problem, indeed we have that
\begin{equation}
    (\hat Z_A\hat Z_B\hat Z_E)=(\hat Z_C\hat Z_F \hat Z_G)\hat S_e\hat S_d,
\end{equation}
 where we defined
the stabilisers
\begin{equation}
    \hat S_e=\hat Z_B\hat Z_E\hat Z_D\hat Z_G,\quad \text{and}\quad
    \hat S_d=\hat Z_A\hat Z_C\hat Z_D\hat Z_F.
\end{equation}
Therefore, the two errors are related by two stabilisers. This means, that recovery operator $\hat R_k$ that corrects $\hat Z_A\hat Z_B\hat Z_E$ can correct also for $\hat Z_C\hat Z_F \hat Z_G$. This has been discussed in Sec.~\ref{sec.stabilisercode}. Thus, every time we can form closed loops, the error can be accounted properly. These are harmless errors.

\begin{figure}[h]
    \centering
    \includegraphics[width=0.5\linewidth]{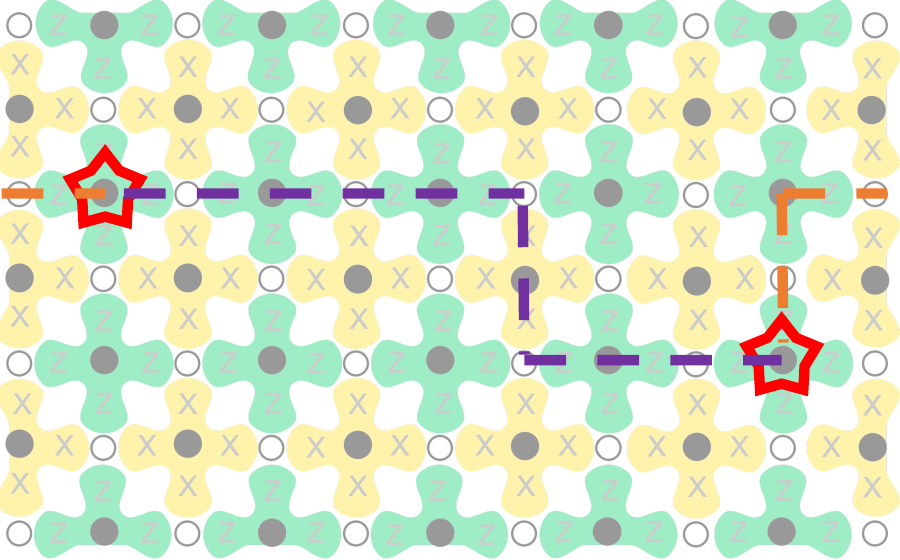}
    \caption{Graphical representation of an error path that crosses the boundary of the array (orange path). If corrected with the purple path, it leads to a logical operation, thus not correcting for the error.}
    \label{fig:array erro surface other errors}
\end{figure}

\begin{figure}[h]
    \centering
    \includegraphics[width=0.7\linewidth]{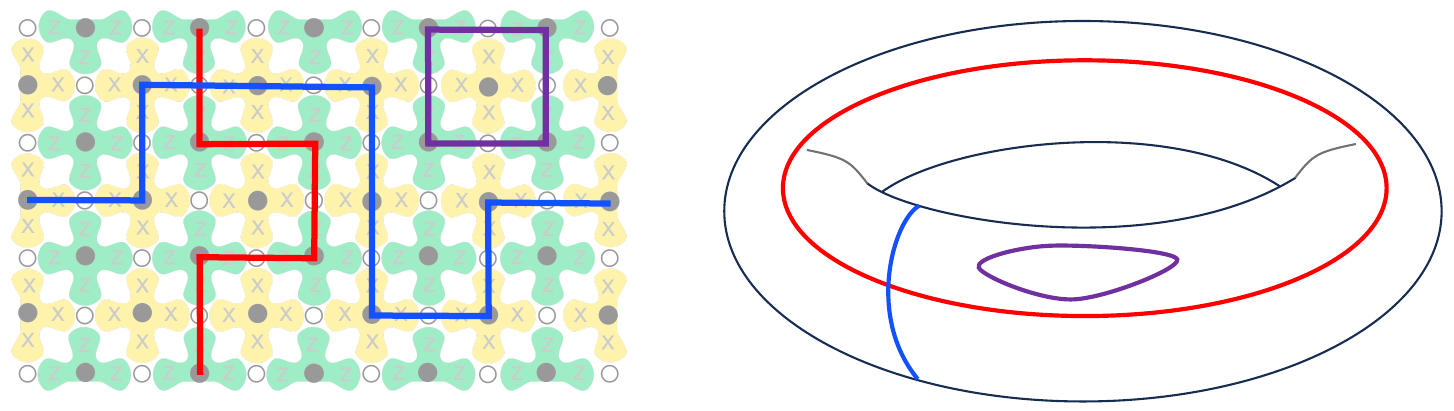}
    \caption{Graphical representation of the mapping of the array on a torus surface. The red path corresponds to a logical $X$, while the blue path to a logical $Z$ operation. These are harmful error. The purple path is an error that can be corrected. }
    \label{fig.torus}
\end{figure}

Conversely, consider now the case shown in Fig.~\ref{fig:array erro surface other errors}. 
Here the error path (shown in orange) crosses the boundary of the array, and due to the periodic boundary conditions only two syndrome qubits highlight the error path. In such a case, one would still be tempted to connect directly the syndrome qubits with a path fully in the array (shown in purple). However, such a correction is not the proper one. Indeed, the two paths would form a logical operation. 
To visualise harmless and harmful paths, one maps the array on a torus. If the path can be closed, then it is harmless. If the path cannot be closed, then it is harmful. Figure \ref{fig.torus} shows the mapping between the array and the torus, and highlights the harmless and harmful paths.

\section{Fault-tolerant computation}

Let us suppose we want to implement the following logical circuit 
\begin{equation}
    \begin{tikzcd}
        \ket0&\gate{H}&\ctrl{1}&\meter{}\\
        \ket0&&\targ{}&\meter{}
    \end{tikzcd}
\end{equation}
so that if an error occurs it can be successfully corrected via a QEC code. The error can essentially occur in any of the components of the circuit. Namely,
\begin{itemize}
    \item[-] in the state preparation,
    \item[-] in the logic quantum gates,
    \item[-] in the measurement,
    \item[-] in the simple transition of the quantum information along the quantum wires.
\end{itemize}
To combat the effect of the noise, one encodes the logical qubits in blocks of physical ones, using QEC codes. However, one needs also to replace the logical operations with encoded gates. Performing QEC periodically on the encoded states prevents the accumulation of errors in the state. However, it is not sufficient to prevent the build-up of errors, even if QEC is applied after each encoded gate.

There are two main  reasons:
\begin{itemize}
    \item[1)] The encoded gates can cause the propagation of errors.\\
    Let us consider a specific example, where two qubits are connected by a CNOT gate and a X error occurs on  the control qubit before the CNOT. Then, the error propagates also to the target qubit. This can be easily computed by considering that the CNOT is implemented by the unitary operator $\hat U_\text{\tiny CNOT}$. Then, one has
    \begin{equation}
        \hat U_\text{\tiny CNOT}\hat \sigma_x^{(1)}=\hat U_\text{\tiny CNOT}\hat \sigma_x^{(1)}\hat U_\text{\tiny CNOT}^\dag\hat U_\text{\tiny CNOT}=\hat \sigma_x^{(1)}\hat \sigma_x^{(2)}\hat U_\text{\tiny CNOT},
    \end{equation}
which can be graphically represented with
\begin{equation}
    \begin{tikzcd}
        &\gate{X}&\ctrl{1}&&\\
        &&\targ{}&\ghost{H}&
    \end{tikzcd}
    \rightarrow
        \begin{tikzcd}
        &&\ctrl{1}&\gate{X}&\\
        &&\targ{}&\gate{X}&
    \end{tikzcd}
\end{equation}
Then, one needs to design the encoded gates carefully, so that errors do not propagate on the entire block, but are limited to some physical qubits. In such a case the QEC code can remove these errors.\\
Performing encoded gates in such a way is a fault-tollerant (FT) procedure.

\item[2)] Also QEC can introduce errors.\\
An example is that graphically represented in Fig.~\ref{fig:array erro surface other errors}, where an error is not correctly recovered.

\end{itemize}
To showcase the FT procedure, we introduce the Steane code.

\subsection{Stean code or 7-qubit code}
\begin{figure}[h]
    \centering
    \includegraphics[width=0.6\linewidth]{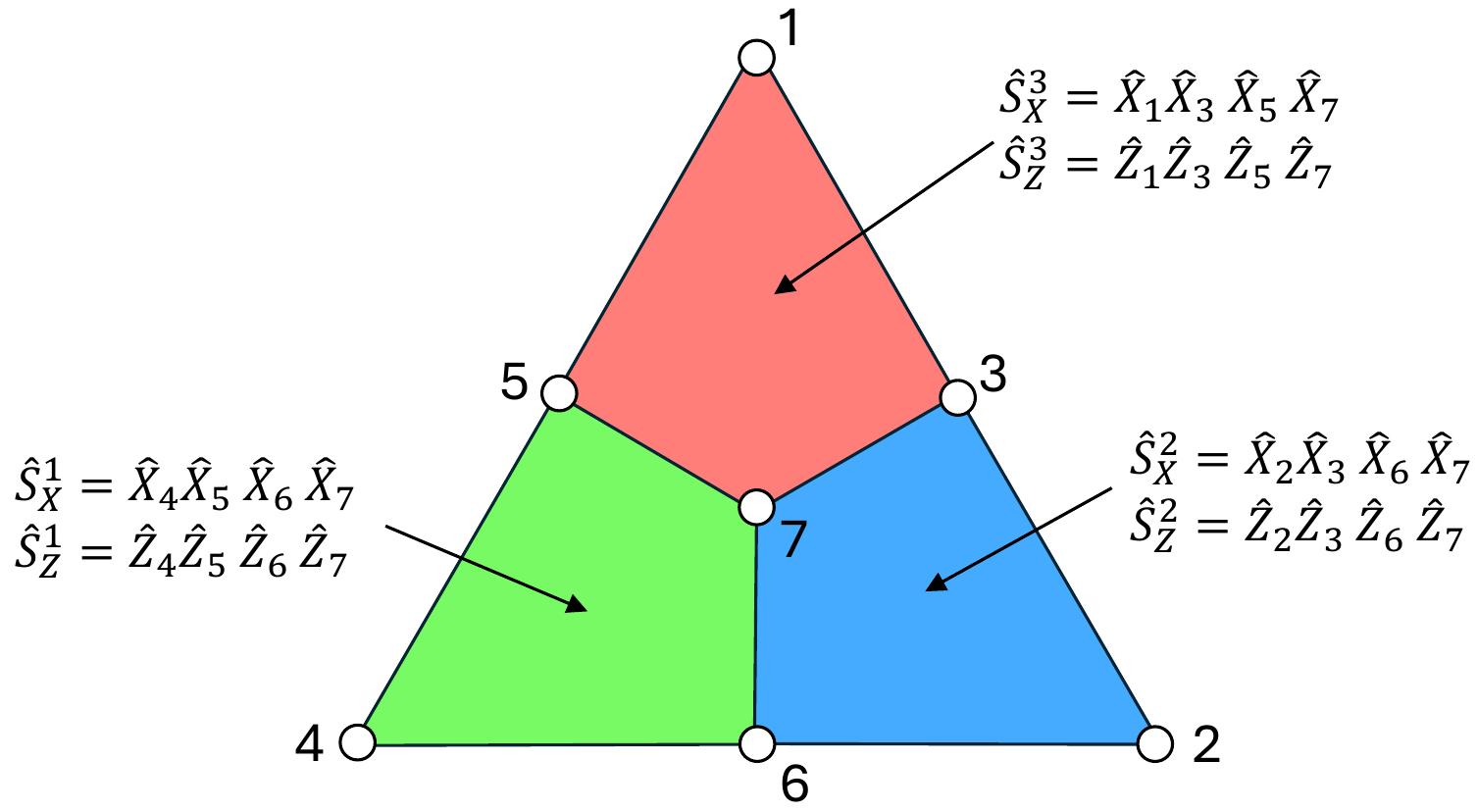}
    \caption{Graphical representation of the Steane code with the data qubit represented as open circles and the corresponding stabilisers.}
    \label{fig.steanecode}
\end{figure}

The Steane code is a stabiliser code employing 7 data qubit and 6 syndrome qubits for each logical qubit. The total Hilbert space $\mathbb H'$ has 128 dimensions that are divided in 64 subspaces of two dimensions. The graphical representation is given in Fig.~\ref{fig.steanecode}.
The generators of the stabilisers are
\begin{equation}
    \set{\hat S_k}_{k=1}^6
    =\left\{
    \begin{matrix}
 \hat{\mathbb 1}&\hat{\mathbb 1}&\hat{\mathbb 1}&\hat X&\hat X  &\hat X&\hat X , \\
 \hat{\mathbb 1}&\hat X&\hat X&\hat{\mathbb 1}&\hat{\mathbb 1}&\hat X&\hat X,\\
 \hat X&\hat{\mathbb 1}&\hat X&\hat{\mathbb 1}&\hat X&\hat{\mathbb 1}  &  \hat X,\\
 \hat{\mathbb 1}&\hat{\mathbb 1}&\hat{\mathbb 1}&\hat Z&\hat Z & \hat Z&\hat Z , \\
 \hat{\mathbb 1}&\hat Z&\hat Z&\hat{\mathbb 1}&\hat{\mathbb 1}&\hat Z&\hat Z,\\
 \hat Z&\hat{\mathbb 1}&\hat Z&\hat{\mathbb 1}&\hat Z&\hat{\mathbb 1}  &  \hat Z
 \end{matrix}
    \right\}=
    \left\{
    \begin{matrix}
    \hat X_4\hat X_5\hat X_6\hat X_7,\\
    \hat X_2\hat X_3\hat X_6\hat X_7,\\
    \hat X_1\hat X_3\hat X_5\hat X_7,\\
    \hat Z_4\hat Z_5\hat Z_6\hat Z_7,\\
    \hat Z_2\hat Z_3\hat Z_6\hat Z_7,\\
    \hat Z_1\hat Z_3\hat Z_5\hat Z_7,
    \end{matrix}
    \right\}.
\end{equation}
With the Steane code, any single-qubit error can be correctly recovered as it send the code space $\mathbb H_\text{C}$ in one of the other subspaces $\mathbb H_i$. Specifically, the code space $\mathbb H_\text{C}$ is given by the span of two logical states $\set{\ket{0_\text{L}},\ket{1_\text{L}}}$. These are encoded as
\begin{equation}
    \begin{aligned}
        \ket{0_\text{L}}=&\frac{1}{\sqrt{8}}\left[\ket{0000000}+\ket{1010101}+\ket{0110011}+\ket{1100110}+\ket{0001111}+\ket{1011010}+\ket{0111100}+\ket{1101001}\right],\\
        \ket{1_\text{L}}=&\frac{1}{\sqrt{8}}\left[\ket{1111111}+\ket{0101010}+\ket{1001100}+\ket{0011001}+\ket{1110000}+\ket{0100101}+\ket{1000011}+\ket{0010110}\right].
    \end{aligned}
\end{equation}
The logical operations are given by the normalisers. In the Pauli group, we have the logical X-gate and the logical Z-gate. These are constructed by applying the corresponding single qubit operators to each physical qubit. Namely,
\begin{equation}
\begin{aligned}
    \hat X_\text{L}=&\hat X_1\hat X_2\hat X_3\hat X_4\hat X_5\hat X_6\hat X_7,\\
    \hat Z_\text{L}=&\hat Z_1\hat Z_2\hat Z_3\hat Z_4\hat Z_5\hat Z_6\hat Z_7.
\end{aligned}
\end{equation}
These operations are allowed, as there are still two free degrees of freedom. 
There are also other logical operations that one can construct. These are normalisers that do not appartain to the Pauli group. An example is the Hadamard gate, which can be also implemented as the application of single qubit Hadamards:
\begin{equation}
        \hat H_\text{L}=\hat H_1\hat H_2\hat H_3\hat H_4\hat H_5\hat H_6\hat H_7.
\end{equation}
At this point, one would naively extrapolate that any logical operation  can be constructed via the application of the corresponding gate on the single data qubits where the logical state is encoded. However, this is not the case. An example is the phase gate $\hat S$, whose representation in the computation basis reads
\begin{equation}
    S=\begin{pmatrix}
        1&0\\0&i
    \end{pmatrix}.
\end{equation}
The corresponding logical gate is constructed as
\begin{equation}
        \hat S_\text{L}=\hat S^\dag_1\hat S^\dag_2\hat S^\dag_3\hat S^\dag_4\hat S^\dag_5\hat S^\dag_6\hat S^\dag_7.
\end{equation}
Indeed, when applied to the logical computational basis we find
\begin{equation}
\begin{aligned}
    \hat S_\text{L}\ket{0_\text{L}}&=\ket{0_\text{L}},\\
    \hat S_\text{L}\ket{1_\text{L}}&=i\ket{1_\text{L}}.
\end{aligned}
\end{equation}
Conversely, if one would have defined a $\hat S'_\text{L}$ as just simply applying the S-gate to each physical qubit, one would have obtained
\begin{equation}
\begin{aligned}
    \hat S'_\text{L}\ket{0_\text{L}}&=\ket{0_\text{L}},\\
    \hat S'_\text{L}\ket{1_\text{L}}&=-i\ket{1_\text{L}}.
\end{aligned}
\end{equation}
Thus, the construction of logical gates needs to be done with care and it depends on the QEC code that one is employing.\\

Let us consider the case of a CNOT gate. Here, we have two logical qubits, each corresponding to 7 data and 6 syndrome qubits. One can see that the logical CNOT gate can be implemented pairwise as described by the following ciurcuit
\begin{equation}
    \begin{tikzcd}
        &\ctrl{7}&&&&&&&\\
        &&\ctrl{7}&&&&&&\\
        &&&\ctrl{7}&&&&&\\
        &&&&\ctrl{7}&&&&\\
        &&&&&\ctrl{7}&&&\\
        &&&&&&\ctrl{7}&&\\
        &&&&&&&\ctrl{7}&\\
        &\targ{}&&&&&&&\\
        &&\targ{}&&&&&&\\
        &&&\targ{}&&&&&\\
        &&&&\targ{}&&&&\\
        &&&&&\targ{}&&&\\
        &&&&&&\targ{}&&\\
        &&&&&&&\targ{}&\\
    \end{tikzcd}
\end{equation}
It follows that one can implement the logical circuit
\begin{equation}
    \begin{tikzcd}
        &\gate{H}&\ctrl{1}&\\
        &&\targ{}&
    \end{tikzcd}
\end{equation}
as
\begin{equation}\label{cir.7qubit Hcnot}
    \begin{tikzcd}
        &\gate{H}&\ctrl{7}&&&&&&&\\
        &\gate{H}&&\ctrl{7}&&&&&&\\
        &\gate{H}&&&\ctrl{7}&&&&&\\
        &\gate{H}&&&&\ctrl{7}&&&&\\
        &\gate{H}&&&&&\ctrl{7}&&&\\
        &\gate{H}&&&&&&\ctrl{7}&&\\
        &\gate{H}&&&&&&&\ctrl{7}&\\
        &&\targ{}&&&&&&&\\
        &&&\targ{}&&&&&&\\
        &&&&\targ{}&&&&&\\
        &&&&&\targ{}&&&&\\
        &&&&&&\targ{}&&&\\
        &&&&&&&\targ{}&&\\
        &&&&&&&&\targ{}&\\
    \end{tikzcd}
\end{equation}
This way of implementing logical gates is called transversal construction. This is a really easy and straightforward construction, but most importantly allows to confine errors, as they are unable to propagate. Let us take as an example the circuit in Eq.~\eqref{cir.7qubit Hcnot} and suppose we have initially a X error on the first qubit of the first block. In such a case, the error propagates accordingly to 
\begin{equation}
    \begin{tikzcd}
        \gate{X}&\gate{H}&\ctrl{7}&&&&&&&\\
        &\gate{H}&&\ctrl{7}&&&&&&\\
        &\gate{H}&&&\ctrl{7}&&&&&\\
        &\gate{H}&&&&\ctrl{7}&&&&\\
        &\gate{H}&&&&&\ctrl{7}&&&\\
        &\gate{H}&&&&&&\ctrl{7}&&\\
        &\gate{H}&&&&&&&\ctrl{7}&\\
        &&\targ{}&&&&&&&\ghost{X}\\
        &&&\targ{}&&&&&&\\
        &&&&\targ{}&&&&&\\
        &&&&&\targ{}&&&&\\
        &&&&&&\targ{}&&&\\
        &&&&&&&\targ{}&&\\
        &&&&&&&&\targ{}&\\
    \end{tikzcd}
    \rightarrow
        \begin{tikzcd}
        &\gate{H}&\ctrl{7}&&&&&&&\gate{X}\\
        &\gate{H}&&\ctrl{7}&&&&&&\\
        &\gate{H}&&&\ctrl{7}&&&&&\\
        &\gate{H}&&&&\ctrl{7}&&&&\\
        &\gate{H}&&&&&\ctrl{7}&&&\\
        &\gate{H}&&&&&&\ctrl{7}&&\\
        &\gate{H}&&&&&&&\ctrl{7}&\\
        &&\targ{}&&&&&&&\gate{X}\\
        &&&\targ{}&&&&&&\\
        &&&&\targ{}&&&&&\\
        &&&&&\targ{}&&&&\\
        &&&&&&\targ{}&&&\\
        &&&&&&&\targ{}&&\\
        &&&&&&&&\targ{}&\\
    \end{tikzcd}
\end{equation}
Thus, the error has been copied to the first qubit of the second block, but all the other qubits are not affected. The Steane code can correct for such an error. This holds true for any transverse construction.\\

Till now, we have covered the case for the normalisers $\set{\hat X_\text{L},\hat Z_\text{L},\hat H_\text{L},\hat S_\text{L},\text{CNOT}_\text{L}}$, which is the so-called Clifford group. Notably, the Gottesman-Knill theorem indicates that operations performed using only elements of this group can be simulated classically. Thus, one cannot have a real quantum advantage over a classical computer. Moreover, the Clifford group is not universal, meaning that the composition of elements of this group is not sufficient to implement any arbitrary gate. This is essentially the argument of the Solovay-Kitaev theorem. One needs to extend the Clifford group with the addition of at least an extra gate not appartaining to the group. This can be the T-gate or the Toffoli gate. 

Unfortunately, one does not know how to implement through unitary operations the T-gate in a transversal way using the Steane code. This is a code-related problem. One could consider a different QEC code and implement the T-gate, but they will have problems with another gate, e.g.~the Hadarmard gate. The Eastin-Knill theorem indicates that it is not possible to construct a fault-tolerant universal set with unitary operations.\\

Notably, the latter theorem applies only to unitary operations. One can still have  the entire universal set by substituting the problematic gate with an effective implementation. Below, we focus on the derivation of an effective T-gate for the Steane code. \\

Consider the following circuit
\begin{equation}
    \begin{tikzcd}
        \ket\psi&&\ctrl{1}&\gate{H}&\meter{y}\\
        \ket{0}&\gate{H}&\ctrl{-1}&&&\ket\varphi
    \end{tikzcd}
\end{equation}
where one obtains that the state $\ket\varphi$ is given by
\begin{equation}
    \ket\varphi=\hat X^y\hat H\ket\psi,
\end{equation}
with $y=0,1$.
Notably, the three operations performed here, the two Hadamards and the control-Z gates, are all fault-tolerant.

Suppose we slightly modify the circuit by adding a T-gate on the second qubit as follows
\begin{equation}
    \begin{tikzcd}
        \ket\psi&&\ctrl{1}&\gate{H}&\meter{y}\\
        \ket{0}&\gate{H}&\ctrl{-1}&\gate{T}&&\ket{\varphi'}
    \end{tikzcd}
\end{equation}
where the representation of the T-gate on the computational basis reads
\begin{equation}
    T=\begin{pmatrix}
        1&0\\0&e^{-i\tfrac{\pi}{4}}
    \end{pmatrix}
\end{equation}
and the final state of the second qubit now is
\begin{equation}
    \ket{\varphi'}=\hat T\hat X^y\hat H\ket{\psi}.
\end{equation}
Since, a part from a total phase, the T-gate can be written as
\begin{equation}
    \hat T=e^{i\tfrac{\pi}{8}\hat Z},
\end{equation}
one obtains
\begin{equation}
    \ket{\varphi'}=\hat X^y\hat T^{(1-2y)}\hat H\ket\psi.
\end{equation}
But, $\hat T^{(1-2y)}$ can be expressed as
\begin{equation}
   \hat T^{(1-2y)}=\hat T  e^{-i\tfrac{\pi y}{4}\hat Z}=e^{-i\tfrac{\pi y}{4}}\hat S^y\hat T,
\end{equation}
where we can safely omit the phase in the last expression. Thus, we get
\begin{equation}
    \ket{\varphi'}=\hat X^y\hat S^y\hat T\hat H\ket\psi=(\hat X\hat S)^y\hat T\hat H\ket\psi.
\end{equation}
Here, the last equality holds since the value of $y$ can be only 0 or 1.
Moreover, we also find that the T-gate commutes with the control-Z, since it commutes with $\hat Z$. Thus, by adding an extra Hadamard gate on the first qubit, we obtain
\begin{equation}
    \begin{tikzcd}
        \ket\psi&\gate{H}&&\ctrl{1}&\gate{H}&\meter{y}\\
        \ket{0}&\gate{H}&\gate{T}&\ctrl{-1}&&&\ket{\varphi''}
    \end{tikzcd}
\end{equation}
where
\begin{equation}
    \ket{\varphi''}=(\hat X\hat S)^y\hat T\ket\psi.
\end{equation}
Specifically, the latter implies the following one
\begin{equation}
    \begin{tikzcd}
        \ket\psi&\gate{H}&&\ctrl{1}&\gate{H}&\meter{y}\wire[d]{c}\\
        \ket{\phi}&&&\ctrl{-1}&&\gate{(\hat S^\dag \hat X)^y}&\ket{\chi}
    \end{tikzcd}
\end{equation}
where
\begin{equation}
    \begin{aligned}
        \ket\chi&=\hat T\ket\psi,\\
        \ket \phi&=\hat T\hat H\ket0.
    \end{aligned}
\end{equation}
This means that, if one is able to prepare the second qubit in the state $\ket\phi$, then they also can apply the T-gate to an arbitrary state $\ket\psi$ that was initially embedded in the first qubit. This is an effective application of the T-gate. Clearly, it requires the preparation of $\ket\phi$. Specifically, this is given by
\begin{equation}
    \ket\phi=\hat T\hat H\ket0=e^{-i\tfrac{\pi}{8}\hat Z}\ket+.
\end{equation}
Now, the state $\ket+$ is eigeinstate of $\hat X $ with eigeinvalue $(+1)$, and in the same way $\ket\phi$ is eigeinstate of the operator
\begin{equation}
    e^{-i\tfrac{\pi}{8}\hat Z}\hat Xe^{i\tfrac{\pi}{8}\hat Z}=e^{-i\tfrac{\pi}{4}\hat Z}\hat X=\hat S\hat X,
\end{equation}
with the eigeinvalue $(+1)$.
Here, the last two equalities follow from the commutaion of $\hat T$ and $\hat X$ and the definition of the S-gate. Thus, $\ket\phi$ is eigeinstate of $\hat S\hat X$ with eigeinvalue $(+1)$. Similarly, the ortogonal state $\ket{\phi_\perp}$ is associated to the eigenvalue $(-1)$. Now, this implies that given an arbitrary state $\ket\nu$ by measuring the operator $\hat S\hat X$, the state will collapse in $\ket\phi$ or $\ket{\phi_\perp}$ with corresponding outcomes respectively given by $(+1)$ and $(-1)$.

The complete circuit thus becomes
\begin{equation}
    \begin{tikzcd}
\ket{\psi}&&\gate{H}&\ctrl{1}&\gate{H}&\meter{y}\wire[d]{c}\\
\ket{\nu}&\meter[2][1.3cm]{\hat S\hat X}\gateoutput{$+1$}&\push{\ket{\phi}}&\ctrl{-1}&&\gate{(\hat S^\dag\hat X)^y}&\ket\chi\\
        &\wireoverride{n}\gateoutput{$-1$}&\push{\text{ discarded}}
    \end{tikzcd}
\end{equation}
and effectively applies the T-gate to an arbitrary state $\ket\psi$. If the measurement is also FT (this can be proven, but it will not be tackled here), then one has that the entire circuit is FT.

\chapter{Dynamical Decoupling and Quantum Error  Mitigation}

The greatest problem in the development of quantum computers are the presence of errors and noises. QEC works in theory: the threshold theorem guarantees it. However, it requires a number of physical qubits ($\sim10^3$ to $\sim10^6$) that is often beyond what possible with the current technology.
In this chapter, we introduce two possible routes to tackle the problem. These are the Dynamical Decoupling and the Quantum Error Mitigation.

\section{Dynamical Decoupling}
\index{Dynamical decoupling}
\label{sec.dynamicaldecoupling}

The dynamical decoupling (DD) approach leverage on averaging out the unwanted effects of the surrounding environment by applying a control on the system. \\

\begin{figure}[ht]
    \centering
    \includegraphics[width=0.6\linewidth]{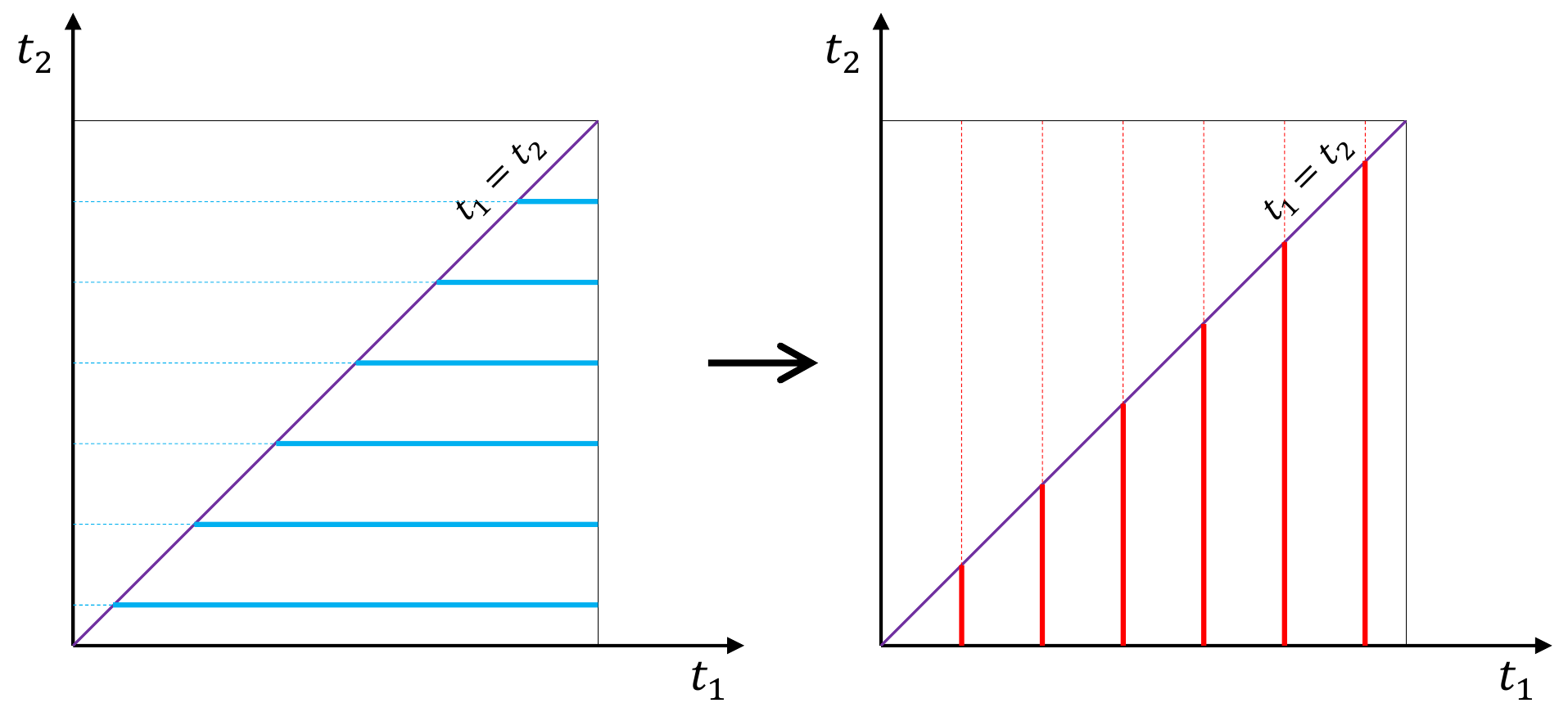}
    \caption{Representation of different   but equivalent ways of how to perform the integral in Eq.~\eqref{eq.integraltriangle}.}
    \label{fig.triangleintegral}
\end{figure}

Two introduce the idea, we focus on the specific model of a qubit coupled to a thermal bath of harmonic oscillators, with $\hat b_k$ being the annihilation operator of the $k$ oscillator. The total Hamiltonian reads
\begin{equation}
    \hat H_0=\hat H_\text{S}+\hat H_\text{B}+\hat H_\text{SB}=\tfrac12\hbar \omega_0\hat \sigma_z+\sum_k \hbar\omega_k \hat b^\dag_k\hat b_k+\sum_k \hbar\hat \sigma_z(g_k\hat b^\dag_k+g^*_k\hat b_k),
\end{equation}
where the first and second contributions are the free Hamiltonian of the qubit system and thermal bath respectively, while the last term describes their interaction being weighted by the constants $g_k$. Eventually, one focuses on the dynamics of the qubit alone. Thus, the state of interest is the reduced density matrix, which is obtained via
\begin{equation}
    \hat \rho_\text{S}(t)=\TR{\text{B}}{e^{-i\hat H_0 t/\hbar}\hat \rho_\text{T}(0)e^{i\hat H_0 t/\hbar}},
\end{equation}
where $\hat \rho_\text{T}(0)$ is the total state at time $t=0$. The latter can be decomposed on the computational basis $\set{\ket0,\ket1}$ as
\begin{equation}
    \hat \rho_\text{S}(t)=\sum_{i,j=0,1}\rho_{ij}(t)\ket i\bra j.
\end{equation}
Now, given the total Hamiltonian $\hat H_0$, one finds that the populations $\rho_{ii}$ are conserved. Indeed, $\com{\hat \sigma_z}{\hat H_0}=0$ and thus the model describes a purely decohering mechanism, where no energy exchange between the system and the bath is present. Specifically, one can focus on the dynamics of the coherences $\rho_{01}(t)$ alone, and we do it in the interaction picture. Thus, we have that the total state is given by
\begin{equation}
    \hat \rho_\text{T}^{\text{\tiny (I)}}(t)=e^{i(\hat H_\text{S}+\hat H_\text{B})t/\hbar}\hat \rho_\text{T}(t)e^{-i(\hat H_\text{S}+\hat H_\text{B})t/\hbar},
\end{equation}
with the effective Hamiltonian reading
\begin{equation}\label{eq.H0DDinteraction}
    \hat H_0^{\text{\tiny{(I)}}}(t)=\hbar\hat \sigma_z\sum_k\left(g_k\hat b_k^\dag e^{i\omega_k t}+g_k^*\hat b_k e^{-i\omega_k t}\right).
\end{equation}
Correspondingly, the unitary operator determining the time evolution in the interaction picture from time $t_0$ to $t$  is
\begin{equation}
    \hat U^{\text{\tiny (I)}}(t_0,t)=T \exp\left\{ -\frac i\hbar\int_{t_0}^t\D s\,\hat H_0^{\text{\tiny{(I)}}}(s)\right\},
\end{equation}
where $T$ indicates the time-ordering operator. The corresponding Dyson expansion, which is effectively a Taylor expansion accounting also for the time-ordering, reads
\begin{equation}
      \hat U^{\text{\tiny (I)}}(t_0,t)=\hat{\mathbb 1}-\frac i\hbar\int_{t_0}^t\D t_1\,\hat H_0^{\text{\tiny{(I)}}}(t_1)-\frac{1}{\hbar^2}\int_{t_0}^t\D t_2\int_{t_0}^{t_2}\D t_1\,\hat H_0^{\text{\tiny{(I)}}}(t_2)\hat H_0^{\text{\tiny{(I)}}}(t_1)+\dots
\end{equation}
Let us focus on the second order term, which can be rewritten in term of the integral from $t_0$ to $t$ for both variables as 
\begin{equation}\label{eq.triangle1st}
    \int_{t_0}^t\D t_2\int_{t_0}^{t_2}\D t_1\,\hat H_0^{\text{\tiny{(I)}}}(t_2)\hat H_0^{\text{\tiny{(I)}}}(t_1)=\int_{t_0}^t\D t_2\int_{t_0}^{t}\D t_1\,\hat H_0^{\text{\tiny{(I)}}}(t_2)\hat H_0^{\text{\tiny{(I)}}}(t_1)-\int_{t_0}^t\D t_2\int_{t_2}^{t}\D t_1\,\hat H_0^{\text{\tiny{(I)}}}(t_2)\hat H_0^{\text{\tiny{(I)}}}(t_1).
\end{equation}
The last term corresponds to the integral over the area highlighted by the blue lines in Fig.~\ref{fig.triangleintegral}: for each value of $t_2\in[t_0,t]$, the $t_1$ integral runs from $t_2$ to $t$. Equivalently, this area can be described by the red lines: for each value of $t_1\in[t_0,t]$, one performs the $t_2$ integral from $t_0$ to $t_1$. 

Mathematically, this implies that the following equality holds
\begin{equation}\label{eq.integraltriangle}
    \int_{t_0}^t\D t_2\int_{t_2}^{t}\D t_1\,\hat H_0^{\text{\tiny{(I)}}}(t_2)\hat H_0^{\text{\tiny{(I)}}}(t_1)=\int_{t_0}^t\D t_1\int_{t_0}^{t_1}\D t_2\,\hat H_0^{\text{\tiny{(I)}}}(t_2)\hat H_0^{\text{\tiny{(I)}}}(t_1).
\end{equation}
Moreover, we can recast the right-hand-side of the latter equation as
\begin{equation}
    \begin{aligned}
        \int_{t_0}^t\D t_1\int_{t_0}^{t_1}\D t_2\,\hat H_0^{\text{\tiny{(I)}}}(t_2)\hat H_0^{\text{\tiny{(I)}}}(t_1)&=\int_{t_0}^t\D t_1\int_{t_0}^{t_1}\D t_2\,\com{\hat H_0^{\text{\tiny{(I)}}}(t_2)}{\hat H_0^{\text{\tiny{(I)}}}(t_1)}+\int_{t_0}^t\D t_1\int_{t_0}^{t_1}\D t_2\,\hat H_0^{\text{\tiny{(I)}}}(t_1)\hat H_0^{\text{\tiny{(I)}}}(t_2),\\
        &=\int_{t_0}^t\D t_1\int_{t_0}^{t_1}\D t_2\,\com{\hat H_0^{\text{\tiny{(I)}}}(t_2)}{\hat H_0^{\text{\tiny{(I)}}}(t_1)}+\int_{t_0}^t\D t_2\int_{t_0}^{t_2}\D t_1\,\hat H_0^{\text{\tiny{(I)}}}(t_2)\hat H_0^{\text{\tiny{(I)}}}(t_1),
    \end{aligned}
\end{equation}
where we swapped the variables $t_1\leftrightarrow t_2$ in the second term. Namely, such second term is identical to that in the left-hand-side of Eq.~\eqref{eq.triangle1st}. 
Here, the non-equal time Hamiltonians do not commute, but give
\begin{equation}
    \begin{aligned}
        \com{\hat H_0^{\text{\tiny{(I)}}}(t_2)}{\hat H_0^{\text{\tiny{(I)}}}(t_1)}&=\hbar^2\hat\sigma_z^2\sum_{kk'}\left(
       g_kg_{k'}^*e^{i \omega_k t_2-i\omega_{k'}t_1}\com{\hat b^\dag_k}{\hat b_{k'}}+ g_k^*g_{k'}e^{-i \omega_k t_2+i\omega_{k'}t_1}\com{\hat b_k}{\hat b^\dag_{k'}}\right),\\
       &=-2i\hbar^2\sum_k|g_k|^2\sin[\omega_k(t_2-t_1)],
    \end{aligned}
\end{equation}
which is an imaginary number.
Now, by merging the last four equations, we find
\begin{equation}
    \begin{aligned}
        \int_{t_0}^t\D t_2\int_{t_0}^{t_2}\D t_1\,\hat H_0^{\text{\tiny{(I)}}}(t_2)\hat H_0^{\text{\tiny{(I)}}}(t_1)&=\frac12\int_{t_0}^t\D t_2\int_{t_0}^{t}\D t_1\,\hat H_0^{\text{\tiny{(I)}}}(t_2)\hat H_0^{\text{\tiny{(I)}}}(t_1)-\frac12\int_{t_0}^t\D t_1\int_{t_0}^{t_1}\D t_2\,\com{\hat H_0^{\text{\tiny{(I)}}}(t_2)}{\hat H_0^{\text{\tiny{(I)}}}(t_1)},\\
        &=\frac12\left(\int_{t_0}^t\D t_1\,\hat H_0^{\text{\tiny{(I)}}}(t_1)\right)^2+i\hbar^2\int_{t_0}^t\D t_1\int_{t_0}^{t_1}\D t_2\,\sum_k|g_k|^2\sin[\omega_k(t_2-t_1)].
    \end{aligned}
\end{equation}
This allows to recast the Dyson expansion as
\begin{equation}
     \hat U^{\text{\tiny (I)}}(t_0,t)=\hat{\mathbb 1}-\frac i\hbar\int_{t_0}^t\D t_1\,\hat H_0^{\text{\tiny{(I)}}}(t_1)+\frac{1}{2}\left(-\frac{i}{\hbar}\int_{t_0}^t\D t_1\,\hat H_0^{\text{\tiny{(I)}}}(t_1)\right)^2-i\phi(t_0,t)+\dots,
\end{equation}
where 
\begin{equation}
    \phi(t_0,t)=\frac{1}{2}\int_{t_0}^t\D t_1\int_{t_0}^{t_1}\D t_2\,\sum_k|g_k|^2\sin[\omega_k(t_2-t_1)].
\end{equation}
By summing all the terms, one gets
\begin{equation}
    \hat U^{\text{\tiny (I)}}(t_0,t)=e^{-i\varphi(t_0,t)}\exp\left\{-\frac i\hbar\int_{t_0}^t\D s\,\hat H_0^{\text{\tiny{(I)}}}(s)\right\},
\end{equation}
where one has an extra global phase, which is hiowever unimportant since
\begin{equation}
    \hat U^{\text{\tiny (I)}}(t_0,t)\hat \rho\left[\hat U^{\text{\tiny (I)}}(t_0,t)\right]^\dag=\exp\left\{-\frac i\hbar\int_{t_0}^t\D s\,\hat H_0^{\text{\tiny{(I)}}}(s)\right\}\hat \rho \exp\left\{\frac i\hbar\int_{t_0}^t\D s\,\hat H_0^{\text{\tiny{(I)}}}(s)\right\},
\end{equation}
and there is not time-ordering operator.

Specifically, one can perform explicitly the time integral in the exponential of $\hat U^{\text{\tiny (I)}}(t_0,t)$, which reads
\begin{equation}
    -\frac i\hbar\int_{t_0}^t\D s\,\hat H_0^{\text{\tiny{(I)}}}(s)=\frac{1}{2}\hat \sigma_z\sum_k\left( \hat b_k^\dag e^{i \omega_k t_0}\xi_k(t-t_0)-\hat b_k e^{-i \omega_k t_0}\xi_k^*(t-t_0)\right),
\end{equation}
where
\begin{equation}
    \xi_k(t-t_0)=\frac{2g_k}{\omega_k}(1-e^{i\omega_k(t-t_0)}).
\end{equation}

Now, the quantity of interest is the coherence, which in the interaction picture is given by
\begin{equation}\label{eq.coherenceddinteraction}
    \rho_{01}^{\text{\tiny (I)}}(t)=\braket{0|\hat \rho_\text{S}^{\text{\tiny (I)}}(t)|1}=\braket{0|\TR{B}{\hat U^{\text{\tiny (I)}}(t_0,t)\hat \rho_\text{T}(t_0)\left(   \hat U^{\text{\tiny (I)}}(t_0,t)\right)^\dag}|1},
\end{equation}
and it can be computed analytically under the following assumptions:
\begin{itemize}
    \item[1)] The total initial state is separable, namely
        \begin{equation}
            \hat \rho_\text{T}(t_0)=\hat \rho_\text{S}(t_0)\otimes\hat \rho_\text{B}(t_0);
        \end{equation}
        \item[2)] The initial state of the bath is a thermal state of the form
        \begin{equation}
            \hat \rho_\text{B}(t_0)=\prod_k(1-e^{-\beta \hbar\omega_k})\exp\left( - \beta\hbar\omega_k\hat b_k^\dag\hat b_k\right),
        \end{equation}
        with $\beta=(\kB T)^{-1}$ being the inverse temperature.
\end{itemize}

By going back to the Schr\"odinger picture, Eq.~\eqref{eq.coherenceddinteraction} reads
\begin{equation}
\begin{aligned}
    \rho_{01}(t)&=e^{-i\omega_0(t-t_0)}\rho_{01}^{\text{\tiny (I)}}(t),\\
    &=e^{-i\omega_0(t-t_0)}\bra{0}\operatorname{Tr}^\text{\tiny (B)}\left[\exp\left[\frac{1}{2}\hat \sigma_z\sum_k\left( \hat b_k^\dag e^{i \omega_k t_0}\xi_k(t-t_0)-\hat b_k e^{-i \omega_k t_0}\xi_k^*(t-t_0)\right)\right]\hat \rho_\text{T}(t_0)\right.\\
    &\quad\quad\times\left.\exp\left[-\frac{1}{2}\hat \sigma_z\sum_k\left( \hat b_k^\dag e^{i \omega_k t_0}\xi_k(t-t_0)-\hat b_k e^{-i \omega_k t_0}\xi_k^*(t-t_0)\right)\right]\right]\ket{1}.
\end{aligned}
\end{equation}
By applying with $\hat \sigma_z$ on the $\bra 0$ and $\ket 1$ states, we get
\begin{equation}\label{eq.coherencedd1}
    \rho_{01}(t)=
    e^{-i\omega_0(t-t_0)}\TR{B}{\exp\left[ \sum_k\left( c_k\hat b_k^\dag -c_k^*\hat b_k \right)\right]\hat \rho_\text{B}(t_0)}\rho_{01}(t_0),
\end{equation}
where we exploited the cyclicity of the partial trace with respect to the bath operators, the assumption of separability of the initial state, the definition of the initial coherence $\rho_{01}(t_0)=\braket{0|\hat \rho_\text{S}(t_0)|1}$, and defined
\begin{equation}
    c_k=c_k(t_0,t)=e^{i\omega_k t_0}\xi_k(t-t_0).
\end{equation}
In Eq.~\eqref{eq.coherencedd1}, one can recognise the displacement operator. Namely, the latter equation can be recasted as
\begin{equation}\label{eq.rho01dd1}
    \rho_{01}(t)=
    e^{-i\omega_0(t-t_0)}\TR{B}{\prod_k\hat D_k(c_k)\hat \rho_\text{B}(t_0)}\rho_{01}(t_0),
\end{equation}
where 
\begin{equation}
    \hat D(\beta)=\exp(\beta\hat b^\dag_k-\beta^*\hat b)
\end{equation}
is the displacement operator.

\begin{myrecall}[Coherent states]
Given the ground state $\ket0$, one can construct a coherent state via the application of a displacement operator to $\ket0$. Namely,
\begin{equation}\label{eq-def.displacement}
    \hat D(\beta)\ket0=\ket\beta,
\end{equation}
where $\beta\in\mathbb C$.
The coherent states form an overcomplete basis of the Hilbert space $\mathbb H$, for which one has
\begin{equation}
    \hat {\mathbb 1}=\int\frac{\D^2z}{\pi}\ket z\bra z,
\end{equation}
where $\D^2 z=\D(\Re z)\D(\Im z)$. The combination of two displacement operators is governed by
\begin{equation}\label{eq.displacementalphabeta}
    \hat D(\alpha)\hat D(\beta)=e^{\tfrac12(\alpha\beta^*-\beta\alpha^*)}\hat D(\alpha+\beta).
\end{equation}
Finally, one can express the coherent states in terms of the Fock basis $\set{\ket n}$, where
\begin{equation}\label{eq.zton}
    \braket{n|z}=\frac{z^n}{\sqrt{n!}}e^{-|z|^2/2},
\end{equation}
determines the weights between the two basis.
\end{myrecall}
Let us then consider the partial trace over the $k$-th mode of the displacement operator and the thermal state, which explicitly reads
\begin{equation}
    \begin{aligned}
        \TR{B}{\hat D_k(c_k)\hat \rho_{\text{B}_k}(t_0)}&=  (1-e^{-\beta\hbar\omega_k})\TR{B}{\hat D_k(c_k)e^{-\beta\hbar\omega_k\hat b_k^\dag\hat b_k}},\\
        &=(1-e^{-\beta\hbar\omega_k})\int\frac{\D^2z}{\pi}\braket{z|e^{-\beta\hbar\omega_k\hat b_k^\dag\hat b_k}\hat D_k(c_k)|z},
    \end{aligned}
\end{equation}
where we exploited the coherent basis to compute the partial trace and exploit its cyclicity. By applying the composition of displacement opertors (namely, Eq.~\eqref{eq.displacementalphabeta} merged with Eq.~\eqref{eq-def.displacement}), and introducing an identity in the Fock basis, i.e.~$\hat{\mathbb 1}=\sum_{n=0}^{+\infty}\ket n  \bra n$, we get
\begin{equation}
    \begin{aligned}
        \TR{B}{\hat D_k(c_k)\hat \rho_{\text{B}_k}(t_0)}&=(1-e^{-\beta\hbar\omega_k})\int\frac{\D^2z}{\pi}\sum_n\braket{z|n}e^{-\beta\hbar\omega_kn}e^{\tfrac12(c_kz^*-zc_k^*)}\braket{n|z+c_k},\\
        &=(1-e^{-\beta\hbar\omega_k})\int\frac{\D^2z}{\pi}\sum_ne^{-\beta\hbar\omega_kn}e^{\tfrac12(c_kz^*-zc_k^*)}\frac{(z^*)^n}{\sqrt{n!}}e^{-|z|^2/2}\frac{(z+c_k)^n}{\sqrt{n!}}e^{-|z+c_k|^2/2},
    \end{aligned}
\end{equation}
where we applied Eq.~\eqref{eq.zton}. By putting together the exponentials, we obtain
\begin{equation}
    \TR{B}{\hat D_k(c_k)\hat \rho_{\text{B}_k}(t_0)}=(1-e^{-\beta\hbar\omega_k})\int\frac{\D^2z}{\pi}e^{-|z|^2}e^{-|c_k|^2/2}e^{-zc_k^*}S(z,c_k)
\end{equation}
where
\begin{equation}
\begin{aligned}
    S(z,c_k)&=\sum_ne^{-\beta\hbar\omega_kn}\frac{(z^*)^n}{\sqrt{n!}}\frac{(z+c_k)^n}{\sqrt{n!}},\\
    &=\sum_n\frac{1}{n!}\left[e^{-\beta\hbar\omega_k}(|z|^2+c_kz^*)\right]^n,\\
    &=\exp\left[e^{-\beta\hbar\omega_k}(|z|^2+c_kz^*)\right].
    \end{aligned}
\end{equation}
Thus,
\begin{equation}\label{eq.guass.integral}
    \TR{B}{\hat D_k(c_k)\hat \rho_{\text{B}_k}(t_0)}=(1-e^{-\beta\hbar\omega_k})\int\frac{\D^2z}{\pi}e^{-|z|^2\alpha_k}e^{-|c_k|^2/2}e^{-zc_k^*}\exp[c_kz^*e^{-\beta\hbar\omega_k}],
\end{equation}
where
\begin{equation}
    \alpha_k=1-e^{-\beta\hbar\omega_k},
\end{equation}
is a positive quantity. Then, the Gaussian integral in  Eq.~\eqref{eq.guass.integral} can be safely implemented and gives
\begin{equation}
\begin{aligned}
    \TR{B}{\hat D_k(c_k)\hat \rho_{\text{B}_k}(t_0)}&=(1-e^{-\beta\hbar\omega_k})\frac{1}{\alpha_k}\exp\left[-\frac{|c_k|^2}{\alpha_k}e^{-\beta\hbar\omega_k}\right]e^{-|c_k|^2/2},\\
    &=\exp\left[-|c_k|^2\left(\frac12+\frac{e^{-\beta\hbar\omega_k}}{(1-e^{-\beta\hbar\omega_k})}\right)\right],\\
    &=\exp\left[-\frac{|c_k|^2}{2}\coth\left(\frac{\beta\hbar\omega_k}{2}\right)\right].
    \end{aligned}
\end{equation}
 Finally, by merging together the latter equation with Eq.~\eqref{eq.rho01dd1}, we obtain
 \begin{equation}\label{eq.rho01.Gamma}
     \rho_{01}(t)=e^{-i\omega_0(t-t_0)}e^{-\Gamma(t_0,t)}\rho_{01}(t_0),
 \end{equation}
where we defined
\begin{equation}\label{eq-def.GammaDD}
\begin{aligned}
    \Gamma(t_0,t)&=\sum_k \frac{|c_k^2|}{2}\coth \left(\frac{\beta\hbar\omega_k}{2}\right),\\
    &=\sum_k\frac{|e^{i\omega_kt_0}\xi_k(t-t_0)|^2}{2}\coth \left(\frac{\beta\hbar\omega_k}{2}\right),\\
    &=\sum_k\frac{|\frac{2g_k}{\omega_k}(1-e^{i\omega_k(t-t_0)})|^2}{2}\coth \left(\frac{\beta\hbar\omega_k}{2}\right),\\
    &=\sum_k\frac{4|g_k|^2}{\omega_k^2}\left(1-\cos[\omega_k(t-t_0)]\right)\coth \left(\frac{\beta\hbar\omega_k}{2}\right)
    \end{aligned}
\end{equation}
where we simply substituted the definitions of $c_k$ and $\xi_k(t-t_0)$. Now, by introducing the spectral density $I(\omega)$ as
\begin{equation}
    I(\omega)=\sum_k\delta(\omega-\omega_k)|g_k|^2,
\end{equation}
which determines the strength of the coupling between the system and each bath's modes, we can rewrite $\Gamma(t_0,t)$ as
\begin{equation}
\begin{aligned}
    \Gamma(t_0,t)&=4\int_0^{+\infty}\D \omega\,I(\omega) \coth \left(\frac{\beta\hbar\omega}{2}\right)\frac{\left(1-\cos[\omega(t-t_0)]\right)}{\omega^2},\\
    &=4\int_0^{+\infty}\D \omega\,I(\omega) \left(2\bar n(\omega,T)+1\right)\frac{\left(1-\cos[\omega(t-t_0)]\right)}{\omega^2},
    \end{aligned}
\end{equation}
where $\bar n(\omega, T)$ is the mean number of excitations of the mode $\omega$ at the temperature $T=(\kB \beta )^{-1}$. Notably, $\Gamma(t_0,t)$ is positive. This can be seen explicitly from the first line of Eq.~\eqref{eq-def.GammaDD}. This means, that --- as expected --- the interaction with the environment reduces the coherences [cf.~Eq.~\eqref{eq.rho01.Gamma}].

Suppose now that we want to perturb the system so the induce a spin-flip transition. Physically, since the interaction Hamiltonian $\hat H_\text{SB}$ is proportional to $\hat \sigma_z$, then opposite contributions arise when the system is in $\ket0$ and $\ket1$. Thus, by making the system change fast between $\ket0$ and $\ket1$, one can average out the contributions from $\hat H_\text{SB}$, effectively decoupling the system from the environment.

Specifically, we will consider a modified Hamiltonian reading
\begin{equation}
    \hat H_0\rightarrow\hat H(t)=\hat H_0+\hat H_\text{P}(t),
\end{equation}
where the Hamiltonian perturbation $\hat H_\text{P}(t)$ can be implemented via a monocromatic alternating magnetic field applied at the resonance. Its explicit form we consider is
\begin{equation}
\begin{aligned}
    \hat H_\text{P}(t)&=\sum_{n=1}^{n_\text{\tiny P}}V^{(n)}(t)\left\{
   \hat \sigma_x \cos[\omega_0(t-t_\text{\tiny P}^{(n)})] +
   \hat \sigma_y \sin[\omega_0(t-t_\text{\tiny P}^{(n)})] 
    \right\},    \\
    &=\sum_{n=1}^{n_\text{\tiny P}}V^{(n)}(t)\left(
    \hat \sigma_+ e^{i\omega_0(t-t_\text{\tiny P}^{(n)})}+
    \hat \sigma_- e^{-i\omega_0(t-t_\text{\tiny P}^{(n)})}
    \right),
\end{aligned}
\end{equation}
with $n_\text{\tiny P}$ being the number of pulses, $t_\text{\tiny P}^{(n)}$ is the time at which the pulse is switched on every $\Delta t$, namely
\begin{equation}
    t_\text{\tiny P}^{(n)}=t_0+n \Delta t, \quad \text{with}\quad n\in\set{1,\dots,n_\text{\tiny P}}.
\end{equation}
Finally, the switch of the impulse is determined by $V^{(n)}(t)$, which is defined as
\begin{equation}\label{eq.def.DD.V}
    V^{(n)}(t)=
    \begin{cases}
        V,&\text{for}\  t\in[t_\text{\tiny P}^{(n)},t_\text{\tiny P}^{(n)}+\tau_\text{\tiny P}],\\
        0,&\text{otherwise},
    \end{cases}
\end{equation}
where $\tau_\text{\tiny P}$ is the duration time of the pulses.\\

The exact dynamics with respect to the modified Hamiltonian $\hat H(t)$ cannot be solved. However, we can assume that during the pulses the contribution of $\hat H_\text{SB}$ is negligible and we completely neglect it. Then, the dynamics becomes piecewise, alternating $\hat H_\text{SB}$ to $\hat H_\text{P}$.

As for the unperturbed case, we tackle the problem in the interaction picture. Namely, the effective Hamiltonian becomes
\begin{equation}
    \hat H^\text{\tiny (I)}(t)=    \hat H_0^\text{\tiny (I)}(t)+    \hat H_\text{P}^\text{\tiny (I)}(t),
\end{equation}
where $ \hat H_0^\text{\tiny (I)}(t)$ is shown in \eqref{eq.H0DDinteraction} and 
\begin{equation}
\begin{aligned}
    \hat H_\text{P}^\text{\tiny (I)}(t)&=\exp\left[ \frac i\hbar\left( \hat H_\text{S}+\hat H_\text{B}\right)\right] \hat H_\text{P}(t)\exp\left[ -\frac i\hbar\left( \hat H_\text{S}+\hat H_\text{B}\right)\right], \\
    &=e^{ i{\omega_0}\hat \sigma_z t/2}\sum_{n=1}^{n_\text{\tiny P}}V^{(n)}(t)\left(
    \hat \sigma_+ e^{i\omega_0(t-t_\text{\tiny P}^{(n)})}+
    \hat \sigma_- e^{-i\omega_0(t-t_\text{\tiny P}^{(n)})}
    \right)e^{- i{\omega_0}\hat \sigma_z t/2}.
\end{aligned}
\end{equation}
However, one has that
\begin{equation}
    \begin{aligned}
        e^{ i{\omega_0}\hat \sigma_z t/2}\hat \sigma_-e^{- i{\omega_0}\hat \sigma_z t/2}&=e^{ i{\omega_0}\hat \sigma_z t/2}\ket0\bra1e^{- i{\omega_0}\hat \sigma_z t/2},\\
        &=e^{i\omega_0t }\ket0\bra1,\\
        &=e^{i\omega_0 t}\hat \sigma_-,
    \end{aligned}
\end{equation}
and similarly
\begin{equation}
        e^{ i{\omega_0}\hat \sigma_z t/2}\hat \sigma_+e^{- i{\omega_0}\hat \sigma_z t/2}=e^{-i\omega_0 t}\hat \sigma_+.
\end{equation}
Then, we obtain
\begin{equation}
    \begin{aligned}
        \hat H_\text{P}^\text{\tiny (I)}(t)&=\sum_{n=1}^{n_\text{\tiny P}}V^{(n)}(t)\left(
    \hat \sigma_+ e^{-i\omega_0t_\text{\tiny P}^{(n)}}+
    \hat \sigma_- e^{i\omega_0t_\text{\tiny P}^{(n)}}
    \right),\\
    &=\sum_{n=1}^{n_\text{\tiny P}}V^{(n)}(t)e^{ i{\omega_0}\hat \sigma_z t_\text{\tiny P}^{(n)}/2}
    \hat \sigma_x e^{- i{\omega_0}\hat \sigma_z t_\text{\tiny P}^{(n)}/2},
    \end{aligned}
\end{equation}
where we exploited that $\hat \sigma_++\hat\sigma_-=\hat \sigma_x$. Notably, the only time dependence is in $V^{(n)}(t)$, but it is only formal as one can see from Eq.~\eqref{eq.def.DD.V}. Then, when considering the corresponding unitary,  we have
\begin{equation}
\begin{aligned}
    \hat {\mathcal V}_n^\text{\tiny (I)}(\tau_\text{\tiny P})&=\exp\left(-\frac i \hbar    \int_{t_\text{\tiny P}^{(n)}}^{t_\text{\tiny P}^{(n)}+\tau_\text{\tiny P}}\D s\,\hat H_\text{P}^\text{\tiny (I)}(s)\right),\\
    &=\exp\left(-\frac i \hbar Ve^{ i{\omega_0}\hat \sigma_z t_\text{\tiny P}^{(n)}/2}
    \hat \sigma_x e^{- i{\omega_0}\hat \sigma_z t_\text{\tiny P}^{(n)}/2} \tau_\text{\tiny P}\right).
\end{aligned}
\end{equation}
By Taylor expanding
\begin{equation}
\begin{aligned}
    \hat {\mathcal V}_n^\text{\tiny (I)}(\tau_\text{\tiny P})
    &=\sum_k\frac{1}{k!}\left(-\frac i \hbar Ve^{ i{\omega_0}\hat \sigma_z t_\text{\tiny P}^{(n)}/2}
    \hat \sigma_x e^{- i{\omega_0}\hat \sigma_z t_\text{\tiny P}^{(n)}/2} \tau_\text{\tiny P}\right)^k,\\
    &=e^{ i{\omega_0}\hat \sigma_z t_\text{\tiny P}^{(n)}/2}\sum_k\frac{1}{k!}\left(-\frac i \hbar V
    \hat \sigma_x \tau_\text{\tiny P} \right)^ke^{- i{\omega_0}\hat \sigma_z t_\text{\tiny P}^{(n)}/2} ,\\
    &=e^{ i{\omega_0}\hat \sigma_z t_\text{\tiny P}^{(n)}/2} e^{-\tfrac i \hbar V \hat \sigma_x\tau_\text{\tiny P}}
    e^{- i{\omega_0}\hat \sigma_z t_\text{\tiny P}^{(n)}/2} .
\end{aligned}
\end{equation}

We finally fix $V$ and $\tau_\text{\tiny P}$ so to have an actual bit-flip. This is provided by setting
\begin{equation}\label{eq.TV.DD}
    \frac{V \tau_\text{\tiny P}}{\hbar}=\frac \pi2,
\end{equation}
which gives
\begin{equation}
    e^{-\tfrac{i}{\hbar}V\hat \sigma_x\tau_\text{\tiny P}}=e^{-i\tfrac{\pi}{2}\hat \sigma_x}=-i\hat \sigma_x.
\end{equation}
Notably, we can consider the limit of the time pulses that go to zero, i.e.~$\tau_\text{\tiny P}\to0$, as long as  $V\to\infty$  and Eq.~\eqref{eq.TV.DD} holds. Since from here $V$ does not appear explicitly, this will only simplify the calculations.

Then, we have that
\begin{equation}
    \hat {\mathcal V}_n^\text{\tiny (I)}(\tau_\text{\tiny P})=\hat {\mathcal V}_n^\text{\tiny (I)}=-ie^{ i{\omega_0}\hat \sigma_z t_\text{\tiny P}^{(n)}/2} \hat \sigma_x
    e^{- i{\omega_0}\hat \sigma_z t_\text{\tiny P}^{(n)}/2}.
\end{equation}
By considering that the following relation holds
\begin{equation}
    e^{-i\omega_0\hat \sigma_z t/2}=\cos(\omega_0t/2)\hat {\mathbb 1}-i\sin(\omega_0t/2)\hat \sigma_z,
\end{equation}
and the anticommutation relation $\acom{\hat \sigma_x}{\hat \sigma_z}=0$, we have that 
\begin{equation}
    \hat \sigma_xe^{-i\omega_0\hat \sigma_z t/2}=e^{i\omega_0\hat \sigma_z t/2}\hat \sigma_x.
\end{equation}
It follows that one can write the operator $\hat {\mathcal V}_n^\text{\tiny (I)}$ in two equivalent ways:
\begin{equation}
    \hat {\mathcal V}_n^\text{\tiny (I)}=-i e^{ i{\omega_0}\hat \sigma_z t_\text{\tiny P}^{(n)}} \hat \sigma_x=
    -i \hat \sigma_xe^{-i{\omega_0}\hat \sigma_z t_\text{\tiny P}^{(n)}}. 
\end{equation}

Let us now consider the time evolution of the first entire cycle of spin-flips: this is from time $t_0$
 through  time $t_\text{\tiny P}^{(1)}$ when the spin flips the first time, to time 
$t_\text{\tiny P}^{(2)}$ when the spin flips back to the original spin state. In particular, we define this latter time as $t_1=t_0+2\Delta t$. The unitary dynamics from $t_0$ to $t_1$ is given by
\begin{equation}\label{eq.DD.Ut0t1}
\begin{aligned}
        \hat {\mathcal U}_\text{P}^\text{\tiny (I)}(t_0,t_1)&=\hat  {\mathcal V}_2^\text{\tiny (I)}\hat U^{\text{\tiny (I)}}(t_\text{\tiny P}^{(1)},t_\text{\tiny P}^{(2)})\hat  {\mathcal V}_1^\text{\tiny (I)}\hat U^{\text{\tiny (I)}}(t_0,t_\text{\tiny P}^{(1)}),\\
        &=\left[\hat  {\mathcal V}_2^\text{\tiny (I)}\hat  {\mathcal V}_1^\text{\tiny (I)}\right]\left[\left(\hat  {\mathcal V}_1^\text{\tiny (I)}\right)^{-1}\hat U^{\text{\tiny (I)}}(t_\text{\tiny P}^{(1)},t_\text{\tiny P}^{(2)})\hat  {\mathcal V}_1^\text{\tiny (I)}\right]\hat U^{\text{\tiny (I)}}(t_0,t_\text{\tiny P}^{(1)}),
\end{aligned}
\end{equation}
where
\begin{equation}\label{eq.dd.U.nopulsealphabeta}
    \hat U^\text{\tiny (I)}(t_\alpha,t_\beta)=\exp\left[\frac{1}{2}\hat \sigma_z\sum_k\left( \hat b_k^\dag e^{i \omega_k t_\alpha}\xi_k(t_\beta-t_\alpha)-\hat b_k e^{-i \omega_k t_\alpha}\xi_k^*(t_\beta-t_\alpha)\right)\right].
\end{equation}
The first square parenthesis in the last line of Eq.~\eqref{eq.DD.Ut0t1} is given by
\begin{equation}
\begin{aligned}
    \hat  {\mathcal V}_2^\text{\tiny (I)}\hat  {\mathcal V}_1^\text{\tiny (I)}&=\left(-i e^{ i{\omega_0}\hat \sigma_z t_\text{\tiny P}^{(2)}} \hat \sigma_x\right)\left(-i \hat \sigma_x e^{ -i{\omega_0}\hat \sigma_z t_\text{\tiny P}^{(1)}} \right),\\
    &=-e^{ i{\omega_0}\hat \sigma_z (t_\text{\tiny P}^{(2)}-t_\text{\tiny P}^{(1)})}.
\end{aligned}
\end{equation}
Similarly, the second square parenthesis in the last line of Eq.~\eqref{eq.DD.Ut0t1} can be rewritten as
\begin{equation}
        \left(\hat  {\mathcal V}_1^\text{\tiny (I)}\right)^{-1}\hat U^{\text{\tiny (I)}}(t_\text{\tiny P}^{(1)},t_\text{\tiny P}^{(2)})\hat  {\mathcal V}_1^\text{\tiny (I)}=e^{ i{\omega_0}\hat \sigma_z t_\text{\tiny P}^{(1)}} \hat \sigma_x \exp\left[\frac{1}{2}\hat \sigma_z\hat B(t_\text{\tiny P}^{(1)},t_\text{\tiny P}^{(2)})\right]\hat \sigma_xe^{ -i{\omega_0}\hat \sigma_z t_\text{\tiny P}^{(1)}}
\end{equation}
where
$B(t_\text{\tiny P}^{(1)},t_\text{\tiny P}^{(2)})=\sum_k\left( \hat b_k^\dag e^{i \omega_k t_\text{\tiny P}^{(1)}}\xi_k(\Delta t)-\hat b_k e^{-i \omega_k t_\text{\tiny P}^{(1)}}\xi_k^*(\Delta t)\right)$. Here, we Taylor expand the central exponential, which gives
    \begin{equation}
\exp\left[\frac{1}{2}\hat \sigma_z\hat B(t_\text{\tiny P}^{(1)},t_\text{\tiny P}^{(2)})\right]= \sum_l\frac{1}{l!}\left(\frac{1}{2}\hat \sigma_z\hat B(t_\text{\tiny P}^{(1)},t_\text{\tiny P}^{(2)})\right)^l.
    \end{equation}
We divide the contributions to the sum in those with even and odd values of $l$. For even values, we have $\hat\sigma_z^l=\hat\sigma_z^{2l'}$, where $l=2l'$; then $\hat\sigma_z^l=\hat{\mathbb 1}=(-\hat\sigma_z)^l$. For odd values of $l$, we  have $\hat\sigma_z^l=\hat\sigma_z^{2l'+1}$, where $l=2l'+1$; then $\hat\sigma_z^l=\hat\sigma_z$. But more specifically, we also have  that $\hat \sigma_x\hat\sigma_z^l\hat \sigma_x=\hat \sigma_x\hat\sigma_z\hat \sigma_x=-\hat \sigma_z=-\hat \sigma_z^l$. Thus, it follows that 
\begin{equation}
        \left(\hat  {\mathcal V}_1^\text{\tiny (I)}\right)^{-1}\hat U^{\text{\tiny (I)}}(t_\text{\tiny P}^{(1)},t_\text{\tiny P}^{(2)})\hat  {\mathcal V}_1^\text{\tiny (I)}= \exp\left[-\frac{1}{2}\hat \sigma_z\hat B(t_\text{\tiny P}^{(1)},t_\text{\tiny P}^{(2)})\right].
\end{equation}
Thus, we have that Eq.~\eqref{eq.DD.Ut0t1} reads
\begin{equation}
    \hat {\mathcal U}_\text{P}^\text{\tiny (I)}(t_0,t_1)=-e^{ i{\omega_0}\hat \sigma_z (t_\text{\tiny P}^{(2)}-t_\text{\tiny P}^{(1)})}\exp\left[-\frac{1}{2}\hat \sigma_z\hat B(t_\text{\tiny P}^{(1)},t_\text{\tiny P}^{(2)})\right]\exp\left[\frac{1}{2}\hat \sigma_z\hat B(t_0,t_\text{\tiny P}^{(1)})\right],
\end{equation}
and can be recasted as
\begin{equation}
    \hat {\mathcal U}_\text{P}^\text{\tiny (I)}(t_0,t_1)=\exp\left[i{\omega_0}\hat \sigma_z (t_\text{\tiny P}^{(2)}-t_\text{\tiny P}^{(1)})+
    \frac12\hat\sigma_z\sum_k\left( \hat b_k^\dag e^{i \omega_k t_0}\eta_k(\Delta t)-\hat b_k e^{-i \omega_k t_0}\eta_k^*(\Delta t)\right)
    \right],
\end{equation}
where we neglected the overall unimportant phase and  we defined
\begin{equation}\label{eq.dd.def.eta}
    \eta_k(\Delta t)=\xi(\Delta t)\left(1-e^{i\omega_k\Delta t}\right).
\end{equation}

Now, the full evolution from time $t_0$ to time $t_N$ after $N$ entire cycles of spin-flip is simply given by
\begin{equation}\label{eq.dd.pulsesN}
        \prod_{n=1}^N\hat {\mathcal U}_\text{P}^\text{\tiny (I)}(t_{n-1},t_n)  
    =\exp\left[i{\omega_0}\hat \sigma_z (t_N-t_0)+
    \frac12\hat\sigma_z\sum_k\left( \hat b_k^\dag\sum_n e^{i \omega_k t_0}\eta_k(N,\Delta t)-\hat b_k \sum_ne^{-i \omega_k t_{0}}\eta_k^*(N,\Delta t)\right)
    \right],
\end{equation}
where we introduced
\begin{equation}
\begin{aligned}
        \eta_k(N,\Delta t)&=e^{-i\omega_k t_0}\sum_n e^{i \omega_kt_{n-1}}\eta_k(\Delta t),\\
        &=\eta_k(\Delta t)\sum_{n=1}^N e^{2i\omega_k\Delta t(n-1)}.
\end{aligned}
\end{equation}
Such an evolution is to be compared to that with no pulses on the same time period. This is given by Eq.~\eqref{eq.dd.U.nopulsealphabeta} where one substitutes $t_\alpha\to t_0$ and $t_\beta\to t_N$. Then, since $t_N-t_0=2N \Delta t$, we have
\begin{equation}\label{eq.dd.nopulsesN}
    \hat U^\text{\tiny (I)}(t_0,t_N)=\exp\left[\frac{1}{2}\hat \sigma_z\sum_k\left( \hat b_k^\dag e^{i \omega_k t_0}\xi_k(2N\Delta t)-\hat b_k e^{-i \omega_k t_0}\xi_k^*(2N\Delta t)\right)\right].
\end{equation}
Notably, the expressions in Eq.~\eqref{eq.dd.pulsesN} and Eq.~\eqref{eq.dd.nopulsesN} have a similar structure, with the important difference being the factor $\eta_k(N,\Delta t)$ substituted with $\xi_k(2N\Delta t)$. Thus, the decohering factor $\Gamma(t_0,t_N)$ will take a suitably modified expression as that in Eq.~\eqref{eq-def.GammaDD}, namely
\begin{equation}
\Gamma(t_0,t_N)=\sum_k\frac{|e^{i\omega_kt_0}\eta_k(N,\Delta t)|^2}{2}\coth \left(\frac{\beta\hbar\omega_k}{2}\right).
\end{equation}
We now compare the difference between these two factors:
\begin{equation}
        \eta_k(N,\Delta t)-\xi_k(2N\Delta t)=\eta_k(\Delta t)\sum_{n=1}^N e^{2i\omega_k\Delta t(n-1)}-\xi_k(2\Delta t)\sum_{n=1}^N e^{2i\omega_k\Delta t(n-1)},
\end{equation}
where we exploited the composition of the $\xi_k$ terms. Then, by considering that
\begin{equation}
\begin{aligned}
        \xi_k(\Delta t)(1+e^{i\omega_k \Delta t})&=\frac{2g_k}{\omega_k}(1-e^{i\omega_k \Delta t})(1+e^{i\omega_k \Delta t}),\\
        &=\frac{2g_k}{\omega_k}(1-e^{2i\omega_k \Delta t}),\\
        &=\xi_k(2\Delta t),
\end{aligned}
\end{equation}
and the definition of $\eta_k(\Delta t)$ in Eq.~\eqref{eq.dd.def.eta},
we obtain
\begin{equation}
    \eta_k(N,\Delta t)-\xi_k(2N\Delta t)=-2\xi_k(\Delta t)e^{i\omega_k \Delta t}\sum_{n=1}e^{2i\omega_k\Delta t(n-1)}.
\end{equation}
Equivalently, we have
\begin{equation}
    \eta_k(N,\Delta t)=\xi_k(2N\Delta t)\left(1-f_k(N,\Delta t)\right),
\end{equation}
where
\begin{equation}
        f_k(N,\Delta t)=2\frac{\xi_k(\Delta t)}{\xi_k(2N\Delta t)}e^{i\omega_k \Delta t}\sum_{n=1}e^{2i\omega_k\Delta t(n-1)}.
\end{equation}
By exploiting the geometric series and the definition of $\xi_k$, we get
\begin{equation}
\begin{aligned}
        f_k(N,\Delta t)&=2\frac{(1-e^{i\omega_k \Delta t})}{(1-e^{2i\omega_k N \Delta t})}e^{i\omega_k \Delta t}\frac{(1-e^{2i\omega_kN\Delta t})}{(1-e^{2i\omega_k \Delta t})},\\
        &=2\frac{(1-e^{i\omega_k \Delta t})}{(1-e^{2i\omega_k  \Delta t})}e^{i\omega_k \Delta t}.
\end{aligned}
\end{equation}
Finally, by taking the limit of dense pulses, i.e.~$\Delta t\to0$, we obtain
\begin{equation}
    \lim_{\Delta t\to 0}f_k(N,\Delta t)=1,
\end{equation}
which means that under the same limit we have
\begin{equation}
    \lim_{\Delta t\to 0}\eta_k(N,\Delta t)=0.
\end{equation}
As a consequence, the decoherence factor vanishes: $\Gamma(t_0,t_N)\to0$. Namely, the decohering effect of the environment on the system is cancelled. Effectively, one has a (dynamical) decoupling of the system from its environment.

\section{Quantum Error Mitigation}
\index{Quantum error  mitigation}
\label{ch.errormitigation}

Quantum Error Mitigation (QEM) wants to translate the improvements of quantum hardware in those of quantum information and computation. Namely, it is an algorithmic scheme that reduces noise-induced bias in the expectation value of an observable of interest by post-processing outputs from an ensemble of circuit runs. To do this, it employs circuits at the same depth as the original unmitigated circuit or above. QEM applies post-processing directly from the hardware outputs. Thus, if the circuit size, being the product of the circuit depth times the number of qubits, becomes too large then QEM loses its usefulness.

A good QEM approach should employ a limited number of qubits, while still providing a guaranteed accuracy. This converts in a formal error bound, which indicates how well the QEM code works. Moreover, it should employ only a few (or better none) assumptions about the final state. For example, assuming that the final state is factorised is not a good assumption. Indeed, it would  strongly limit the applicability of the corresponding QEM algorithm.\\

Before dwelling in two, among various, algorithms in the QEM context, we provide the general idea of the QEM approach.
We defined  the primary circuit as that  process that would ideally produce the perfect output state $\hat \rho_0$. Due to the presence of noises, the primary circuit produces the noisy state $\hat \rho$. To account how a circuit works, we consider an observable of interests $\hat O$ whose expectation value is the output information we seek. To compute this, we will run the circuit $N_\text{sample}$ times, which is the number of circuit executions. Also in the noiseless case, a finite value of $N_\text{sample}$ implies a finite inaccuracy of the estimated average. This is the so-called shot noise. However, in such a case,  there will be no systematic shift, i.e.~bias, in the expectation value of $\hat O$ due to the noise. QEM aims to reduce such a bias. Often, this implies that the corresponding variace increases. Then, one needs to increase the number of circuit runs $N>N_\text{sample}$ to compensate. The sampling overhead is the cost, in terms of number of repetitions, of the QEM method when compared to the noiseless circuit.

We underline that, conversely to QEC, in QEM there is no monitoring of the errors occurring during the run of the circuit.

\subsection{Zero noise extrapolation}\index{Zero noise extrapolation}

The Zero noise extrapolation (ZNE) method extracts the zero-noise expectations from a fitting of the circuit ran at different values of the noise. 
We define a time dependent Hamiltonian $\hat H(t)$ that embeds action of the noiseless circuit. It can be written as
\begin{equation}
    \hat H(t)=\sum_\alpha J_\alpha(t) \hat P_\alpha,
\end{equation}
where $J_\alpha(t)$ are some time dependent couplings that switch on and off the gates of the circuit, which are implemented by the corresponding $N$-qubit Pauli operators $\hat P_\alpha$. The full dynamics, including the action of the noise, is given by the following master equation
\begin{equation}\label{eq.master.zeronoise}
    \frac{\D \hat \rho_\lambda(t)}{\D t}=-\frac{i}{\hbar}\com{\hat H(t)}{\hat \rho_\lambda(t)}+\lambda\mathcal L[\hat \rho_\lambda(t)],
\end{equation}
where $t\in[0,T]$, with $T$ being the time at which the circuit ends. We assume here that the noise coupling $\lambda$ is small. Moreover, we assume that the noise dissipator $\mathcal L$ is invariant under time rescaling and it is independent from $\hat J_\alpha(t)$.

Now, given the observable of interest $\hat A$, we compute the corresponding expectation value on the noisy circuit as $E(\lambda)=\Tr{\hat A\hat \rho_\lambda(T)}$, where $\hat \rho_\lambda(T)$ is the solution of Eq.~\eqref{eq.master.zeronoise}. What we want to do is to estimate $E(\lambda)$ for $\lambda\to0$. 
Since one cannot reduce the value of $\lambda$, to construct a series of measurement from where extrapolate the estimate $E(0)$, we increase the value of $\lambda$. This can be done by considering the following rescaling.
We dilate the time $T$ at which the circuit is ran and, due to the time invariance of $\mathcal L$, this is equivalent to let the noise act more on the circuit. Then, one applies this idea with different values of $\lambda$ and can perform a fit and deduce the value of $E$ for $\lambda\to0$. Practically, we perform the circuit $N_\text{cir}$ times, at different values of the noise rate $\lambda_j=c_j\lambda$, $j=0,\dots,N_\text{cir}-1$ with $c_0=1<c_1<\dots<c_{N_\text{cir}-1}$.
For each value of $\lambda_j$, we run the circuit with the following rescaled Hamiltonian
\begin{equation}\label{eq.rescaledH.zeronoise}
    \hat H^{(j)}(t)=\sum_\alpha J_\alpha^{(j)}(t)\hat P_\alpha, \quad\text{where}\quad J_\alpha^{(j)}(t)=c_j^{-1}J_\alpha(c_j^{-1} t),
\end{equation}
for a time $T_j=c_j T$. The rescaled dynamics gives
\begin{equation}\label{eq.rescaledmaster.zeronoise}
\frac{\D \hat \rho_\lambda^{(j)}(t)}{\D t}=-\frac{i}{\hbar}\com{\hat H^{(j)}(t)}{\hat \rho_\lambda^{(j)}(t)}+\lambda\mathcal L[\hat \rho_\lambda^{(j)}(t)].
\end{equation}
By merging the latter with Eq.~\eqref{eq.rescaledH.zeronoise}, we obtain
\begin{equation}
\frac{\D \hat \rho_\lambda^{(j)}(t)}{\D t}=-\frac{i}{\hbar}\sum_\alpha c_j^{-1}J_\alpha(c_j^{-1} t)\com{\hat P_\alpha}{\hat \rho_\lambda^{(j)}(t)}+\lambda\mathcal L[\hat \rho_\lambda^{(j)}(t)].
\end{equation}
By defining $s=c_j^{-1} t$, which runs in the interval $s\in[0,T]$ since $t\in[0,T_j]$, we rewrite the above master equation as
\begin{equation}
    \frac{\D \hat \rho_\lambda^{(j)}(t)}{\D t}=\frac{\D \hat \rho_\lambda^{(j)}(c_js)}{c_j\D s}=-\frac{i}{\hbar}\sum_\alpha c_j^{-1}J_\alpha(s)\com{\hat P_\alpha}{\hat \rho_\lambda^{(j)}(c_js)}+\lambda\mathcal L[\hat \rho_\lambda^{(j)}(c_js)].
\end{equation}
By multiplying the left and right hand side by $c_j$ we obtain
\begin{equation}
    \frac{\D \hat \rho^{(j)}_\lambda(c_js)}{\D s }=-\frac{i}{\hbar}\com{\hat H(s)}{\hat \rho^{(j)}_\lambda(c_js)}+c_j\lambda\mathcal L[\hat \rho^{(j)}_\lambda(c_js)],
\end{equation}
which is Eq.~\eqref{eq.master.zeronoise} with $\lambda$ substituted with $c_j\lambda$. Its solution at time $s=T$ is given by $\hat \rho_{c_j\lambda}(T)=\hat\rho_\lambda^{(j)}(T_j)$.
Correspondingly, we compute the  expectation value $E(\lambda_j)=\Tr{\hat A\hat \rho_\lambda^{(j)}(T_j)}=\Tr{\hat A\hat \rho_{c_j\lambda}(T)}$.
Experimentally, for each $c_j$, one performs $N_\text{sample}$ runs of the circuit and obtains an estimator $\tilde E(\lambda_j)$, which converges to the true value $E(\lambda_j)$ only in the asymptotic limit $N_\text{sample}\to\infty$. Specifically, one has
\begin{equation}
    \tilde E(\lambda_j)=E(\lambda_j)+\tilde \delta,
\end{equation}
where $\tilde \delta$ is a random variable with zero mean and variance $\mathbb E[\tilde\delta^2]=\sigma_0^2/N_\text{sample}$, with $\sigma_0^2$ corresponding to the single-shot variance. Here, $\mathbb E$ is to the mean over the sampling. 

\begin{figure}[h]
    \centering
    \includegraphics[width=0.7\linewidth]{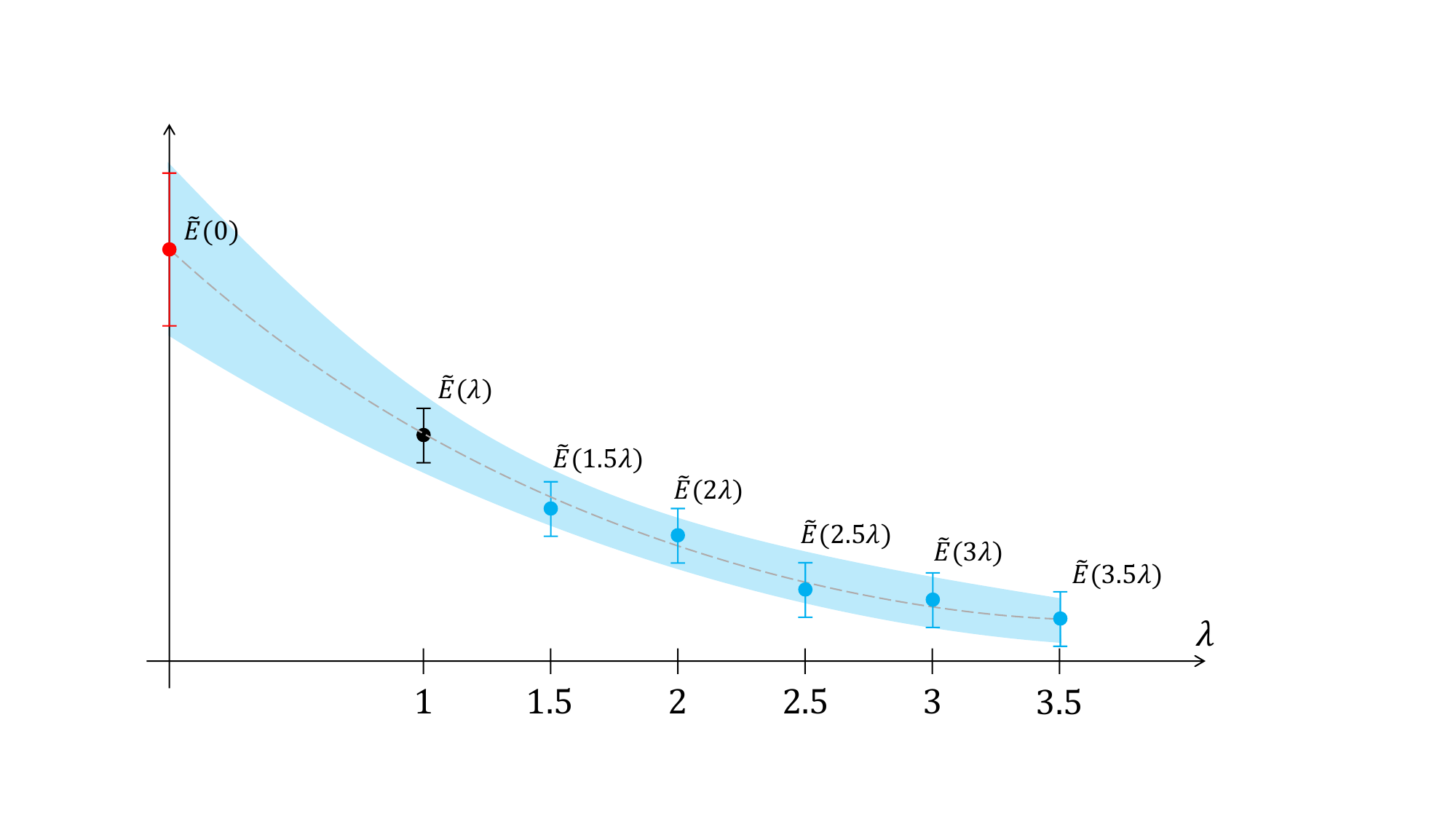}
    \caption{Graphical representation of the Zero Error Extrapolation. Given the set of estimators $\tilde E(\lambda)$ at different values of the noise (black and blue dots and error bars), one performs a fit assuming a specific model (gray dashed line) with corresponding confidence region (light blue region). In such a way, one extrapolates the value of $\tilde E(0)$ with its corresponding error bar (red point and error bar).}
    \label{fig:zero-error}
\end{figure}

Now, the ZNE problem is to construct a good estimator $\tilde E(0)$ for the expectation value $E(\lambda=0)=\Tr{\hat A\hat  \rho_0(T)}$ from the set of estimators $\tilde E(\lambda_j)$. Figure  \ref{fig:zero-error} represents the problem.
To be a good estimator, we want that its bias
\begin{equation}\label{def.bias,l0}
    \text{Bias}(\tilde E(0))=\mathbb E[\tilde E(0)-E(0)],
\end{equation}
and its variance
\begin{equation}
    \text{Var}(\tilde E(0))=\mathbb E[\tilde E(0)^2]-\mathbb E[\tilde E(0)]^2,
\end{equation}
are both small. We employ the mean squared error (MSE) are a figure of merit with respect to the true unknown parameter
\begin{equation}
\begin{aligned}
    \text{MSE}(\tilde E(0))&=\mathbb E[(\tilde E(0)-E(0))^2],\\
    &=\text{Var}(\tilde E(0))+(\text{Bias}(\tilde E(0)))^2.
    \end{aligned}
\end{equation}
If the expectation value $E(\lambda)$ can be an arbitrary function of $\lambda$ without any regularity assumption, then ZNE is impossible.  However, from physical considerations, it is reasonable to have a model for it, for example we can assume a linear, a polynomial or an exponential dependence with respect to $\lambda$. 
\begin{itemize}
    \item[1]
If we assume a linear dependence on $\lambda$, the corresponding linear model is given by
\begin{equation}
    E_\text{linear}(\lambda)=a_0+a_1\lambda.
\end{equation}
In such a case, a simple analytic solution exists, which is that of the ordinary least squared estimator of the intercept parameter. Namely, we have
\begin{equation}
    \tilde E_\text{linear}(0)=\bar E(\lambda)-\frac{S_{\lambda E}}{S_{\lambda\lambda}}\bar \lambda,
\end{equation}
where
\begin{equation}
    \begin{aligned}
        \bar \lambda&=\frac{1}{N_\text{cir}}\sum_{j=0}^{N_\text{cir}-1}\lambda_j,\\
        \bar E(\lambda)&=\frac{1}{N_\text{cir}}\sum_{j=0}^{N_\text{cir}-1}\tilde E(\lambda_j),\\
        S_{\lambda E}&=\sum_{j=0}^{N_\text{cir}-1}(\lambda_j-\bar\lambda)(\tilde E(\lambda_j)-\bar E(\lambda)),\\
        S_{\lambda \lambda}&=\sum_{j=0}^{N_\text{cir}-1}(\lambda_j-\bar\lambda)^2.
    \end{aligned}
\end{equation}
With respect to the zero noise value $E_\text{linear}(0)$, the estimator $\tilde E_\text{linear}(0)$ is unbiased. Its variance, under the assumption that the statistical uncertainty is the same for each $\lambda_j$, reads
\begin{equation}
    \text{Var}(\tilde E_\text{linear}(0))=\frac{\sigma_0^2}{N_\text{sample}}\left(\frac{1}{N_\text{cir}}+\frac{\bar \lambda^2}{S_{\lambda\lambda}}\right).
\end{equation}

\item[2] The Richardson's extrapolation is a special case of the polynomial extrapolation, which is limited  at  order $N_\text{cir}-1$. The corresponding model is given by
\begin{equation}
    E_\text{Rich}(\lambda)=a_0+a_1\lambda+\dots+c_{N_\text{cir}-1}\lambda^{N_\text{cir}-1}.
\end{equation}
This is the only case in which the fitted polynomial perfectly interpolates the $N_\text{cir}$ data points such that, in the ideal limit of an infinite number of samples $N_\text{sample}\to\infty$, the error with respect to the true expectation value is by construction $\mathcal O(N_\text{cir})$. Using the Lagrange polynomial, the estimator can be expressed explicitly as
\begin{equation}
    \tilde E_\text{Rich}(0)=\sum_{j=0}^{N_\text{cir}-1}\tilde E(\lambda_j)\gamma_j,
\end{equation}
where
\begin{equation}
    \gamma_j=\prod_{m\neq j}\frac{c_m}{c_j-c_m}.
\end{equation}
The error of the estimator is $\mathcal O(N_\text{cir})$ only in the asymptotic limit $N_\text{sample}\to \infty$. In other words, $\mathcal O(N_\text{cir})$ corresponds to the bias term in Eq.~\eqref{def.bias,l0}. In a real scenario, $N_\text{sample}$ is finite, and the variance term in Eq.~\eqref{def.bias,l0} grows exponentially as we increase $N_\text{cir}$.

\end{itemize}

\subsection{Probabilistic error cancellation}\index{Probabilistic error cancellation}

The Probabilistic error cancellation (PEC) method cancels the effects of the noise employing a  map that acts as the inverse of the noise map under suitable average.

Suppose the ideal circuit is performed by a unitary CPTP map $\mathcal U$ being the consecutive application of unitary gates: $\hat U_\text{circuit}=\hat U_d\dots\hat U_1$, where $d$ id the depth of the circuit. One can represent the corresponding noisy circuit by substituting each unitary operation with its noisy counterpart, namely $\hat U_i \hat \rho \hat U_i^\dag\to\hat U_i\Lambda_i[  \hat \rho ]\hat U_i^\dag$, where $\Lambda_i$ is a CPTP noisy map and  $\hat \rho$ is the $N$-qubit state. If we focus on a single gate $\hat U_i$, the two corresponding circuits are represented as
\begin{equation}
    \begin{tikzcd}
        &\gate[5]{U_i}&\\
        &&\\
        &&\\
        &&\\
        &&
    \end{tikzcd}
    \to
        \begin{tikzcd}
        &\gate[5]{\Lambda_i}&\gate[5]{U_i}&\\
        &&&\\
        &&&\\
        &&&\\
        &&&
    \end{tikzcd}
\end{equation}
Now, the point is if we can invert the CPTP map $\Lambda_i$ via the application of its inverse $\Lambda_i^{-1}$. In general, this is not possible. Indeed, typically,  $\Lambda_i^{-1}$ is not a CPTP map and thus such an inverse operation of the noise cannot be implemented. Nevertheless, such an operation can be implemented on average.

\begin{figure}[h]
    \centering
    \includegraphics[width=0.45\linewidth]{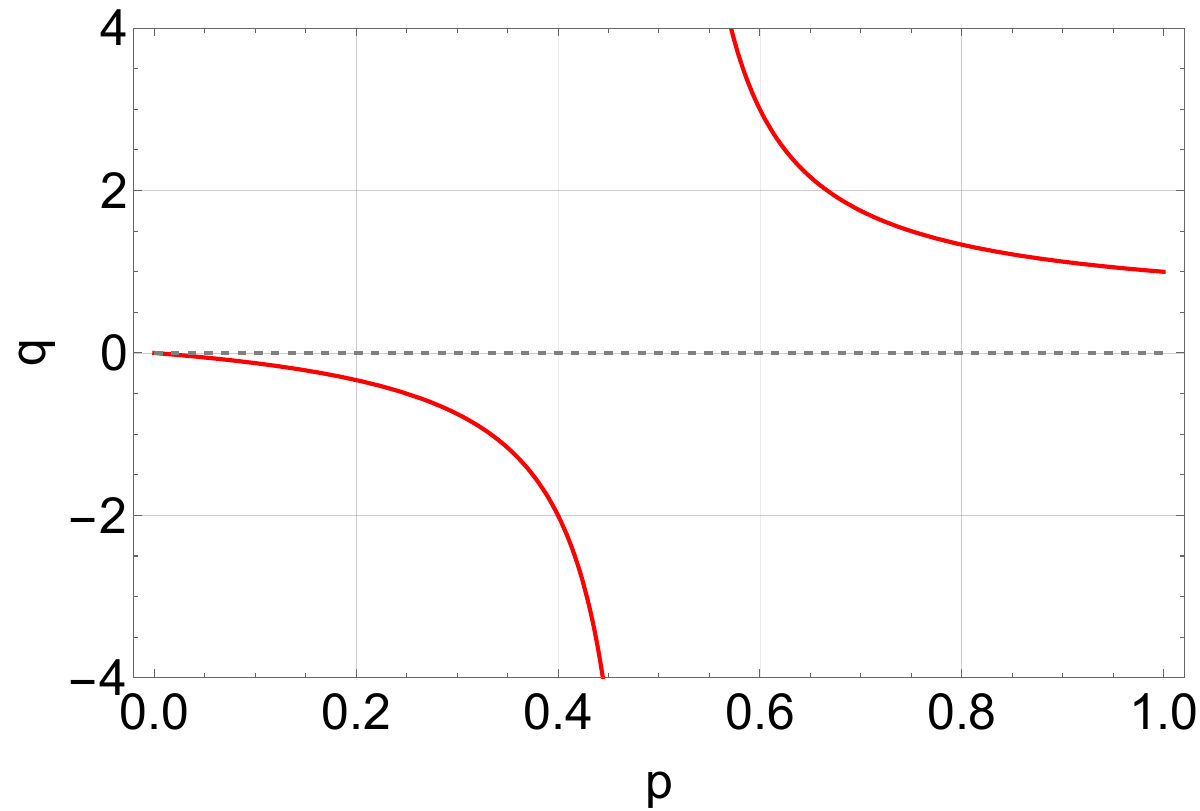}    \includegraphics[width=0.45\linewidth]{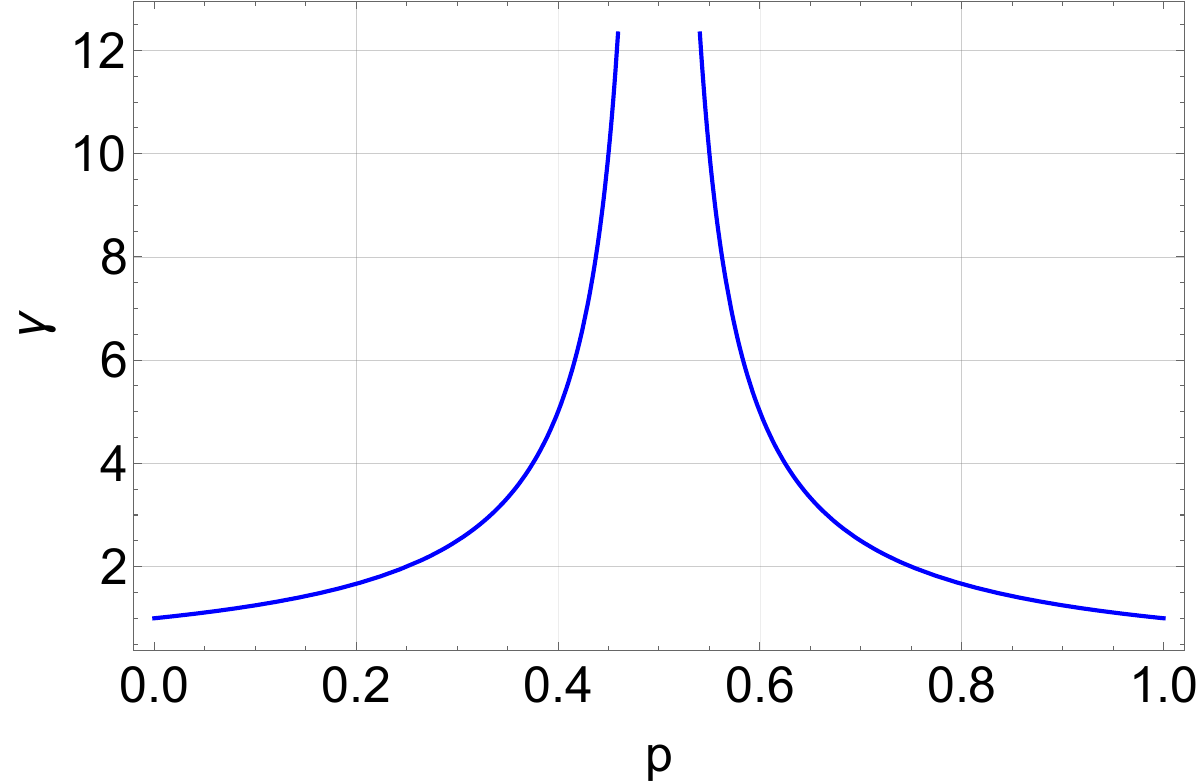}  
    \caption{Quasi-probability $q$ (left panel) and renormalisation constant $\gamma$ (right panel) as a function of the probability $p$ of having an error.}
    \label{fig:q-gamma_cancellation}
\end{figure}

Consider the toy model of a single qubit, where the unitary noiseless operation is the identity: $\hat U=\hat {\mathbb 1}$, and the noise channel is the bit-flip with a probability $p$. Thus, the corresponding map is
\begin{equation}
\Lambda(\hat\rho)=(1-p)\hat {\mathbb 1}\hat \rho\hat {\mathbb 1}+p\hat \sigma_x\hat \rho\hat\sigma_x.
\end{equation}
This map corresponds to the unravelling with two components: with a probability $p$ one applies an extra gate $X$, and with probability $(1-p)$ one does nothing, i.e.~applies the gate $\mathbb 1$. Notably, both these gates have an inverse. Indeed, $\hat {\mathbb 1}^{-1}=\hat {\mathbb 1}$ and $\hat \sigma_x^{-1}=\hat \sigma_x$. Then, we construct the inverse noise map $\Lambda^{-1}$ as having two components: with a probability $q$ we apply an $X$ gate, and with a probability $(1-q)$ we apply an $\mathbb 1$ gate. The corresponding total circuit can be then decomposed in the four components:
\begin{equation}
        \begin{tikzcd}
        &\gate{\Lambda^{-1}}&\gate{\Lambda}&\gate{\mathbb 1}&\meter{}
    \end{tikzcd}
=
\begin{cases}
            \begin{tikzcd}
        &\gate{\mathbb 1}&\gate{\mathbb 1}&\gate{\mathbb 1}&\meter{}
    \end{tikzcd}
    &\text{a), with prob = $(1-p)(1-q)$,}\\
            \begin{tikzcd}
        &\gate{X}&\gate{\mathbb 1}&\gate{\mathbb 1}&\meter{}
    \end{tikzcd}
    &\text{b), with prob = $(1-p)q$,}\\
            \begin{tikzcd}
        &\gate{\mathbb 1}&\gate{X}&\gate{\mathbb 1}&\meter{}
    \end{tikzcd}
    &\text{c), with prob = $p(1-q)$,}\\
            \begin{tikzcd}
        &\gate{X}&\gate{X}&\gate{\mathbb 1}&\meter{}
    \end{tikzcd}
    &\text{d), with prob = $pq$.}
    \end{cases}
\end{equation}
Now, we want to fix $q$ such that, under the ensemble average, the circuit b) occurs with a probability being the opposite value of that of circuit c) occuring, and that the sum of the probabilities of having the circuit a) and d) gives 1. This implies the following system of equations
\begin{equation}
    P_aP_d=(1-p)(1-q)+pq=1,\quad\text{and}\quad P_bP_c=(1-p)q+p(1-q)=0.
\end{equation}
The solution is given by
\begin{equation}
    q=\frac{-p}{1-2p},
\end{equation}
which is a quasi-probability, since it can take negative values, and it is shown in the left panel of Fig.~\ref{fig:q-gamma_cancellation}.
Now, the inverse noise map $\Lambda^{-1}$ is given by
\begin{equation}\label{eq.map.pec}
    \begin{aligned}
        \Lambda^{-1}(\hat \rho)&=(1-q)\hat {\mathbb 1}\hat \rho\hat{\mathbb 1}+q\hat \sigma_x\hat \rho\hat \sigma_x,\\
    &=\text{sgn}(1-q)|1-q|\hat {\mathbb 1}\hat \rho\hat{\mathbb 1}+\text{sgn}(q)|q|\hat \sigma_x\hat \rho\hat \sigma_x,\\
    &=\gamma\left[S_{\mathbb 1}P_{\mathbb 1}\hat {\mathbb 1}\hat \rho\hat{\mathbb 1}+S_{X}P_{X}\hat \sigma_x\hat \rho\hat \sigma_x\right],
    \end{aligned}
\end{equation}
where in the first line we used $x=\text{sgn}(x)|x|$, with sgn indicating the sign function, and in the second line we introduced
\begin{equation}
        \gamma=|1-q|+|q|,
\end{equation}
which is represented in the right panel of Fig.~\ref{fig:q-gamma_cancellation}. Finally, we defined
\begin{equation}
\begin{aligned}
S_{\mathbb 1}=\text{sgn}(1-q),&\quad P_{\mathbb 1}=\frac{|1-q|}{\gamma},\\
S_{X}=\text{sgn}(q),&\quad P_{X}=\frac{|q|}{\gamma}.
    \end{aligned}
\end{equation}
Thus, independently from the unravelling of the noise map, i.e.~without knowing if the bit-flip noise is applied or not, we apply the map $\Lambda^{-1}$ in the last line of Eq.~\eqref{eq.map.pec}. This can be implemented with the following circuit:
\begin{equation}
        \begin{tikzcd}
        &\gate{\Lambda^{-1}}&\gate{\Lambda}&\gate{\mathbb 1}&\meter{\braket{M}}&\setwiretype{c}&\gate[style={rounded
    corners}]{\text{CPP}}
    \end{tikzcd}
    =
\begin{cases}
            \begin{tikzcd}
        &\gate{\mathbb 1}&\gate{\Lambda}&\gate{\mathbb 1}&\meter{M_{\mathbb 1}}&\setwiretype{c}&\gate{ S_{\mathbb 1}}
    \end{tikzcd}
    &\text{a), with prob = $P_{\mathbb 1}$,}\\
            \begin{tikzcd}
        &\gate{X}&\gate{\Lambda}&\gate{\mathbb 1}&\meter{M_X}&\setwiretype{c}&\gate{ S_X}
    \end{tikzcd}
    &\text{b), with prob = $P_X$.}
    \end{cases}
\end{equation}
A classical post-processing (CPP) is applied to multiplicate the outcome of the result by the proper sign factor. Eventually, the mitigated result  is given by
\begin{equation}
    \braket{M}=\gamma\left(S_{\mathbb 1} P_{\mathbb 1}M_{\mathbb 1}+S_XP_XM_X \right).
\end{equation}
This is an unbiased estimator. The cost of the mitigation procedure goes in the variance, which grows by a factor $\gamma^2$ compared to the unmitigated one.\\

Consider a more general case of the noise map $\Lambda$ acting on a single qubit, which reads
\begin{equation}
    \Lambda(\hat \rho)=\lambda_0\hat \rho+\lambda_1\hat \sigma_x\hat \rho\hat \sigma_x+\lambda_2\hat \sigma_y\hat \rho\hat \sigma_y+\lambda_3\hat \sigma_z\hat \rho\hat \sigma_z,
\end{equation}
where
\begin{equation}
    \lambda_\alpha\geq0,\quad\text{and}\quad\sum_{\alpha=0}^3\lambda_\alpha=1.
\end{equation}
Such a map is a CPTP map. Similarly as done above, we construct the inverse map $\Lambda^{-1}$ as
\begin{equation}
    \Lambda^{-1}(\hat \rho)=q_0\hat \rho+q_1\hat \sigma_x\hat \rho\hat \sigma_x+q_2\hat \sigma_y\hat \rho\hat \sigma_y+q_3\hat \sigma_z\hat \rho\hat \sigma_z,
\end{equation}
where we require that
\begin{equation}
    \sum_{\alpha=0}^3q_\alpha=1,
\end{equation}
but we do not add any restriction on the sign of $q_\alpha$. Then, in terms of unravellings, we have 4 possible evolutions provided by $\Lambda$ and 4 by $\Lambda^{-1}$ for a total of 16 possible mappings. Explicitly they give
\begin{equation}
    \begin{array}{c|c|c|c}
        \Lambda & \Lambda^{-1}&\hat\rho\to\hat\rho'&\text{probability }P_{\alpha\beta}\\
        \midrule
        \hat{\mathbb 1}&\hat{\mathbb 1}&\hat \rho&\lambda_0q_0\\
        \hat{\mathbb 1}&\hat \sigma_x&\hat\sigma_x\hat \rho\hat \sigma_x&\lambda_0q_1\\
        \hat{\mathbb 1}&\hat \sigma_y&\hat\sigma_y\hat \rho\hat \sigma_y&\lambda_0q_2\\
        \hat{\mathbb 1}&\hat \sigma_z&\hat\sigma_z\hat \rho\hat \sigma_z&\lambda_0q_3\\
        \hat\sigma_x&\hat{\mathbb 1}&\hat\sigma_x\hat \rho\hat\sigma_x&\lambda_1q_0\\
        \hat\sigma_x&\hat \sigma_x&\hat\sigma_x^2\hat \rho\hat \sigma_x^2&\lambda_1q_1\\
        \hat\sigma_x&\hat \sigma_y&\hat\sigma_x\hat\sigma_y\hat \rho\hat \sigma_y\hat\sigma_x&\lambda_1q_2\\
        \hat\sigma_x&\hat \sigma_z&\hat\sigma_x\hat\sigma_z\hat \rho\hat \sigma_z\hat\sigma_x&\lambda_1q_3\\
        \hat\sigma_y&\hat{\mathbb 1}&\hat\sigma_y\hat \rho\hat\sigma_y&\lambda_2q_0\\
        \hat\sigma_y&\hat \sigma_x&\hat\sigma_y\hat\sigma_x\hat \rho\hat \sigma_x\hat\sigma_y&\lambda_2q_1\\
        \hat\sigma_y&\hat \sigma_y&\hat\sigma_y^2\hat \rho\hat \sigma_y^2&\lambda_2q_2\\
        \hat\sigma_y&\hat \sigma_z&\hat\sigma_y\hat\sigma_z\hat \rho\hat \sigma_z\hat\sigma_y&\lambda_2q_3\\
        \hat\sigma_z&\hat{\mathbb 1}&\hat\sigma_z\hat \rho\hat\sigma_z&\lambda_3q_0\\
        \hat\sigma_z&\hat \sigma_x&\hat\sigma_z\hat\sigma_x\hat \rho\hat \sigma_x\hat\sigma_z&\lambda_3q_1\\
        \hat\sigma_z&\hat \sigma_y&\hat\sigma_z\hat\sigma_y\hat \rho\hat \sigma_y\hat\sigma_z&\lambda_3q_2\\
        \hat\sigma_z&\hat \sigma_z&\hat\sigma_z^2\hat \rho\hat \sigma_z^2&\lambda_3q_3
    \end{array}
\end{equation}
However, we can exploit that $\hat\sigma_\alpha^2=\hat{\mathbb 1}$ and that $\hat\sigma_i\hat \sigma_j=i\epsilon_{ijk}\hat \sigma_k$. Thus, the above table becomes
\begin{equation}
    \begin{array}{c|c|c|c}
        \Lambda & \Lambda^{-1}&\hat\rho\to\hat\rho'&\text{probability }P_{\alpha\beta}\\
        \midrule
        \hat{\mathbb 1}&\hat{\mathbb 1}&\hat \rho&\lambda_0q_0\\
        \hat{\mathbb 1}&\hat \sigma_x&\hat\sigma_x\hat \rho\hat \sigma_x&\lambda_0q_1\\
        \hat{\mathbb 1}&\hat \sigma_y&\hat\sigma_y\hat \rho\hat \sigma_y&\lambda_0q_2\\
        \hat{\mathbb 1}&\hat \sigma_z&\hat\sigma_z\hat \rho\hat \sigma_z&\lambda_0q_3\\
        \hat\sigma_x&\hat{\mathbb 1}&\hat\sigma_x\hat \rho\hat\sigma_x&\lambda_1q_0\\
        \hat\sigma_x&\hat \sigma_x&\hat \rho&\lambda_1q_1\\
        \hat\sigma_x&\hat \sigma_y&\hat\sigma_z\hat \rho\hat \sigma_z&\lambda_1q_2\\
        \hat\sigma_x&\hat \sigma_z&\hat\sigma_y\hat \rho\hat \sigma_y&\lambda_1q_3\\
        \hat\sigma_y&\hat{\mathbb 1}&\hat\sigma_y\hat \rho\hat\sigma_y&\lambda_2q_0\\
        \hat\sigma_y&\hat \sigma_x&\hat\sigma_z\hat \rho\hat \sigma_z&\lambda_2q_1\\
        \hat\sigma_y&\hat \sigma_y&\hat \rho&\lambda_2q_2\\
        \hat\sigma_y&\hat \sigma_z&\hat\sigma_x\hat \rho\hat \sigma_x&\lambda_2q_3\\
        \hat\sigma_z&\hat{\mathbb 1}&\hat\sigma_z\hat \rho\hat\sigma_z&\lambda_3q_0\\
        \hat\sigma_z&\hat \sigma_x&\hat\sigma_y\hat \rho\hat \sigma_y&\lambda_3q_1\\
        \hat\sigma_z&\hat \sigma_y&\hat\sigma_x\hat \rho\hat \sigma_x&\lambda_3q_2\\
        \hat\sigma_z&\hat \sigma_z&\hat \rho&\lambda_3q_3
    \end{array}
\end{equation}
Finally, we impose that the sum of the probabilities of getting $\hat \rho'=\hat \rho$ should be 1, and those such $\hat \rho'\neq\hat \rho$ should be 0. Namely
\begin{equation}
    \begin{aligned}
        P_{00}+P_{11}+P_{22}+P_{33}&=\lambda_0q_0+\lambda_1q_1+\lambda_2q_2+\lambda_3q_3=1,\\
        P_{01}+P_{10}+P_{23}+P_{32}&=\lambda_0q_1+\lambda_1q_0+\lambda_2q_3+\lambda_3q_2=0,\\
        P_{02}+P_{20}+P_{13}+P_{31}&=\lambda_0q_2+\lambda_2q_0+\lambda_1q_3+\lambda_3q_1=0,\\
        P_{03}+P_{30}+P_{12}+P_{21}&=\lambda_0q_3+\lambda_3q_0+\lambda_1q_2+\lambda_2q_1=0.
    \end{aligned}
\end{equation}
The  solution to this system of linear equations gives
\begin{equation}
\begin{aligned}
    q_0&=\frac14\left(
    1+\frac{1}{1-2\lambda_1-2\lambda_2}+\frac{1}{1-2\lambda_1-2\lambda_3}+\frac{1}{1-2\lambda_2-2\lambda_2}\right),\\
    q_1&=\frac14\left(
    1-\frac{1}{1-2\lambda_1-2\lambda_2}-\frac{1}{1-2\lambda_1-2\lambda_3}+\frac{1}{1-2\lambda_2-2\lambda_2}
    \right),\\
    q_2&=\frac14\left(
    1-\frac{1}{1-2\lambda_1-2\lambda_2}+\frac{1}{1-2\lambda_1-2\lambda_3}-\frac{1}{1-2\lambda_2-2\lambda_2}\right),\\
    q_3&=\frac14\left(
    1+\frac{1}{1-2\lambda_1-2\lambda_2}-\frac{1}{1-2\lambda_1-2\lambda_3}-\frac{1}{1-2\lambda_2-2\lambda_2}
    \right).
\end{aligned}
\end{equation}
The inverse map can be rewritten as
\begin{equation}
    \begin{aligned}
        \Lambda^{-1}(\hat \rho)&=\sum_{\alpha=0}^3q_\alpha\hat \sigma_\alpha\hat\rho\hat \sigma_\alpha,\\
        &=\sum_{\alpha=0}^3 \text{sgn}(q_\alpha)|q_\alpha|\hat \sigma_\alpha\hat\rho\hat \sigma_\alpha,\\
        &=\gamma \sum_{\alpha=0}^3 S_\alpha P_\alpha\hat \sigma_\alpha\hat\rho\hat \sigma_\alpha,
    \end{aligned}
\end{equation}
where
\begin{equation}
    \gamma=\sum_{\alpha=0}^3|q_\alpha|, \quad S_\alpha=\text{sgn}(q_\alpha),\quad\text{and}\quad P_\alpha=\frac{|q_\alpha|}{\gamma}.
\end{equation}
Then, the mitigated result is given by
\begin{equation}
    \braket{M}=\gamma\sum_{\alpha=0}^3S_\alpha P_\alpha M_\alpha,
\end{equation}
where $M_\alpha$ is the outcome obtained from the measurement at the end of the circuit at whose beginning we applied $\hat \sigma_\alpha$.\\

The application of PEC mitigation works if one has an almost perfect knowledge of the noise. However, for such a characterisation for $N$ qubits, one needs to quantify $4^N-1$ parameters, where 4 is the dimensions of the single-qubit algebra and 1 degree of freedom is fixed as it corresponds to the map given by $\hat{\mathbb 1}^{\otimes N}$ whose associated probability is given by the unity minus the sum of all the other probabilities. To be quantitative, for 2 qubits one needs 15 parameters, for 10 qubits these become $\sim10^{6}$, and for 50 qubits we have $\sim 10^{30}$ parameters. Therefore, it is an approach that requires too many classical processing to be used for a large number of qubits.

\newpage
\appendix
\chapter{Solutions of the exercises}
\label{ch.solutions}

\section{Solution to Exercise \ref{ex.Hadamardasrot}}

To express the Hadamard gate $H$ as a rotation, we proceed as follows.
We consider the general rotation 
\begin{equation}\label{sol.rotation}
    \hat R^{\bf n}(\theta)=\cos(\theta/2)\hat{\mathbb 1}-i\sin{(\theta/2)}{\bf n}\cdot\hat {\boldsymbol \sigma},
\end{equation}
of an angle $\theta$ around ${\bf n}$, where the Pauli matrices are
\begin{equation}
    X=\begin{pmatrix}
        0&1\\1&0
    \end{pmatrix},
    \quad   
    Y=\begin{pmatrix}
    0&-i\\i&0
    \end{pmatrix},
    \quad
    Z=\begin{pmatrix}
    1&0\\0&-1
    \end{pmatrix}.
\end{equation}
From such a rotation, we want to obtain
\begin{equation}
    H=\frac{1}{\sqrt{2}}\begin{pmatrix}
        1&1\\1&-1
    \end{pmatrix}.
\end{equation}
First thing, we highlight that the sum $X+Z$ gives
\begin{equation}
    X+Z=\begin{pmatrix}
        1&1\\1&-1
    \end{pmatrix},
\end{equation}
then it follows that
\begin{equation}
    H=\frac{X+Z}{\sqrt{2}}=\{\tfrac{1}{\sqrt{2}},0,\tfrac{1}{\sqrt{2}}\}\cdot\hat {\boldsymbol \sigma}.
\end{equation}
Such an expression recalls the last term in Eq.~\eqref{sol.rotation}. Finally, we need to set the angle $\theta$ so that the first term in Eq.~\eqref{sol.rotation} vanishes. This is $\theta=\pi$. Then
\begin{equation}
    H=i\hat R^{{\bf n}}(\pi),\quad\text{where}\quad{\bf n}=\{\tfrac{1}{\sqrt{2}},0,\tfrac{1}{\sqrt{2}}\},
\end{equation}
gives the solution.

\section{Solution to Exercise \ref{ex.3rotations}}

To prove that,  given two fixed non-parallel  normalised vectors ${\bf n}$ and ${\bf m}$, any unitary $\hat U$ can be expressed as
    \begin{equation}
        \hat U=e^{i\alpha}\hat R^{\bf n}(\beta)\hat R^{\bf m}(\gamma)\hat R^{\bf n}(\delta),
    \end{equation}
    with $\alpha, \beta, \gamma, \delta \in \mathbb R$, then one needs to recast $\hat U$ in the form
    \begin{equation}
        \hat U=e^{i\alpha}\hat R^{\bf t}(\omega),
    \end{equation}
with $\alpha\in \mathbb R$ and ${\bf t}\in\mathbb R^3$ suitably chosen.

The first step of the proof is to write $    ({\bf m}\cdot\hat{\boldsymbol \sigma})({\bf n}\cdot\hat{\boldsymbol \sigma})$ in terms of a single Pauli matrix vector $\hat {\boldsymbol \sigma}$. We have
\begin{equation}\label{sol.m.n}
\begin{aligned}
    ({\bf m}\cdot\hat{\boldsymbol \sigma})({\bf n}\cdot\hat{\boldsymbol \sigma})&=(m_1\hat \sigma_x+m_2\hat\sigma_y+m_3\hat\sigma_z)(n_1\hat \sigma_x+n_2\hat\sigma_y+n_3\hat\sigma_z),\\
    &={\bf m}\cdot{\bf n}+m_1n_2\hat \sigma_x\hat \sigma_y+m_1n_3\hat \sigma_x\hat\sigma_z+m_2n_1\hat\sigma_y\hat \sigma_x+m_2n_3\hat \sigma_y\hat\sigma_z+m_3n_1\hat\sigma_z\hat\sigma_x+m_3n_2\hat\sigma_z\hat\sigma_y.
\end{aligned}
\end{equation}
By applying 
\begin{equation}
    \hat\sigma_i\hat\sigma_j=\delta_{ij}\hat {\mathbb 1}+i\epsilon_{ijk}\hat \sigma_k,
\end{equation}
we have that Eq.~\eqref{sol.m.n} becomes
\begin{equation}\label{sol.m.n.2}
\begin{aligned}
    ({\bf m}\cdot\hat{\boldsymbol \sigma})({\bf n}\cdot\hat{\boldsymbol \sigma})&=({\bf m}\cdot{\bf n})\hat {\mathbb 1}+i({\bf m}\times{\bf n})\cdot \hat {\boldsymbol{\sigma}}.
\end{aligned}
\end{equation}

The second step is to consider the composition of two rotations:
\begin{equation}\begin{aligned}
\hat R^{\bf m}(\gamma)\hat R^{\bf n}(\delta)=&\left(\cos(\gamma/2)\hat{\mathbb 1}-i\sin{(\gamma/2)}{\bf m}\cdot\hat {\boldsymbol \sigma}\right)\left(\cos(\delta/2)\hat{\mathbb 1}-i\sin{(\delta/2)}{\bf n}\cdot\hat {\boldsymbol \sigma}\right),\\
=&\cos(\gamma/2)\cos(\delta/2)\hat{\mathbb 1}-i\cos(\gamma/2)\sin{(\delta/2)}{\bf n}\cdot\hat {\boldsymbol \sigma}-i\cos(\delta/2)\sin{(\gamma/2)}{\bf m}\cdot\hat {\boldsymbol \sigma}\\
&-\sin{(\gamma/2)}\sin{(\delta/2)}({\bf m}\cdot\hat{\boldsymbol \sigma})({\bf n}\cdot\hat{\boldsymbol \sigma}).
\end{aligned}\end{equation}
Substituting Eq.~\eqref{sol.m.n.2} in the last expression, we find that 
\begin{equation}
    \hat R^{\bf m}(\gamma)\hat R^{\bf n}(\delta)=\hat R^{\bf h}(\epsilon)=\cos(\epsilon/2)\hat{\mathbb 1}-i\sin{(\epsilon/2)}{\bf h}\cdot\hat {\boldsymbol \sigma},
\end{equation}
where $\epsilon$ and ${\bf h}$ are taken such that
\begin{equation}
\begin{aligned}
    \cos(\epsilon/2)=&\cos(\gamma/2)\cos(\delta/2)-\sin{(\gamma/2)}\sin{(\delta/2)}{\bf m}\cdot{\bf n},\\
    \sin(\epsilon/2){\bf h}=&\cos(\gamma/2)\sin{(\delta/2)}{\bf n}+\cos(\delta/2)\sin{(\gamma/2)}{\bf m}+\sin{(\gamma/2)}\sin{(\delta/2)}({\bf m}\times{\bf n}).
\end{aligned}
\end{equation}
Then, we have
\begin{equation}
    \hat R^{\bf n}(\beta)\hat R^{\bf m}(\gamma)\hat R^{\bf n}(\delta)=\hat R^{\bf n}(\beta)\hat R^{\bf h}(\epsilon)=\hat R^{\bf t}(\omega),
\end{equation}
where we applied again the composition of two rotations, which ends the proof.

\section{Solution to Exercise \ref{ex.CNOT.entanglement}}

Consider two qubits, where the first is prepared in the superposition
\begin{equation}
    \ket{\psi_1}=\frac{\ket 0 +\ket 1}{\sqrt{2}},
\end{equation}
while the second is initialised in the ground state $\ket{\psi_2}=\ket 0$. The total state is
\begin{equation}
    \ket{\psi_{12}}=\frac{\ket 0 +\ket 1}{\sqrt{2}}\ket0=\frac{\ket {00} +\ket {10}}{\sqrt{2}}.
\end{equation}
From the first expression, one clearly sees that the state is separable.
By appling the CNOT gate, we find that the state becomes
\begin{equation}
    \ket{\psi_{12}}=\frac{\ket {00} +\ket {11}}{\sqrt{2}},
\end{equation}
which is a fully entangled state.

\section{Solution to Exercise \ref{ex.S.im.U}}

Consider the circuit
\begin{equation}
    \begin{quantikz}
    \ket0&\gate{H}&\gate{S^\dag}&\ctrl{1}&         \gate{H}   &\meter{}\\
    \ket\psi&\qwbundle{}&&\gate{U}&              &
\end{quantikz}    
\end{equation}
Its action is the following
\begin{equation}
\begin{aligned}
    \ket0\ket\psi&\xrightarrow[]{\hat H\otimes\hat {\mathbb 1}}
    \tfrac{1}{\sqrt{2}}(\ket0+\ket1)\ket\psi\\
    &\xrightarrow[]{\hat S^\dag\otimes\hat {\mathbb 1}}
    \tfrac{1}{\sqrt{2}}(\ket0-i\ket1)\ket\psi\\
    &\xrightarrow[]{C(U)}
   \tfrac{1}{\sqrt{2}}(\ket0\ket\psi-i\ket1\hat U\ket\psi)\\
   & \xrightarrow[]{\hat H\otimes\hat {\mathbb 1}}
    \tfrac12\left[(\ket0+\ket1)\ket\psi-i(\ket0-\ket1)\hat U\ket\psi\right]\\
    &=\tfrac12\left[\ket0(\hat{\mathbb 1}-i\hat U)\ket\psi+\ket1(\hat{\mathbb 1}+i\hat U)\ket\psi\right].
    \end{aligned}
\end{equation}
Finally, one measures qubit 0, and the probability of finding the qubit in $\ket 0$ is
\begin{equation}
    P(\ket0)=\tfrac14\braket{\psi|\left(   \hat{\mathbb 1}+i\hat U^\dag \right)\left(   \hat{\mathbb 1}-i\hat U \right)|\psi}=\tfrac12\left(1+\Im\braket{\psi|\hat U|\psi}\right),
\end{equation}
which ends the exercise.

\printindex

\end{document}